\documentclass[12pt, openright, twoside, a4paper, english]{abntex2}

\usepackage{lmodern}			
\usepackage[T1]{fontenc}		
\usepackage[utf8]{inputenc}		
\usepackage{lastpage}			
\usepackage{indentfirst}		
\usepackage{color}				
\usepackage{graphicx}			
\usepackage{microtype} 			
\usepackage{amsthm}
\usepackage{amsmath}
\usepackage{mathtools}
\usepackage{amsfonts}
\usepackage{amssymb}
\usepackage{caption}
\usepackage{subcaption}
\usepackage{tabularx}
\usepackage{csquotes}
\usepackage{makecell}
\usepackage{hyperref}
\usepackage{amsmath}
\usepackage[ruled,lined,linesnumbered]{algorithm2e}
\usepackage{multirow}
\usepackage[brazil]{babel}
\addto\captionsbrazil{
\def\bibname{References}
\renewcommand{\figurename}{Figure}
\renewcommand{\tablename}{Table}
\renewcommand{\listfigurename}{List of figures}
\renewcommand{\listtablename}{List of tables}
\renewcommand{\contentsname}{Contents}

}
\usepackage{lipsum}
\usepackage[alf]{abntex2cite}

\renewcommand{\orientadorname}{Supervisor:}

\newcommand{\hmsun}{h^{-1}\mathrm{M}_\odot}
\newcommand{\hmpc}{h^{-1}\mathrm{Mpc}}
\newcommand{\kms}{\mathrm{km}~\mathrm{s}^{-1}}
\newcommand{\xispo}{\xi(\theta,\zeta)}

\newcommand{\xisp}{\xi(\sigma,\pi)}
\newcommand{\xiposo}{\xi_\mathrm{pos}(\theta)}
\newcommand{\xiloso}{\xi_\mathrm{los}(\zeta)}
\newcommand{\xipos}{\xi_\mathrm{pos}(\sigma)}
\newcommand{\xilos}{\xi_\mathrm{los}(\pi)}
\newcommand{\zsim}{z}
\newcommand{\zbox}{z_\mathrm{box}}
\newcommand{\posv}{\textbf{X}_\mathrm{v}}
\newcommand{\posvx}{X_\mathrm{v1}}
\newcommand{\posvy}{X_\mathrm{v2}}
\newcommand{\posvz}{X_\mathrm{v3}}
\newcommand{\disp}{\textbf{d}_\mathrm{v}}
\newcommand{\dispx}{d_\mathrm{v1}}
\newcommand{\dispy}{d_\mathrm{v2}}
\newcommand{\dispz}{d_\mathrm{v3}}
\newcommand{\velv}{\textbf{V}_\mathrm{v}}
\newcommand{\velvx}{V_\mathrm{v1}}
\newcommand{\velvy}{V_\mathrm{v2}}
\newcommand{\velvz}{V_\mathrm{v3}}
\newcommand{\rrs}{R_\mathrm{v}^\mathrm{rs}}
\newcommand{\rzs}{R_\mathrm{v}^\mathrm{zs}}
\newcommand{\q}{R_\mathrm{v}^\mathrm{zs}/R_\mathrm{v}^\mathrm{rs}}
\newcommand{\qrsd}{q_\mathrm{RSD}}
\newcommand{\qrsds}{q^s_\mathrm{RSD}}
\newcommand{\qrsdl}{q^l_\mathrm{RSD}}
\newcommand{\qap}{q_\mathrm{AP}}
\newcommand{\dR}{\delta R_\mathrm{v}}
\newcommand{\dRs}{\delta R_\mathrm{v}^s}
\newcommand{\dRl}{\delta R_\mathrm{v}^l}
\newcommand{\zxz}{z  \times z}
\newcommand{\rxz}{r \times z}
\newcommand{\rxr}{r \times r}
\newcommand{\did}{\Delta_\mathrm{id}}

\newcommand\aap{A$\&$A}              
\newcommand\aj{AJ}                   
\newcommand\apj{ApJ}                 
\newcommand\apjl{ApJ}                
\newcommand\baas{BAAS}               
\newcommand\jcap{J.~Cosmology Astropart. Phys.} 
\newcommand\jcp{J.~Chem.~Phys.}      
\newcommand\mnras{MNRAS}             
\newcommand\nat{Nature}              
\newcommand\prd{Phys. Rev.~D}        
\newcommand\prl{Phys. Rev.~Lett.}    
\newcommand\physrep{Phys.~Rep.}      

\titulo{Cosmic voids as cosmological laboratories}
\autor{Carlos Mauricio Correa}

\local{FaMAF - UNC

\begin{figure}[t]
    \centering
    \includegraphics[width=40mm]{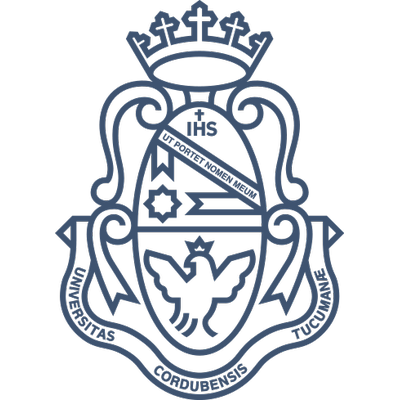}
 \end{figure}
}

\data{
2021

\begin{figure}[b]
    \centering
    \includegraphics[width=25mm]{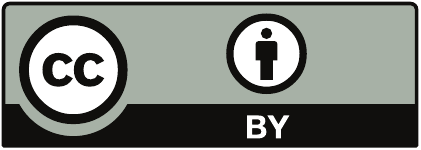}\\
    \href{https://creativecommons.org/licenses/by/4.0/}{This work is under a Creative Commons Attribution 4.0 International License.}
\end{figure}
}

\orientador{Dr. Dante Javier Paz}

\instituicao{
 Facultad de Matemática, Astronomía, Física y Computación
  \par
 Universidad Nacional de Córdoba - República Argentina
 }
 
\tipotrabalho{Doctorate Thesis}

\preambulo{
}

\definecolor{blue}{RGB}{41,5,195}

\makeatletter
\hypersetup{
		pdftitle={\@title}, 
		pdfauthor={\@author},
    	pdfsubject={\imprimirpreambulo},
	    pdfcreator={LaTeX with abnTeX2},
		pdfkeywords={abnt}{latex}{abntex}{abntex2}{trabalho acadêmico}, 
		colorlinks=true,       		
    	linkcolor=blue,          	
    	citecolor=blue,        		
    	filecolor=magenta,      	
		urlcolor=blue,
		bookmarksdepth=4
}
\makeatother

\makeindex

\begin{document}

\frenchspacing 

\pretextual

\imprimircapa

\cleardoublepage

\imprimirfolhaderosto


\newpage

\begin{center}
{\ABNTEXchapterfont\Large\textsc{Authorship declaration}}
\end{center}

\vspace*{1cm}

Thesis presented at the Faculty of Mathematics, Astronomy, Physics and Computer Science\footnote{In Spanish, \textit{Facultad de Matemática, Astronomía, Física y Computación.}} (FaMAF) of the National University of Córdoba\footnote{In Spanish, \textit{Universidad Nacional de Córdoba.}} (UNC), in the Argentine Republic, as a requirement to obtain the PhD degree in Astronomy.

I declare that I am the author of this thesis.
Any source used directly or indirectly is appropriately cited in the References section.
I also declare that I have not presented this thesis in any other institution in order to obtain another degree.

This is a faithful English translation of the official version of this thesis, which was written in Spanish.
It is archived in the Daniel Oscar Sonzini Library of FaMAF.
Also available in digital format: \url{https://rdu.unc.edu.ar/handle/11086/21041}.

\cleardoublepage


\newpage

\begin{dedicatoria}
  \vspace*{\fill}
  \noindent
  \textit{
This thesis is dedicated to my family.
To my mom, Susana, who has always been my emotional support, and taught me through her example how a good person should be.
To my dad, Carlos, who instilled in me a love for science, the importance of working hard and honestly, and fundamentally, to never give up.
To my brother, José, who is also my best friend.
To my grandparents, Sarita, Marta, Carlos and Mauricio, who have been my second parents.
To my girlfriend, Daniela, the love of my life.
Thank you all for your unconditional support.
  } \vspace*{\fill}
\end{dedicatoria}


\begin{agradecimentos}

My PhD research was carried out at the Institute for Theoretical and Experimental Astronomy\footnote{In Spanish, \textit{Instituto de Astronomía Teórica y Experimental.}} (IATE - CONICET - UNC) in the city of Córdoba, Argentina, and was financially supported by the Argentine National Scientific and Technical Research Council\footnote{In Spanish, \textit{Consejo Nacional de Investigaciones Científicas y Técnicas.}} (CONICET).
This project has also received a partial financial support from the European Union's Horizon 2020 Research and Innovation programme under the Marie Sklodowska-Curie grant agreement number 734374 - project acronym: LACEGAL, and from the Munich Institute for Astro- and Particle Physics, which is funded by the Deutsche Forschungsgemeinschaft (German Research Foundation) under Germany's Excellence Strategy – EXC-2094 – 390783311 and the Technical University of Munich.
Numerical calculations were performed on the computer clusters from the UNC - High Performance Computing Centre (\url{https://ccad.unc.edu.ar/}).
Plots were made with the \textsc{ggplot2} package \cite{ggplot2} of the \textsc{R} software \cite{R}.
Figures~\ref{fig:rsd2}, \ref{fig:binning} and \ref{fig:effects} were designed with the help of Daniela Taborda, a special thanks to her.
This thesis was structured from a template developed by the abnTeX2 team (\url{http://www.abntex.net.br/}).
I want to specially acknowledge the hospitality of the Max Planck Institute for Extraterrestrial Physics (MPE) and the Pontifical Catholic University of Chile (PUC), where part of this work has been done.

A special appreciation to my PhD examiners: Dr. María Laura Ceccarelli (IATE - CONICET - UNC), Dr. Yan-Chuan Cai (Institute for Astronomy - University of Edinburgh) and Dr. Emanuel Gallo (Enrique Gaviola Institute for Physics - CONICET - UNC), who have reviewed my work thoroughly and have made valuable contributions to improve it.
It has been an honour for me to have been evaluated by such prestigious professionals.

I would like to thank my collaborators and also coauthors of my papers: Dr. Dante Paz (IATE - CONICET - UNC), Dr. Nelson Padilla (PUC), Dr. Ariel Sánchez (MPE), Dr. Andrés Ruiz (IATE - CONICET - UNC) and Dr. Raúl Angulo (Donostia International Physics Center), for their invaluable contribution, help and fruitful discussions.
A special thanks to Dante, Ariel and Nelson for giving me the opportunity to travel abroad to visit the MPE and PUC academically, to present my work at various conferences and workshops, and to contact other colleagues working on the same topic. 

Finally, I would like to thank Dante for being an excellent supervisor, not only academically, but also as a person.

\end{agradecimentos}


\newpage

\begin{epigrafe}
    \vspace*{\fill}
	
	    \noindent
		\textit{No te des por vencido, ni aun vencido, no te sientas esclavo, ni aun esclavo; trémulo de pavor, piénsate bravo, y arremete feroz, ya mal herido.
		       }
		
		\begin{flushright}
		Almafuerte
		\end{flushright}
\end{epigrafe}


\newpage

\setlength{\absparsep}{18pt}
\begin{resumo}[Abstract]
 \begin{otherlanguage*}{english}

One of the major challenges of modern Cosmology is to understand the nature of dark energy, which drives cosmic acceleration.
In this sense, cosmic voids are promising cosmological probes.
Voids are vast underdense regions that encode key information about the expansion history and geometry of the Universe.
They are receiving special attention nowadays in view of the new generation of galaxy spectroscopic surveys, which will cover a volume and redshift range without precedents.
Furthermore, voids offer two advantages over traditional clustering studies: their dynamics is less affected by non-linearities, and Modified Gravity theories predict that potential deviations from the predictions of General Relativity must be more pronounced in this type of environments.

There are two primary statistics in void studies: the void size function, which characterises their abundance, and the void-galaxy cross-correlation function, which contains information about the density and velocity fields in their surrounding regions.
Both statistics are affected by distortions in the observed spatial distribution of galaxies, which are originated by two main sources: the dynamics of galaxies (redshift-space distortions, RSD), and the fiducial cosmology assumed to establish a physical distance scale (Alcock-Paczy\'nski effect, AP).
However, both of them are of high dynamical and cosmological content.

We present a new cosmological test based on two perpendicular projections of the correlation function with respect to the line-of-sight direction.
We treat correlations directly in terms of angular distances and redshift differences between void-galaxy pairs, without the need to assume a fiducial cosmology.
The method is based on a physical model that reproduces the coupled RSD and AP distortions on the correlation function, and also the mixture of scales due to the projection ranges.
The model primarily depends on two cosmological parameters: the present-day matter fraction of the Universe and the growth rate of cosmic structures.
The covariance matrices associated with the method are relatively small and their inversion is numerically more stable, which reduces the noise in the likelihood analysis.
This aspect allows us to significantly reduce the number of mock catalogues needed to estimate them.
We tested the method with the Millennium XXL simulation for different redshifts and projection ranges.

There is, however, an obstacle that has prevented the use of voids as reliable tools.
Our standard picture of distortions around voids is incomplete.
Traditionally, we have focused only on the spatial distribution of galaxies.
It turns out that the RSD and AP effects also affect global properties of voids, such as their number, size and spatial distribution.
This generates additional distortion patterns on observations that lead to biased cosmological constraints if they are not taken into account appropriately.
Moreover, given the precision achievable nowadays with modern surveys, it is imperative a full description of these systematicities.
Using a spherical void finder and the Millennium XXL simulation, we addressed this problematic by describing physically the connection between the identification in real space (unaffected by distortions) and in redshift space (affected by them).
We found three important results.
First, redshift-space voids above the shot-noise level have a unique real-space counterpart spanning the same region of space, hence void number conservation can be assumed.
Second, the volume of voids is affected by three sources: (i) the RSD induced by tracer dynamics at scales around the void radius (t-RSD), which manifests as a systematic expansion, (ii) the AP effect, which manifests as an overall expansion or contraction depending on the fiducial cosmology, and (iii) the intrinsic ellipticity of voids (e-RSD).
Third, void centres are systematically shifted along the line-of-sight direction, a by-product of a different class of RSD induced by the global dynamics of the whole region containing the void (v-RSD).
We also present an evaluation of the impact that these redshift-space effects in voids have on the void size function and the void-galaxy cross-correlation function.
In this way, we lay the foundations for improvements in current models for these statistics in order to obtain unbiased cosmological constraints.

Finally, we present preliminary results from observational data, namely, the Baryon Oscillation Spectroscopic Survey Data Release 12.
The analysis is focused on understanding the origin of the anisotropic patterns observed on the correlation function by applying the methodology developed throughout the work.
We conclude that, in addition to its practical cosmological importance, the analysis carried out in this thesis also has profound implications on studies of the structure and dynamics of the Universe at the largest cosmic scales.

\textbf{Key words:} large-scale structure of the Universe -- cosmic voids -- dark energy -- cosmological parameters -- distance scale -- galaxies: distances, redshifts and velocities 

\end{otherlanguage*}
\end{resumo}


\newpage

\pdfbookmark[0]{\listfigurename}{lof}
\listoffigures*
\cleardoublepage


\pdfbookmark[0]{\listtablename}{lot}
\listoftables*
\cleardoublepage


\begin{siglas}

    \item[2dFGRS] Two Degree Field Galaxy Redshift Survey
    \item[AP] Alcock-Paczyński effect/test
    \item[BAO] Baryon acoustic oscillations
    \item[BBN] Big Bang Nucleosynthesis
    \item[BOSS] Baryon Oscillation Spectroscopic Survey
    \item[BOSS DR12] BOSS Data Release 12
    \item[CAMB] \textsc{Code for Anisotropies in the Microwave Background}
    \item[CDM] Cold dark matter
    \item[CMB] Cosmic microwave background
    \item[DE] Dark energy
    \item[DESI] Dark Energy Spectroscopic Instrument survey
    \item[eBOSS] BOSS extension
    \item[EdS] Einstein - de Sitter cosmological model
    \item[Eq.] Equation or formula
    \item[e-RSD] Ellipticity effect
    \item[FC-l] Fiducial-cosmology catalogue of voids: $\Omega^l_m = 0.2$-cosmology version
    \item[FC-u] Fiducial-cosmology catalogue of voids: $\Omega^u_m = 0.3$-cosmology version
    \item[flat-$\Lambda$CDM] Standard cosmological model
    \item[FRW] Friedmann-Robertson-Walker
    \item[GS] Gaussian streaming model
    \item[HETDEX] Hobby-Eberly Telescope Dark Energy Experiment survey
    \item[IQR] Interquartile range
    \item[LOS] Line of sight
    \item[MCMC] Markov Chain Monte Carlo
    \item[MXXL] Millennium XXL cosmological simulation
    \item[POS] Plane of the sky
    \item[PR] Projection range
    \item[$\rxr$-space] Configuration: $r$-space voids, $r$-space tracers
    \item[RSD] Redshift-space distortions
    \item[$r$-space] Real space
    \item[$\rxz$-space] Configuration: $r$-space voids, $z$-space tracers
    \item[SDSS] Sloan Digital Sky Survey
    \item[SNe Ia] Supernovae Ia
    \item[SvdW] Sheth $\&$ van de Weygaert model for the abundance of voids
    \item[TC-rs-b] True-cosmology catalogue of voids: MXXL-cosmology, $r$-space, bijective version
    \item[TC-rs-f] True-cosmology catalogue of voids: MXXL-cosmology, $r$-space, full version
    \item[TC-zs-b] True-cosmology catalogue of voids: MXXL-cosmology, $z$-space, bijective version
    \item[TC-zs-f] True-cosmology catalogue of voids: MXXL-cosmology, $z$-space, full version
    \item[t-RSD] Expansion effect
    \item[Vdn] Volume-conserving model for the abundance of voids
    \item[v-RSD] Off-centring effect
    \item[VSF] Void size function
    \item[ZOBOV] \textsc{ZOnes Bordering On Voidness} (void finder)
    \item[$z$-space] Redshift space
    \item[$\zxz$-space] Configuration: $z$-space voids, $z$-space tracers
  
\end{siglas}


\pdfbookmark[0]{\contentsname}{toc}
\tableofcontents*
\cleardoublepage


\textual

\chapter{Introduction}
\label{chp:intro}
Modern Cosmology rests on six observational pillars.

\begin{figure}
    \centering
    \includegraphics[width=70mm]{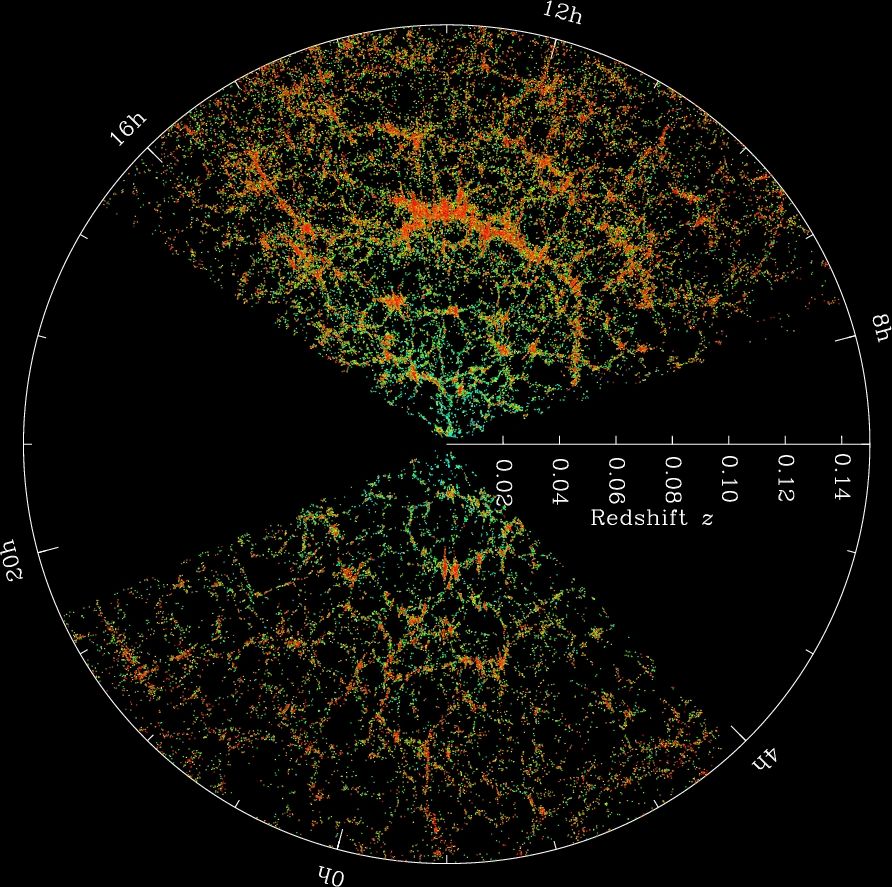}
    \includegraphics[width=\textwidth/2]{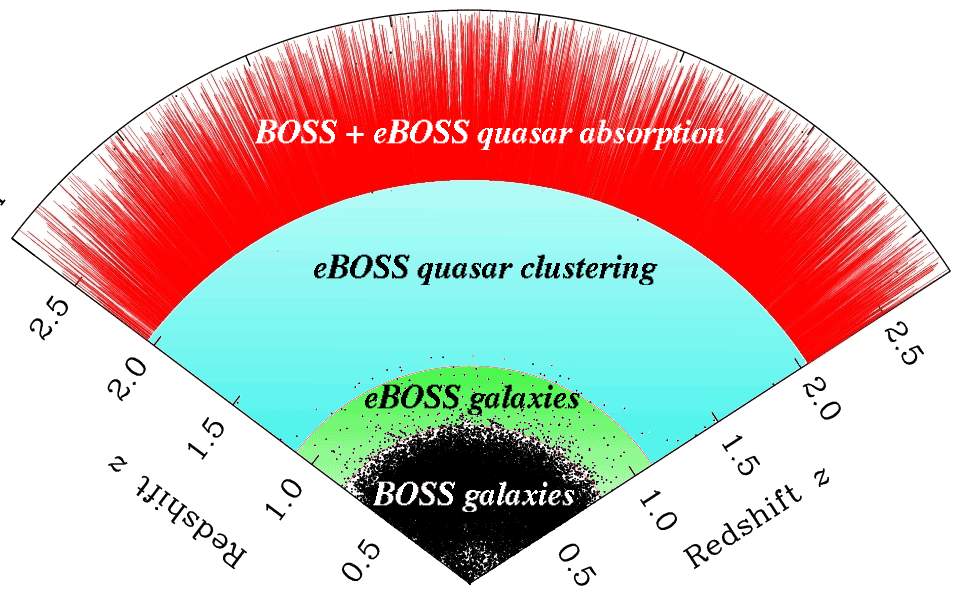}
    \caption[
    Maps of the SDSS, BOSS and eBOSS spectroscopic surveys.
    ]{
    \textit{Left-hand panel.}
    The SDSS map of the Universe.
    Each dot is a galaxy; the colour bar shows the local density.
    The darkest zones are the underdense regions of the Universe, and hence, potential cosmic voids.
    \textit{Right-hand panel.}
    BOSS and eBOSS coverage of the Universe.
    \textit{Figure credit:} M. Blanton $\&$ SDSS \url{https://www.sdss.org/}.
    }
    \label{fig:lss_surveys}
\end{figure}

\begin{enumerate}

\item 
\textit{The expansion of the Universe.}
The radial velocity of galaxies, measured by means of the shift of spectral lines, is positive for nearly all of them (redshift), i.e. they appear to be moving away from us.
Moreover, this velocity increases linearly with the distance.
This is Hubble's law, interpreted such that the Universe is expanding.
We are able to trace back the cosmic expansion under the assumption that the known laws of Physics were also valid in the past. 
From that we get the Big Bang model of the Universe, according to which our Universe has evolved out of a very dense and hot state.

\item
\textit{The cosmic microwave background} (CMB).
This is a background electromagnetic radiation coming from all directions with an almost perfect black-body spectrum, confirming the smooth, hot and dense past of the Universe.
The temperature of this radiation is \cite{cmb_temperature}
\begin{equation}
    T_\mathrm{CMB} = 2.72548 \pm 0.00057 ~ \mathrm{K}.
    \label{eq:tcmb}
\end{equation}

\item
\textit{The abundance of light elements.}
The observed abundance of the chemical elements shows that the Universe is composed of about $75\%$ of hydrogen-1 and $25\%$ of helium-4, with small traces of deuterium, helium-3 and lithium, and negligible heavier elements, in good agreement with the predictions of the Big Bang Nucleosynthesis theory (BBN).
Moreover, BBN indicates that ordinary matter represents only a $15\%$ of total matter, and hence provides a compelling argument for the existence of non-baryonic dark matter \cite{bbn}.

\item
\textit{The supernova experiment.}
Hubble's law is inaccurate to explain the redshift-distance relation of Supernovae Ia (SNe Ia), the farthest standard candles that can be observed, for which the notion of distance must be redefined according to the cosmological model used to explain the evolution of the Universe.
The surprising result is that data are not compatible with a flat matter-dominated universe, nor are they with an open universe.
Instead, data favour a flat Universe with $30\%$ matter and $70\%$ of an enigmatic component known as dark energy.
The existence of this component means that the Universe is currently in a phase of accelerated expansion, a completely unexpected result \cite{sn_riess,sn_perlmutter}.
This astonishing result means that dark matter and dark energy make up $95\%$ of the total energy budget of the Universe, whose true nature we ignore.

\item
\textit{The large-scale spatial distribution of galaxies.}
The last decades saw a number of large spectroscopic surveys of galaxies designed to measure structure in the Universe.
These culminated in two large surveys, the Two Degree Field Galaxy Redshift Survey \cite[2dFGRS]{2df} and the Sloan Digital Sky Survey \cite[SDSS]{sdss}, which compiled the redshift of a million galaxies.
The left-hand panel of Figure~\ref{fig:lss_surveys} shows the spatial distribution of the galaxies in the SDSS.
Each dot is a galaxy, the colour gradient shows the local density.
This map shows that the galaxies are clearly not distributed randomly, the Universe has structure on large scales.
For instance, by the action of gravity, galaxies form groups, clusters and even larger structures like filaments and walls.
Complementarily, they leave on their way vast underdense regions in this process, the so-called cosmic voids (darkest zones in the figure).
However, no evidence of structures with dimensions larger than $150~\hmpc$ have been found.
Hence, the Universe seems to be basically homogeneous if averaged over these scales.
The existence of dark matter and dark energy plays a fundamental role when trying to explain this large-scale distribution from a theoretical point of view.

\item
\textit{Anisotropies in the CMB.}
Although the CMB is almost homogeneous and isotropic, there are small anisotropies in the temperature distribution at the level between $10^{-4}$ and $10^{-5}$, the seeds of structure.
Present-day measurements of the anisotropies in the CMB are the best example of the accuracy attainable by current cosmological experiments.
For this reason, it is said that we have entered an era of high-precision Cosmology.
These observations provide tight constraints on the basic set of parameters of cosmological models, and have been decisive in establishing the current standard model of Cosmology \cite{planck}.

\end{enumerate}

In this scenario, one of the major challenges of modern Cosmology is to understand the nature of dark energy.
The standard model rests on two fundamental hypotheses.
One is the cosmological principle, which states that the Universe is spatially homogeneous and isotropic on large scales.
The other is that gravitational interactions are described by Einstein’s General Theory of Relativity.
One plausible explanation for the dark-energy phenomenon is that it is a manifestation of the cosmological constant present in Einstein's field equations, which can be also associated with the vacuum energy.
Alternatively, this could be an indication that General Relativity must be reviewed, leading to the field of Modified Gravity theories.

Cosmic voids are the largest observable structures.
Therefore, they encode key information about the expansion history and geometry of the Universe, emerging in this way as natural cosmological probes to test the standard model.
Moreover, they do not only constitute a complementary view of the large-scale structure, but they offer two main advantages over galaxy clustering studies.
On the one hand, void dynamics can be treated linearly, hence it is easier to model systematicities such as redshift-space distortions.
On the other hand, Modified Gravity theories predict that deviations from General Relativity should be more pronounced in these low density environments.
In view of this, cosmic voids are ideal for testing different dark-energy models.
We will tackle the study of voids in Chapter~\ref{chp:voids}.

The cosmological potential of voids has increased recently with the development of the new generation of spectroscopic surveys.
Some of them are the SDSS Baryon Oscillation Spectroscopic Survey \cite[BOSS]{boss}, its extension eBOSS \cite{eboss}, the Hobby-Eberly Telescope Dark Energy Experiment \cite[HETDEX]{hetdex}, the Dark Energy Spectroscopic Instrument \cite[DESI]{desi} and the Euclid mission \cite{euclid}, which will probe our Universe covering a volume and redshift range without precedents.
This will allow us to obtain rich samples of voids at different redshifts, and in this way, to test the evolution of the Universe with high precision.
The right-hand panel of Figure~\ref{fig:lss_surveys} shows the coverage planned for BOSS and eBOSS.

There are two primary statistics in void studies: the void size function and the void-galaxy cross-correlation function.
We will develop these tools in Chapter~\ref{chp:voids}.
The void size function, on the one hand, describes the abundance of voids.
It is analogous to the mass function of dark-matter haloes, hence it can be modelled using the excursion set theory combined with the spherical evolution of matter fluctuations derived from the cosmological perturbation theory.
The void-galaxy cross-correlation function, on the other hand, characterises the void density field when considered at small to intermediate scales.
Both statistics are affected by distortions in the observed spatial distribution of galaxies, which translate into anisotropic patterns in the measurements.
There are two main sources of distortions: the redshift-space distortions \cite[hereinafter RSD]{rsd_kaiser}, a dynamical effect caused by the peculiar velocities of the galaxies, and the Alcock-Paczyński effect \cite[hereinafter AP]{ap}, a geometrical effect caused by the selection of a fiducial cosmology necessary to transform the angles and redshifts provided by a survey into distances expressed in physical units.
These systematicities can be modelled from physical grounds, hence they encode valuable information about the dynamics of the large-scale structure and the cosmology in which it is embedded.

The present thesis can be divided into five parts.
In the first part, we will present the theoretical background and necessary tools to develop the work, and also to discuss the main conclusions.
This part will be developed in Chapters~\ref{chp:lcdm}, \ref{chp:lss}, \ref{chp:voids} and \ref{chp:data}.
In Chapter~\ref{chp:data}, in particular, we will present the data set used throughout the work.
Specifically, we will describe the numerical simulations and mock catalogues used to calibrate the different tests and models, the void finding method, and finally, the spectroscopic survey used to compare observational results with the corresponding theoretical predictions.

Next, in the second part, we will present a new cosmological test based on the void-galaxy cross-correlation function.
This will be developed in Chapter~\ref{chp:cosmotest}.
Our method provides two novel aspects.
In the first place, we treat correlations directly in terms of void-centric angular distances and redshift differences between void-galaxy pairs, hence it is not necessary to assume a fiducial cosmology.
In this way, the AP effect is taken into account naturally.
Secondly, we model a third type of systematicity besides the RSD and AP effects that also affects the comparison between predictions and observations.
This is the mixture of scales, which is due to the binning scheme used in the measurement process.
In particular, this allows us to work with two perpendicular projections of the correlation function with respect to the line-of-sight direction.
This variant of measuring correlations offers some advantages over the traditional case.
For instance, it represents two simple one-dimensional profiles affected by distortions in a different way, with associated covariance matrices easier to manipulate.
The first aspect implies that any degeneration in the parameter space due to the coupled RSD and AP distortions can be easily broken, whereas the last aspect means that the covariance matrices are smaller and numerically more stable, which reduces the noise in the likelihood analysis and allows us to use a smaller number of mock catalogues to estimate them.

The statistical properties of voids depend on the method used to identify them from the spatial distribution of galaxies.
There are different classes of void finders.
Despite the intrinsic differences between them, though, there is a consensus on their basic statistical properties.
However, the RSD and AP distortions have a direct impact on the void identification process itself, affecting global properties of voids, such as their number, size and the position of their centres.
This is true for all methods, and translates into additional anisotropic patterns in the measurements of the void size function and the correlation function.
Given the precision achievable nowadays with modern surveys, it is extremely important to detect and model all these systematicities in order to obtain unbiased cosmological constraints.
One solution is to use the reconstruction technique, an algorithm to approximately recover the real position of the galaxies based on the Zel'dovich approximation \cite{rsd_nadathur, reconstruction_nadathur}.
This technique has been recently applied to data \cite{aprsd_nadathur, aprsd_nadathur2020}, showing a robust power in constraining the cosmological parameters.
However, it also presents some disadvantages, particularly, it loses the physical information about the structure and dynamics of voids encoded in these additional anisotropic patterns, which could otherwise be extracted when voids are directly identified from the observed galaxies.
In the present work, we propose an alternative approach.
With the help of a numerical simulation, we will analyse the void finding method in order to find a physical connection between the identification with and without distortions, and in this way, be able to detect the physical effects responsible for these additional systematic effects.
This is the third part of this work, developed in Chapter~\ref{chp:zeffects}.

In the fourth part, we will analyse the impact that these additional void systematicities have on the measurements of the void size function and the void-galaxy cross-correlation function.
This will be developed in Chapters~\ref{chp:impact_vsf} and \ref{chp:impact_vgcf}.
We will focus the analysis on the theoretical framework developed in Chapter~\ref{chp:zeffects} to model them.
The importance of this framework resides in the fact that it depends strongly on cosmological considerations, hence it must be incorporated in current models for the abundance of voids and the correlation function.
In this way, we lay the foundations for a full treatment that will allow us to obtain unbiased cosmological constraints from modern spectroscopic surveys.
But besides this practical cosmological importance, this also has profound implications in our understanding of the large-scale structure of the Universe, since these effects encode key information about the structure and dynamics of voids.

Finally, in Chapter~\ref{chp:boss}, we will present a preliminary analysis of the void-galaxy cross-correlation function measured from observational data taken from the BOSS DR12 survey.
The main goal will be to understand the origin of the observed anisotropic patterns in the context of the dynamical and cosmological effects that affect voids studied in the previous chapters.

\chapter{The standard model of Cosmology}
\label{chp:lcdm}
Before we start\footnote{This chapter and the next one are mainly based on the lectures on \textit{Formation and evolution of cosmic structures} given by Dr. \citeonline{sanchez_book} at the Ludwig Maximilians University in Munich, and on the books of \citeonline{dodelson_book} and \citeonline{schneider_book}.}, a comment about a few important conventions to be followed.
In the first place, the word \textit{universe} will refer to a generic space-time, whereas our particular \textit{Universe} will be denoted with a capital letter.
We will use units in which the speed of light is $c = 1$, although we will include it explicitly when relevant to compare with results from the literature.
We will also use the convention in which Latin indices represent spatial coordinates, running from $1$ to $3$ (e.g. $i = 1, 2, 3$), whereas Greek letters represent space-time coordinates, running from $0$ to $3$ (e.g. $\mu = 0, 1, 2, 3$).
Moreover, we will use the summation convention in which identical lower and upper indices are summed over all their possible values.
A dot over a variable will represent a derivative with respect to the time coordinate $t$, whereas an apostrophe next to a variable, a derivative with respect to conformal time $\tau$.
Finally, the covariant derivative will be denoted by $\nabla_\mu$, and the brackets $\langle \rangle$ will denote an average over an assembly of universes with the same statistical properties.


\section{The cosmological principle}
\label{sec:lcdm_cosmoprinciple}

The standard model of Cosmology rests on the following two fundamental hypotheses.
\begin{enumerate}
    \item 
    \textit{Validity of General Relativity.}
    Gravitational interactions, which shape the global evolution of the Universe, are described by Einstein's General Theory of Relativity.
    \item
    \textit{Cosmological principle.}
    This principle assumes that the Universe is spatially homogeneous and isotropic on large scales.
    This means that there are no preferred locations nor directions.
    Of course, this hypothesis must be understood in a statistical sense.
    For instance, the Universe exhibits large density fluctuations, such as galaxies, galaxy clusters, filaments and voids.
\end{enumerate}

Under these hypotheses, the most general form that the metric tensor $g_{\mu \nu}$ can adopt is the Friedmann-Robertson-Walker (FRW) metric:
\begin{equation}
    ds^2 = -dt^2 + a^2(t) [ \gamma_{ij}(x^k) dx^i dx^j ],
    \label{eq:frw_metric1}
\end{equation}
where $\gamma_{ij}$ is the spatial part of the metric, and $x^i$ denotes comoving coordinates.
In these coordinates, the change in position of a given point of space as the universe evolves is absorbed into the scale factor $a(t)$.
In terms of spherical coordinates, the FRW metric can be written as
\begin{equation}
    ds^2 = -dt^2 + a^2(t) [ d\chi^2 + S^2_K(\chi) ( d\theta^2 + \mathrm{sin}^2\theta d\phi^2 ) ],
    \label{eq:frw_metric2}
\end{equation}
where we used the convention for the radial coordinate given by
\begin{equation}
    d\chi := \frac{dr}{\sqrt{1 - Kr^2}},
    \label{eq:frw_metric_chi}
\end{equation} 
and 
\begin{equation}
	S_K(\chi) :=
	\left\{
		\begin{array}{lll}
			\frac{1}{\sqrt{K}} \mathrm{sin}(\sqrt{K}\chi) & \mathrm{if} ~ K > 0\\
			\chi & \mathrm{if} ~ K = 0\\
            \frac{1}{\sqrt{|K|}} \mathrm{sinh}(\sqrt{|K|}\chi) & \mathrm{if} ~ K < 0.
		\end{array}
	\right.
	\label{eq:frw_metric_s}
\end{equation}
This metric is completely determined by two quantities: $a(t)$, which determines the kinematics of the universe, and $K$, which characterises its spatial curvature.
Depending on the value of $K$, FRW universes can be classified as open $(K < 0)$, flat $(K = 0)$ or closed $(K > 0)$.
These terms refer to the topological properties of the spatial hypersurfaces of the universe.

In some cases, it will be convenient to express the metric in terms of conformal time $\tau$, defined by the relation
\begin{equation}
    dt = a(\tau) d\tau,
\end{equation}
so that
\begin{equation}
    ds^2 = a^2(\tau) [ -d\tau^2 + \gamma_{ij}(x^k) dx^i dx^j ].
\end{equation}
In this case, the FRW metric can be factorised into a static metric, for which $K = 0$ corresponds to the Minkowski metric, multiplied by a time-dependent factor.


\section{Dynamics of Friedmann-Robertson-Walker universes}
\label{sec:lcdm_frw}

The relation between the geometry of space-time, characterised by the metric tensor $g_{\mu \nu}$, and the energetic components that it contains, encompassed by the energy-momentum tensor $T_{\mu \nu}$, is governed by Einstein's field equations:
\begin{equation}
    G_{\mu \nu} = 8 \pi G T_{\mu \nu},
    \label{eq:einstein}
\end{equation}
where $G$ is the Newtonian constant of gravitation, and $G_{\mu \nu}$ is the Einstein tensor, defined as
\begin{equation}
    G_{\mu \nu} := R_{\mu \nu} - \frac{1}{2} g_{\mu \nu} R.
    \label{eq:einstein_tensor}
\end{equation}
Here, $R_{\mu \nu}$ is the Ricci tensor, and $R := g^{\mu \nu} R_{\mu \nu}$, the scalar curvature.

When the FRW metric is introduced into Einstein's field equations, we obtain equations for $a(t)$ and $K$.
To do this, we need to first assume a form for the energy-momentum tensor.
The most general form of $T_{\mu \nu}$ that is compatible with the cosmological principle is that of a perfect fluid:
\begin{equation}
    T^{\mu \nu} = (p + \rho) U^\mu U^\nu + p g^{\mu \nu},
    \label{eq:tensor_pfluid}
\end{equation}
where $U^\mu$  is the four-velocity of the fluid, and $\rho$ and $p$ correspond to its energy density and pressure, respectively.
In this way, the field equations are reduced to two independent equations, the so-called Friedmann's equations:
\begin{equation}
    \left( \frac{\dot{a}}{a} \right)^2 = \frac{8 \pi G}{3} \rho - \frac{K}{a^2}
    \label{eq:frw1}
\end{equation}
and 
\begin{equation}
    \frac{\ddot{a}}{a} = -\frac{4 \pi G}{3} (\rho + 3p).
    \label{eq:frw2}
\end{equation}

Einstein's field equations implicitly contain the energy conservation equation: $\nabla_\nu T^{\mu \nu} = 0$.
This equation can also be obtained by differentiating Eq.~(\ref{eq:frw1}) and combining it with Eq.~(\ref{eq:frw2}), which leads to the following relation:
\begin{equation}
    \dot{\rho} + 3 \frac{\dot{a}}{a}(\rho + p) = 0.
    \label{eq:energy_consv}
\end{equation}

To derive explicit solutions for $a(t)$ and $K$, we can use Eq.~(\ref{eq:frw1}) in combination with Eq.~(\ref{eq:energy_consv}).
To do so, it is also necessary to assume a relation between the pressure and density of the ideal fluid in the form of an equation of state: $p = p(\rho)$.
Using this relation it is possible to integrate Eq.~(\ref{eq:energy_consv}) to obtain $\rho = \rho(a)$.
In Cosmology, we are mostly interested in barotropic fluids, for which the pressure is proportional to the density: $p = w\rho$, and are then characterised by a constant sound speed: $c^2_s = dp/d\rho$.
The parameter $w$ is known as the equation-of-state parameter.
For barotropic fluids, the energy conservation equation implies that
\begin{equation}
    \rho \propto a^{-3(w+1)}.
    \label{eq:state_eq}
\end{equation}
If the universe contains $N$ different components with equations of state given by $w_i$, this result will hold for each species separately as long as they do not interact.
If we denote as $\rho_{i,0}$ the present-day density of a given species $i$, then the total energy density of the universe at a time $t$ is given by
\begin{equation}
    \rho = \sum_{i=1}^N \rho_{i,0} a^{-3(w_i+1)},
    \label{eq:density_species}
\end{equation}
where we have set $a_0 := a(t_0) = 1$ for simplicity, and $t_0$ denotes the present time.
In this way, Eq.~(\ref{eq:frw1}) can be written as
\begin{equation}
    \left( \frac{\dot{a}}{a} \right)^2 = \frac{8 \pi G}{3} \sum_{i=1}^N \rho_{i,0} a^{-3(w_i+1)} - \frac{K}{a^2}.
    \label{eq:frw3}
\end{equation}
This is the desired equation for $a(t)$.

We define now the Hubble parameter:
\begin{equation}
    H := \frac{\dot{a}}{a}.
    \label{eq:hubble_parameter}
\end{equation}
Note that $H>0$, since we are interested in expanding universes ($\dot{a}>0$).
Therefore, this parameter quantifies the expansion rate of the universe at a time $t$.
The present-day value of $H$ is denoted as $H_0$, and referred to as the Hubble constant.
This constant is usually parametrised in terms of a dimensionless quantity $h$ in the following way:
\begin{equation}
    H_0 = 100~h~\mathrm{km}~\mathrm{s}^{-1}~\mathrm{Mpc}^{-1}.
    \label{eq:hubble_cte}
\end{equation}
Using this definition, and evaluating Eq.~(\ref{eq:frw3}) at present time, we get the following expression:
\begin{equation}
    K = H_0 \left( \frac{8 \pi G}{3 H_0^2} \sum_{i=1}^N \rho_{i,0} - 1 \right).
    \label{eq:K}
\end{equation}
This is the desired equation for $K$.

In Eqs.~(\ref{eq:frw3}) and (\ref{eq:K}), there are $N+1$ unknowns: $H_0$ and the densities $\rho_{i,0}$.
They are not variables.
They are measurable quantities that must be obtained from observational experiments.
In this way, it is clear that the geometry of space-time and its evolution are determined by the energy content of the universe.
In particular, it will be flat (K = 0) if the total energy density is equal to a critical density, $\rho_c$, given by
\begin{equation}
    \rho_c = \frac{3 H_0^2}{8 \pi G} = 2.773 \times 10^{11} \hmsun (\hmpc)^{-3} = 11.26~h^2~\frac{\mathrm{protons}}{\mathrm{m}^3}.
    \label{eq:density_cr}
\end{equation}

Instead of working with the quantities $\rho_{i,0}$, it is a common practice to define a dimensionless density parameter for each species by expressing their energy density in units of the critical density:
\begin{equation}
    \Omega_i := \frac{\rho_{i,0}}{\rho_c} = \frac{8 \pi G}{3 H_0^2} \rho_{i,0}.
    \label{eq:density_species2}
\end{equation}
A similar expression is defined for the curvature:
\begin{equation}
    \Omega_K := -\frac{K}{H_0^2}.
    \label{eq:density_K}
\end{equation}
In this way, it is possible to express the equation for the curvature (Eq.~\ref{eq:K}) in a more compact form:
\begin{equation}
    \sum_{i=1}^N \Omega_i + \Omega_K = 1.
    \label{eq:K2}
\end{equation}
Also, with this notation, the equation governing $a(t)$ (Eq.~\ref{eq:frw3}) can be written in the following way:
\begin{equation}
    \dot{a}^2 = H_0^2 \left[ \sum_{i=1}^N \Omega_i a^{-(3w_i + 1)} + \Omega_K \right].
    \label{eq:frw4}
\end{equation}

As an example, consider a flat universe with a unique energy component characterised by an equation of state with $w \ne -1$.
Then, the time evolution of the scale factor is given by
\begin{equation}
    a(t) \propto t^{ \frac{2}{3(1+w)} }.
    \label{eq:scale_factor}
\end{equation}


\section{Matter and radiation}
\label{sec:lcdm_matrad}

In this section, we analyse two main components of the universe: matter and radiation, reviewing the consequences that each of them has on the evolution of the scale factor.

Matter (m) behaves as a fluid with vanishing pressure, i.e. $w_m = 0$, which according to Eq.~(\ref{eq:state_eq}), this implies that
\begin{equation}
    \rho_m \propto a^{-3}.
    \label{eq:density_mat}
\end{equation}
This simply states that the density of matter decreases in proportion to the volume of the universe as it expands.
From Eq.~(\ref{eq:scale_factor}), we see that the evolution of the scale factor in a flat matter-dominated universe ($\Omega_m = 1$ and $\Omega_K = 0$) is given by
\begin{equation}
    a(t) \propto t^{2/3}.
    \label{eq:scale_factor_mat}
\end{equation}
This solution is also known as the Einstein-de Sitter (EdS) universe.

Radiation ($\gamma$), on the other hand, behaves as a relativistic fluid with an equation of state given by $w_\gamma = 1/3$, which according to Eq.~(\ref{eq:state_eq}), this implies that
\begin{equation}
    \rho_\gamma \propto a^{-4}.
    \label{eq:density_rad}
\end{equation}
This dependence can be understood in the following way: as the universe expands, a factor $a^{-3}$ can be associated with a change in the volume element, whereas an additional factor $a^{-1}$ can be associated with the stretching of the radiation wavelength.
As in the case of matter, we can study the behaviour of a flat radiation-dominated universe by inspecting Eq.~(\ref{eq:scale_factor}).
This leads to
\begin{equation}
    a(t) \propto t^{1/2}.
    \label{eq:scale_factor_rad}
\end{equation}

Analysing the dependences with the scale factor in Eqs.~(\ref{eq:density_mat}) and (\ref{eq:density_rad}), it is straightforward to realise that matter dominates the evolution and energy budget of the universe at later times, but radiation is the dominant component at early times.
Furthermore, a comparison between Eqs.~(\ref{eq:scale_factor_mat}) and (\ref{eq:scale_factor_rad}) shows that the universe expands more slowly in the radiation-dominated era than in the matter-dominated era.


\section{Dark energy}
\label{sec:lcdm_de}

In recent years, a variety of precise cosmological observations have shown that our Universe is currently undergoing a phase of accelerated expansion ($\ddot{a}>0$).
According to Eq.~(\ref{eq:frw2}), this requires the presence of a new and strange component with an equation of state $w_\mathrm{DE} < -1/3$, known as dark energy (DE).
Fluids for which $\rho + 3p \geq 0$ are said to satisfy the strong energy condition.
This implies that dark energy must violate this condition.
There are several models to explain this phenomenon.
Here, we will only focus on the cosmological constant and its association with the vacuum energy.

Given the fact that the field equations lead to dynamic universes, Einstein originally modified them in order to find static solutions by introducing an additional term proportional to the metric tensor, as
\begin{equation}
    G_{\mu \nu} - \Lambda g_{\mu \nu} = 8 \pi G T_{\mu \nu}.
    \label{eq:einstein_lambda}
\end{equation}
With the introduction of the constant $\Lambda$, known as the cosmological constant, the Friedmann's equations (Eqs.~\ref{eq:frw1} and \ref{eq:frw2}) are modified as
\begin{equation}
    \left( \frac{\dot{a}}{a} \right)^2 = \frac{8 \pi G}{3} \rho - \frac{K}{a^2} + \frac{\Lambda}{3}
    \label{eq:frw1_lambda}
\end{equation}
and
\begin{equation}
    \frac{\ddot{a}}{a} = -\frac{4 \pi G}{3} (\rho + 3p) + \frac{\Lambda}{3}.
    \label{eq:frw2_lambda}
\end{equation}
Later, due to the solid evidence about the expansion of the Universe, the cosmological constant was left aside.
However, models with $\Lambda > 0$ can also lead to solutions with $\ddot{a} > 0$, hence the $\Lambda$-term appeared again as a possible explanation for this phenomenon.

It is common to find an association between the cosmological constant and the vacuum energy.
Zero-point fluctuations must satisfy Lorentz invariance, and therefore the relation $\langle T_{\mu \nu} \rangle = \rho_\mathrm{vac} g_{\mu \nu}$, which corresponds to a barotropic fluid with an equation of state given by $w = -1$.
In this way, the contribution from the vacuum must be included in the right-hand side of Einstein's field equations:
\begin{equation}
    G_{\mu \nu} - \Lambda g_{\mu \nu} = 8 \pi G ( T_{\mu \nu} + \rho_\mathrm{vac} g_{\mu \nu} ).
\end{equation}
This clearly shows that the effect of the vacuum energy cannot be distinguished from that of a cosmological constant.
The net effect is that of an effective cosmological constant given by
\begin{equation}
    \Lambda_\mathrm{eff} := \Lambda + 8 \pi G \rho_\mathrm{vac},
\end{equation}
with an effective energy density given by
\begin{equation}
    \rho_\mathrm{eff} := \frac{\Lambda}{8 \pi G} + \rho_\mathrm{vac}.
\end{equation}

Even when both phenomena produce the same effect, they are very different in nature.
A cosmological constant implies that the gravitational constant $G$ is not enough to describe the behaviour of gravity correctly, so that $\Lambda$ must be taken into account as a second constant of nature.
The vacuum energy, on the other hand, is related to the energy content of the universe, associated with zero-point fluctuations, which, even in the absence of matter, are capable of curving space-time.
As these two cases cannot be distinguished, we will not include explicit terms for both in our equations, but rather we will treat this case simply as a dark-energy component characterised by the parameter $w_\mathrm{DE} = -1$.

Setting $w_\mathrm{DE} = -1$ in Eq.~(\ref{eq:state_eq}) shows that the corresponding energy density, $\rho_\mathrm{DE}$, remains constant as the universe expands.
Introducing this behaviour into Friedmann's equations, we obtain that, in this case, the time evolution of the scale factor is given by
\begin{equation}
    a(t) \propto \mathrm{exp}(H_0 t),
\end{equation}
i.e. the universe expands exponentially.

Before ending this section, we mention two alternative models for dark energy.
One is quintessence, in which dark energy is given by a scalar field.
The other is known as phantom energy, which not only violates the strong energy condition, but also the so-called weak energy condition: $\rho + p \geq 0$.
In this model, $\rho_\mathrm{DE}$ increases as the universe expands, and it is characterised by a strange property, the Big Rip, in which $a(t)$ diverges at a finite time.


\section{Distances in Cosmology}
\label{sec:lcdm_distance}

In this section, we study different distance measurements than can be used as means to describe the Universe.
These distance definitions only rely on the assumption that the metric takes the FRW form.
So far, we have used natural units in which $c = 1$, convenient for theoretical purposes.
However, they are not the most practical choice when dealing with observations.
Therefore, we will go back to traditional units in which $c$ appears explicitly, remembering that we must replace the time coordinate in Eq.~(\ref{eq:frw_metric2}) by the product $ct$.


\subsection{The comoving distance}
\label{subsec:lcdm_distance_com}

The expansion of the universe is contained in the scale factor.
The distance between two objects fixed in comoving coordinates, known as proper distance, $D_p$, changes according to $a(t)$.
The proper distance to an object with coordinates $(r, \theta, \phi)$, can be computed from the spatial part of the FRW metric given by Eq.~(\ref{eq:frw_metric2}), taking our position as the origin ($r=0$) and pointing the coordinate system so that $d\phi = d\theta = 0$: 
\begin{equation}
    D_p(t) = a(t) \int_0^r \frac{dr'}{\sqrt{1 - K r'^2}} = a(t) \chi.
    \label{eq:proper_distance}
\end{equation}
We can clearly see the physical meaning of the radial coordinate $\chi$, defined from Eq.~(\ref{eq:frw_metric_chi}), which represents the comoving distance to the object.
We will be mostly interested in the present-day proper distance between two points in space, which is given then by the corresponding comoving distance, since we have set $a_0 = 1$.


\subsection{The cosmological redshift}
\label{subsec:lcdm_distance_redshift}

The cosmological redshift, $z$, represents the change in the wavelength of radiation due to the expansion of the universe.
It is defined as
\begin{equation}
    z := \frac{\lambda_0 - \lambda_e}{\lambda_e},
    \label{eq:redshift}
\end{equation}
where $\lambda_0$ is the wavelength observed at a time $t_0$, and $\lambda_e$ is the wavelength emitted at a time $t_e < t_0$.
The redshift is a very important physical quantity as it is directly accessible through spectroscopic observations.

We can find a relation between the redshift and the scale factor at the time a source emits a light beam by studying the propagation of the light ray in an expanding universe described by an FRW metric.
For this, let us consider a ray propagating radially from $r = 0$ to $r = r_0$, and let us orient the coordinate system so that $d\phi = d\theta = 0$.
The origin is now at the emitting source.
As light travels along null geodesics ($ds = 0$), we find that
\begin{equation}
    \frac{c~dt}{a(t)} = d \chi = \frac{dr}{ \sqrt{ 1 - Kr^2 } }.
    \label{eq:dif_chi}
\end{equation}
Integrating this relation from $r=0$ to $r=r_0$, and considering the emission and reception times between two contiguous peaks of the light wave, it can be seen that
\begin{equation}
    \frac{\lambda_e}{\lambda_0} = \frac{a(t_e)}{a(t_0)}.
\end{equation}
In an expanding universe, $a(t_0) > a(t_e)$, which means that the wavelength increases.
This demonstrates that the cosmic expansion leads to a redshift of spectral lines.
If $t_0$ represents the present time, so that  $a(t_0) = 1$, we get the following relation between the cosmological redshift and the scale factor:
\begin{equation}
    a = \frac{1}{1 + z}.
    \label{eq:a_z}
\end{equation}
This means that the light we receive from a distant galaxy at a redshift $z$ left this galaxy when the scale factor was a fraction $1/(1 + z)$ of its present value.

As the redshift $z$ is a direct observable, it is often a more convenient variable than the time $t$.
From Eq.~(\ref{eq:dif_chi}), and the following relation:
\begin{equation}
    H(z) = - \frac{1}{(1+z)} \frac{dz}{dt},
\end{equation}
obtained in turn from Eqs.~(\ref{eq:hubble_parameter}) and (\ref{eq:a_z}), we can rewrite the comoving distance in terms of $z$:
\begin{equation}
    \chi(z) = c \int_0^z \frac{dz'}{H(z')},
    \label{eq:dcom_z}
\end{equation}
with
\begin{equation}
    H(z) = H_0 \left[ \sum_{i=1}^N \Omega_i (1+z)^{3(w_i+1)} + \Omega_K (1+z)^2 \right]^{1/2}.
    \label{eq:hubble_z}
\end{equation}
Note that for small redshifts, $z \ll 1$,
\begin{equation}
    \chi(z) \approx \frac{c~z}{H_0}.
    \label{eq:hubble_law}
\end{equation}
We recovered Hubble's law.
Here, $cz$ is the apparent receding velocity of a source.

The redshift is a direct observable quantity, but the distance is not.
Nevertheless, the redshift is also a distance indicator.
In fact, the comoving distance to a source can be calculated from its redshift by means of Eq.~(\ref{eq:dcom_z}).
However, this calculation depends on the Hubble parameter, given by Eq.~(\ref{eq:hubble_z}), which in turn depends on the parameters $H_0$ (or $h$), $\Omega_i$ and $\Omega_K$, the so-called cosmological parameters.
Therefore, these parameters must be known in order to estimate distances.
We will now see that there are different observational definitions of distance that are closely related to $\chi$, and that can be used to design tests in order to estimate the value of these parameters.
These are not truly distances, but recipes to compute different physical aspects of a source, like its light or its size.


\subsection{The angular-diameter distance}
\label{subsec:lcdm_distance_ang}

Let us first consider an extended object at a redshift $z$ with an angular diameter $\Delta \phi$.
Using Eq.~(\ref{eq:frw_metric2}), and setting $d\chi = d\theta = 0$, we can compute the physical (proper) size of the object as
\begin{equation}
    \Delta l = a(t) S_K(\chi) \Delta \phi.
    \label{eq:disang}
\end{equation}
In an Euclidean universe, the corresponding relation is
\begin{equation}
    \Delta l = D \Delta \phi,
    \label{eq:disang2}
\end{equation}
where $D$ denotes the distance to the object.
A comparison between these two expressions leads to the definition of the angular-diameter distance:
\begin{equation}
    D_A(z) := \frac{\Delta l}{\Delta \phi} = \frac{ S_K(\chi(z)) }{ 1 + z }.
    \label{eq:dist_ang}
\end{equation}
For a flat universe, $S_K(\chi(z)) = \chi(z)$, which corresponds to the comoving distance (as we have set $a_0 = 1$, it also corresponds to the true physical distance).

\begin{figure}
    \centering
    \includegraphics[width=\textwidth/2]{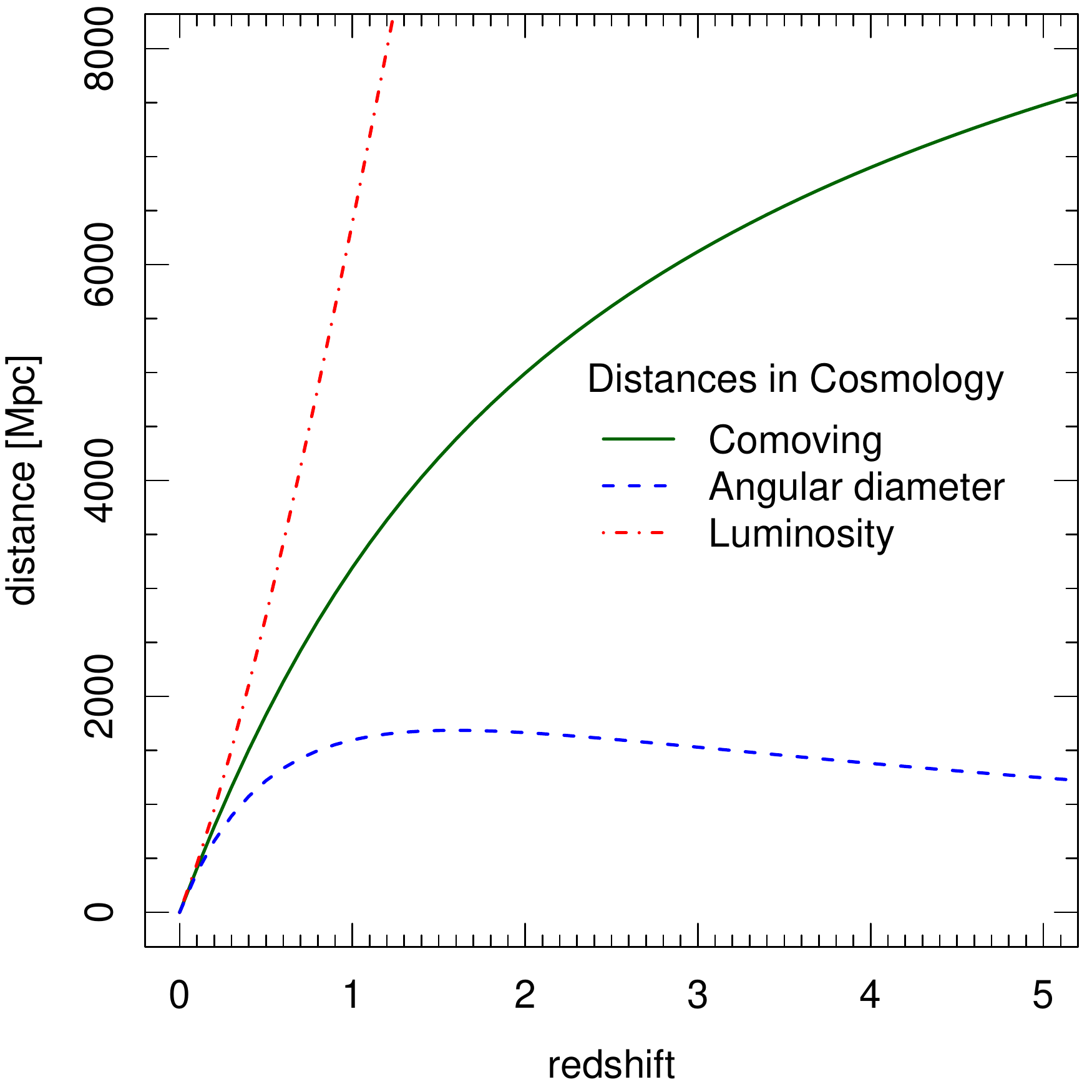}
    \caption[
    Redshift evolution of the different notions of distance used in Cosmology. 
    ]{
    Redshift evolution of the comoving distance (green solid curve, Eq.~\ref{eq:dcom_z}), the angular-diameter distance (blue dashed curve, Eq.~\ref{eq:dist_ang}), and the luminosity distance (red dot-dashed curve, Eq.~\ref{eq:dist_lum}) for a universe similar to ours with parameters $h=0.72$, $\Omega_K=0$, $\Omega_m=0.31$ and $\Omega_\Lambda=0.69$.
    }
    \label{fig:distances_cosmology}
\end{figure}

Figure~\ref{fig:distances_cosmology} shows the redshift evolution of all types of distances defined in this work for a universe similar to ours with parameters $h=0.72$, $\Omega_K=0$, $\Omega_m=0.31$ and $\Omega_\Lambda=0.69$.
Note, in particular, that the angular-diameter distance of an object is systematically smaller than its actual distance.
In fact, although for nearby objects both are similar, $D_A$ goes to zero as $z$ tends to infinity.
This means that, beyond a turning point, objects that are further away have larger angular sizes.
This apparently counter-intuitive behaviour can be understood if we notice that we are using as a reference an object of fixed proper size, which corresponds to a larger comoving size as $z$ increases.

\subsection{The luminosity distance}
\label{subsec:lcdm_distance_lum}

Let us now consider a source of light at a redshift $z$.
For this object, we define the luminosity distance, denoted by $D_L$, by means of the relation between the observed bolometric flux, $S$, and its intrinsic bolometric luminosity, $L$, in an Euclidean space-time:
\begin{equation}
    D_L := \sqrt{ \frac{L}{4 \pi S} }.
    \label{eq:dislum1}
\end{equation}

Two effects must be taken into account to understand the redshift evolution of the luminosity distance.
First, individual photons lose energy due to the cosmological redshift, and second, photons arrive less frequently as the universe expands.
In more detail, the bolometric luminosity of a source at a time $t_1$ is given by
\begin{equation}
    L = \int N_1(\nu_1) h_P \nu_1 d \nu_1,
\end{equation}
where $N_1(\nu_1)$ is the total number of photons of frequency $\nu_1$ emitted from the source per unit time, and $h_P$ is the Planck constant.
If these photons are emitted isotropically, the flux observed at a time $t_0$ at a comoving distance $\chi$ from the source will be given by
\begin{equation}
    S = \int n_0(\nu_0) h_P \nu_0 d \nu_0,
\end{equation}
where $n_0(\nu_0)$ is the received number of photons of frequency $\nu_0$ per unit area per unit time, related to $N_1(\nu_1)$ by
\begin{equation}
    n_0(\nu_0) d\nu_0 = \frac{ N_1(\nu_1) }{ 4 \pi a_0^2 \chi^2} \frac{d\nu_0}{d\nu_1} d\nu_1,
\end{equation}
with $\nu_1 =  \nu_0 a(t_0)/a(t_1)$.
Putting these ingredients together into Eq.~(\ref{eq:dislum1}), we finally get
\begin{equation}
    D_L(z) = (1 + z) \chi(z).
    \label{eq:dist_lum}
\end{equation}
This means that, as the redshift reduces the apparent luminosity of distant objects, they appear to be further away than they really are (see Figure~\ref{fig:distances_cosmology}).


\section{The redshift-distance relation}
\label{sec:lcdm_tests}

Let us assume that we could simultaneously measure the distances and redshifts of some objects of the same type distributed throughout the Universe.
Then, we could build a redshift-distance relation, also known as a Hubble diagram.
Measuring redshifts from the shift of spectral lines is straightforward, the hard part is determining distances from an independent method.
However, based on the definitions of the luminosity and angular-diameter distances given before, together with their dependence on the comoving distance, the redshift-distance relation can then be used to derive constraints on the cosmological parameters involved, and in this way, to quantify the expansion rate and energy budget of the Universe.


\subsection{Standard candles}
\label{subsec:lcdm_tests_candle}

If we could identify a type of objects of the same intrinsic luminosity $L$, and measure their fluxes $S$ at different redshifts $z$, then it would be possible to build a Hubble diagram using the luminosity distance: $D_L(z)$.
Such an object is known as a standard candle.

In practice, what is relevant is that any difference between the apparent brightness (measured in terms of astronomical magnitudes) of two such objects is a direct result of their different distances from us.
In this way, the Hubble diagram is represented in terms of a redshift-magnitude relation.
In some cases, the method is generalised to find a correlation between an observable and intrinsic brightness.
For example, Cepheid variables are stars for which their intrinsic brightness is tightly related to their period.
The Hubble Space Telescope was able to measure the periods of thousands of Cepheid variables in galaxies as far away as $20~\mathrm{Mpc}$.
With distances to these galaxies fixed, several secondary distance indicators could be calibrated to go much further, as far away as $400~\mathrm{Mpc}$.
In this way, it was possible to calibrate the Hubble constant \cite{hst_project}.

The standard candle that can be seen at largest distances is SN Ia.
Since they are so bright, SNe Ia can be used to extend the Hubble diagram to very large redshifts ($z > 1$), a regime where the simple linear Hubble's law (Eq.~\ref{eq:hubble_law}) ceases to work, and therefore they can be used to test the redshift evolution of the luminosity distance.
SNe Ia have provided key measurements for the discovery of the current phase of accelerated expansion of the Universe \cite{sn_riess, sn_perlmutter, supernova_suzuki}.


\subsection{Standard rulers: the Alcock-Paczyński test}
\label{subsec:lcdm_tests_ap}

We turn now to the case in which we could identify a type of objects of the same physical size $\Delta l$, and measure their angular sizes $\Delta \phi$ at different redshifts $z$.
Then, it would be possible to build a Hubble diagram using the angular-diameter distance: $D_A(z)$.
Such an object is known as a standard ruler.
One example is the signal of the baryon acoustic oscillations (BAO) imprinted on the clustering of galaxies, a relic of the early Universe, whose pattern is due to the acoustic waves experimented by the primordial photo-baryonic fluid.
See for instance the work of \citeonline{bao_sanchez}.

We can do even more with a standard ruler.
Let us assume that, instead of being extended across the sky, it is located along the line of sight, and it is large enough for us to measure the redshift difference between its two endpoints, $\Delta z$.
Then, according to Eqs.~(\ref{eq:proper_distance}), (\ref{eq:a_z}) and (\ref{eq:dcom_z}), and assuming that the scale factor and the Hubble parameter remain almost constant between the endpoints, we find that the physical size of the object is given by
\begin{equation}
    \Delta l = \frac{c \Delta z}{(1+z) H(z)}.
    \label{eq:ap0}
\end{equation}
Therefore, measurements of this kind can be used to probe the redshift evolution of the Hubble factor as $H(z) = c \Delta z / (1+z) \Delta l$, providing a direct measurement of the expansion rate of the Universe.

These tests require the knowledge of the true physical size $\Delta l$ of our standard ruler.
Even if this is not the case, we can still infer cosmological information from these objects.
What we need is to identify them in both directions, i.e. some of them across the sky and some of them along the line of sight.
In other words, we can exploit the measurements of the angular and redshift extent of the ruler to obtain cosmological information.
By equating Eqs.~(\ref{eq:disang}) and (\ref{eq:ap0}), we obtain the following relation:
\begin{equation}
    F_\mathrm{AP}(z) := \frac{\Delta z}{\Delta \phi} = \frac{(1+z)}{c} D_A(z) H(z).
    \label{eq:ap_parameter}
\end{equation}
This is the Alcock-Paczyński test (hereinafter AP test), first proposed by \citeonline{ap}.
The combination $F_\mathrm{AP}$ is referred to as the Alcock-Paczyński parameter.
This test will be fundamental for our study of voids.


\section{Our Universe: flat-\texorpdfstring{$\Lambda$}{lambda}CDM}
\label{sec:lcdm_lcdm}

In this section, we summarise the main characteristics of our Universe, like its energy content and geometry, provided by cosmological experiments.

\begin{enumerate}

\item[1.]
\textit{Expansion rate.}
Current observations show that $h = 0.74 \pm 0.01$ \cite{hubblecons_riess}.
This means that the critical density is approximately 6 protons per cubic metre (Eq.~\ref{eq:density_cr}).
Nevertheless, a tension has been detected between this value, obtained from observations of the local Universe, and that obtained from studies of the early Universe, which suggest a lower value of $h = 0.674 \pm 0.005$ \cite{planck}.
In view of this uncertainty in the value of $h$, it is a common practice to express distances and masses in units of $\hmpc$ and $\hmsun$, respectively.

\item[2.]
\textit{Radiation.}
We can estimate the current energy density of radiation from the temperature of CMB photons, Eq.~(\ref{eq:tcmb}):
\begin{equation}
    \rho_\mathrm{\gamma,0} = \frac{ \pi^2 k_B^4 }{ 15 \hbar_P^3 c^3 } T_\mathrm{CMB}^4 = 4.5 \times 10^{-34} \frac{\mathrm{g}}{\mathrm{cm}^3},
\end{equation}
where $k_B$ is the Boltzmann constant.
This implies that
\begin{equation}
    \Omega_\gamma = 2.4 \times 10^{-5} ~ h^{-2}.
\end{equation}
In principle, three massless neutrino species would contribute a similar amount.
Beyond this aspect, it is then possible to neglect the contribution of radiation to the present-day energy budget of the Universe.

\end{enumerate}

\begin{figure}
    \centering
    \includegraphics[width=\textwidth/2]{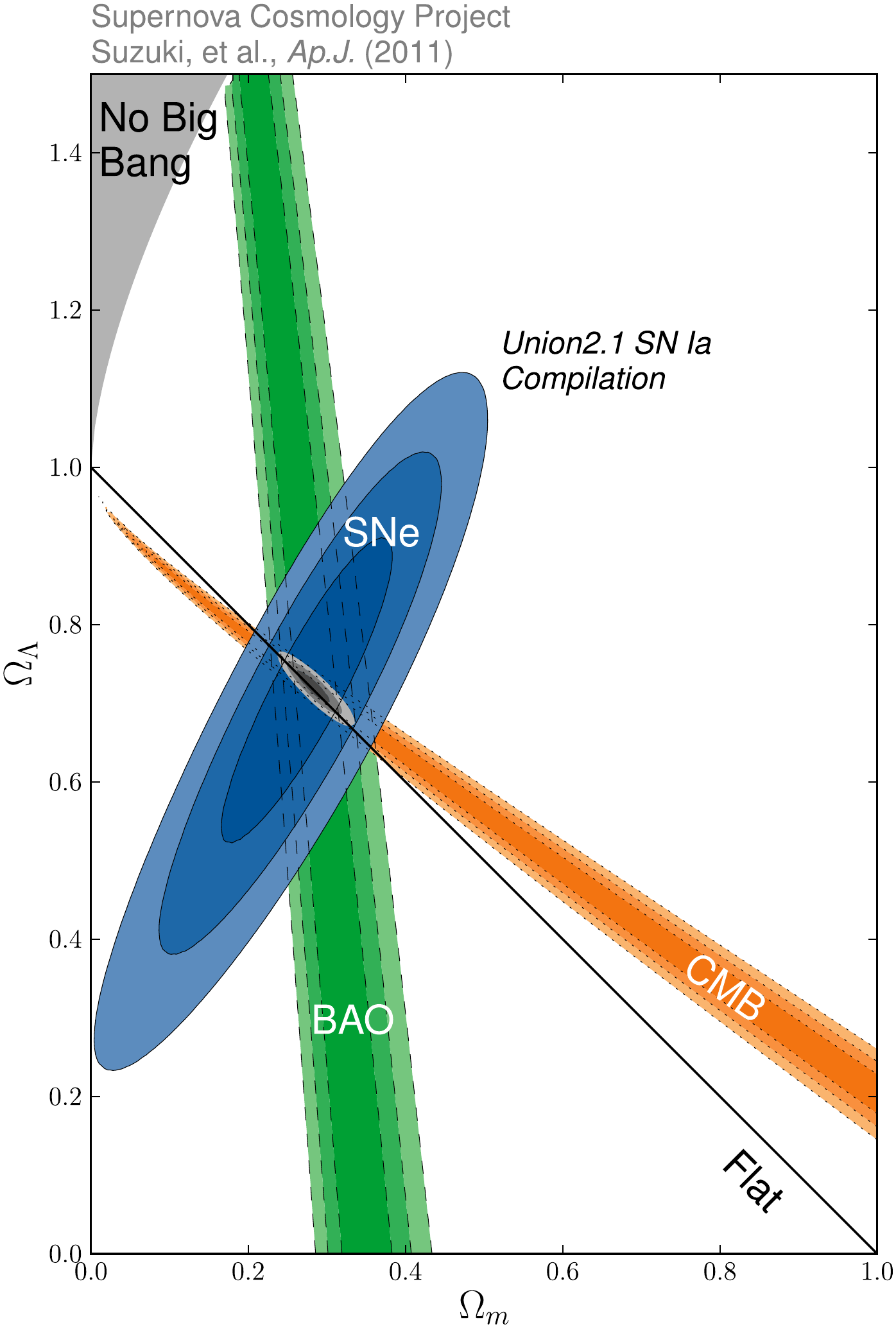}
    \caption[
    Fundamental cosmological experiments for studying dark energy: SNe Ia, CMB and BAO.
    ]{
    Allowed regions of the parameter space spanned by $\Omega_m$ and $\Omega_\Lambda$, as derived from three fundamental cosmological experiments: the study of CMB anisotropies, the extended Hubble diagram using SNe Ia, and the AP test using the BAO signal on the galaxy clustering.
    Since the confidence regions overlap and have substantially different orientations in this plane, their combination provides a much better constraint than each method by itself.
    Void-based experiments are promising cosmological probes that will contribute to these constraints even further.
    \textit{Figure credit:} \citeonline{supernova_suzuki}.
    }
    \label{fig:cosmoconstraints}
\end{figure}

In view of this, we will hereinafter focus on models containing only matter and dark energy, and also assume that $w_\mathrm{DE} = -1$.
In this case, Eq.~(\ref{eq:frw4}) can be rewritten as
\begin{equation}
    \dot{a}^2 = H_0^2[ \Omega_m a^{-1} + \Omega_\mathrm{DE} a^2 + (1 - \Omega_m - \Omega_\mathrm{DE} ) ].
    \label{eq:frw5}
\end{equation}
This relation also provides an expression for the age of the Universe:
\begin{equation}
    t_0 = \frac{1}{H_0} \int_0^1 \frac{ da }{ \sqrt{ \Omega_m a^{-1} +  \Omega_\mathrm{DE} a^2 + (1 - \Omega_m - \Omega_\mathrm{DE}) } }.
    \label{eq:age}
\end{equation}
It is useful to study the behaviour of various models in the $\Omega_m-\Omega_\mathrm{DE}$ plane, as the one shown in Figure~\ref{fig:cosmoconstraints}, which can be divided into regions with different properties.
This figure contains the parameter constraints from three fundamental cosmological experiments: the study of CMB anisotropies, the extended Hubble diagram using SNe Ia, and the AP test using the BAO signal on the galaxy clustering.
The importance of these results resides in the fact that the three methods are completely independent from each other, and yet they are consistent.
But more important is that the confidence regions have substantially different orientations, hence their combination provides a much tighter constraint than each method by itself.

\begin{enumerate}

\item [3.]
\textit{Curvature.}
Current observations suggest that the Universe is flat to a high precision: $|\Omega_K \leq 10^{-5} |$ \cite{planck}.

\item[4.]
\textit{Baryonic matter.}
This is ordinary matter, the one that interacts with light.
There are four established ways of measuring the baryon density: (i) observing the baryon content directly (the greatest contribution to the density comes, not from stars in galaxies, but rather from gas in groups and clusters of galaxies), (ii) looking at spectra of distant quasars (the amount of light absorbed from these beacons is a measure of the intervening hydrogen), (iii) the analysis of anisotropies in the CMB, and (iv) the analysis of the light-elements abundance according to the predictions of BBN.
Remarkably, these estimates with very different techniques all agree.
They all place the baryon density at $5\%$ of the critical density.

\item[5.]
\textit{Dark matter.}
All of the methods mentioned above involve the interaction of matter and radiation.
The methods that do not rely on this interaction usually involve measuring the gravitational field in a given system, thereby inferring information about the total mass responsible for that field.
We mention four methods: (i) studying the spatial distribution of galaxies by means of the power spectrum, (ii) measuring the cosmic velocity field and its relation with the observed galaxy distribution, (iii) picking out observations sensitive to $\Omega_b/\Omega_m$, such as the ratio of the mass of gas in clusters to the total mass by looking for X-ray emission, and (iv) the analysis of anisotropies in the CMB.
All these experiments favour a value $\Omega_m = 0.31$, which means that the total matter density is equal to approximately $31\%$ of the critical density.
We therefore have an enormous amount of evidence telling us that the total matter density is approximately five times larger than the baryon density.
Most of the matter in the Universe must be in an unknown form referred to as dark matter.
Moreover, it is said that dark matter must be cold (CDM) in order to explain the observed large-scale structure, in contrast to hot or warm dark matter.
The differences between them lie in the thermal velocities of their constituents when radiation and matter had equal density.
The particles of CDM were non-relativistic at that time.
Currently, the most plausible candidate for dark matter is a weakly interacting massive particle.

\item[6.]
\textit{Dark energy.}
In recent years, three methods have provided a solid evidence about the existence of a dark-energy component responsible for the current phase of accelerated expansion: (i) the extended Hubble diagram using distant SNe Ia, (ii) the analysis of anisotropies in the CMB, and (iii) the AP test using the BAO signal imprinted on the galaxy clustering.
They all suggest a value $\Omega_\mathrm{DE} = 0.69$ for the density, and $w_\mathrm{DE} = -1$ for the equation of state.
Hereinafter, we will consider that dark energy is described by a cosmological constant, and hence we will use the notation $\Omega_\Lambda$ to refer to its density.

\item[7.]
\textit{Age of the Universe.}
According to Eq.~(\ref{eq:age}), the present age of the Universe is estimated to be $t_0 = 13.8~\mathrm{Gyr}$, a time by which the Universe is entering into the dark-energy dominated era, and hence, approaching the exponential expansion.

\end{enumerate}

\begin{figure}
    \centering
    \includegraphics[width=79mm]{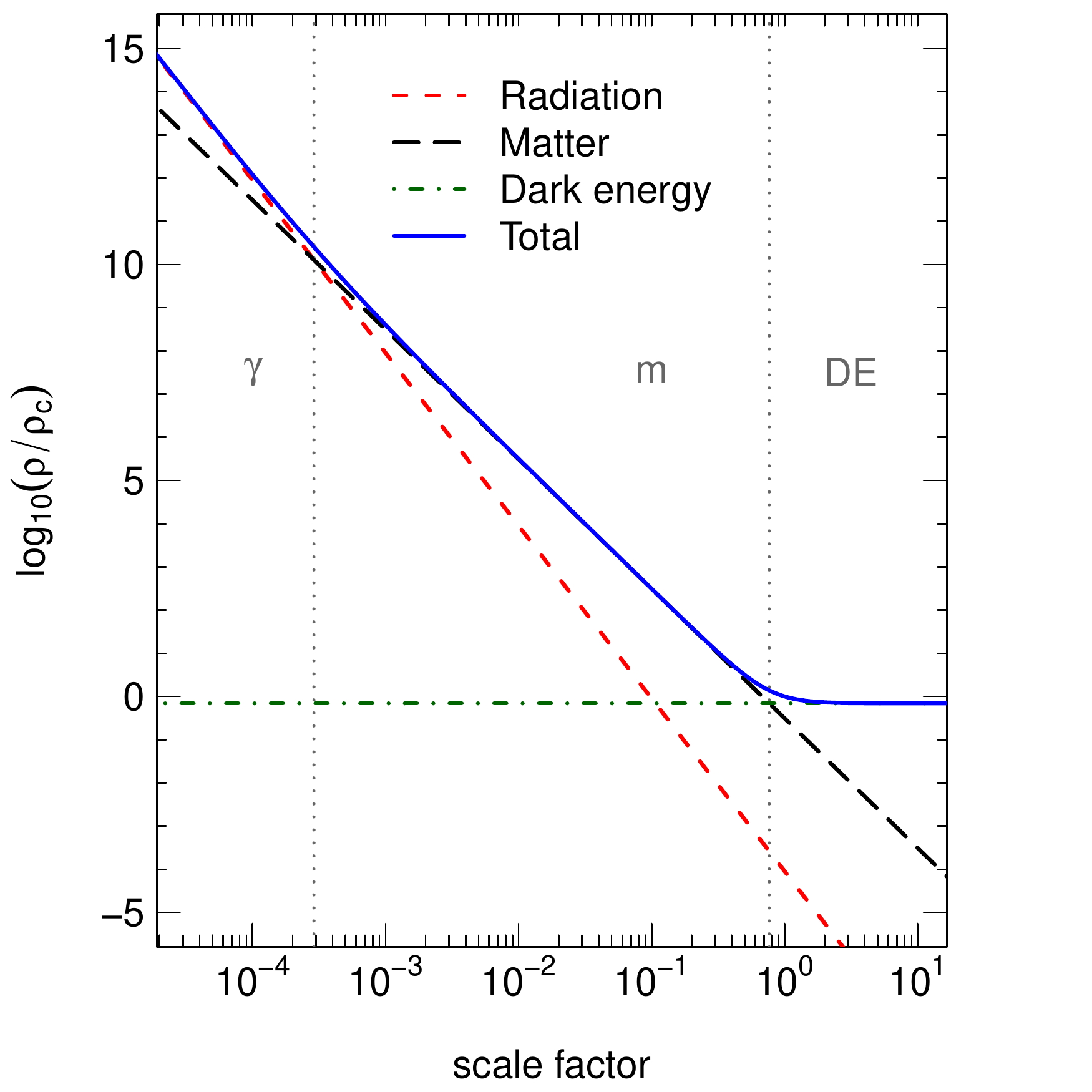}
    \includegraphics[width=79mm]{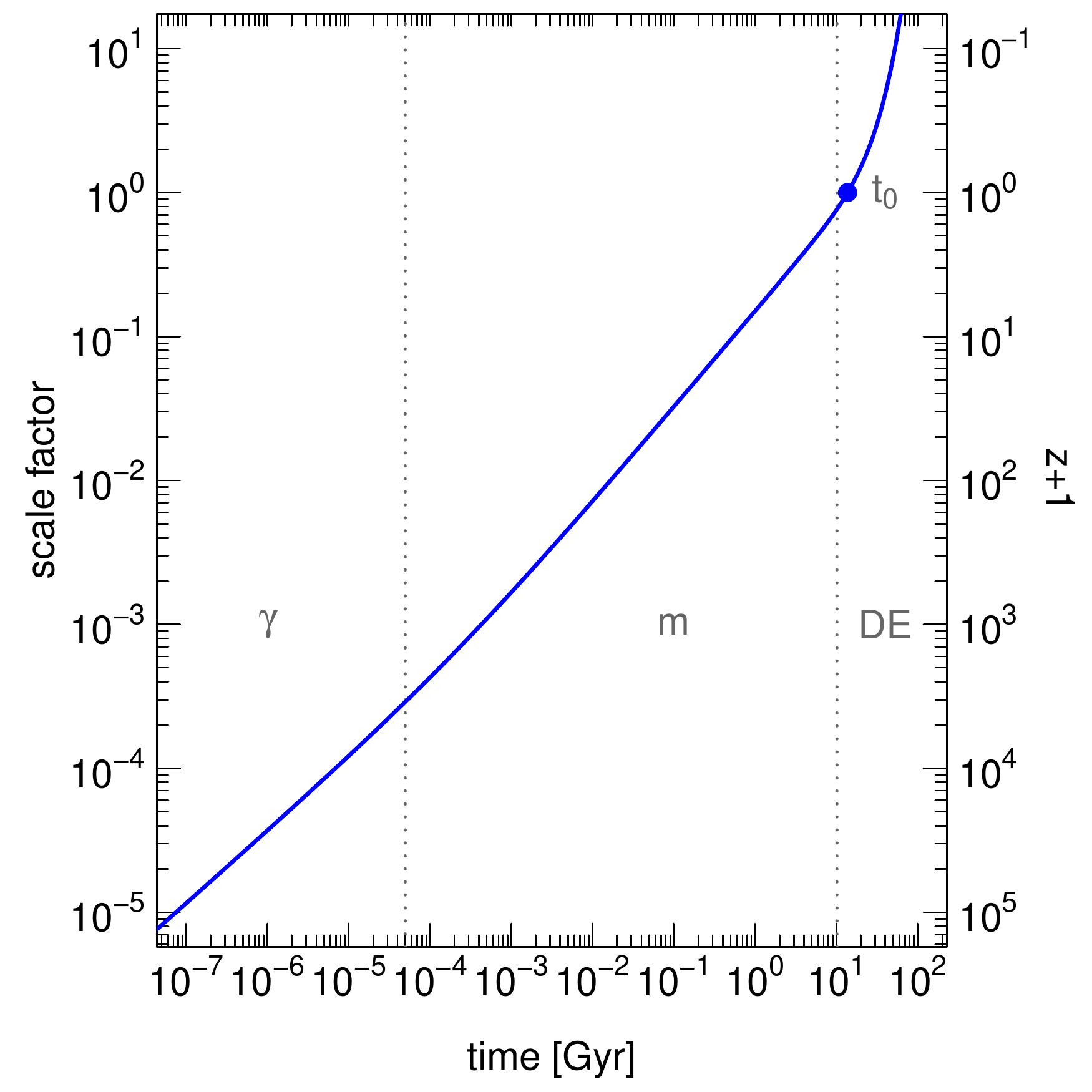}
    \caption[
    Evolution of the Universe.
    ]{
    \textit{Left-hand panel.}
    Evolution of the energy density as a function of the scale factor for the different constituents of our Universe: radiation (red dashed line), matter (black long-dashed line) and dark energy in the form of a cosmological constant (green dot-dashed line). 
    The total contribution is represented by the blue solid curve.
    \textit{Right-hand panel.}
    Evolution of the scale factor (or equivalently, the redshift) as a function of time.
    In both panels, it can be appreciated the different cosmological eras.
    Radiation dominated at early times, whereas matter dominated at later times.
    Apparently, dark energy is the dominant component nowadays.
    }
    \label{fig:evolution_rho_a}
\end{figure}

The cosmological model presented in this chapter, with all the characteristics describing our Universe, is called the flat-$\Lambda$CDM model.
The left-hand panel of Figure~\ref{fig:evolution_rho_a} shows the evolution of the densities of radiation, total matter and dark energy as a function of the scale factor, which follow the relations described in Sections~\ref{sec:lcdm_matrad} and \ref{sec:lcdm_de}.
It is also shown the evolution of the total density.
The right-hand panel shows the corresponding evolution of the scale factor as a function of time.
In both panels, three different epochs in the history of the Universe can be seen, in which its energy budget is dominated by radiation during the early phase, matter at later times, and dark energy today.

\chapter{Structures in the Universe}
\label{chp:lss}
In this chapter, we develop the relevant statistical tools for the analysis of the large-scale structure of the Universe, and in this way, be able to tackle the study of cosmic voids in the next chapter.
The main aspect to consider is to avoid scales dominated by non-linearities.
While perturbations to matter on small scales, less than $10~\hmpc$, have grown non-linear, larger-scale perturbations are still small, hence linear theory can be assumed.


\section{The density field and the power spectrum}
\label{sec:lss_density}

The statistical analysis of the large-scale structure is based on the density contrast, $\delta(\mathbf{x})$, which describes the density fluctuation field of the Universe, and is defined as
\begin{equation}
    \delta(\mathbf{x}) := \frac{\rho(\mathbf{x}) - \bar{\rho}}{\bar{\rho}},
    \label{eq:delta}
\end{equation}
where $\bar{\rho}$ represents the mean density of the Universe, and $\rho(\mathbf{x})$, the local density at position $\mathbf{x}$.
In order to distinguish the density contrast of the total matter (dark plus baryonic) from that of galaxies, we will use the notations $\delta_m$ and $\delta_g$, respectively.
The same will apply to any other quantity to be defined.
If no index is specified, an arbitrary distribution of objects will be implicitly assumed.

The analysis of small density fluctuations in linear theory is simpler in Fourier space, since different Fourier modes evolve independently.
Our convention will be
\begin{equation}
    \delta(\mathbf{x}) = \frac{1}{(2 \pi)^3} \int \delta(\mathbf{k}) e^{i \mathbf{k}.\mathbf{x}} d^3k.
    \label{eq:delta_fourier}
\end{equation}
Here, $\mathbf{k}$ is a wave vector whose modulus, $k = |\mathbf{k}|$, characterises a given scale of space.

A useful tool to characterise fluctuations is the power spectrum, $P(k)$, defined via the following relation:
\begin{equation}
    \langle \delta(\mathbf{k}) \delta^*(\mathbf{k}') \rangle = (2 \pi)^3 \delta_D(\mathbf{k} - \mathbf{k}') P(k),
    \label{eq:powspec}
\end{equation}
where the angular brackets denote an average over the whole distribution, and $\delta_D$, the Dirac delta function.
Qualitatively speaking, the power spectrum is the spread or variance of the amplitude distribution of fluctuations on a given scale.
If there are a lot of prominent underdense and overdense regions, the power spectrum will be large, whereas it will be small if the distribution is smooth.


\section{The two-point correlation function}
\label{sec:lss_correlation}

The power spectrum is a useful statistical tool for theoretical grounds.
However, in practice, it is often simpler to use other methods and relate them to the power spectrum.
Of particular importance is the two-point correlation function, defined as
\begin{equation}
    \xi(\mathbf{r}) := \langle \delta(\mathbf{x}) \delta(\mathbf{x} - \mathbf{r}) \rangle.
    \label{eq:corfunc_def}
\end{equation}
As clustering is statistically isotropic, the correlation function must solely depend on the separation $r = |\mathbf{r}|$ between two points, i.e. $\xi(\mathbf{r}) = \xi(r)$.
Expressing the density fluctuations in terms of their Fourier transform, we find that
\begin{equation}
    \xi(r) = \frac{1}{(2 \pi)^3} \int P(k) e^{i \mathbf{k} . \mathbf{r}} d^3k,
    \label{eq:xi_pk}
\end{equation}
which shows that the correlation function is simply the Fourier transform of the power spectrum.
As the power spectrum is also isotropic, we can perform the angular part of the integral independently.
Using the fact that $\xi(r)$ must be a real function, the correlation function can be finally written as
\begin{equation}
    \xi(r) = \frac{1}{2 \pi^2} \int P(k) \frac{\mathrm{sin}(kr)}{kr} k^2 dk.
\end{equation}

Although the general shape of the correlation function is complex, it can be empirically described by a power law over a limited range of scales.
Indeed, measurements of the correlation function from early galaxy surveys were correctly described by fitting functions of the form
\begin{equation}
    \xi_g(r) = \left( \frac{r}{r_0} \right)^\alpha.
\end{equation}
Typical values of these parameters are $\alpha \approx -1.8$ and $r_0 \approx 5 \hmpc$.
As larger galaxy surveys provided more accurate measurements of $\xi_g(r)$, deviations from this simple behaviour were detected, mostly caused by non-linear effects on small scales.


\subsection{Estimators}
\label{subsec:lss_correlation_estimators}

An alternative definition of the correlation function for a discrete set of point particles is related to the probability of finding pairs of such particles as a function of their separation.
For a completely homogeneous distribution, the probability of finding a pair of particles in two volume elements $dV_1$ and $dV_2$ separated by a distance $r_{12}$ is given by
\begin{equation}
    dP_{12} = \bar{\rho}^2 dV_1 dV_2.
    \label{eq:dP_homo}
\end{equation}
For a general distribution, this probability can be computed as
\begin{equation}
    dP_{12} = \langle \rho_1 \rho_2 \rangle dV_1 dV_2.
\end{equation}
Expressing the densities according to Eq.~(\ref{eq:delta}) as $\rho_i = \bar{\rho} (1 + \delta_i)$, this expression can be written as
\begin{equation}
    dP_{12} = \bar{\rho}^2 (1 + \langle \delta_1 \delta_2 \rangle) dV_1 dV_2 = \bar{\rho}^2 (1 + \xi(r_{12})) dV_1 dV_2.
    \label{eq:dP_lss}
\end{equation}
This means that the correlation function, $\xi(r)$, gives the excess probability of finding pairs of particles separated by a distance $r$ with respect to a homogeneous distribution.

The estimation of the correlation function for the galaxy distribution depends then on the probability of finding galaxy pairs separated by distances between $r$ and $r + dr$, $DD(r)$.
That is,
\begin{equation}
    DD(r) = \frac{N_\mathrm{pairs}(r)}{N_\mathrm{tot}},
    \label{eq:data_data}
\end{equation}
where $N_\mathrm{pairs}(r)$ is the number of galaxy pairs inside the volume element defined by $r$ and $r + dr$, a spherical shell, and $N_\mathrm{tot}$ is the total number of pairs.
In turn, $N_\mathrm{tot} = N_g(N_g - 1)/2$, where $N_g$ is the total number of galaxies.
The probability of finding pairs of galaxies in a homogeneous distribution, $RR(r)$, can be computed in an analogous way by means of a random catalogue.
From Eqs.~(\ref{eq:dP_homo}) and (\ref{eq:dP_lss}), we can see that both probabilities are related by $DD(r) = RR(r) (1 + \xi_g(r))$.
We can then estimate the two-point correlation function for the galaxy distribution as
\begin{equation}
    \hat{\xi}_g(r) = \frac{DD(r)}{RR(r)} - 1,
    \label{eq:estimator_natural}
\end{equation}
which is known as the natural estimator.

A number of alternative estimators based on pair counts have been developed.
A few examples are:
\begin{equation}
    \hat{\xi}_g(r) = \frac{DD(r)}{DR(r)} - 1,
    \label{eq:estimator_dp}
\end{equation}
from \citeonline{estimator_davis}, and
\begin{equation}
    \hat{\xi}_g(r) = \frac{DD(r) - 2DR(r) + RR(r)}{RR(r)},
    \label{eq:estimator_ls}
\end{equation}
from \citeonline{estimator_landy}.
These estimators depend on the quantity $DR(r)$, which measures the probability of finding pairs with one element in the galaxy sample and the other element in the random catalogue (in which case $N_\mathrm{tot} = N_g N_\mathrm{ran}$, with $N_\mathrm{ran}$ the total number of random particles).
The first two estimators are more sensitive to edge effects in the sample.
The performance of the last one is the most common choice in current clustering analyses.


\section{The bias parameter}
\label{sec:lss_bias}

Measurements of the correlation function of galaxy samples of varying colour or luminosity exhibit different amplitudes.
This is just an indication that the galaxy clustering does not necessarily match that of the underlying matter distribution.
In general, we can assume that the galaxy density fluctuation field is a function of the local matter density fluctuation field, which can be expanded as
\begin{equation}
    \delta_g(\mathbf{x}) = \sum_{k=1}^\infty \frac{b_k}{k!} \delta_m^k(\mathbf{x}),
\end{equation}
where the coefficients $b_k$ are known as the bias parameters.

In the linear regime, where $\delta_m \ll 1$, we can then expect a simple linear relation of the form
\begin{equation}
    \delta_g(\mathbf{x}) = b_1 \delta_m(\mathbf{x}),
    \label{eq:bias_linear}
\end{equation}
in which case the linear bias parameter is simply denoted as $b$.
In this case, the galaxy and matter correlation functions are related by
\begin{equation}
    \xi_g(r) = b^2 \xi_m(r).
    \label{eq:bias_relation}
\end{equation}

Galaxies form in dark-matter haloes.
It is possible to derive analytical and semianalytical recipes for the dependence of the bias parameter with halo mass: $b(M)$.
Then, assuming a given value for the amplitude of the scalar power spectrum (see Section~\ref{subsec:lss_evolution_pk}), the amplitude of the correlation function of different galaxy samples can provide information about the mass of the dark-matter haloes in which they live, or more generally, the way in which they populate haloes of different masses.
This information is valuable for galaxy formation models, but is of little cosmological interest.
Our current understanding of the physical processes that control the formation and evolution of galaxies is not detailed enough to predict the bias of a given galaxy population with high accuracy.
This means that no cosmological information can be extracted from the amplitude of the correlation function or the power spectrum.
Only the shape of these statistics is used when constraining the cosmological parameters.


\section{The evolution of density fluctuations}
\label{sec:lss_evolution}

The structure that we observe today evolved from small density fluctuations in the early Universe.
The process that describes this evolution is gravitational instability.
As time evolves, matter accumulates in initially overdense regions.
Complementarily, initially underdense regions become emptier.
These are the seeds of cosmic voids.
Gravitational instability can be schematically depicted in the following way:
\begin{equation}
    \ddot{\delta} + [\mathrm{pressure} - \mathrm{gravity}] \delta = 0.
\end{equation}
The two basic forces, gravity and pressure, act in opposite directions.
There are different manifestations of this basic idea, where different ambient cosmological conditions alter the growth rate.
For example, in the matter-dominated era, $\delta$ grows as a power of time, whereas in the radiation-dominated era, the growth is logarithmic.


\subsection{The evolution equations}
\label{subsec:lss_evolution_eqs}

In order to find the equations that govern the evolution of small departures from the smooth FRW cosmological models, we will assume that the metric tensor can be written as
\begin{equation}
    g_{\mu \nu} = \bar{g}_{\mu \nu} + \delta g_{\mu \nu},
\end{equation}
where $\bar{g}_{\mu \nu}$ represents the FRW metric, and $\delta g_{\mu \nu}$ characterises the small deviations in this metric.
In the same way, we must include small deviations in the energy-momentum tensor from that of an ideal fluid, $\delta T_{\mu \nu}$.
Perturbations in the metric and the energy-momentum tensors are related by Einstein's field equations, which, after subtracting the terms corresponding to the homogeneous background model, and keeping only linear terms in the metric perturbations, they result in the linearised field equations:
\begin{equation}
    \delta G_{\mu \nu} = 8 \pi G \delta T_{\mu \nu}.
\end{equation}

Working in the longitudinal gauge, also known as the conformal Newtonian gauge, the metric perturbations are characterised by the gauge invariant variables $\Phi$ and $\Psi$, so that the metric takes the form
\begin{equation}
    ds^2 = a^2(\tau) [ -(1 + 2\Phi) d\tau^2 + (1 - 2\Psi) \gamma_{ij} dx^i dx^j ].
    \label{eq:frw_pert_metric}
\end{equation}
In the absence of $\Phi$ and $\Psi$, this expression is simply the FRW metric.
Similarly, in the absence of expansion $(a=1)$, this metric describes a weak gravitational field.
The perturbations have the following interpretation: $\Phi$ corresponds to the Newtonian potential, whereas $\Psi$, the perturbation to the spatial curvature.
A comment worth mentioning.
Eq.~(\ref{eq:frw_pert_metric}) only contains scalar perturbations.
In principle, it is possible that the metric also has vector or tensor perturbations.
If so, $g_{\mu \nu}$ would require other functions besides $\Phi$ and $\Psi$ to fully describe all perturbations.
However, scalar, vector and tensor perturbations are decoupled: each one evolves independently of the others.
Therefore, we will solely focus on scalar perturbations, these are the only ones that couple to matter perturbations.

In this context, the linearised field equations are given by
\begin{equation}
    \nabla^2 \Phi - 3\mathcal{H} (\Phi' + \mathcal{H}\Phi) = 4\pi G a^2 \delta \rho,
    \label{eq:poisson}
\end{equation}
\begin{equation}
    \nabla_i (\Phi' + \mathcal{H}\Phi) = 4\pi G a^2 (\bar{\rho} + \bar{p}) v_i
\end{equation}
and
\begin{equation}
    \Phi'' + 3\mathcal{H}\Phi' + (2\mathcal{H}' + \mathcal{H}^2)\Phi = 4\pi G a^2 \delta p.
\end{equation}
Here, the fluctuation in density, $\delta \rho$, the fluctuation in pressure, $\delta p$, and the peculiar velocity of the fluid,
\begin{equation}
    v^i := \frac{dx^i}{d\tau},
\end{equation}
are the perturbative variables relevant in $\delta T_{\mu \nu}$.
Moreover, $\mathcal{H}:=a'/a$ is the Hubble parameter defined in terms of conformal time $\tau$.
Note that $\Psi$ does not appear in these equations.
This is because we have assumed that $\Phi \approx \Psi$, an approximation valid in the limit when there are no quadrupole moments in the photon distribution.
On subhorizon scales, the second term of Eq.~(\ref{eq:poisson}) can be neglected, and the equation takes then the form of Poisson's equation.

Rather than the full linearised field equations, it is more convenient to use the energy conservation equations: $\nabla^\mu T_{0,\mu}=0$ and $\nabla^\mu T_{i,\mu}=0$, which correspond to the general relativistic versions of the continuity and Euler's equations, respectively.
More specifically, we need their linearised versions.
Using the relation between pressure and density in the homogeneous case, $\bar{p} = w\bar{\rho}$, and relating their perturbations by means of the sound speed, $c_s^2 = \delta p/\delta \rho$, we find that
\begin{equation}
    \delta' + 3\mathcal{H}(c_s^2 - w)\delta + (1 + w)(\nabla_i v^i - 3\Phi') = 0
\end{equation}
and 
\begin{equation}
    v'_i + \mathcal{H}(1 - 3c_s^2)v_i + \frac{c_s^2}{1 + w} \nabla_i \delta + \nabla_i \Phi = 0.
\end{equation}
Let us consider the case of a universe containing only dark matter ($w_m = 0$) and radiation ($w_\gamma = 1/3$) for this analysis, and let us focus on the scalar part of the velocity field: $v^i = \nabla^i v$.
The system is then completely specified by the density fluctuations, $\delta_m$ and $\delta_\gamma$, their respective velocity potentials, $v_m$ and $v_\gamma$, and the potential $\Phi$.
After Fourier-transforming, the continuity and Euler's equations for the matter and radiation components take the form
\begin{equation}
    \delta'_m - k^2 v_m = 3 \Phi',
    \label{eq:continuity}
\end{equation}
\begin{equation}
    \frac{3}{4} \delta'_\gamma - k^2 v_\gamma = 3 \Phi',
\end{equation}
\begin{equation}
    v'_m + \mathcal{H}v_m = - \Phi
\end{equation}
and
\begin{equation}
    v'_\gamma + \frac{1}{4} \delta_\gamma = - \Phi.
\end{equation}

In order to find solutions for the five perturbative variables $\delta_m$, $v_m$, $\delta_\gamma$, $v_\gamma$ and $\Phi$, we need to complement these equations with one of the linearised field equations.


\subsection{The growth factor}
\label{subsec:lss_evolution_growth}

Using the continuity and Euler's equations for the matter component, and Poisson's equation in the limit $k \gg \mathcal{H}$ (neglecting the contribution $\delta_\gamma$ from radiation), we get the following second order equation:
\begin{equation}
    \delta''_m + 2 \mathcal{H} \delta'_m - 4 \pi G a^2 \rho \delta_m = 0.
\end{equation}
It is remarkable that neither does this equation contain derivatives with respect to the spatial coordinates, nor do the coefficients depend on them.
Therefore, this equation has solutions of the form
\begin{equation}
    \delta_m(k, a) = D(a) \Tilde{\delta}_m(k),
    \label{eq:growth_solution}
\end{equation}
i.e. the spatial and temporal dependences factorise.
This differential equation has two linearly independent solutions.
One of them increases with time, whereas the other decreases.
If, at some early time, both functional dependences were present, the increasing solution dominates at later times, whereas the decreasing solution becomes irrelevant.
Therefore, we will only consider the increasing solution, characterised by the so-called growth factor denoted by $D_+ (a)$.

This mathematical aspect allows us to draw two important conclusions.
First, the type of solution indicates that, in linear perturbation theory, the spatial shape of the density fluctuations is frozen in comoving coordinates, only their amplitude increases.
Second, the growth factor of the amplitudes, $D_+ (a)$, follows a simple ordinary differential equation that is easily solvable for any cosmological model.
This is important because it allows us to incorporate cosmological models with dark energy.
In fact, it can be shown that for arbitrary values of the density parameter of matter and dark energy, the growth factor has the form
\begin{equation}
    D_+ (a) = \frac{5 \Omega_m H(a)}{2 H_0} \int_0^a \frac{ d\Tilde{a} }{ (\Tilde{a} H(\Tilde{a})/H_0)^3 },
    \label{eq:growth_factor}
\end{equation}
which is normalised to give $D_+ (a) = a$ during matter domination.


\subsection{The matter power spectrum}
\label{subsec:lss_evolution_pk}

According to the law of linear structure growth of Eq.~(\ref{eq:growth_solution}), the corresponding linear evolution of the matter power spectrum can also be factorised into a spatial and temporal dependence:
\begin{equation}
    P_m(k,a) = D^2_+(a) P_p(k) T^2(k).
\end{equation}
The spatial part, in turn, is made up of two components: the primordial power spectrum, $P_p(k)$, and the transfer function, $T(k)$.

Today, the most accepted theory that explains the origin of density fluctuations in the Universe is Inflation.
This theory predicts that quantum-mechanical perturbations in the very early Universe were first produced when the relevant scales were casually connected.
Then, these scales were whisked outside the horizon by inflation, a brief period of exponential expansion, only to reenter much later to serve as initial conditions for the growth of structure.
Multiple inflationary models differing on the details of how inflation starts and ends have been proposed, but almost all models where inflation is caused by a single scalar field lead to some generic predictions: (i) the amplitude of the density fluctuations follow a Gaussian probability distribution, (ii) the density fluctuations are adiabatic, (iii) inflation generates a background of gravitational waves, and (iv) the density fluctuations are characterised by a nearly scale-invariant power spectrum.
This last aspect implies that
\begin{equation}
    P_p(k) \propto k^{n_s},
\end{equation}
with small deviations from the value $n_s=1$.
This parameter is known as the spectral index, and a power spectrum of the form $P_p(k) \propto k$ is also known as a Harrison-Zeldovich-Peebles spectrum, crediting the people who first proposed it, a proposal that predates inflation by many years.

As we mentioned at the beginning of this section, different ambient cosmological conditions alter the growth rate, and hence, the shape of the power spectrum.
Initially, the wavelength of a fluctuation is larger than the horizon and its amplitude remains constant.
Once a given mode enters the horizon, it begins to grow.
This growth is logarithmic during radiation domination, and linear with the scale factor during matter domination.
The transfer function describes the evolution during these epochs, particularly, during horizon-crossing and the radiation-matter transition.
It is not easy to describe the behaviour of $T(k)$ analytically, except for two limit cases: when perturbations enter the horizon well inside the radiation-dominated era, for which $k \gg k_\mathrm{eq}$, and for those that enter well inside the matter-dominated era, for which $k \ll k_\mathrm{eq}$.
Here, $k_\mathrm{eq}$ represents the characteristic scale that crosses the horizon exactly in the matter-radiation equality ($\rho_m = \rho_\gamma$), usually referred to as the turn-over scale.
Specifically,
\begin{equation}
	T(k) \propto 
	\left\{
		\begin{array}{ll}
			1, & \mathrm{if}~  k \ll k_\mathrm{eq} \\
            k^{-2} \mathrm{ln}(k), & \mathrm{if}~ k \gg k_\mathrm{eq}.
		\end{array}
	\right.    
\end{equation}
In turn, the turn-over scale is given by
\begin{equation}
    k_\mathrm{eq} = 0.073 \Omega_m h^2 \mathrm{Mpc}^{-1}.
\end{equation}
Several more detailed approximations to the transfer function have been developed by \citeonline{bbks} and \citeonline{pk_eishu}, but these recipes are not accurate enough for the analysis of data from current galaxy surveys.
More accurate transfer functions can be numerically computed using codes such as the software \textsc{Code for Anisotropies in the Microwave Background} \cite[CAMB]{camb}.

We can see then that the evolution of the Universe through the radiation-matter transition imprints a characteristic scale $k_\mathrm{eq}$ on the power spectrum, even though the primordial power spectrum is scale invariant.
As the power spectrum is usually expressed in units of $h~\mathrm{Mpc}^{-1}$, the parameter that sets the position of the turn over scale is the shape parameter: $\Gamma := \Omega_m h$.
This means that the value of $\Omega_m$ is imprinted on the large-scale distribution of matter, hence the importance of dark matter in the formation of structures in the Universe.

In terms of the growth factor, the transfer function and the spectral index, the matter power spectrum can be written as
\begin{equation}
    P_m(k, a) = A \left[ \frac{D_+(a)}{D_+(1)} \right]^2 \left( \frac{k}{k_0} \right)^{n_s} T^2(k),
\end{equation}
where $k_0$ is known as the pivot scale, and $A$ is a parameter setting the global amplitude of the fluctuations at present epoch.


\subsection{The peculiar velocity field}
\label{subsec:lss_evolution_vel}

The evolution of the peculiar velocity field is closely related to the growth of density fluctuations.
The continuity equation for matter on subhorizon scales can be written as
\begin{equation}
    \delta'_m  - k^2 v_m = 0,
\end{equation}
where we used the fact that the gravitational potential is constant during matter domination.
In this regime, density fluctuations grow as $\delta_m \propto D_+(a)$, which then gives
\begin{equation}
    v_m = \frac{ \delta_m(k,\tau) }{ k^2 D_+(\tau) } D'_+(\tau).
\end{equation}
We define now the logarithmic growth rate factor:
\begin{equation}
    f := \frac{a}{D_+} \frac{dD_+}{da} = \frac{ d \mathrm{ln} D_+ }{ d \mathrm{ln} a }.
\end{equation}
Using this definition, and the relation $d/d\tau = a^2 H d/da$, we can express the velocity potential in terms of the scale factor and the parameter $f$:
\begin{equation}
    v_m(k, a) = \frac{1}{k^2} a H f(a) \delta_m(k, a).
\end{equation}
The velocity field is given by the gradient of the potential, so that
\begin{equation}
    v_{m_j}(k, a) = i \frac{k_j}{k^2} a H f(a) \delta_m(k, a).
    \label{eq:velocity_linear}
\end{equation}
This relation is the starting point of the study of redshift-space distortions.
The growing mode of the velocity field represents matter flowing to overdense regions and from underdense regions, increasing in this way the amplitude of perturbations.


\subsection{Non-linear evolution of density fluctuations}
\label{subsec:lss_evolution_nonlin}

The linear regime in the cosmological perturbation theory has a limited range of applicability.
For example, the evolution of structures like clusters of galaxies cannot be treated within this framework.
For instance, the fluid approximation is no longer valid if gravitationally bound systems form because multiple streams of matter occur.
One possibility is to evolve the system of equations to higher orders in the small variables $\delta$ and $v^i$, and so consider a non-linear perturbation theory.
However, the achievements of this theory do not in general justify the large mathematical effort.

There is, though, one case of non-linear evolution that can be treated analytically: the spherical evolution model.
Spherical collapse allows us to understand how dark-matter haloes form, and hence to characterise some of their main statistical properties, such as their associated mass function.
Analogously, spherical expansion helps us to understand some aspects of the evolution of voids.
We will review its main characteristics in the next chapter.

In general, studying the non-linear structure evolution requires the use of numerical methods to simulate structure formation.
The results of these simulations, when compared to observations, have substantially contributed to establishing the standard cosmological model, allowing us to make accurate quantitative comparisons between different models.
The enormous development in computing power rendered corresponding progress in simulations possible.
In addition, the continuous improvement of numerical algorithms and computer hardware has made it possible to constantly improve the spatial resolution of simulations.

Since the Universe is mainly dominated by dark matter and dark energy, it is often sufficient to compute only the behaviour of these two components, and thus to consider gravitational interactions solely.
Recently, however, computing power increased to a level where hydrodynamic processes can also be taken into account approximately, so that the baryonic component of the Universe can be traced as well.
In addition, radiative transfer can be included in such simulations, hence the influence of radiation on the heating and cooling of the baryonic component can also be examined.
One of the biggest simulations ever built to date is the Millennium simulation \cite{millennium}.
In this work, we use an extension of it.
We will review its main properties in Chapter~\ref{chp:data}.


\section{Redshift-space distortions}
\label{sec:lss_rsd}

In spectroscopic surveys, distances are inferred from the measured redshifts (Section~\ref{sec:lcdm_distance}).
In a homogeneous universe, the redshift of a galaxy would be purely given by the cosmological expansion, providing an accurate estimate of its distance.
In reality, the presence of inhomogeneities induce peculiar velocities to the galaxies, which introduce an extra component in the measured redshifts, and give rise to a difference between the real and apparent positions of them.
This leads to a change in the shape and amplitude of the measured two-point statistics.
This effect, known as redshift-space distortions (RSD), complicates the cosmological interpretation of galaxy clustering measurements.
However, as peculiar velocities are gravitationally induced by the inhomogeneities in the density field, the pattern of RSD contains information on the underlying matter distribution that can be used to probe the rate at which cosmic structures grow, offering a test to detect potential deviations from the predictions of General Relativity.


\subsection{The observed redshift}
\label{subsec:lss_rsd_redshift}

The total observed redshift of a galaxy, $z_\mathrm{obs}$, is given by the combination of two contributions:
\begin{equation}
    1 + z_\mathrm{obs} = (1 + z_\mathrm{cos})(1 + z_\mathrm{pec}),
    \label{eq:z_obs}
\end{equation}
where $z_\mathrm{cos}$ represents the cosmological component due to the expansion of the Universe, and $z_\mathrm{pec} = v_{g\parallel}/c$, the contribution due to the Doppler effect caused by $v_{g\parallel}$, the component of the peculiar velocity of the galaxy along the line-of-sight direction (hereinafter LOS).
This additional component changes the position of the galaxies along the LOS, introducing an anisotropic pattern in their spatial distribution.

\begin{figure}
    \centering
    \includegraphics[width=\textwidth]{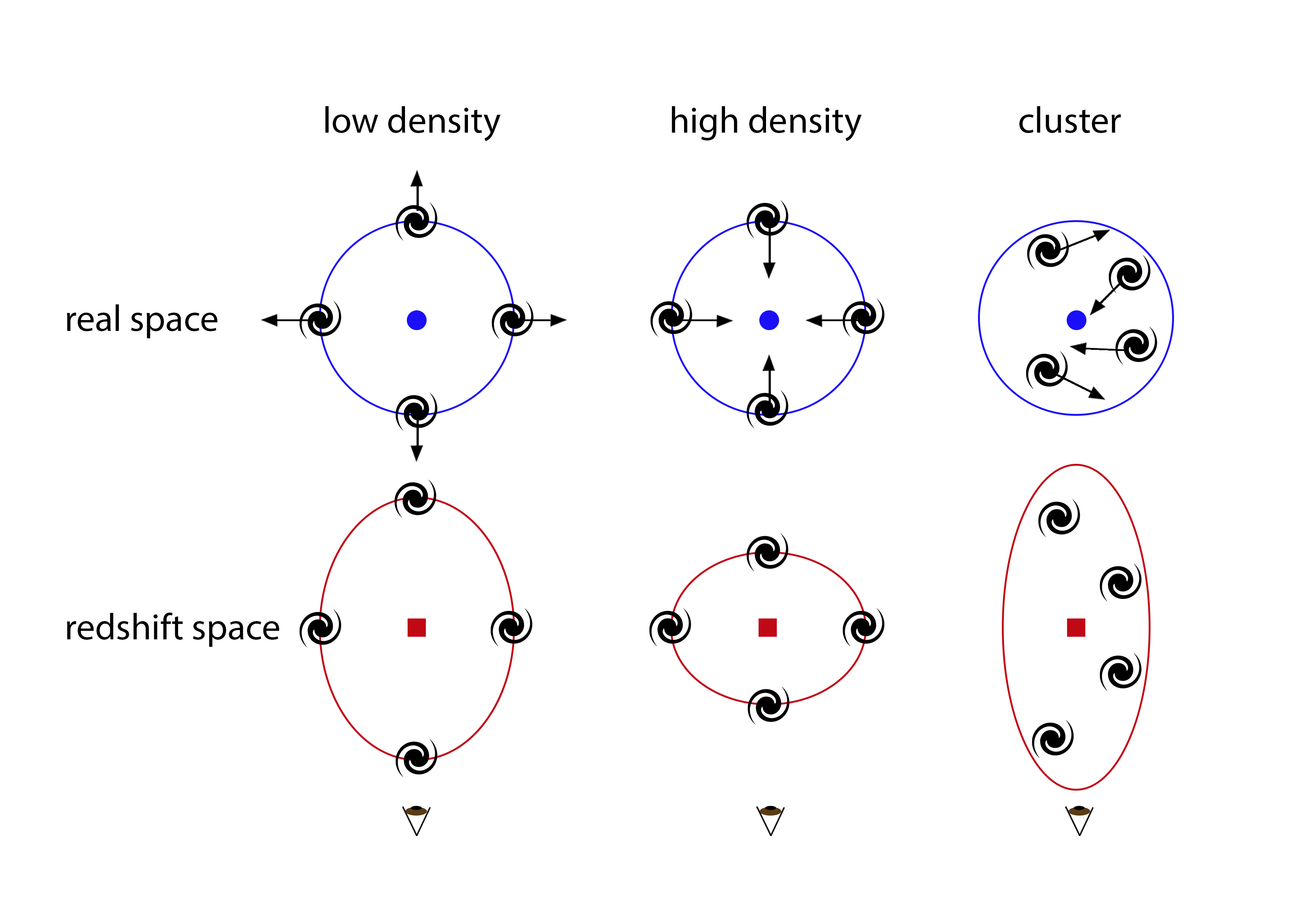}
    \caption[
    Dynamical distortions in redshift space.
    ]{
    \textit{Left-hand and central panels.}
    Schematic representation of the effect of redshift-space distortions on large and linear scales.
    As galaxies move away from underdense regions, and are directed towards overdense regions, these regions appear elongated and flattened along the line of sight when seen in redshift space, respectively.
    \textit{Right-hand panel.}
    Representation of the RSD effect in a smaller and virialised region like a cluster.
    Here, the galaxies are spread out in redshift space due to the large velocity dispersion, yielding large radial patterns in the wedge diagrams, the so-called fingers-of-God effect.
    }
    \label{fig:rsd2}
\end{figure}

Figure~\ref{fig:rsd2} shows a schematic representation of the effect of RSD on large and linear scales (left-hand and central panels).
In the upper panels, we see that the velocity field (represented by arrows) is mostly given by the bulk flow of matter towards overdense regions.
In the linear regime, the divergence of the velocity field is proportional to the density fluctuations, but with an opposite sign.
This implies that galaxies move away from underdense regions (left-hand panel) and are directed towards overdense regions (central panel).
In the lower panels, we see how these regions are actually observed.
They appear elongated and flattened along the LOS, respectively, introducing an anisotropic pattern in the clustering of galaxies.
By way of comparison, the right-hand panel shows the effect of RSD in a smaller virialised region like a cluster.
Here, the galaxies are spread out in observations due to the large velocity dispersion, yielding large radial patterns also known as the fingers-of-God effect.

In observations, the line of sight is a preferred direction, hence the correlation function and the power spectrum are no longer functions of $r$ and $k$ alone, respectively.
The full characterisation of the anisotropies caused by RSD requires us to express the correlation function in terms of the components of the observational separation vector, $\mathbf{s}$, which can be decomposed in the directions perpendicular and parallel to the LOS\footnote{We will use the notation $\sigma$ and $\pi$, instead of the more intuitive $s_\perp$ and $s_\parallel$, as they have been traditionally used in the literature, and also to be in agreement with the works of \citeonline{aprsd_correa} and \citeonline{zvoids_correa} on which part of this thesis is based.}, $\sigma$ and $\pi$.
That is, $\xi^s_g(\mathbf{s}) = \xi^s_g(\sigma,\pi)$.
The spatial configuration spanned by these coordinates is called redshift space, and we will refer to quantities in this space with the superscript $s$.
The same applies to the power spectrum, which needs to be expressed as a function of the components of the wave vector $\mathbf{k}$ in the transverse and LOS directions: $P^s_g(k_\perp, k_\parallel)$.
Alternatively, the correlation function can be expressed in polar coordinates as $\xi^s_g(s,\mu)$, where $s = |\mathbf{s}|$ is the radial coordinate, and $\mu$, the cosine of the angle between $\mathbf{s}$ and the LOS direction.

By contrast, the spatial configuration spanned by the true positions of the galaxies is called real space, free of RSD.
Sometimes, we will need to compare $\xi^s_g(\sigma,\pi)$ or $\xi^s_g(s,\mu)$ with their two-dimensional real-space counterpart, for which we will use the notation $\xi_g(\mathbf{r}) = \xi_g(r_\perp,r_\parallel) = \xi_g(r)$, where $\mathbf{r}$, $r_\perp$ and $r_\parallel$ are the real-space counterparts of the redshift-space quantities $\mathbf{s}$, $\sigma$ and $\pi$, respectively.
The last equality holds since $\xi_g$ must exhibit spherical symmetry in real space.
With this notation, $|\mathbf{r}| = r = \sqrt{r_\perp^2 + r_\parallel^2}$, $|\mathbf{s}| = s = \sqrt{\sigma^2 + \pi^2}$ and $\mu = \pi/s$.


\subsection{The impact of peculiar velocities}
\label{subsec:lss_rsd_vel}

The first step to model the effect of RSD on two-point clustering measurements is to study the mapping between the real-space position $\mathbf{r}$ and its redshift-space counterpart $\mathbf{s}$.
The following analysis will be applied to the total matter distribution.
From Eq.~(\ref{eq:z_obs}), we see that the observed redshift is given by
\begin{equation}
    z_\mathrm{obs} = z + (1+z) \frac{v_{m\parallel}}{c},
\end{equation}
where we have dropped the ``cos'' subscript of the cosmological redshift for simplicity.
When using the observed redshift to infer a distance, we must replace the true comoving distance, $\chi(z)$, by $\chi(z_\mathrm{obs})$.
Assuming that the peculiar redshifts are small, we can write
\begin{equation}
    \chi(z_\mathrm{obs}) = \chi(z) + \frac{d\chi}{dz}(z) \Delta z = \chi(z) + \frac{1+z}{H(z)} v_{m\parallel}.
    \label{eq:dcom_zobs}
\end{equation}
The observed position in redshift space corresponds to the true position with an additional term proportional to the peculiar velocity.

We are now in a position to study the impact of RSD on the correlation function.
We will work in the linear regime and assume the distant observer approximation, the assumption that the objects we are considering cover a small region of the sky and are far away.
In consequence, we can treat the LOS direction as one of the Cartesian directions.
In this context, Eq.~(\ref{eq:dcom_zobs}) implies that
\begin{equation}
    \mathbf{s} = \mathbf{r} + \frac{1+z}{H(z)} v_{m\parallel} ~ \mathbf{\hat{x}_\parallel},
    \label{eq:rsd}
\end{equation}
where $\mathbf{\hat{x}_\parallel}$ is a unit vector oriented along the LOS direction.
Equivalently, in terms of the components,
\begin{equation}
    \sigma = r_\perp
    \label{eq:sigma0}
\end{equation}
and
\begin{equation}
    \pi = r_\parallel + \frac{1+z}{H(z)} v_{m\parallel}.
    \label{eq:pi0}
\end{equation}
Note that only $\pi$ is affected by RSD.
It is worth pointing out that, expressed in this way, $\sigma$ and $\pi$ correspond to comoving distances.

The next step is to consider that the total amount of matter in a given volume element is conserved. 
In terms of the density contrast, this relation implies that
\begin{equation}
    (1 + \delta^s_m(\mathbf{s})) d^3s = (1 + \delta_m(\mathbf{r})) d^3 r,
    \label{eq:gnumconsv}
\end{equation}
where we have assumed that RSD do not change the mean density of the sample: $\bar{\rho}^s_m(\mathbf{s})=\bar{\rho}_m(\mathbf{r})$.
The volume elements in real and redshift space are related by the Jacobian of the transformation, given by Eq.~(\ref{eq:rsd}): $J = |\partial r_i / \partial s_j|$, in such a way that
\begin{equation}
    d^3 r = J d^3 s.
\end{equation}
Explicitly, this Jacobian is
\begin{equation}
    J = \left( 1 + \frac{1}{\mathcal{H}(z)} \frac{\partial v_{m\parallel}}{\partial r_\parallel} \right)^{-1}
    \approx \left( 1 - \frac{1}{\mathcal{H}(z)} \frac{\partial v_{m\parallel}}{\partial r_\parallel} \right).
    \label{eq:jacobian}
\end{equation}
Inserting this result into Eq.~(\ref{eq:gnumconsv}) leads to
\begin{equation}
    1 + \delta^s_m(\mathbf{s}) = (1 + \delta_m(\mathbf{r})) \left( 1 - \frac{1}{\mathcal{H}(z)} \frac{\partial v_{m\parallel}}{\partial r_\parallel} \right).
\end{equation}
Distributing and keeping only linear terms in the perturbations, we find that
\begin{equation}
    \delta^s_m(\mathbf{s}) = \delta_m(\mathbf{r}) - \frac{1}{\mathcal{H}(z)} \frac{\partial v_{m\parallel}}{\partial r_\parallel}.
\end{equation}
Fourier-transforming this relation, and using Eq.~(\ref{eq:velocity_linear}), we finally find that
\begin{equation}
    \delta^s_m(\mathbf{k}) = \delta_m(\mathbf{k}) (1 + f(z) \mu_k^2),
    \label{eq:kaiser_delta}
\end{equation}
where $\mu_k := k_\parallel/k$ represents the cosine of the angle between $\mathbf{k}$ and the LOS direction.
With this relation, the anisotropic power spectrum is given by
\begin{equation}
    P_m^s(k,\mu_k) = (1 + f(z) \mu_k^2)^2 P_m(k).
    \label{eq:kaiser_pk}
\end{equation}
This result was first derived by \citeonline{rsd_kaiser}.

Eq.~(\ref{eq:kaiser_delta}) shows that, in the linear regime, the density fluctuations observed in redshift space correspond to those in real space with a correction factor.
Since $f(z) \mu_k^2 \geq 0$, this additional factor leads to an enhancement of the density fluctuations.
This enhancement depends on $\mu_k$: it is maximum when $\mu_k = 1$, which corresponds to the LOS direction, and it is zero when $\mu_k = 0$, which corresponds to the transverse directions, which define what is called the plane of the sky (hereinafter POS).


\subsection{The RSD parameter}
\label{subsec:lss_rsd_beta}

Eqs.~(\ref{eq:kaiser_delta}) and (\ref{eq:kaiser_pk}) imply that, by studying the anisotropic patterns caused by RSD, it is possible to measure the logarithmic growth rate, $f(z)$.
Within the context of General Relativity, the growth rate can be accurately described by the following empirical relation:
\begin{equation}
    f(z) = \Omega_m^\gamma(z),
    \label{eq:fz}
\end{equation}
where $\Omega_m(z) := \rho_m(z)/\rho_c(z)$ is the generalisation of the parameter $\Omega_m$ for any redshift $z$, and $\gamma = 0.55$ (with a weak dependence on other parameters such as the dark-energy equation of state).
This means that a measurement of $f(z)$ provides additional information on the matter density parameter.
Perhaps more interesting is the fact that such a measurement also provides us with a test of General Relativity.
The exponent $\gamma$ can be treated as a free parameter to be constrained. 
A measurement of $\gamma \ne 0.55$ would be a signature of a deviation from the predictions of General Relativity.

Nevertheless, there is a drawback here.
Eqs.~(\ref{eq:kaiser_delta}) and (\ref{eq:kaiser_pk}) are only valid for the total matter distribution.
However, what we observe are galaxies.
Therefore, we need to take into account the bias relation studied in Section~\ref{sec:lss_bias}.
For simplicity, we will assume the linear bias relation of Eq.~(\ref{eq:bias_linear}).
Moreover, we will consider that the velocity field of the galaxies corresponds exactly to that of the total matter distribution: $\mathbf{v}_g = \mathbf{v}_m$, i.e. there is no velocity bias.
For this reason, we will hereinafter drop the subindices in the velocity vector.
In this case, Eq.~(\ref{eq:kaiser_delta}) implies that
\begin{equation}
    \delta_g^s(\mathbf{k}) = \delta_g(\mathbf{k}) (1 + \beta(z) \mu_k^2),
    \label{eq:kaiser_delta2} 
\end{equation}
where we have defined the quantity
\begin{equation}
    \beta(z) := \frac{f(z)}{b},
    \label{eq:beta}
\end{equation}
known as the RSD parameter.
The redshift-space galaxy power spectrum will be then given by
\begin{equation}
    P_g^s(k,\mu_k) = (1 + \beta(z) \mu_k^2)^2 b^2 P_m(k).
\end{equation}
This means that, by measuring the anisotropic patterns in the two-point statistics caused by RSD, it is possible to constrain the parameter $\beta$, a combination of $f$ and $b$.
This will be a fundamental cosmological parameter in our study of voids.


\section{The fiducial cosmology}
\label{sec:lss_ap}

An important fact is that a measurement of the two-point statistics requires the assumption of a fiducial cosmology to transform the observed angles and redshifts of galaxies provided by a spectroscopic survey into distances expressed in physical units, like the Megaparsec.
This is evident in Eqs.~(\ref{eq:dcom_z}), (\ref{eq:dist_ang}) and (\ref{eq:dist_lum}), for instance.
These expressions need to assume certain values for the cosmological parameters $H_0$, $\Omega_m$ and $\Omega_\Lambda$.
In this selection, any deviation between the chosen parameters and the true underlying ones will skew the distance estimations, inducing in this way, additional distortion patterns in the spatial distribution of the galaxies.
This has an impact on the clustering measurements.
This phenomenon is a manifestation of the AP effect studied in Section~\ref{subsec:lcdm_tests_ap}.
AP distortions due to the fiducial cosmology must be taken into account in order to constrain unbiased cosmological parameters.

Given two galaxies, their spatial separation can be estimated by two direct measurable quantities: an angle $\Delta \phi$ measured across the POS, and a redshift difference $\Delta z$, an indicator of their separation along the LOS.
Then, according to Eqs.~(\ref{eq:dist_ang}) and (\ref{eq:ap0}), it is possible to estimate their separation vector $\mathbf{s}$ by means of its components:
\begin{equation}
    \sigma = D_\mathrm{M}^\mathrm{fid}(z) \Delta \phi
    \label{eq:sigma_ap}
\end{equation}
and
\begin{equation}
    \pi = \frac{c}{H_\mathrm{fid}(z)} \Delta z,
    \label{eq:pi_ap}
\end{equation}
where $D_M(z) := (1+z)D_A(z)$ is the comoving angular-diameter distance\footnote{Note that $D_M(z) = \chi(z)$ for a flat universe.}.
Here, the legend ``fid'' refers to quantities calculated with the fiducial cosmology.
Analogue equations follow using the (unknown) true cosmology for the true separations $r_\perp$ and $r_\parallel$ (using the legend ``true'').
Therefore, a simple comparison leads to
\begin{equation}
    \sigma = q_\mathrm{AP}^\perp r_\perp
    \label{eq:s_perp}
\end{equation}
and
\begin{equation}
    \pi = q_\mathrm{AP}^\parallel r_\parallel,
    \label{eq:s_parallel}
\end{equation}
where the AP factors $q_\mathrm{AP}^\perp$ and $q_\mathrm{AP}^\parallel$ are defined by the following relations:
\begin{equation}
    q_\mathrm{AP}^\perp := \frac{D_\mathrm{M}^\mathrm{fid}(z)}{D_\mathrm{M}^\mathrm{true}(z)}
    \label{eq:qap_perp}
\end{equation}
and
\begin{equation}
    q_\mathrm{AP}^\parallel := \frac{H_\mathrm{true}(z)}{H_\mathrm{fid}(z)}.
    \label{eq:qap_parallel}
\end{equation}

\begin{figure}
    \centering
    \includegraphics[width=\textwidth]{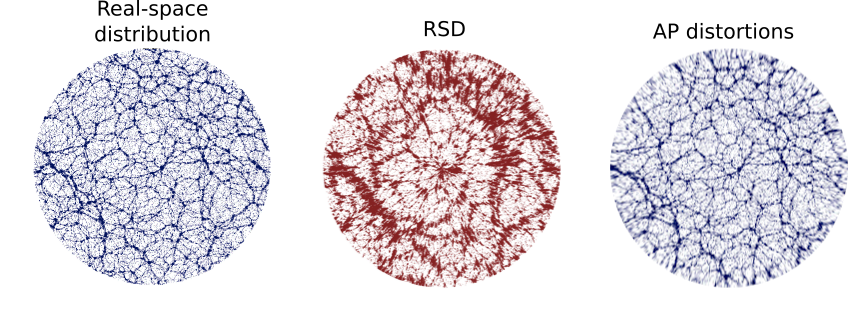}
    \caption[
    Visual and comparative impression of the RSD and AP effects.
    ]{
    Visual impression of the types of distortions in the spatial distribution of galaxies.
    \textit{Left-hand panel.}
    Real-space distribution of the matter in a simulation.
    A hypothetical observer is located at the centre.
    \textit{Central panel.}
    Redshift-space distortions induced by the peculiar velocity of the galaxies, a dynamical effect.
    \textit{Right-hand panel.}
    Alcock-Paczyński distortions induced by the selection of a fiducial cosmology, a geometrical effect.
    Both effects are coupled in observations.
    \textit{Figure credit:} B. Thomas \url{https://www.washburn.edu/faculty/bthomas/work.html}.
    }
    \label{fig:rsd_ap}
\end{figure}

Figure~\ref{fig:rsd_ap} shows a visual impression of the two types of spatial distortions studied in this chapter.
The left-hand panel shows the real-space distribution of the matter in a simulation, where a hypothetical observer is located at the centre.
The central panel shows how RSD induce radial patterns in the distribution due to the peculiar velocity of galaxies.
It is particularly evident the fingers-of-God effect.
The right-hand panel, on the other hand, shows the AP distortions induced by the selection of a fiducial cosmology.
In reality, both effects are coupled in observations.

\chapter{Cosmic voids}
\label{chp:voids}
Cosmic voids are vast underdense regions of the Universe.
Since their discovery \cite{voids_gregory_thompson,voids_kirshner,voids_lapparent}, they have been recognised as powerful cosmological laboratories \cite{cosmovoids1, cosmovoids2}.
Most of the cosmic volume is occupied by voids \cite{clues1}, making them the largest observable structures.
Hence, they encode key information about the expansion history and geometry of the Universe.

Large-scale underdensities arise naturally as the result of structure growth.
As the Universe evolves, galaxies progress towards matter concentrations by the action of gravity, forming diverse structures like groups, clusters and filaments.
In this process, galaxies also dissipate from underdense regions, which become even emptier, forming in this way cosmic voids.
This interplay in the formation of voids and structures allows us to think of them as complementary.
Moreover, cosmic voids can be related to the small density fluctuations seen in the early Universe, manifested in the small anisotropies in the CMB \cite{cmbvoids_cai2014, cmbvoids_cai2017}.

The statistical properties of voids depend on two conditioning factors: (i) the matter tracers used to map the large-scale structure, namely, galaxies, dark-matter haloes, dark-matter particles, etc., and (ii) the method used to identify them from the spatial distribution of these tracers.
There are different classes of void finders.
For a comparison of different techniques see \citeonline{voidID_colberg}.
In this work, we use the so-called spherical void finder, originally developed by \citeonline{voids_padilla}, which is based on the integrated density contrast of underdense regions assuming spherical symmetry.
We leave for Chapter~\ref{chp:data} the details of the identification process using this method.
As examples of different techniques, we can mention the \textsc{ZOBOV}\footnote{Acronym for \textsc{ZOnes Bordering On Voidness}.} void finder \cite{zobov}, based on the watershed algorithm, and the method of \citeonline{voidID_elyiv}, based on the dynamical properties of tracers.

Despite the intrinsic differences between all the methods, there is a consensus on the basic statistical properties of voids.
For instance, and roughly speaking, voids are underdense regions with densities as low as $10-20\%$ of the average of the Universe, with several tens of Megaparsecs in diameter.
It is also important to stress the fact that galaxies and haloes trace the void distribution in a similar fashion.
\citeonline{voids_padilla} found that they have comparable statistical and dynamical properties, such as their abundance, correlation function and velocity field.


\section{Void evolution}
\label{sec:voids_evolution}

Initial overdense regions expand slightly less rapidly than the homogeneous background, reach a maximum size, and then turn around to finally collapse (this is strictly true only in an EdS or closed universe).
In contrast, initial underdense regions do not turn around, they undergo simple expansion (more rapidly than the background).
This expansion is simple and linear until matter from the interior overtakes the initially outer shells.
The generic characteristics of these evolutionary paths are best appreciated in terms of the evolution of isolated spherically symmetric density perturbations, either overdense or underdense, in an otherwise homogeneous and expanding background universe.
This spherical model provides a key reference for understanding and interpreting more complex situations.
As a result of the spherical symmetry, the problem is essentially one-dimensional, allowing a full analytical treatment.

\citeonline{svdw} explained that the most basic and universal properties of evolving spherical voids are the following.

\begin{enumerate}

\item 
\textit{Expansion.}
Voids expand, in contrast to overdense regions, which collapse.

\item
\textit{Evacuation.}
As they expand, the density within them decreases continuously.
To first order, the density decrease is a consequence of the redistribution of mass over the expanding volume.

\item
\textit{Spherical shape.}
Outward expansion makes voids evolve towards a spherical geometry.

\item
\textit{Reverse top-hat density profile.}
The effective repulsion of the matter interior to the void decreases with the distance from the centre, so the matter distribution evolves into a reverse top-hat.

\item
\textit{Super-Hubble velocity field.}
Consistent with its homogeneous interior density distribution, the peculiar velocity field in voids has a constant Hubble-like interior velocity divergence.
Thus, voids evolve into genuine super-Hubble bubbles.

\item
\textit{Suppressed structure growth.}
Density inhomogeneities in the interior are suppressed and, as the void begins to resemble an underdense universe, structure formation within it gets frozen-in.

\item
\textit{Boundary ridge.}
As matter from the interior accumulates near the boundary, a ridge develops around the void.

\item
\textit{Shell crossing.}
The transition from a quasi-linear towards a mature non-linear stage occurs as inner shells pass across outer shells.
When this process occurs, a void, as a distinctive entity, is considered to have formed.

\end{enumerate}
A detailed study about the spherical evolution of voids in different cosmological contexts can be found in \citeonline{sph_ev_voids}.

In reality, voids are not isolated structures, but rather part of a complex network.
\citeonline{svdw} showed that the formation and evolution of voids is strongly affected by their surrounding large-scale environment, resulting in a more complicated structure and dynamics than the simplistic isolated expansion.
Essentially, the hierarchy of voids arises by the assembly of matter in the growing nearby structures.
There are two scenarios for void evolution.
In the first case, voids embedded in an environment similar to the global background density expand and remain as underdense regions following the so-called void-in-void mode.
This is the case when small voids merge at an early epoch with another void to form a larger void at a later epoch.
It is analogous to the cloud-in-cloud problem associated with the formation of dark-matter haloes and clusters of galaxies.
In the second case, voids surrounded by an overdense shell shrink at later times due to the gravitational collapse of the dense structure surrounding them.
This is the void-in-cloud mode, and this is generally the case of small voids.
Note however that virialised haloes within voids are not likely to be torn apart as the void expands around them.
Thus, the cloud-in-void phenomenon is irrelevant for halo formation.
Therefore, the asymmetry between the void-in-cloud and cloud-in-void processes induce a symmetry breaking between the emerging halo and void populations: although they evolve out of the same symmetric initial conditions, the two populations are expected to have different statistical properties.


\section{The abundance of voids}
\label{sec:voids_abundance}

The excursion set theory provides a useful framework for describing the formation history of dark-matter haloes in scenarios of hierarchical structure formation \cite{press_schechter, zentner}.
It provides analytical approximations for the distribution of halo masses (known as mass function), merger rates and formation times that are quite accurate.
The key ingredient is the assumption that the final virialised state of haloes form from a smooth spherical collapse.
The analogous spherical expansion can likewise be used to make excursion set predictions for voids \cite{svdw}.
Particularly, we are interested in the void size function (hereinafter VSF), a description of the abundance of voids by means of the distribution of their sizes.
This is analogous to the mass function in the halo case.
In this work, voids will be defined as underdense spheres with a well defined radius $R_\mathrm{v}$ (see Chapter~\ref{chp:data} for more details), so that the VSF is essentially a radius distribution.

One of the main features of the spherical model is that the evolution does not depend on the initial size $R$ or enclosed mass $M$ of the region, but only on the amplitude of the initial top-hat density fluctuation.
Therefore, $R$ and $M$ can be treated indistinctly by means of their relation with the volume of the region.
In the case of collapse, it occurs when the linear density fluctuation reaches a critical barrier $\Delta_c$.
The excursion set formalism predicts then the fraction of trajectories $df$ that cross this barrier for the first time within some scale interval $d \mathrm{ln} \sigma$, accounting in this way for the cloud-in-cloud process:
\begin{equation}
    f_{\mathrm{ln} \sigma}(\sigma) := \frac{df}{d \mathrm{ln} \sigma} = \sqrt{ \frac{2}{\pi} } \frac{\Delta_c}{\sigma}
    \mathrm{exp} \left[ - \frac{\Delta_c^2}{2\sigma^2} \right].
\end{equation}
A trajectory is a sequence of overdensities given by subsequent increases in a smoothing scale $\mathcal{R}$.
It is common to relate this smoothing scale to the corresponding variance of the linear density field:
\begin{equation}
    \sigma^2(\mathcal{R}) = \int \frac{k^2}{2 \pi^2} P_m(k) |W(k, \mathcal{R})|^2 ~ dk,
\end{equation}
where $W(k, \mathcal{R})$ is a filter function.
Note the dependence of $\sigma^2(\mathcal{R})$ on the matter power spectrum, $P_m(k)$.
The strong cosmological dependence of calculating abundances is thus evident.
Since both mass and number of particles are conserved during collapse, the linear mapping $\sigma(M)$ carries over to the non-linear regime, so that the mass function expressed by means of the comoving differential number density of haloes is
\begin{equation}
    \frac{dn}{d \mathrm{ln} M} = \frac{\rho_m}{M} f_{\mathrm{ln} \sigma}(\sigma) \frac{d \mathrm{ln} \sigma^{-1}}{d \mathrm{ln} M}.
\end{equation}

We can extend this model to underdense regions in the initial density field.
These are naturally associated with voids in the evolved density field today.
In this case, the critical density threshold is defined to be when the expanding shells cross.
For an EdS universe, this occurs when the non-linear average density contrast within the void reaches the value $-0.8$, which corresponds to a linearly extrapolated value of $\Delta_\mathrm{v} = -2.71$.
\citeonline{abundance_jennings} showed that this value is still valid for alternative cosmologies, such as the flat-$\Lambda$CDM model.
Therefore, similarly to the cloud-in-cloud process, we can then follow the excursion set formalism for determining the fraction of trajectories which pierce the underdense barrier $\Delta_\mathrm{v}$, thus taking into account the void-in-void mode.

Nevertheless, this approach will not give an accurate prediction, since we cannot assume spherical evolution in isolation.
This makes sense for collapsing objects, since the comoving volume occupied shrinks.
In contrast, voids evolve in a more complex way according to both the void-in-void and void-in-cloud evolution modes.
The underdense barrier $\Delta_\mathrm{v}$ only accounts for the former.
In order to consider the void-in-cloud mode as well, \citeonline{svdw} propose that there must be a second overdense barrier $\Delta_c$ as in the halo case.
In calculating the first crossing distribution, we need to determine the largest scale at which a trajectory crosses the barrier $\Delta_\mathrm{v}$ given that it has not crossed $\Delta_c$ on any larger scale.
They posit that this threshold should lie somewhere in between $\Delta_c = 1.06$, the value at turnaround in the spherical collapse model, and $\Delta_c = 1.686$, the value at the point of collapse.
By the same reasoning as applied to haloes, the abundance of voids in the linear regime is given by
\begin{equation}
    \frac{d n_\mathrm{v}}{d \mathrm{ln} R_\mathrm{v}} = \frac{f_{\mathrm{ln} \sigma}(\sigma)}{V(R_\mathrm{v})} 
    \frac{d \mathrm{ln} \sigma^{-1}}{d \mathrm{ln} R_\mathrm{v}},
    \label{eq:vsf_linear}
\end{equation}
where $V(R_\mathrm{v}) = 4/3 \pi R_\mathrm{v}^3$ is the volume of a spherical void of radius $R_\mathrm{v}$, and
\begin{equation}
    f_{\mathrm{ln} \sigma}(\sigma) = 2 \sum_{j=1}^\infty
    \mathrm{exp} \left[ -\frac{(j \pi x)^2}{2} \right] j \pi x^2 \mathrm{sin}(j \pi \mathcal{D}),
\end{equation}
with $\mathcal{D} := |\Delta_v|/(\Delta_c + |\Delta_v|)$ and $x := \mathcal{D}\sigma/|\Delta_v|$.
Eq.~(\ref{eq:vsf_linear}) is known as the linear model.
Actually, the spherical evolution model predicts that a void expands from its linear radius at the epoch of shell crossing by a factor approximately equal to $1.7$.
Hence, the radius in Eq.~(\ref{eq:vsf_linear}) must be replaced with $1.7 R_\mathrm{v}$.
With this replacement, we get the Sheth $\&$ van de Weygaert (SvdW) model.

\begin{figure}
    \centering
    \includegraphics[width=\textwidth/2]{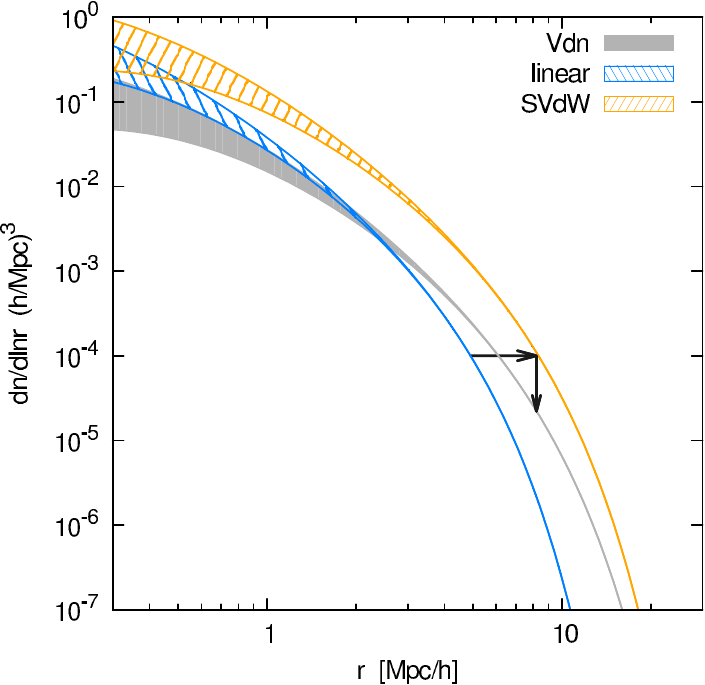}
    \caption[
    Models for the abundance of voids: linear, SvdW and Vdn.
    ]{
    Void  abundance  model  predictions: linear (blue curve), SvdW (orange) and Vdn (grey). 
    In the SvdW model, the comoving number density remains unchanged and only the size changes (arrow to orange curve).
    In the Vdn model, the number density also changes so as to conserve the comoving volume fraction in voids, lowering the amplitude at fixed shape (arrow to grey curve). 
    The shaded regions correspond to a variation in the overdense barrier according to $1.06 \leq \Delta_c \leq 1.686$, which is significant only for small voids.
    \textit{Figure credit:} \citeonline{abundance_jennings}.
    }
    \label{fig:abundance_models}
\end{figure}

The key assumption of the SvdW model is that the comoving number density of voids is conserved during the evolution, so that only the size changes with respect to the linear prediction.
Unfortunately, this assumption is not valid for large voids.
For instance, the cumulative comoving volume fraction in voids exceeds unity.
\citeonline{abundance_jennings} propose a simple fix to this problem.
They require that the comoving volume fraction is fixed during the evolution, instead.
In this picture, when a void expands, it combines with its neighbours to conserve volume and not number.
In this way, the abundance of voids becomes
\begin{equation}
    \frac{d n_\mathrm{v}}{d \mathrm{ln} R_\mathrm{v}} = \frac{f_{\mathrm{ln} \sigma}(\sigma)}{V(R_\mathrm{v})}
    \frac{d \mathrm{ln} \sigma^{-1}}{d \mathrm{ln} R_\mathrm{v}^L}
    \frac{d \mathrm{ln} R_\mathrm{v}^L}{d \mathrm{ln} R_\mathrm{v}},
    \label{eq:vsf_vdn}
\end{equation}
where the superscript $L$ refers to the linear radius.
Eq.~(\ref{eq:vsf_vdn}) is known as the volume-conserving (Vdn) model.
Figure~\ref{fig:abundance_models} shows the general characteristics of the three models described here: linear, SvdW and Vdn.

\section{The density field in voids}
\label{sec:voids_density}

Section~\ref{sec:voids_evolution} provides a useful framework to characterise the density and velocity fields around voids.
For instance, property (3) implies that the density fluctuation field around voids, Eq.~(\ref{eq:delta}), can be characterised with a radial profile $\delta_{\mathrm{v}g}(r)$, where $r$ denotes distance to the void centre.
Actually, this is not a good description for voids taken individually, but it works in a statistical sense when a stacked sample of similar voids is considered, since then individual irregularities are erased.
We will get rid of the subscript ``vg'' for simplicity.

\begin{figure}
    \centering
    \includegraphics[width=0.75\textwidth]{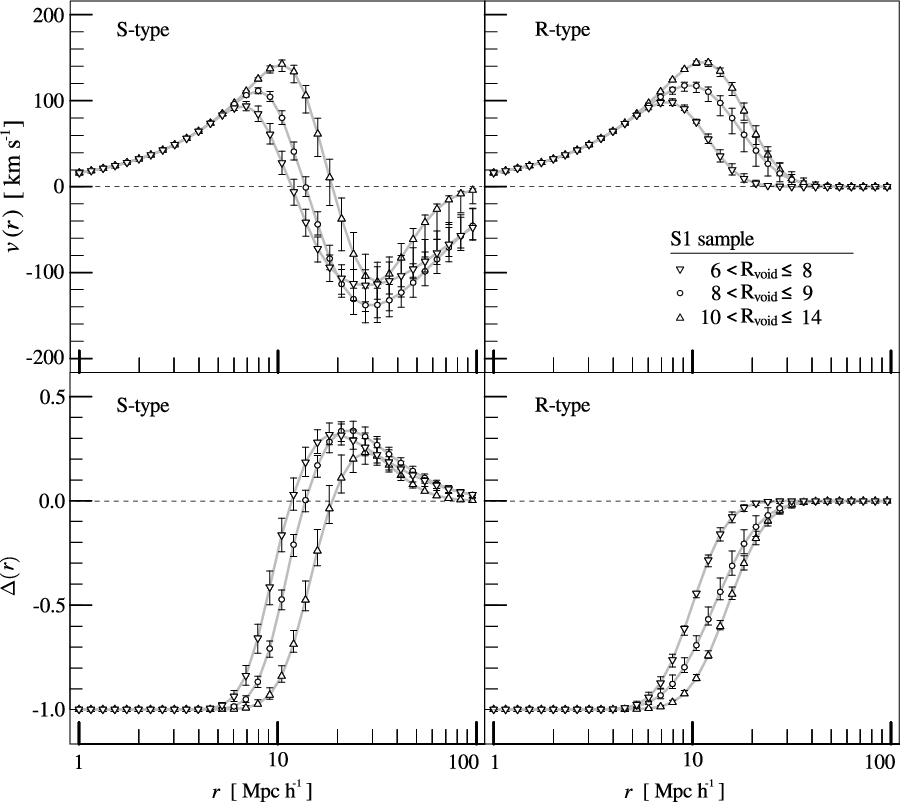}
    \caption[
    Velocity and integrated density contrast profiles for samples of voids selected from the SDSS survey.
    ]{
    Integrated density contrast profiles (\textit{lower panels}) and velocity profiles (\textit{upper panels}) of galaxies in the SDSS for different S-type (\textit{left-hand panels}) and R-type (\textit{right-hand panels}) samples of voids.
    Different void radius ranges are indicated: $6-8~\hmpc$ (downward triangles), $8-9~\hmpc$ (circles) and $10-14~\hmpc$ (upward triangles).
    Error bars indicate the region enclosing all curves within $68.3\%$ uncertainty.
    \textit{Figure credit:} \citeonline{clues2}.
    }
    \label{fig:densvel_voids}
\end{figure}

\citeonline{clues1} showed that the integrated density contrast is a good descriptor of the void environment, since it is tightly connected to the evolution modes.
This profile is obtained by integrating $\delta(r)$ over a volume $V(r)$ given by successive void-centric spheres of radius $r$:
\begin{equation}
    \Delta(r) := \frac{1}{V(r)} \int_V \delta(r') dV = \frac{3}{r^3} \int_0^r \delta(r') r'^2 dr'.
    \label{eq:delta_int}
\end{equation}
These authors examined the distribution of galaxies around voids in the SDSS and obtained two characteristic types of voids according to their integrated density profiles: (i) R-type voids, characterised by an increasing profile that tends to zero (the mean value of the Universe) at large distances, and (ii) S-type voids, characterised by a profile with a noticeable and positive peak near the typical radius of the sample that decreases later tending to zero at large distances.
Moreover, small voids are more frequently of S-type, whereas large voids are more likely classified as R-type.
The key aspect is that R-type voids match with the void-in-void mode, whereas S-type voids, with the void-in-cloud mode.
The lower panels of Figure~\ref{fig:densvel_voids} show the measured integrated density profiles for different samples of voids identified in the SDSS survey with different radii ranges taken from \citeonline{clues2}.
The main characteristics described here for the two types of voids are clearly distinguished.

Property (4) concerning the reverse top-hat density profile gives a hint about how to model $\delta(r)$.
The top-hat profile is a good idealisation to understand the process of void evolution qualitatively, but it is inaccurate to use in practice.
At the moment, there is not a successful model derived from first principles, but it is a common practice to use empirical and parametric approaches.
For instance, \citeonline{clues2} provides a parametric model for the two types of voids.
Actually, they model $\Delta(r)$.
For R-type voids, the provide a functional form based on the error function:
\begin{equation}
    \Delta_\mathrm{R}(r) = \frac{1}{2} [ \mathrm{erf}( S \mathrm{log}_{10}(r/R) ) - 1 ].
    \label{eq:Rtype}
\end{equation}
This model has two parameters: a scale $R$ characterising the void radius, and a steepness coefficient $S$.
S-type voids, on the other hand, are a bit more complex and require an extra term in order to account for the overdense shell surrounding them:
\begin{equation}
    \Delta_\mathrm{S}(r) = \frac{1}{2} [ \mathrm{erf}( S \mathrm{log}_{10}(r/R) ) - 1]
    + P \mathrm{exp} \left[ -\frac{ \mathrm{log}_{10}^2(r/R) }{ 2 \Theta^2(r) } \right],
    \label{eq:Stype}
\end{equation}
where in turn,
\begin{equation}
	\Theta(r) :=
	\left\{
		\begin{array}{ll}
			1/\sqrt{2S} ~ \mathrm{if} ~ r < R\\
            1/\sqrt{2W} ~ \mathrm{if} ~ r > R.
		\end{array}
	\right.
\end{equation}
The peak on density due to this shell is modelled with two semi-Gaussians.
This model has two extra parameters: an amplitude $P$, and a semi-width $W$ (the other semi-width is given by $S$).

It is worth keeping in mind that void properties depend on the void identification method.
For instance, the works of \citeonline{clues1} and \citeonline{clues2} cited before used a spherical void finder, from which the models given by Eqs.~(\ref{eq:Rtype}) and (\ref{eq:Stype}) were tested.
To mention another example, \citeonline{density_hamaus} provide a universal parametric model for $\delta(r)$ (without distinguishing between types of voids) based on a modified version of the ZOBOV void finder that works under the watershed algorithm:
\begin{equation}
    \delta(r) = \delta_{ct} \frac{ 1 - (r/r_s)^\alpha  }{ 1 + (r/r_v)^\beta }.
\end{equation}
This model has five parameters: the central density contrast $\delta_{ct}$, the effective void radius $r_v$, a scale radius $r_s$ where the void density equals the mean density, and two exponents, $\alpha$ and $\beta$, that determine the inner and outer slope of the compensation wall, respectively.


\section{The velocity field in voids}
\label{sec:voids_velocity}

Properties (1) to (3) of Section~\ref{sec:voids_evolution} imply that the peculiar velocity field around voids can also be characterised with a one-dimensional radial profile $v(r)$.
As before, this actually applies to a stacked sample of similar voids.
The upper panels of Figure~\ref{fig:densvel_voids} show the corresponding velocity profiles for the SDSS' voids of \citeonline{clues2}.
Note that R-type voids exhibit expansion velocities only, since $v(r)$ is positive at all scales $r$.
In the inner parts, the velocity increases and reaches a peak near the typical radius of the sample, whereas in the outer parts, it decreases, tending towards zero at large distances, the mean global value of the Universe.
S-type voids, on the other hand, exhibit expansion velocities only in the inner parts and near the void walls.
The velocity increases at first, reaches a peak near the typical radius of the sample, and then decreases in the outer parts becoming negative, a signal of the gradual shrinking of the surrounding overdense shell that characterises this type of voids.
Finally, the velocity increases again, tending to zero at large distances.

In contrast to the density field, it is possible to derive an analytical expression for $v(r)$ following linear theory in the evolution of density perturbations.
The starting point is Eq.~(\ref{eq:velocity_linear}), from which we obtain
\begin{equation}
    v(r) = - \frac{1}{3} \frac{H(z)}{(1+z)} \beta(z) r \Delta(r).
    \label{eq:velocity0}
\end{equation}
Note that the RSD parameter, $\beta=f/b$, appears here because we assumed the linear bias relation of Eq.~(\ref{eq:bias_linear}) between the matter and galaxy void-centric density fields: $\delta_{\mathrm{v}m}(r) = \delta_{\mathrm{v}g}(r)/b$.
Eq.~(\ref{eq:velocity0}) is the basis for studying redshift-space distortions around voids, and hence constitutes one of the most important results on which most of our work is based.


\section{The void-galaxy cross-correlation function}
\label{sec:voids_correlation}

In Section~\ref{subsec:lss_correlation_estimators}, we defined the correlation function for a discrete set of points, which is related to the probability of finding pairs of points as a function of their separation.
We can extend that discussion to consider a cross-correlation between two discrete sets of points.
Particularly, we are interested in cross-correlating void centres and galaxies.
In this case, Eq.~(\ref{eq:dP_lss}) becomes
\begin{equation}
    dP_{\mathrm{v}g} = \bar{\rho}_\mathrm{v} \bar{\rho}_g (1 + \xi_{\mathrm{v}g}(r)) dV_\mathrm{v} dV_g,
\end{equation}
where $\xi_{\mathrm{v}g}(r)$, known as the void-galaxy cross-correlation function, gives the excess probability of finding void-galaxy pairs separated by a distance $r$ with respect to a homogeneous distribution of such pairs.
As before, we will get rid of the subscript ``vg'' for simplicity.

The estimation of the void-galaxy cross-correlation function can still be accounted for with the natural estimator of Eq.~(\ref{eq:estimator_natural}), but their components now have a different meaning.
On the one hand, $DD(r)$ is the probability of finding void-galaxy pairs separated by distances between $r$ and $r + dr$, also given by Eq.~(\ref{eq:data_data}), but this time $N_\mathrm{pairs}$ is the number of void-galaxy pairs inside the volume element defined by $r$ and $r + dr$, and $N_\mathrm{tot}$ is the total number of pairs.
In this case, $N_\mathrm{tot} = N_\mathrm{v} N_g$, where $N_\mathrm{v}$ is the total number of voids.
On the other hand, the probability of finding pairs in a homogeneous distribution of galaxies and voids, $RR(r)$, can be computed in an analogous way by means of a random catalogue.
The alternative estimators given by Eqs.~(\ref{eq:estimator_dp}) and (\ref{eq:estimator_ls}) can also be applied for the case of a cross correlation (see for instance \citeonline{clues2}).

An important relation must be noted.
In the case of a stacked sample of voids, the void-galaxy cross-correlation function is equivalent to the density contrast profile:
\begin{equation}
    \xi(r) = \delta(r),
    \label{eq:xi_delta}
\end{equation}
hence we will refer to them indistinctly.


\section{Redshift-space distortions around voids}
\label{sec:voids_rsd}

The fundamental assumption of spherical symmetry breaks in observations due to the presence of redshift-space distortions, which modify the spatial distribution of the galaxies around voids.
In Section~\ref{sec:lss_rsd}, we showed that the line of sight is a preferred direction in such a way that the spherical symmetry reduces to a cylindrical symmetry along this direction.
Therefore, it is instructive to visualise the stacked density and velocity fields of a void sample in a void-centric reference system whose coordinates are determined by the comoving distances $\sigma$ and $\pi$, which are perpendicular and parallel to the LOS direction, respectively.
In this system, it is obtained the redshift-space void-galaxy cross-correlation function, $\xi^s(\sigma,\pi)$, a function that can be represented as a two-dimensional map.

\begin{figure}
    \centering
    \includegraphics[width=0.8\textwidth]{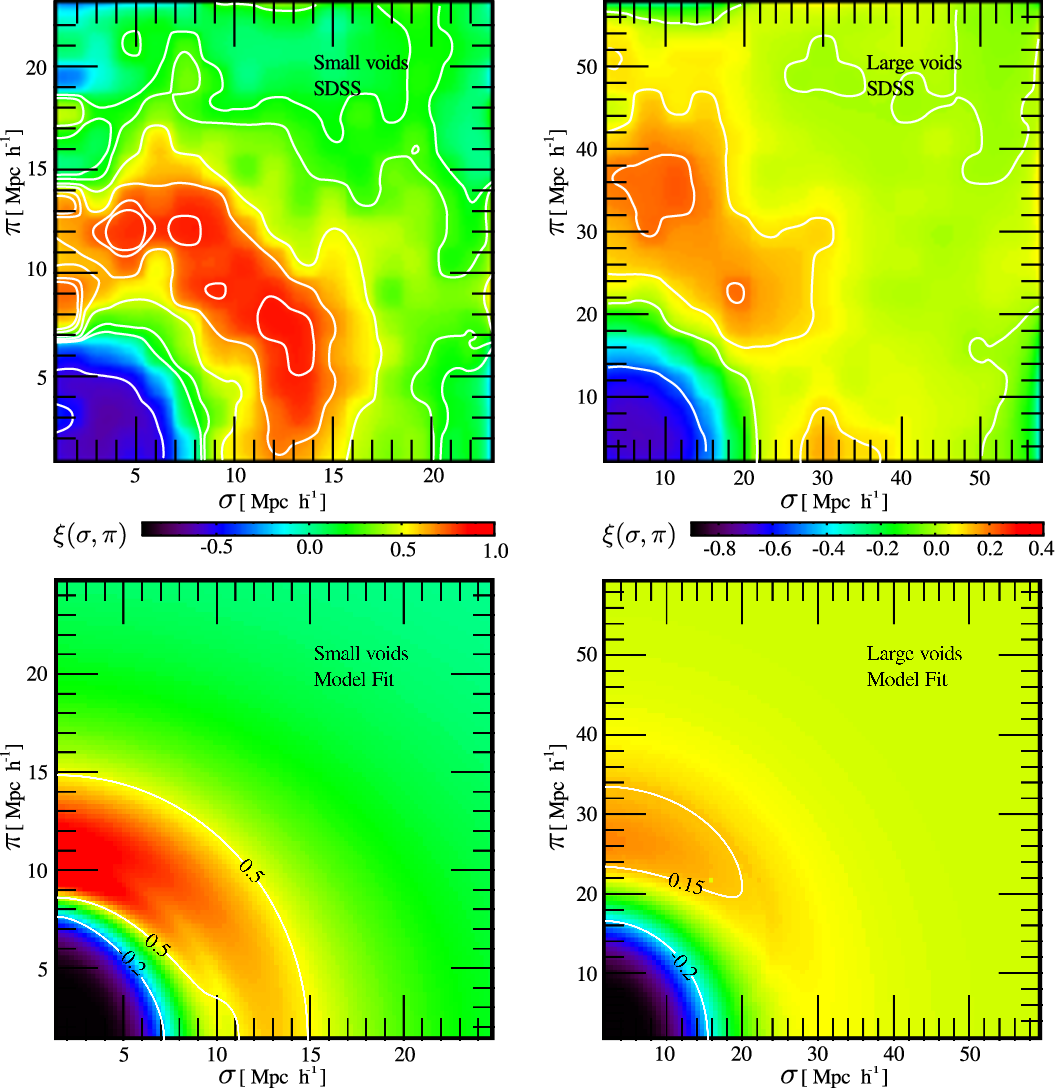}
    \caption[
    Void-galaxy cross-correlation function for samples of voids selected from the SDSS survey.
    ]{
    \textit{Upper panels.}
    Redshift-space distortions exhibited by the void-galaxy cross-correlation function measured for small S-type voids ($6 < R_\mathrm{v}/\hmpc < 8$) and large R-type voids ($10 < R_\mathrm{v}/\hmpc < 20$), both selected from the SDSS.
    \textit{Lower panels.}
    Best-fitting model predictions (Section~\ref{subsec:voids_rsd_gsm}).
    Notice that the spatial and colour scales change for both types of voids.
    \textit{Figure credit:} \citeonline{clues2}.
    }
    \label{fig:rsd_voids}
\end{figure}

Figure~\ref{fig:rsd_voids} shows the redshift-space void-galaxy cross-correlation function for the SDSS' voids of \citeonline{clues2}, distinguishing both types of voids: S-type on the left, R-type on the right.
The upper panels show the measurements, whereas the lower panels are the best-fitting predictions of a model of distortions that these authors developed, which will be explained in Section~\ref{subsec:voids_rsd_gsm}.
The general characteristics of RSD discussed in Section~\ref{subsec:lss_rsd_redshift}, and represented in Figure~\ref{fig:rsd2} schematically, can be observed for both types of voids: elongation of the inner contour levels (low density regime) and flattening of the outer contour levels (high density regime).
In particular, there is a red cloud near the $\pi$-axis, a manifestation of the enhanced density in this direction explained in Section~\ref{subsec:lss_rsd_vel}.

The anisotropic patterns on the correlation function encode valuable information about the dynamics of the galaxies around voids.
Below, we will explain two main approaches to model the RSD effect.
Eq.~(\ref{eq:velocity0}) for the radial velocity profile will be a fundamental component of these models.


\subsection{Linear model}
\label{subsec:voids_rsd_linear}

The first model is equivalent to the Kaiser model for RSD in the galaxy autocorrelation discussed in Section~\ref{subsec:lss_rsd_vel}, and was first derived for the case of voids by \citeonline{rsd_cai}.
We will refer to it as the linear model.
We review here its main assumptions and characteristics.

The starting point is the assumption that the number of void-galaxy pairs is conserved under the redshift-space mapping:
\begin{equation}
    ( 1 + \xi^s(\mathbf{s}) ) d^3s = ( 1 + \xi(\mathbf{r}) ) d^3r,
    \label{eq:number_consv}
\end{equation}
in analogy with Eq.~(\ref{eq:gnumconsv}), but here $\mathbf{r}$ and $\mathbf{s}$ denote the vectors from the void centre to the galaxy position in real and redshift space, respectively.
The difference between the galaxy-galaxy and void-galaxy cases, and also a fundamental assumption, is that we are considering the relative peculiar velocities with respect to one central point, the void centre.
The bulk motion of voids will therefore not affect the correlation function within the scales where the bulk velocity field can be considered to be coherent.
In this context, and within the plane-parallel approximation, it is only the void-centric velocity $\mathbf{v}$ of a galaxy that is relevant in this mapping, recovering Eq.~(\ref{eq:rsd}).
A further key assumption is that the velocity field is isotropic and radially directed, such that
\begin{equation}
    \mathbf{v} = v(r) \mathbf{\hat{r}},
    \label{eq:vel_rad}
\end{equation}
where $v(r)$ is given by Eq.~(\ref{eq:velocity0}), and $\mathbf{\hat{r}}$ is a unit vector oriented radially.
With these assumptions, we can rewrite Eq.~(\ref{eq:number_consv}) as
\begin{equation}
    1 + \xi^s(\mathbf{s}) = ( 1 + \xi(\mathbf{r}) ) \left[ 1 + \frac{v(r)(1+z)}{rH(z)} + \frac{(v'(r)-v(r)/r)(1+z)}{H(z)} \mu^2 \right]^{-1},
\end{equation}
where the prime denotes a derivative with respect to $r$.
Here, the term in square brackets represents the Jacobian of the coordinate transformation under the mapping from $\mathbf{r}$ to $\mathbf{s}$.
This expression can be expanded to linear order in the densities to obtain the basic linear model for the redshift-space correlation function:
\begin{equation}
    1 + \xi^s(s,\mu) = ( 1 + \xi(r) ) \left[ 1 + \frac{1}{3}\beta(z)\Delta(r) + \beta(z)\mu^2[\xi(r)-\Delta(r)] \right],
\end{equation}
where the radial separations in real and redshift space are related by
\begin{equation}
    r = s \left( 1 + \frac{1}{3}\beta(z)\Delta(s)\mu^2 \right).
\end{equation}
A further approximation can be made by dropping terms of order $\xi \Delta$ and $\xi^2$:
\begin{equation}
    \xi(s,\mu) = \xi(r) + \frac{1}{3}\beta(z)\Delta(r) + \beta(z)\mu^2[\xi(r)-\Delta(r)].
\end{equation}


\subsection{Gaussian streaming model}
\label{subsec:voids_rsd_gsm}

The linear model is valid on the assumption that $|\delta| \ll 1$ and that any random dispersion in velocity is small.
These assumptions can be relaxed, leading to a quasi-linear model.

Following \citeonline{xirsdmodel_peebles}, $\xi^s(\sigma, \pi)$ can be computed as the convolution of the real-space correlation, $\xi(r)$, and a pairwise velocity distribution of void-galaxy pairs, $g(\mathbf{r},\mathbf{v})$:
\begin{equation}
	1 + \xi^s(\sigma,\pi) = \int [1 + \xi(r)]~g(\mathbf{r}, \mathbf{v})~d^3v.
	\label{eq:gsm_raw}
\end{equation}
The pairwise velocity distribution can be chosen as a Maxwell-Boltzmann distribution.
Given that only $\pi$ is affected by peculiar velocities, Eq.~(\ref{eq:gsm_raw}) reduces to a one-dimensional integral via the replacements $g(\mathbf{r}, \mathbf{v}) \rightarrow g(r,r_\parallel,v_\parallel)$ and $d^3v \rightarrow dv_\parallel$, where $g$ reduces to a Gaussian distribution centred on the radial velocity profile $v(r)$, with a constant velocity dispersion $\sigma_\mathrm{v}$.
In this way\footnote{$\hat{\pi}$ refers to the irrational number \textit{pi}: 3.14159..., to avoid confusion with the notation of the $\pi$-coordinate.},
\begin{equation}
    1 + \xi^s(\sigma, \pi) = 
    \int_{-\infty}^{\infty} [1 + \xi(r)] \frac{1}{\sqrt{2\hat{\pi}}\sigma_\mathrm{v}}
    \mathrm{exp} \left[- \frac{(v_\parallel - v(r)\frac{r_\parallel}{r})^2}{2\sigma_\mathrm{v}^2} \right]
    \mathrm{d}v_\parallel.
	\label{eq:gsm0}
\end{equation}
The relations between $(r_\perp,r_\parallel)$ and $(\sigma,\pi)$, needed in this equation, are given by Eqs.~(\ref{eq:sigma0}) and (\ref{eq:pi0}).
When $|\delta| \ll 1$ and $\sigma_\mathrm{v}$ is small, the linear model is well recovered.
This model was first derived for the case of voids by \citeonline{clues2}, and we will refer to it as the Gaussian streaming model (hereinafter GS model).


\section{Cosmological relevance of voids}
\label{sec:voids_cosmology}

The physical description developed throughout this chapter of the evolution of voids, their abundance and the RSD around them has a strong cosmological dependence.
Therefore, cosmic voids emerge as natural cosmological probes to test the standard model.
Moreover, voids do not only constitute a complementary view of the large-scale structure, but they offer two main advantages over galaxy clustering studies.
On the one hand, void dynamics can be treated linearly, and hence, it is easier to model systematicities such as redshift-space distortions.
On the other hand, Modified Gravity theories predict that deviations from General Relativity should be more pronounced in voids \cite{modgr_li, modgr_clampitt, modgr_clifton, modgr_barreira, modgr_cai, modgr_lam, modgr_zivick, modgr_achitouv, modgr_cai2, modgr_joyce, modgr_koyama, modgr_cautun, modgr_falck, modgr_sahlen, modgr_davies, modgr_paillas}.
For instance, some of these alternative models postulate that a fifth force is responsible for the accelerated expansion, which is unscreened in low density environments.
In view of this, cosmic voids are ideal for testing different dark-energy models.
The potential of voids has also increased recently with the development of the new generation of spectroscopic surveys, such as BOSS, eBOSS, HETDEX, DESI and Euclid, which will probe our Universe covering a volume and redshift range without precedents.
This will allow us to obtain rich samples of voids at different redshifts, and in this way, to test the expansion history and geometry of the Universe with high precision.

As we have seen, there are two primary statistics in void studies: the void size function and the void-galaxy cross-correlation function.
The basic process consists of measuring these statistics from galaxy spectroscopic surveys, and then comparing them with model predictions, such as the ones developed in Sections~\ref{sec:voids_abundance} and \ref{sec:voids_rsd}.
In this way, we are able to constrain the intervening cosmological parameters that best fit the data, together with the corresponding confidence regions.
The ultimate goal is to add void-based tests to the cosmological toolkit for evaluating dark-energy models, together with the experiments that involve the CMB anisotropies, SNe Ia and the BAO signal.
Even more important is to make joint analyses combining the results of different experiments, such as the one shown in Figure~\ref{fig:cosmoconstraints}, and thus see if even tighter constraints can be achieved.

With respect to the void size function, there are many works in the literature that attempt to lay the foundations for the design of cosmological tests using this statistic \cite{svdw,abundance_furlanetto,abundance_jennings,abundance_achitouv,abundance_pisani,abundance_ronconi1,abundance_bias_contarini,abundance_ronconi2,abundance_verza}.
In general, the community has concentrated their efforts on modelling the VSF with the excursion set formalism and the spherical evolution.
Nevertheless, the RSD and AP distortions have an impact on the VSF, and they lead to biased cosmological constraints if they are not taken into account properly.
In the present work, we will tackle this problematic in Chapter~\ref{chp:impact_vsf}.
This is the first time that redshift-space systematicities on the VSF are treated.

Cosmic voids have been explored as standard rulers as well, since a stacked sample of voids must show a spherical shape in real space, and hence an AP test can be performed by analysing the deviations from this spherical symmetry \cite{apvoids_ryden,apvoids_lavaux,apvoids_sutter,apvoids_mao}.
However, it is not easy to take into account the anisotropic patterns induced by the dynamical RSD effect with this approach.

It is better to use the correlation function.
We know that in real space, the correlation function must manifest spherical symmetry, hence the observed anisotropic patterns are a clear evidence of the presence of geometrical and dynamical distortions.
In this way, the correlation function can be used as an AP test by quantifying these anisotropic patterns.
Moreover, the dynamical models of Section~\ref{sec:voids_correlation} allow us to take into account the RSD contribution to the anisotropic patterns from physical grounds.
We highlight that the fundamental ingredient is Eq.~(\ref{eq:velocity0}) for the void-centric radial velocity profile, which is not only important from a dynamical point of view, but it is also important for cosmology too, since it depends on the parameters $H(z)$ and $\beta(z)$.
Recall that $H(z)$ depends in turn on the background cosmological parameters like $h$, $\Omega_m$ and $\Omega_\Lambda$ (Eq.~\ref{eq:hubble_z}).
This approach has been thoroughly investigated in the literature, both in simulations and with observational data \cite{clues2, aprsd_hamaus1, rsd_cai, aprsd_hamaus2, rsd_achitouv1, rsd_achitouv2, rsd_chuang, rsd_hamaus, rsd_hawken1, rsd_achitouv3, rsd_nadathur, reconstruction_nadathur, aprsd_nadathur, rsd_hawken2, aprsd_hamaus2020, aprsd_nadathur2020}.
The study of the void-galaxy cross-correlation function as a cosmological probe is one of the main topics of the present work.

\chapter{Data set}
\label{chp:data}
In this chapter, we introduce the data set to be used throughout all this work.
Specifically, we describe the numerical simulations and the resulting simulated catalogues used to calibrate the different tests and models.
We also explain the void finding method, and finally, the galaxy spectroscopic catalogue used to compare observational results with the corresponding theoretical predictions.


\section{Millennium XXL simulation}
\label{subsec:data_sim}

For most of the work, we used the Millennium XXL N-body simulation \cite[hereinafter MXXL]{mxxl_angulo}, which extends the previous Millennium \cite{millennium} and Millennium-II \cite{millennium2} simulations, and follows the non-linear evolution of $6720^3$ dark-matter particles inside a periodic cubic box of length $3000~\hmpc$.
The particle mass is $8.456\times10^9~\hmsun$.
It was designed under a flat-$\Lambda$CDM cosmology with the same cosmological parameters as the previous runs: $\Omega_m=0.25$, $\Omega_\Lambda=0.75$, $\Omega_b=0.045$, $\Omega_\nu=0.0$, $h=0.73$, $n_s=1.0$ and $\sigma_8=0.9$.
We used the snapshots belonging to the redshifts $\zsim=0.51$, $0.99$, and $1.50$, which will be assumed as the mean redshifts of the samples selected for the analyses.

\begin{figure}
    \centering
    \includegraphics[width=\textwidth/2]{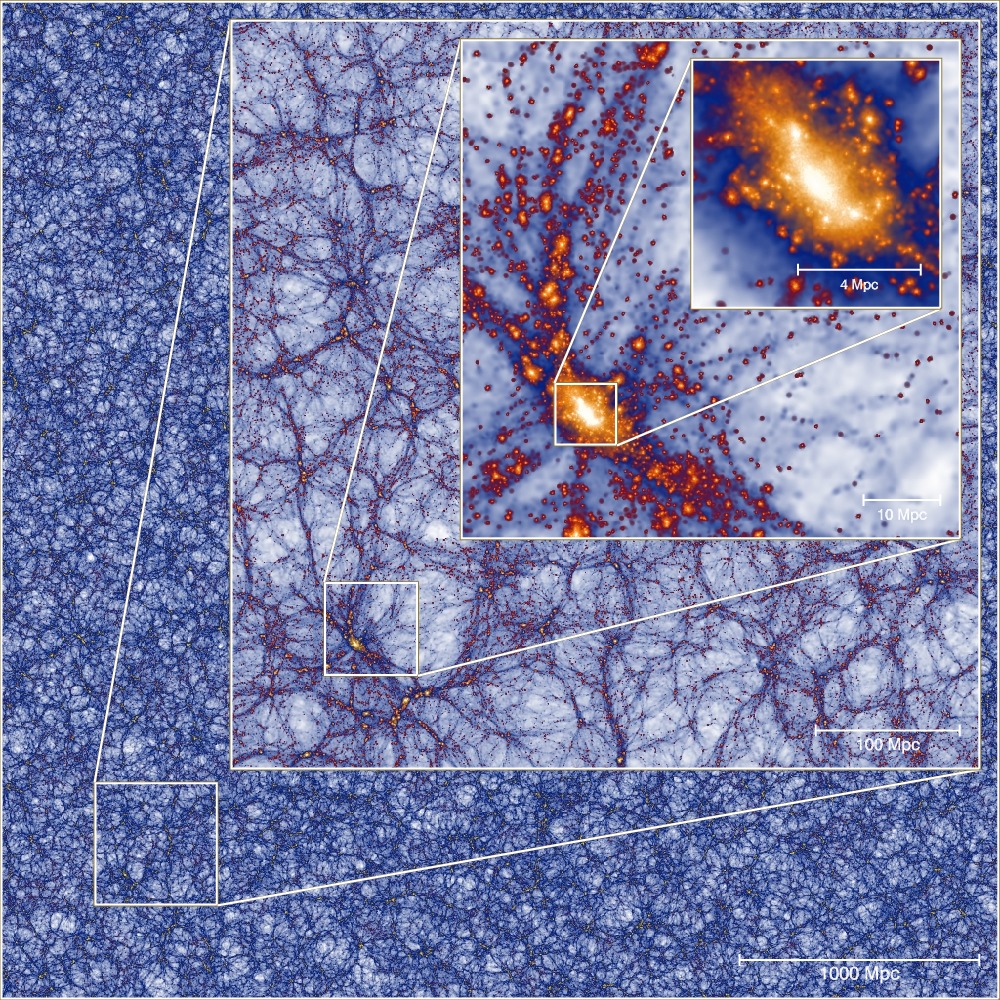}
    \caption[
    Spatial distribution of dark matter in the Millennium XXL simulation.
    ]{
    The mass density field in the Millennium XXL simulation focusing on the most massive halo present at $z=0$.
    Each inset zooms by a factor of $8$ from the previous one; the side-length varies from $4.1~\mathrm{Gpc}$ down to $8.1~\mathrm{Mpc}$.
    All these images are projections of a thin slice through the simulation of thickness $8~\mathrm{Mpc}$. 
    \textit{Figure credit:} R. Angulo \url{https://wwwmpa.mpa-garching.mpg.de/mpa/research/current_research/hl2011-9/hl2011-9-en.html}.
    }
    \label{fig:mxxl}
\end{figure}

The MXXL is one of the largest cosmological simulations ever performed.
The simulated volume is equivalent to that of the whole observable Universe up to redshift $0.72$, is more than $200$ times larger than that of the previous Millennium simulation, and about $7$ times larger than the volume of the Baryon Oscillation Spectroscopic Survey, making it ideal for conducting feasibility studies for cosmological tests.
This is particularly suitable for voids studies.
The mass resolution is sufficient to identify dark-matter haloes hosting galaxies with stellar masses of $1.5 \times 10^{10}~\hmsun$, considerably smaller than the Milky Way, and also to robustly predict the internal properties of haloes corresponding to very massive clusters, which are represented by more than $10^5$ dark-matter particles.
Figure~\ref{fig:mxxl} shows the mass density field in the MXXL at $z=0$, with zooms into different scales.

Dark-matter haloes were chosen as matter tracers, which were identified as groups of more than $60$ particles using a friends-of-friends algorithm with a linking length parameter of $0.2$ times the mean interparticle separation.
In all cases, we selected a lower mass cut of $5\times10^{11}~\hmsun$.

The simulation provides the position $\mathbf{x} = (x_1,x_2,x_3)~[\hmpc]$ and peculiar velocity $\mathbf{v} = (v_1,v_2,v_3)~[\kms]$ of each halo in a Cartesian comoving coordinate system.
In order to simulate the RSD effect, we treated the $x_3$-axis of the simulation box as the line-of-sight direction, assuming the distant observer approximation.
For the purposes of this work, this is a fair assumption for two main reasons.
On the one hand, the redshifts of the snapshots can be considered far enough with respect to the void-centric distances that we treat.
On the other hand, although the box length is large, the simulation box is taken as a simple mock survey, which is periodic, complete in volume and without any complicated selection functions.
The volume of the simulation simply allows us to have a large sample of voids to detect the different redshift-space systematicities with a good signal.
According to Eq.~(\ref{eq:rsd}), we applied the following equation to shift the LOS coordinates of haloes from real to redshift space:
\begin{equation}
    \Tilde{x}_3 = x_3 + \frac{1 + \zsim}{H(\zsim)} v_3,
	\label{eq:halo_zspace}
\end{equation}
where $\Tilde{x}_3$ denotes the shifted $x_3$-coordinate.
According to the MXXL cosmology, the Hubble parameter can be expressed in terms of the cosmological parameters as follows:
\begin{equation}
    H(z) = 100~h~\sqrt{\Omega_m(1+z)^3 + \Omega_\Lambda}.
    \label{eq:hubble}
\end{equation}
For our flat case,
\begin{equation}
    \Omega_\Lambda = 1 - \Omega_m.
    \label{eq:flat}
\end{equation}


\section{Spherical void finder}
\label{sec:data_id}

In order to identify voids from the spatial distribution of the MXXL haloes, we applied the spherical void finder developed by \citeonline{clues3}, based on the original method developed by \citeonline{voids_padilla}.
This is a method based on the integrated density contrast of underdense regions assuming spherical symmetry.
The void identification is a fundamental step for our work, hence we detail below the main steps of this procedure.

\begin{enumerate}

\item
\textit{Voronoi tessellation.}
A Voronoi tessellation is performed to obtain an estimation of the density field: each halo has an associated cell with volume $V_\mathrm{cell}$, and a density given by the inverse of that volume: $\rho_\mathrm{cell} = 1/V_\mathrm{cell}$.
We used a parallel version of the public library \textsc{voro++} \cite{voropp}.

\item
\textit{Selection of candidates.}
A first selection of underdense regions is done by selecting all Voronoi cells that satisfy the following criterion in local density contrast: $\delta_\mathrm{cell} < -0.7$.
Each underdense cell is considered the centre of a potential void.

\item
\textit{Growth of spheres.}
Centred on each candidate, the integrated density contrast, $\Delta(r)$, is computed in spheres of increasing radius $r$ until the overall density contrast satisfies a redshift-dependent threshold of $\Delta_\mathrm{id}$, obtained from the spherical evolution model \cite{sphcollapse1,sphcollapse2} by fixing a final spherical perturbation of $\Delta_\mathrm{id,0} = -0.9$ for $z=0$.

\item
\textit{Optimisation.}
Once these first void candidates are identified, step $3$ is repeated, but now starting in a randomly displaced centre proportional to $0.25$ times the radius of the candidate.
Then, the void centre is updated to a new position if the new radius is larger than the previous one.
This process is repeated iteratively until convergence to a sphere of maximum radius is achieved.
We adopted the criterion that the optimal sphere is obtained if the algorithm can not find a bigger one during a lapse of $50$ iterations.
This normally takes between $200$ and $300$ iterations in total.
In this way, this procedure mimics a random walk around the original centre in order to obtain the largest possible sphere in that local minimum of the density field.

\item
\textit{Overlap filtering.}
Finally, the list of void candidates is cleaned so that each resulting sphere does not overlap with any other.
This cleaning is done by ordering the list of candidates by size and starting from the largest one.
The final result is a catalogue of non-overlapping spherical voids with a well-defined centre, a radius $R_\mathrm{v}$, and overall density contrast $\Delta(R_\mathrm{v}) = \Delta_\mathrm{id}$.

\end{enumerate}

An important comment about step $3$.
The choice of $\Delta_\mathrm{id,0}$ is motivated by previous studies that use voids identified from biased tracers, namely, haloes or galaxies, which keep a bias relation with respect to the total matter.
For dark-matter voids, i.e. those identified from the spatial distribution of dark-matter particles, there is a theoretical threshold: $\Delta_\mathrm{id,0} = -0.8$, which corresponds to the moment of shell-crossing in the expansion process, taken as the moment of void formation \cite[see Section~\ref{sec:voids_evolution}]{svdw,abundance_jennings}.
However, it is not trivial to extrapolate this value for the case of voids identified from biased tracers.
Some studies provide a firm evidence that there is still a linear bias relation between dark matter and tracers in voids \cite{abundance_furlanetto,bias_chan1,bias_pollina1,bias_chan2,abundance_bias_contarini,density_bias_fang,bias_pollina2,bias_schuster,bias_chan3}.
As pointed out by \citeonline{abundance_bias_contarini}, assuming that voids from dark-matter particles and haloes have the same radii when the phenomenon of shell-crossing occurs, implies that the latter have a lower embedded density contrast.
Therefore, if the bias is greater than $1$, then the density contrast can reach values so low that the phenomenon of shell-crossing might not even happen.
Hence, it is a common practice to define voids as empty as possible.
Previous works using the spherical void finder have demonstrated that choosing $\Delta_\mathrm{id,0} = -0.9$ (and extrapolating this value to the corresponding redshift according to the spherical model) give a sample of voids with a well characterised dynamics and suitable for cosmological studies (see for instance \citeonline{clues1} and \citeonline{clues2}).
In the next chapter, we will reinforce this aspect by presenting a new cosmological test based on voids identified in this manner.


\section{Void catalogues I}
\label{sec:data_voids1}

\begin{figure}
    \centering
    \includegraphics[width=\textwidth/2]{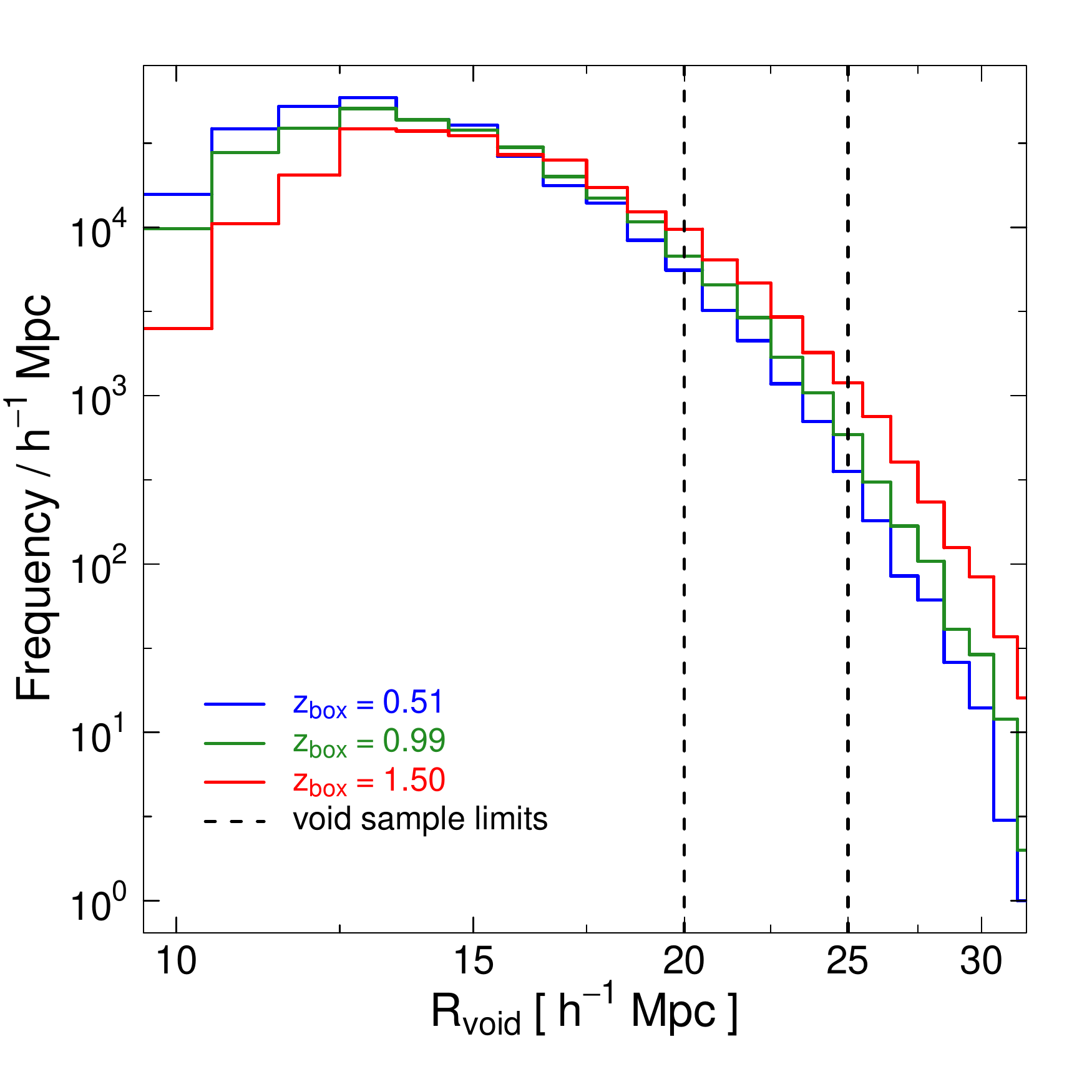}
    \caption[
    Radius distribution of voids from the MXXL simulation (Table~\ref{tab:catalogues1}).
    ]{
    Radius distribution corresponding to the void catalogues of Table~\ref{tab:catalogues1}.
    The vertical dashed lines delimit the void samples used to perform the cosmological test presented in Chapter~\ref{chp:cosmotest}.    
    }
    \label{fig:rad_dist}
\end{figure}

For the analysis of Chapter~\ref{chp:cosmotest}, where we design a new cosmological test, we identified voids using the three snapshots of the MXXL simulation, namely, $z=0.51$, $0.99$ and $1.50$.
To do this, we adopted the same cosmology of the simulation in order to compute distances and densities, needed in void definition.
Moreover, the identification was performed in real space.
The reasons for this choice will be clear in that chapter and the following ones.
Table~\ref{tab:catalogues1} shows the main characteristics of the halo and void catalogues used for this analysis, and Figure~\ref{fig:rad_dist}, the corresponding void radius distribution.

\begin{table}
\centering
\caption[
Main characteristics of the dark-matter halo and void catalogues used to develop the cosmological test of Chapter~\ref{chp:cosmotest}.
]{
Main characteristics of the halo and void catalogues used in Chapter~\ref{chp:cosmotest}.
From left to right: MXXL snapshot, number of dark-matter haloes, density threshold criterion for void identification and number of identified voids.
}
\label{tab:catalogues1}
\begin{tabular}{ccccc}
\hline
Snapshot & Haloes & $\Delta_\mathrm{id}$ & Voids \\
\hline
\hline
0.51 & 136 993 439 & -0.8764 & 333 741 \\ 
0.99 & 133 688 808 & -0.8533 & 305 082 \\ 
1.50 & 118 244 901 & -0.8302 & 254 993 \\ 
\hline
\end{tabular}
\end{table}

In order to perform the cosmological test, we selected a sample of voids for each MXXL snapshot with sizes between $20 \leq R_\mathrm{v}/\hmpc \leq 25$.
This is shown with vertical dashed lines in the figure.
Our results are not sensitive to the size of the voids used, as we have verified.
However, large void sizes ensure that they are mostly of R-type, which simplifies the modelling of the density profiles (see Section~\ref{subsec:cosmotest_model_densdiff} for more details).
The number of voids found in each sample are the following: $10157$ for $z=0.51$, $13703$ for $z=0.99$ and $21034$ for $z=1.50$.


\section{Void catalogues II}
\label{sec:data_voids2}

In Chapters~\ref{chp:zeffects}, \ref{chp:impact_vsf} and \ref{chp:impact_vgcf}, we study the effects of RSD and AP distortions on the void identification process itself, and the subsequent impact on the void size function and the void-galaxy cross-correlation function as cosmological tests.
For this analysis, we focused only on the $z = 0.99$ snapshot as representative, since the remaining snapshots revealed similar results.
We applied the void finder in two ways, resulting in two types of catalogues.

In the first case, we adopted the same cosmology of the MXXL simulation as before.
In turn, in order to study the impact of RSD, we performed the identification in both spatial configurations: real and redshift space.
We will refer to these catalogues as the \textit{true-cosmology (TC) void catalogues}: TC-rs for the real-space version, and TC-zs for the redshift-space one.

In the second case, in order to study the impact of the combined RSD and AP effects, we modified the coordinate system of the simulation according to two different fiducial cosmologies.
Specifically, we fixed all the MXXL global parameters with the exception of $\Omega_m$ and $\Omega_\Lambda$, in such a way that the cosmology is still flat, i.e. Eq.~(\ref{eq:flat}) is still valid.
Concretely, we chose two fiducial values for $\Omega_m$: a lower value with respect to the true one, $\Omega_m^l = 0.20$, and a higher value, $\Omega_m^u = 0.30$.
In this case, the identification was performed only in redshift space for reasons that will be clear in Chapter~\ref{chp:impact_vsf}.
We will refer to these catalogues as the \textit{fiducial-cosmology (FC) void catalogues}: FC-l for the $\Omega_m^l = 0.20$ version, and FC-u for the $\Omega_m^u = 0.30$ one. 

Table~\ref{tab:catalogues2} shows the main characteristics of the void catalogues used for this analysis.
According to Table~\ref{tab:catalogues1}, these catalogues were identified with a density threshold of $\Delta_\mathrm{id}=-0.8533$ for $z=0.99$.
In Chapter~\ref{chp:zeffects}, we will need to define two subcatalogues from the TC ones.
The original catalogues, as they were presented up to here, will be referred to as the \textit{full catalogues}: TC-rs-f for the real-space version, and TC-zs-f for the redshift-space one.
The subcatalogues will be precisely defined in Section~\ref{sec:zeffects_map}, and will be referred to as the \textit{bijective catalogues}: TC-rs-b for the real-space version, and TC-zs-b for the redshift-space one.
Essentially, bijective voids are a subset of the full voids where each void has a unique counterpart in the other spatial configuration spanning the same region of space.

Note that the TC-rs-f catalogue must be then equivalent to that of the second entry of Table~\ref{tab:catalogues1}.
Nevertheless, this catalogue has many more voids.
This is because, in order to build it, we relaxed the criterion of step $2$ in the void identification process to $\delta_\mathrm{cell} < -0.4$.
This allowed us to obtain a more complete sample, an important aspect when analysing void abundances.
However, this modification is not that important, since it only results in an increase of smaller voids, which are generally irrelevant from a statistical point of view (in Section~\ref{sec:zeffects_map} we provide a more detailed discussion).

\begin{table*}
\centering
\caption[
Main characteristics of the void catalogues used to study the AP, t-RSD, v-RSD and e-RSD effects in Chapters~\ref{chp:zeffects}, \ref{chp:impact_vsf} and \ref{chp:impact_vgcf}.
]{
Main characteristics of the void catalogues used in Chapters~\ref{chp:zeffects}, \ref{chp:impact_vsf} and \ref{chp:impact_vgcf}.
These catalogues correspond to the MXXL snapshot $z=0.99$, and were identified with a density threshold of $\Delta_\mathrm{id}=-0.8533$.
From left to right: catalogue name, cosmology used in void definition, $\Omega_m$ choice, spatial configuration where the identification was performed, subcatalogue regarding the bijective condition, number of voids and type of systematic effects considered.
}
\label{tab:catalogues2}
\begin{tabular}{ccccccc}
\hline
Catalogue & Cosmology & $\Omega_m$ & Space & Subcatalogue & Voids & Systematic effects \\
\hline
\hline
TC-rs-f & MXXL     & 0.25 & real     & full      & 463 690 & none \\ 
TC-rs-b & MXXL     & 0.25 & real     & bijective & 318 784 & none \\
TC-zs-f & MXXL     & 0.25 & redshift & full      & 455 482 & RSD \\
TC-zs-b & MXXL     & 0.25 & redshift & bijective & 318 784 & RSD \\
FC-l    & Fiducial & 0.20 & redshift & full      & 375 560 & AP + RSD \\ 
FC-u    & Fiducial & 0.30 & redshift & full      & 526 552 & AP + RSD \\
\hline
\end{tabular}
\end{table*}

\begin{figure}
    \centering
    \includegraphics[width=79mm]{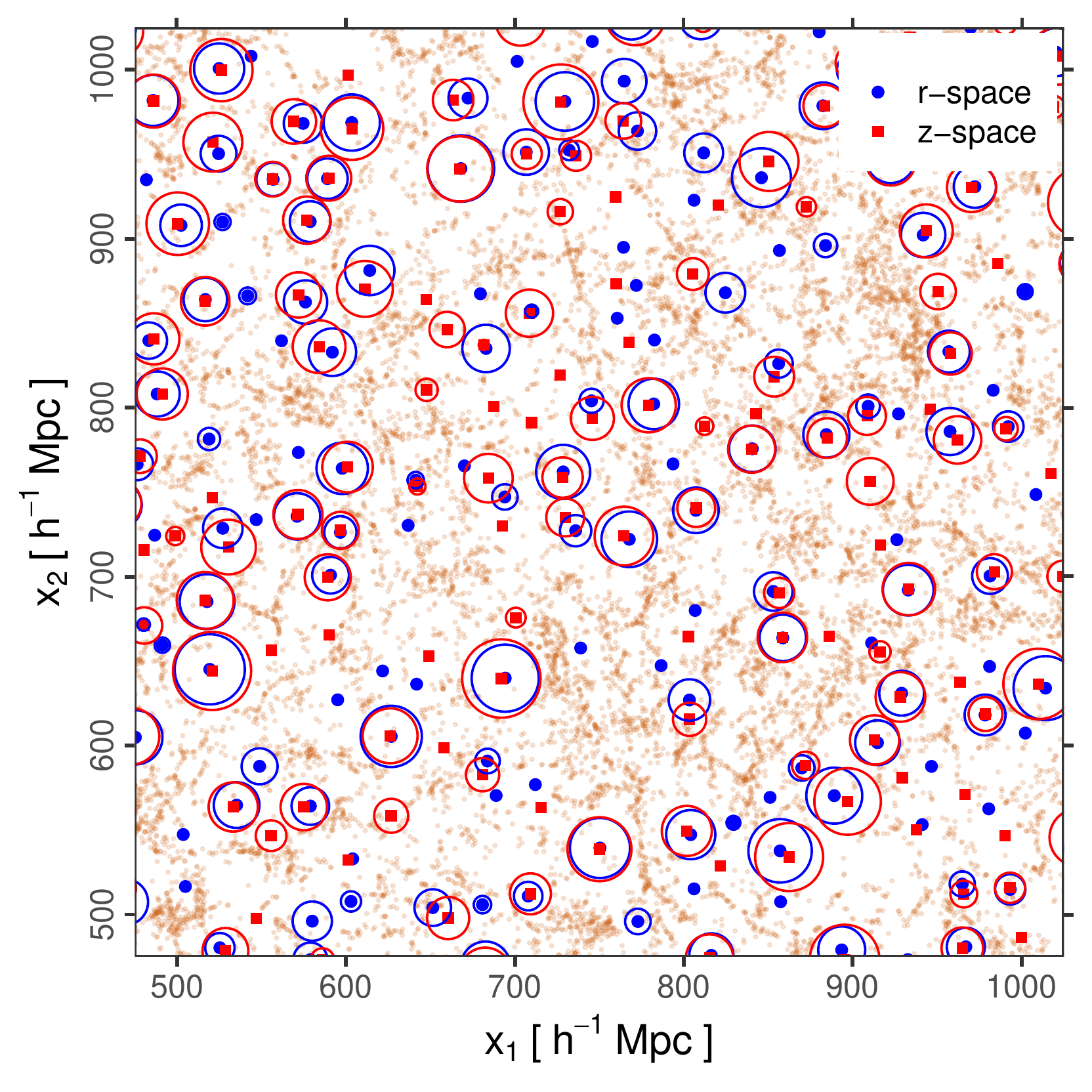}
    \includegraphics[width=79mm]{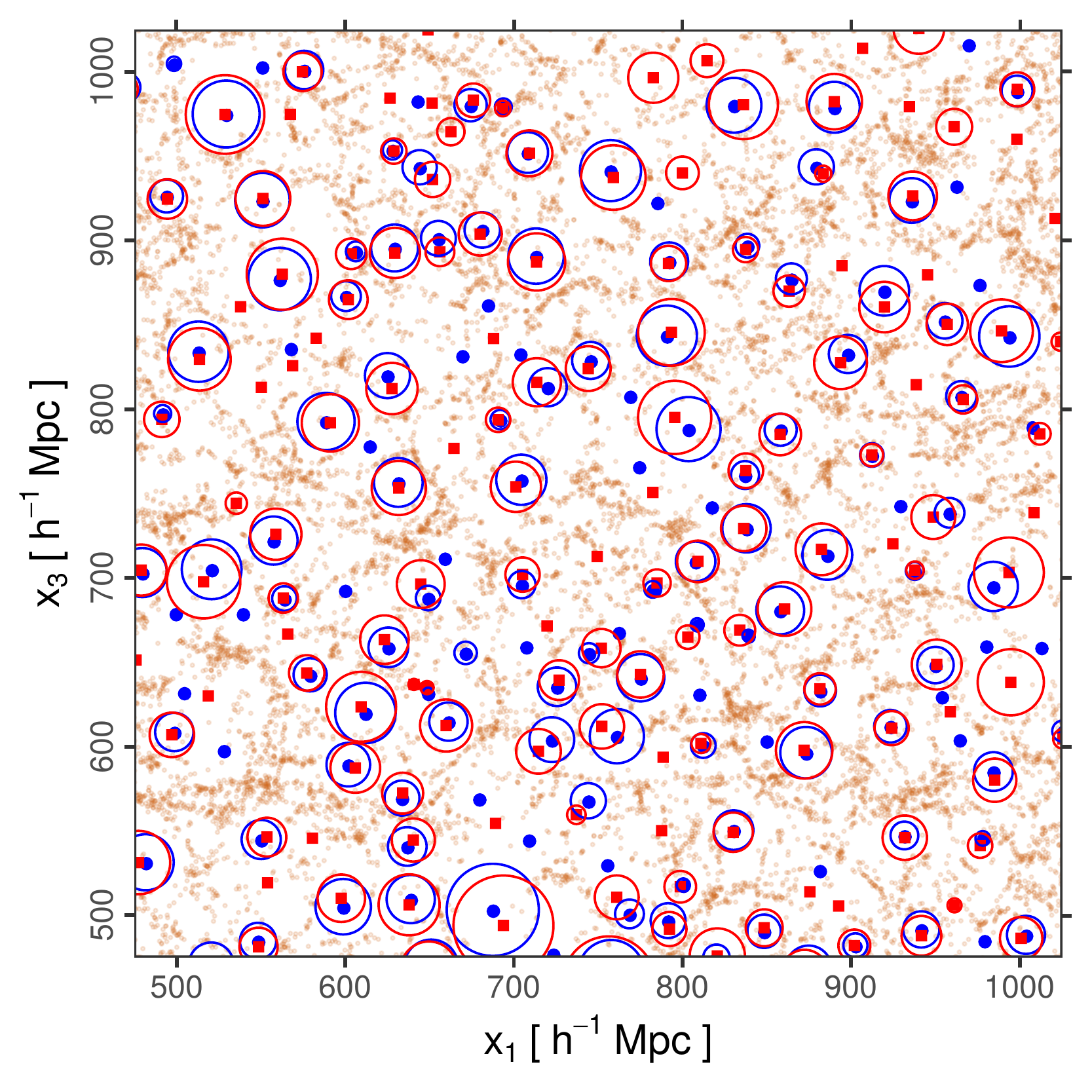}
    \caption[
    Slices of the MXXL simulation showing the spatial distribution of their dark-matter haloes and voids.
    ]{
    Slices of the MXXL simulation box showing the spatial distribution of its haloes and voids, the latter selected from the TC catalogues of Table~\ref{tab:catalogues2}.
    Both slices have a thickness of $10~\hmpc$.
    Eq.~(\ref{eq:halo_zspace}) was used to simulate the RSD effect.
    Real-space voids are represented in blue, whereas the redshift-space ones, in red.
    Bijective voids, in turn, are represented with circles, the intersections of the spherical voids with the midplane of the slice.
    Each pair of bijective voids span the same region of space.
    \textit{Left-hand panel.}
    Representation of the plane-of-sky distribution.
    \textit{Right-hand panel.}
    Representation of the line-of-sight distribution.
    }
    \label{fig:mxxl_slice}
\end{figure}

In order to visualise how the void finder works, Figure~\ref{fig:mxxl_slice} shows two slices of the MXXL simulation box, where the spatial distribution of its haloes and voids can be appreciated, the latter obtained from the TC catalogues of Table~\ref{tab:catalogues2}.
For this representation, we used Eq.~(\ref{eq:halo_zspace}) to simulate the RSD effect.
The left-hand panel shows a slice in the range $500 \leq x_1/\hmpc \leq 1000$, $500 \leq x_2/\hmpc \leq 1000$ and $95 \leq x_3/\hmpc \leq 105$.
Hence, it is a representation of the POS distribution of haloes and voids.
The right-hand panel, on the other hand, shows a slice in the range $500 \leq x_1/\hmpc \leq 1000$, $95 \leq x_2/\hmpc \leq 105$ and $500 \leq x_3/\hmpc \leq 1000$, i.e. it shows the LOS distribution of haloes and voids.
In both cases, the real-space void centres were represented with blue dots, whereas the redshift-space centres, with red squares.
The circles surrounding them represent the intersections of the spherical voids with the midplane of the slice, but they were plotted only for the special class of bijective voids.

\section{The Baryon Oscillation Spectroscopic Survey}
\label{sec:data_boss}

The Baryon Oscillation Spectroscopic Survey \cite[BOSS]{boss} is part of a six-year program of the SDSS-III (mid-2008 to mid-2014) that used the wide-field $2.5~\mathrm{m}$-telescope at Apache Point Observatory in New Mexico, United States of America, to carry out four surveys on three scientific themes: (i) dark energy and cosmological parameters, (ii) the structure, dynamics, and chemical evolution of the Milky Way, and (iii) the architecture of planetary systems.
BOSS was designed for the first purpose.
Specifically, BOSS measured the redshift of $1.5$ million luminous red galaxies and Lyman-$\alpha$ absorption towards $160 000$ high redshift quasars.
The main scientific project is to detect the characteristic scale imprinted by the baryon acoustic oscillations in the early Universe (about $150~\mathrm{Mpc}$) and use it as a physically calibrated ruler to determine the cosmic distance scale with a precision of $1.0\%$ at $z=0.35$, $1.1\%$ at $z=0.6$, and $1.5\%$ at $z=2.5$, achieving tight constraints on the equation of state of dark energy.
The high-precision clustering measurements over a wide range of redshifts and length scales also provides rich insights into the origin of cosmic structure and the matter content of the Universe.
Therefore, besides BAO analyses, the characteristics of this survey make it ideal for similar tests with cosmic voids.
The right-hand panel of Figure~\ref{fig:lss_surveys} shows the coverage of the Universe planned by BOSS and its extended survey, eBOSS.

In Chapter~\ref{chp:boss}, we will present some preliminary results obtained from this survey.
Specifically, we will analyse the void-galaxy cross-correlation function measured from these data.
For this part of the work, we used the CMASS and LOWZ samples from the Data Release 12 (BOSS DR12).
They contain $864464$ and $333082$ galaxies in the northern and southern hemispheres, respectively.


\section{The MultiDark Patchy mock surveys}
\label{sec:data_patchy}

In order to interpret the measurements from BOSS data, we used the MultiDark Patchy mocks \cite{patchy_kitaura, patchy_rodriguez}, a series of mock light cones that reproduce the DR12 galaxy clustering catalogue with high fidelity on all relevant scales in order to allow a robust analysis of BAO and redshift-space distortions.
These mocks test the cosmic evolution in the redshift range from $0.15$ to $0.75$.
They have been calibrated using a reference galaxy catalogue based on the halo abundance matching modelling of the BOSS DR12 galaxy clustering data and on the data themselves.
Their production follows five steps.
First, an accurate reference catalogue is generated, for which a large N-body simulation capable of resolving distinct haloes and the corresponding substructures is used.
This technique is applied at different redshift bins to obtain a detailed galaxy bias evolution spanning the redshift range covered by BOSS DR12 galaxies.
In this way, several mock galaxy catalogues in full cubical volumes of $2500~\hmpc$ side at different redshifts are obtained.
Second, the method is trained to match the two- and three-point clustering of the full mock galaxy catalogues for each redshift bin.
In the third step, stellar masses are assigned to the individual objects.
In the fourth step, a light cone is generated, which includes selection effects, the masking, and combines different boxes at different redshifts.
Finally, the resulting mocks are compared to the observations.
The process is iterated until the desired accuracy for different statistical measures is reached.
For our analysis, we made use of $2048$ mocks for each hemisphere.

The reference catalogues were extracted from one of the Big MultiDark simulations \cite{bigmultidark}, which was performed using the GADGET-2 code with $3840^3$ particles on a volume of a cubic box of side length $2500~\hmpc$ with periodic boundary conditions, assuming a flat-$\Lambda$CDM cosmology from Planck with the following parameters: $\Omega_m=0.307115$, $\Omega_b=0.048206$, $h=0.6777$, $n_s=0.9611$ and $\sigma_8=0.8288$.
Dark-matter haloes were defined based on the BoundDensity Maximum halo finder, and the halo abundance matching technique was used to connect haloes to galaxies.


\section{Void catalogues III}
\label{sec:data_voids_boss}

\begin{figure}
    \centering
    \includegraphics[width=\textwidth/2]{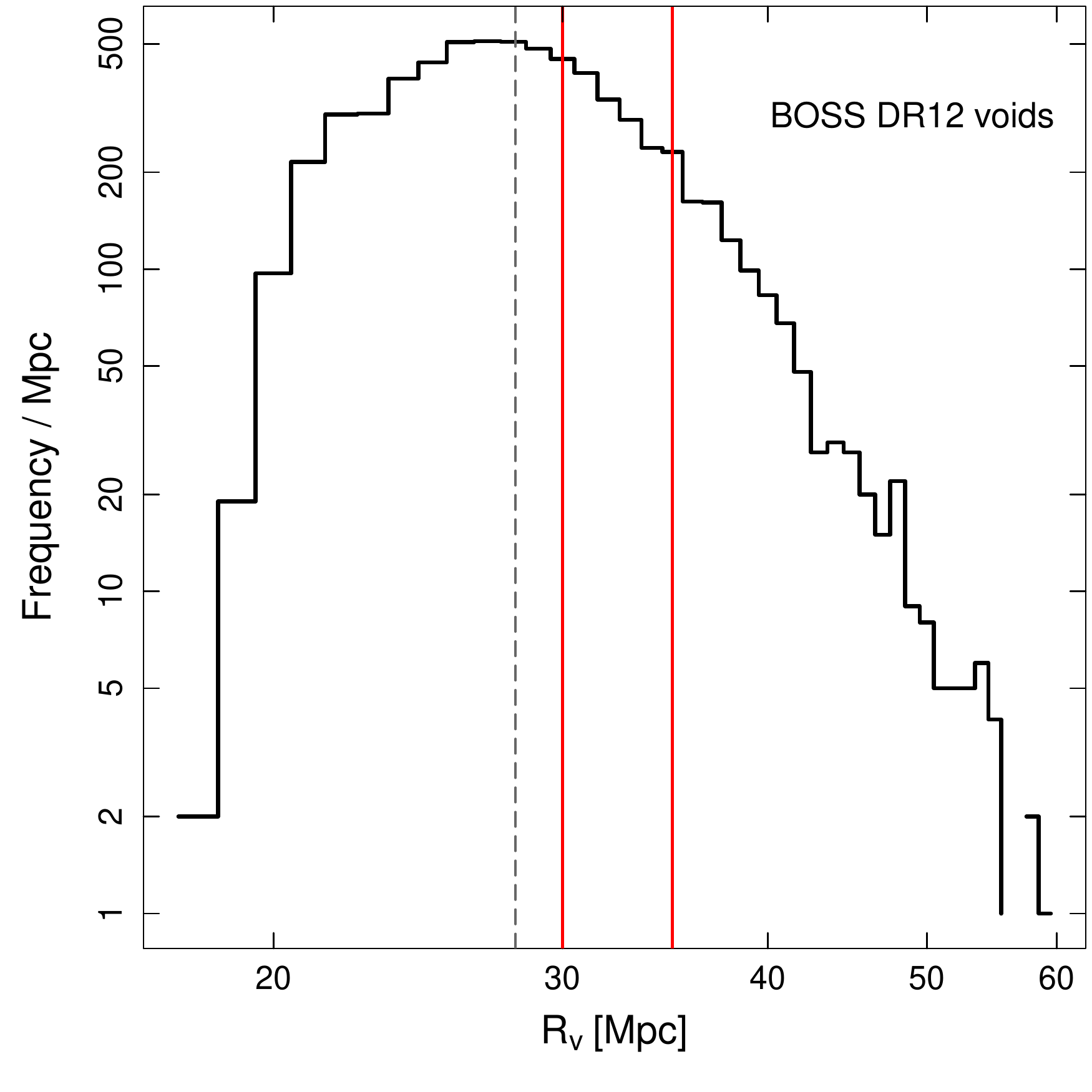}
    \caption[
    Radius distribution of voids from the BOSS DR12 survey.
    ]{
    Radius distribution corresponding to the BOSS DR12 voids.
    The grey dashed line indicates the median of the distribution.
    The red solid lines delimit the samples used for the analysis of the void-galaxy cross-correlation function in Chapter~\ref{chp:boss}.
    }
    \label{fig:boss_abundance}
\end{figure}

In order to identify voids from the Patchy mocks and BOSS DR12 galaxies, we applied our spherical void finder described in Section~\ref{sec:data_id} using, in both cases, the fiducial cosmology of the Patchy mocks and a density threshold of $\did = -0.9$.
From the BOSS survey, we found $4784$ and $1873$ voids in the northern and southern hemispheres, respectively ($6657$ voids in total).
They are distributed in a redshift range between $0.2$ and $0.8$, with a mean redshift of $z_\mathrm{BOSS} = 0.48$.
Figure~\ref{fig:boss_abundance} shows the radius distribution of the BOSS DR12 voids.
The vertical grey dashed line indicates the median of the distribution.
In order to perform the analysis of the void-galaxy cross-correlation function, we selected two samples (delimited by the vertical red solid lines in the figure): one with sizes between $30 \leq R_\mathrm{v}/\mathrm{Mpc} \leq 35$, and the other with sizes $R_\mathrm{v} \geq 30~\mathrm{Mpc}$.

\chapter{Designing a new cosmological test}
\label{chp:cosmotest}
\section*{Abstract}

In this chapter, we present a new cosmological test based on the distribution of galaxies around cosmic voids without assuming a fiducial cosmology.
The test is based on a physical model for the void-galaxy cross-correlation function projected along and perpendicular to the line of sight.
Correlations are measured in terms of angular distances and redshift differences between void-galaxy pairs, hence it is not necessary to assume a fiducial cosmology.
This model reproduces the coupled RSD and AP distortions that affect correlation measurements.
It also takes into account the scale mixing due to the projection range in both directions.
The model is general, it can be applied to an arbitrary binning scheme, not only to the case of the projected correlations.
It primarily depends on two cosmological parameters: $\Omega_m$ (more sensitive to AP distortions) and $\beta$ (more sensitive to RSD).
The test was calibrated using the MXXL simulation for different redshifts.
The method successfully recovers the cosmological parameters.
We studied the effect of measuring with different projection ranges, finding robust results up to wide ranges.
The data covariance matrices associated with the method are relatively small, which reduces the noise in the likelihood analysis, and therefore allows us to use a smaller number of mock catalogues to estimate them.
The performance evaluated in this work indicates that the developed method is a promising test to be applied on observational data.
This part of the work has been published in the Monthly Notices of the Royal Astronomical Society Journal \cite{aprsd_correa}.


\section{Foundations of the test}
\label{sec:cosmotest_foundations}


\subsection{Geometrical and dynamical distortions}
\label{subsec:cosmotest_foundations_aprsd}

Measuring the void-galaxy cross-correlation function involves measuring distances between pairs made up of a void centre and a galaxy.
Three observable quantities characterise the separation between such a pair: an angle $\theta$ subtended by the void centre and the galaxy on the plane of the sky, the redshift $z$ of the galaxy, and the redshift $z'$ of the void centre provided by the void finder.
According to Eqs.~(\ref{eq:sigma_ap}) and (\ref{eq:pi_ap}), the respective POS and LOS comoving separations, $\sigma$ and $\pi$, are related to these observable quantities by
\begin{equation}
    \sigma = D_\mathrm{M}(z') ~ \theta
	\label{eq:sigma_ap2}
\end{equation}
and
\begin{equation}
    \pi    = \left|\chi(z)-\chi(z')\right|.
	\label{eq:pi_ap2}
\end{equation}
Actually, Eq.~(\ref{eq:pi_ap2}) is a generalisation of Eq.~(\ref{eq:pi_ap}), since the void-centric separations can be large.
As we studied in Section~\ref{sec:lss_ap}, it is necessary to assume fiducial values for the cosmological parameters in order to estimate $\sigma$ and $\pi$ from the observable quantities, inducing possible distortions in the spatial distribution of the galaxies around voids due to the AP effect.

Dynamical distortions, on the other hand, are quantified by Eqs.~(\ref{eq:sigma0}) and (\ref{eq:pi0}), which we rewrite here to have the set of equations relevant for the test handy:
\begin{equation}
    \sigma = r_\perp
	\label{eq:sigma_rsd}
\end{equation}
and
\begin{equation}
    \pi = r_\parallel + \frac{1+z}{H(z)} v_\parallel.
	\label{eq:pi_rsd}
\end{equation}
Recall that $r_\perp$ and $r_\parallel$ are the real-space analogues of $\sigma$ and $\pi$, respectively, whereas $v_\parallel$ is the LOS component of the void-centric peculiar velocity of the galaxy under consideration.

It will be convenient to define a new quantity to express the LOS separation of a void-galaxy pair in terms of the difference between their redshifts:
\begin{equation}
    \zeta~:=~|z~-~z'|.
    \label{eq:delta_z}
\end{equation}
Throughout this chapter, we will distinguish between three spatial configurations: (i) the \textit{observable space}\footnote{In this chapter, we will use the notation $\theta$ and $\zeta$, instead of $\Delta \phi$ and $\Delta z$ that we have been using, to be in agreement with the notation used in \citeonline{aprsd_correa}, on which this part of the thesis is based.} defined by the coordinates $(\theta, \zeta)$, where measurements are made; (ii) the \textit{real space} defined by the coordinates $(r_\perp, r_\parallel)$, free of distortions; and (iii) the \textit{redshift space} defined by the coordinates $(\sigma, \pi)$, where the RSD and AP distortions are jointly observed.

A comment worth mentioning about the redshifts.
Strictly speaking, it is not possible to obtain the redshift $z'$ of a void centre directly from the shift of spectral lines.
This is only possible for astrophysical bodies like galaxies.
However, $z'$ can be indirectly obtained by means of the void finder when it localises the corresponding centre.
This is achieved by performing the reversal transformation from comoving coordinates of the centre to observable-space coordinates.
For now, we are only interested in $z'$ for the theoretical modelling.
The question of void identification in a non-fiducial way, namely in observable space, has not been investigated yet.
Whatever the method, this process needs to assume a Mpc-scale.
To make matters worse, the RSD and AP effects have an impact on the identification process itself, generating additional distortion patterns in the measurements of the correlation function.
We will tackle this problematic thoroughly in Chapter~\ref{chp:zeffects}.
For this reason, and in order to design and calibrate the test in this chapter appropriately, we have applied the spherical void finder to the real-space distribution of the MXXL haloes using the cosmology of the simulation, leading to the data set detailed in Section~\ref{sec:data_voids1}.
In this way, we make sure to have genuine voids with a well-characterised structure and dynamics that obey the evolution and physical properties described in Chapter~\ref{chp:voids}.
The additional systematicities observed in the correlation function due to the identification in redshift space will be thoroughly investigated in Chapter~\ref{chp:impact_vgcf}.

\subsection{Fiducial-free test and mixture of scales}
\label{subsec:cosmotest_foundations_scales}

Our method provides two novel aspects.
In the first place, we treat correlations directly in terms of the observable-space quantities $\theta$ and $\zeta$, thus obtaining the observable-space correlation function, $\xi^s(\theta,\zeta)$.
Hence, it is not necessary to assume a fiducial cosmology, and in this way, the AP effect is taken into account naturally.
For simplicity, we will hereinafter drop the superscript $s$ that refers to a redshift-space quantity.

The second aspect is related to a third type of systematicity, besides the RSD and AP effects, that also leads to deviations between observations and theoretical predictions when modelling the correlation function.
Models evaluate the correlation function on a given point of space.
However, in the measuring process, a binning scheme must be assumed, and hence, several scales are mixed.
This is not a problem if we work with almost differential bins, nevertheless, this implies a poor signal.
On the contrary, increasing the bin sizes improves the signal, but so does the mixture of scales.
Therefore, the correlation function must then be modelled taking into account this effect, which finds a description from the volume and geometry of the bins.
Our method provides a treatment for this scale-mixing effect.
This allows us to work with bins of arbitrary sizes, so we can go further and work with fully projected correlation functions.
This variant of measuring correlations constitutes the basis of the test.


\section{The projected correlation functions}
\label{sec:cosmotest_pcorrelations}

As we mentioned previously, the correlation function must be isotropic in real space.
However, this is no longer valid in redshift space due to the presence of the coupled RSD and AP distortions.
In this case, the spherical symmetry breaks into a cylindrical symmetry about an axis oriented along the LOS direction.
Therefore, it is instructive to visualise a void sample as a two-dimensional stack with cylindrical axes.
In view of this symmetry, the correlation function can be represented as a two-dimensional map $\xisp$ in the coordinates $\sigma$ and $\pi$. 

\begin{figure}
    \centering
    \includegraphics[width=50mm]{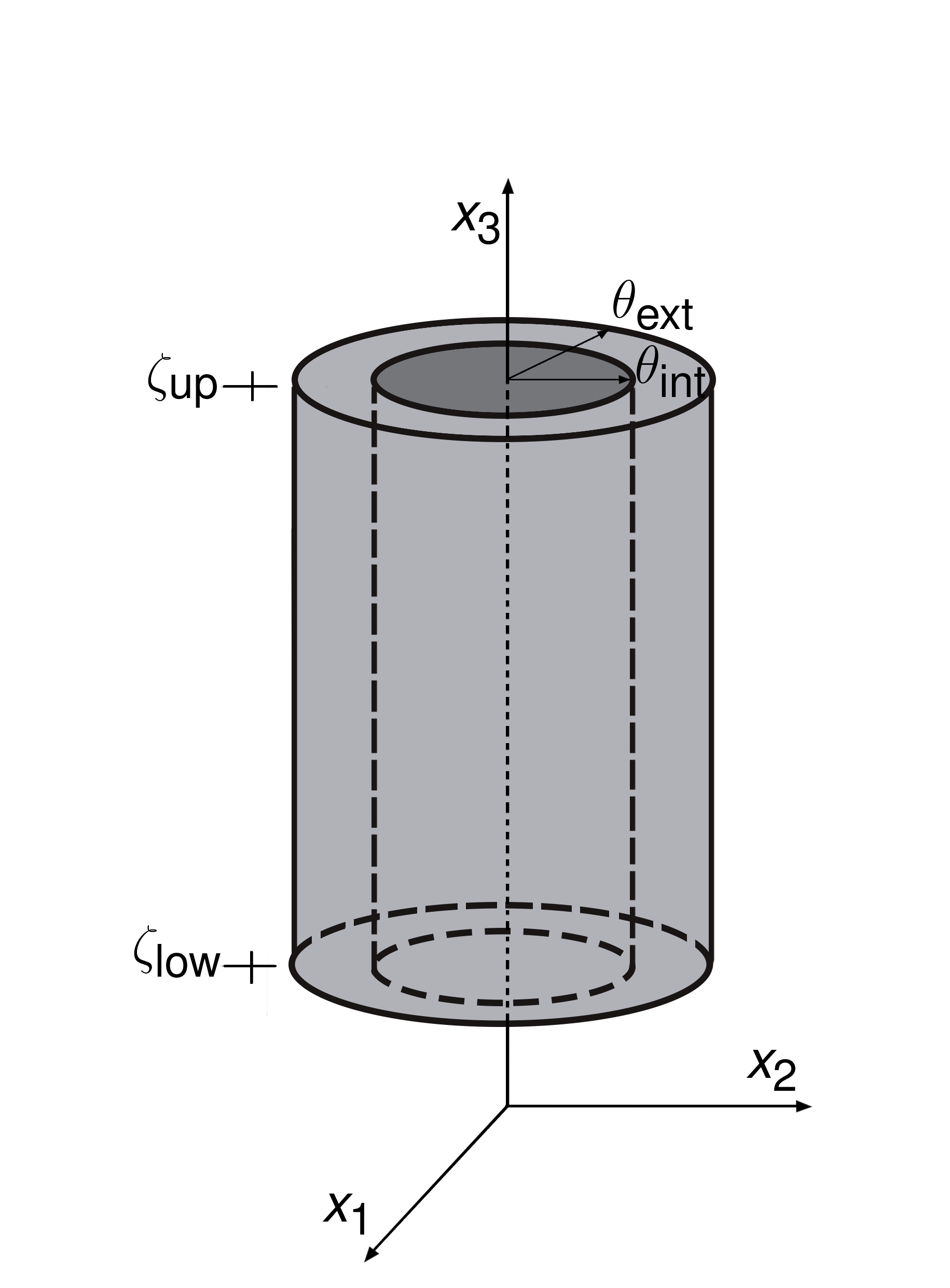}
    \includegraphics[width=50mm]{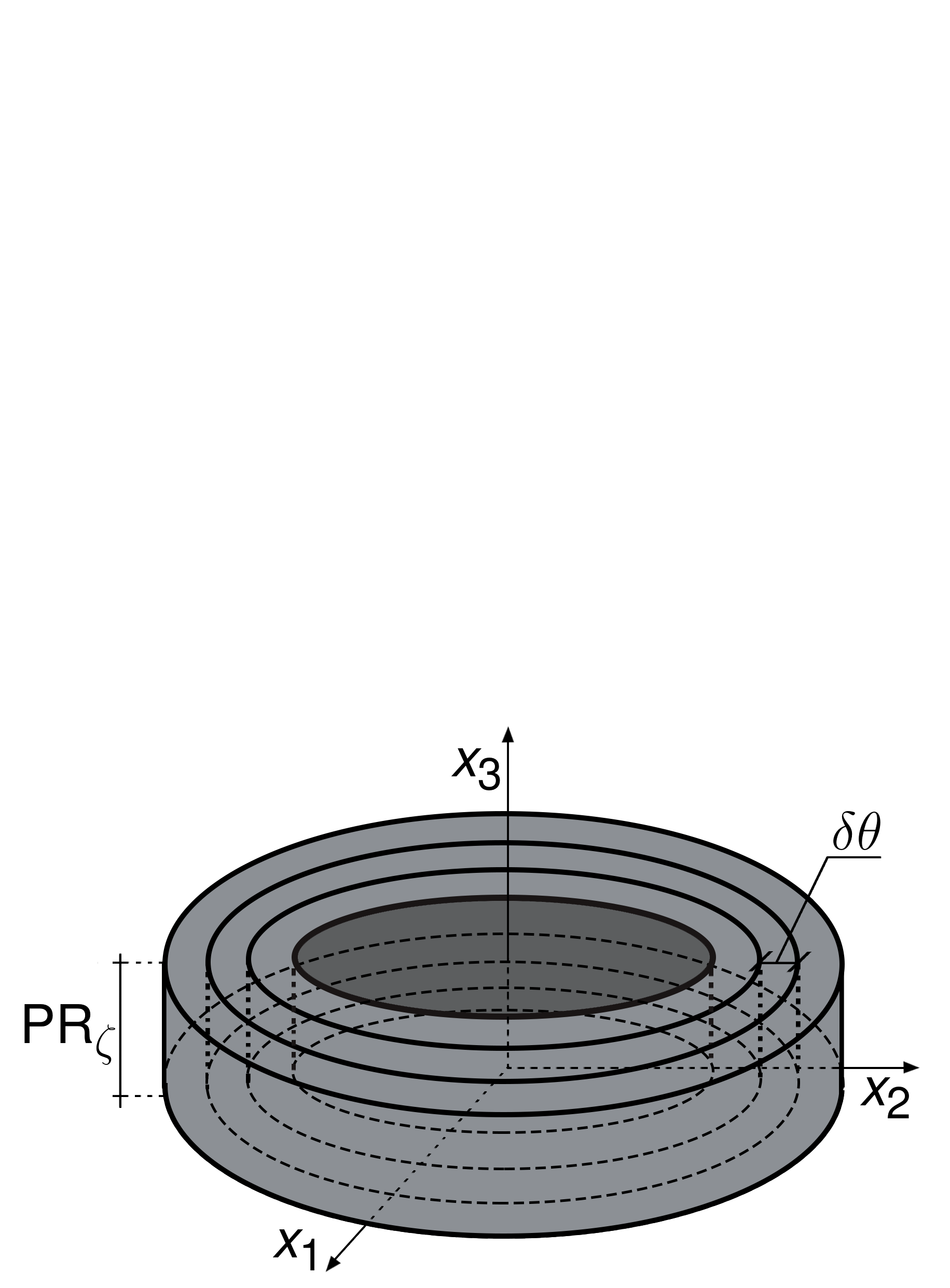}
    \includegraphics[width=50mm]{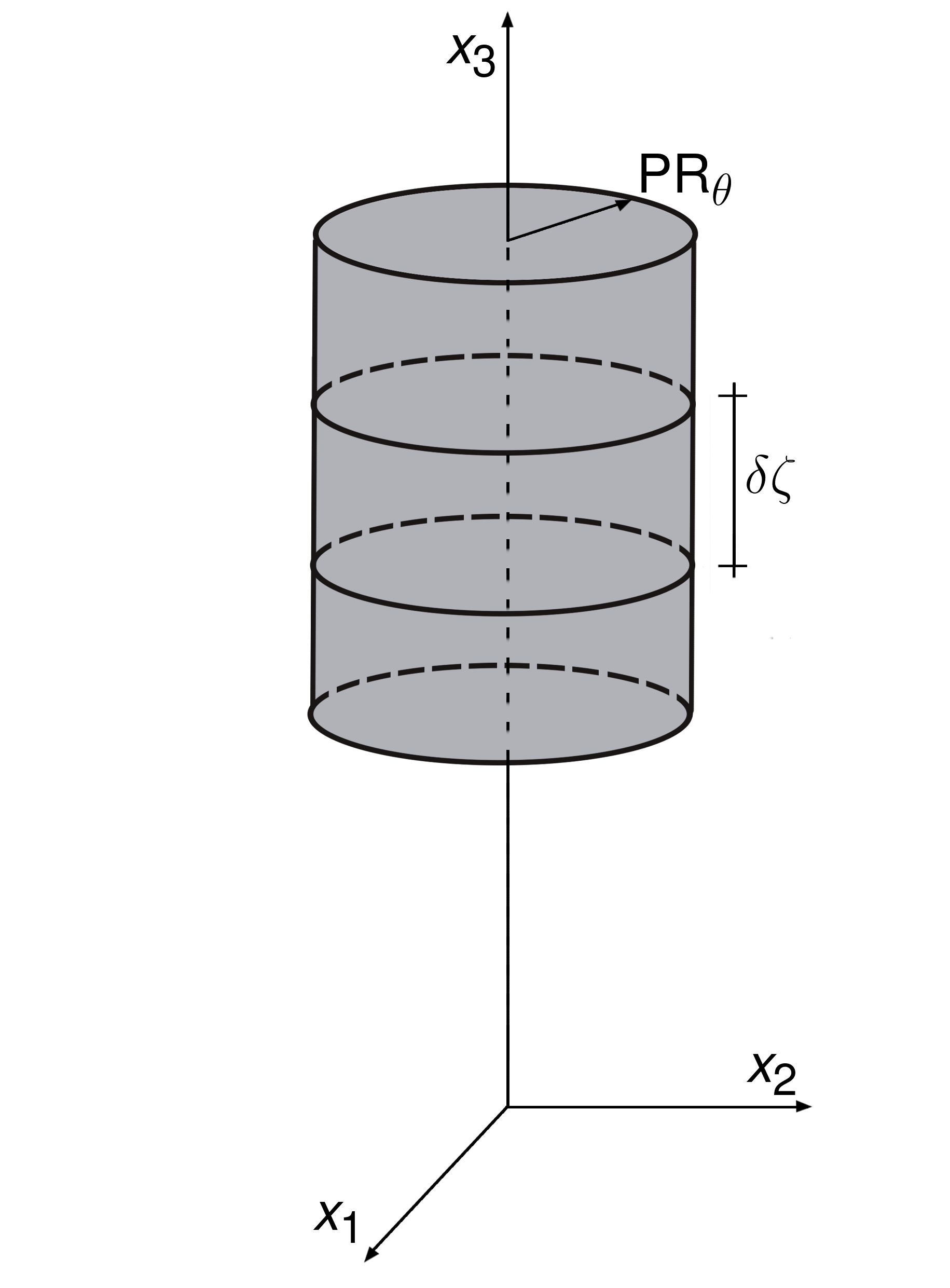}
    \caption[
    Binning scheme used to measure the void-galaxy cross-correlation function in observable space.
    ]{
    Binning scheme used to measure the void-galaxy cross-correlation function in observable space.
    In each panel, the coordinate system of the MXXL in the plane-parallel approximation was represented, but with the origin located at a void centre.
    The $x_3$-axis was chosen as the line-of-sight direction.
    The correlation function is estimated by counting galaxies inside these bins.
    \textit{Left-hand panel.}
    A bin is a cylindrical shell oriented along the LOS, with internal radius $\theta_\mathrm{int}$, external radius $\theta_\mathrm{ext}$, a lower height $\zeta_\mathrm{low}=|z_\mathrm{low}-z'|$, and upper height $\zeta_\mathrm{up}=|z_\mathrm{up}-z'|$.
    \textit{Central panel.}
    Binning scheme for the POS correlation function: a set of nested cylindrical shells distributed across the POS with dimensions $\theta_\mathrm{int}$, $\theta_\mathrm{ext}$, $\zeta_\mathrm{low}=0$ and $\zeta_\mathrm{up}=\mathrm{PR_\zeta}$.
    $\mathrm{PR_\zeta}$ is the redshift projection range, whereas $\delta \theta = \theta_\mathrm{ext} - \theta_\mathrm{int}$, the corresponding binning step.
    \textit{Right-hand panel.}
    Binning scheme for the LOS correlation function: a string of filled cylinders oriented along the LOS with dimensions $\theta_\mathrm{int}=0$, $\theta_\mathrm{ext}=\mathrm{PR_\theta}$, $\zeta_\mathrm{low}$ and $\zeta_\mathrm{up}$.
    $\mathrm{PR_\theta}$ is the angular projection range, whereas $\delta \zeta = \zeta_\mathrm{up} - \zeta_\mathrm{low}$, the corresponding binning step.
    }
    \label{fig:binning}
\end{figure}

Working in observable space, the correlation function is then represented as a two-dimensional map $\xispo$ in the coordinates $\theta$ and $\zeta$.
This statistic is estimated by counting void-galaxy pairs inside cylindrical bins defined in observable space.
In this geometry, a bin is a cylindrical shell oriented along the LOS with dimensions characterised by an internal radius $\theta_\mathrm{int}$, an external radius $\theta_\mathrm{ext}$, a lower height $\zeta_\mathrm{low}=|z_\mathrm{low}-z'|$ and an upper height $\zeta_\mathrm{up}=|z_\mathrm{up}-z'|$.
This is schematically represented in the left-hand panel of Figure~\ref{fig:binning}.

If we project $\xispo$ towards the plane of the sky (towards the $\theta$-axis) in a given range of redshift separations $\mathrm{PR}_\zeta$, we get the \textit{POS correlation function}, denoted by $\xiposo$, which is a function only of the angular distance $\theta$.
Conversely, if we project $\xispo$ towards the line of sight (towards the $\zeta$-axis) in a given angular range $\mathrm{PR}_\theta$, we get the \textit{LOS correlation function}, denoted by $\xiloso$, which is a function only of the redshift difference $\zeta$.
These projections of the correlation function can be considered as special cases in the cylindrical binning scheme.
Specifically, the POS correlation function is measured from a set of nested cylindrical shells distributed across the POS, whereas the LOS correlation function is measured from a string of filled cylinders oriented along the LOS.
Concretely, the scheme for the POS projection involves bins with dimensions $\theta_\mathrm{int}$, $\theta_\mathrm{ext}$, $\zeta_\mathrm{low}=0$ and $\zeta_\mathrm{up}=\mathrm{PR_\zeta}$.
In this way, $\delta \theta := \theta_\mathrm{ext} - \theta_\mathrm{int}$ is the POS binning step, assumed equal for all bins.
Similarly, the scheme for the LOS projection involves bins with dimensions $\theta_\mathrm{int}=0$, $\theta_\mathrm{ext}=\mathrm{PR_\theta}$, $\zeta_\mathrm{low}$ and $\zeta_\mathrm{up}$.
In a similar way, $\delta \zeta := \zeta_\mathrm{up} - \zeta_\mathrm{low}$ is the LOS binning step.
These two binning schemes are schematically represented in the central and right-hand panels of Figure~\ref{fig:binning}.
The goal of our method is to measure and model these two complementary functions by taking into account the RSD and AP distortions, together with the mixture of scales due to the projection range.
In this way, we can perform an AP test to constrain the cosmological parameters $\Omega_m$ and $\beta$, as we will see.

Although the final intention is to apply the method to a galaxy spectroscopic survey, the aim of this chapter is to evaluate its performance and calibrate it, for which we have used the dark-matter haloes of the MXXL as matter tracers.
For this simulation study, in which the box is assumed to be a simple and periodic mock, complete in volume and without any complicated selection functions, there is no need to employ any estimators involving a random comparison sample.
Therefore, the estimation of the correlation function can then be thought as analogous to the natural estimator, Eq.~(\ref{eq:estimator_natural}), in which the quantity $RR$ can be computed analytically.
In this context, given a bin in observable space labelled by $(i,j)$, the estimator for the correlation function can be written as
\begin{equation}
    \hat{\xi}(\theta_i, \zeta_j) = \frac{DD_{ij}}{RR_{ij}} - 1,
    \label{eq:estimator_voids}
\end{equation}
where $DD_{ij}$ is computed by counting void-halo pairs inside the bin, and $RR_{ij}$, the expected number of pairs in a homogeneous distribution, as the product of the density of tracers, the volume of the bin and the total number of voids.
Here, $(\theta_i, \zeta_j)$ denotes the coordinates of the geometrical centre of the bin, merely in order to label it.

\begin{figure}
    \centering
    \includegraphics[width=79mm]{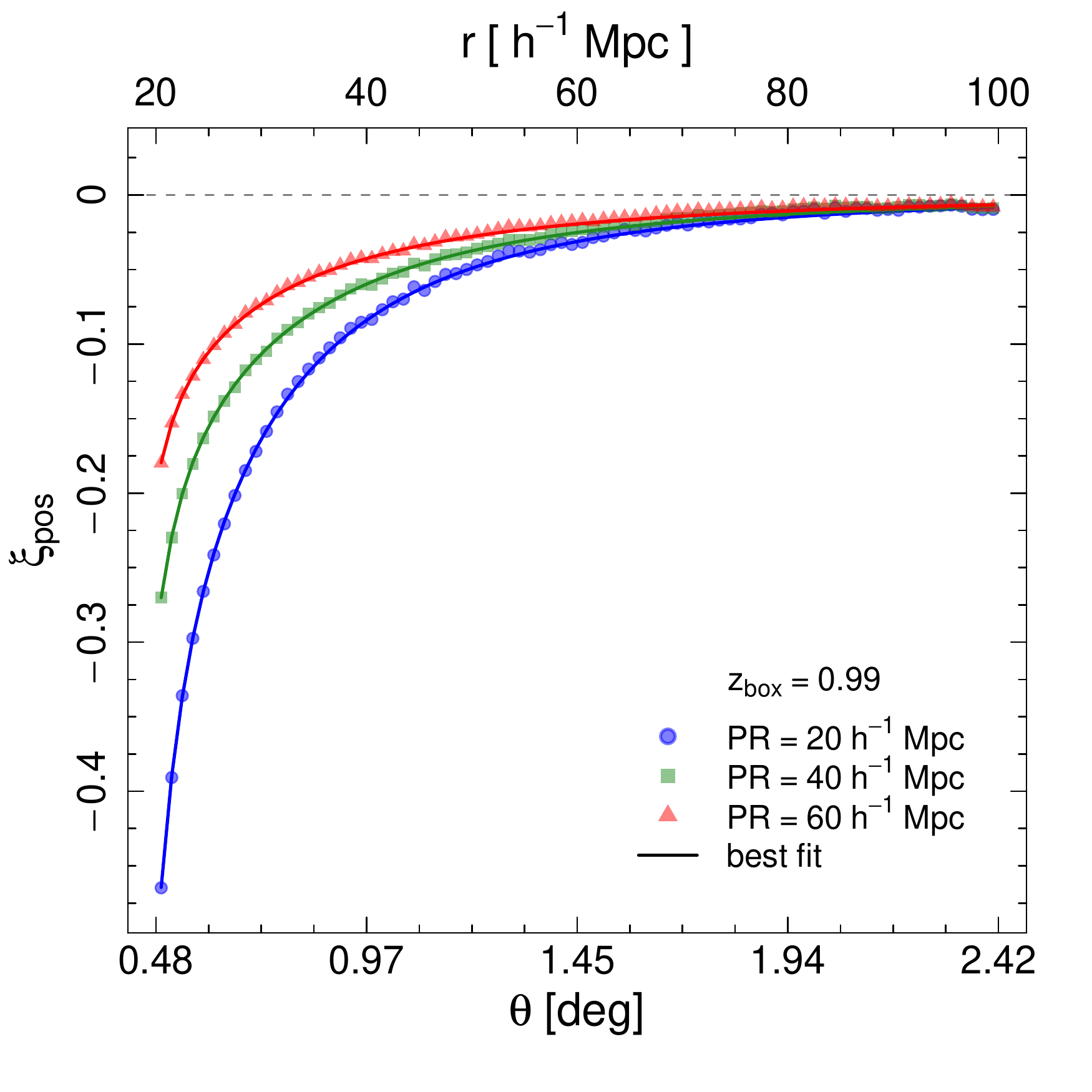}
    \includegraphics[width=79mm]{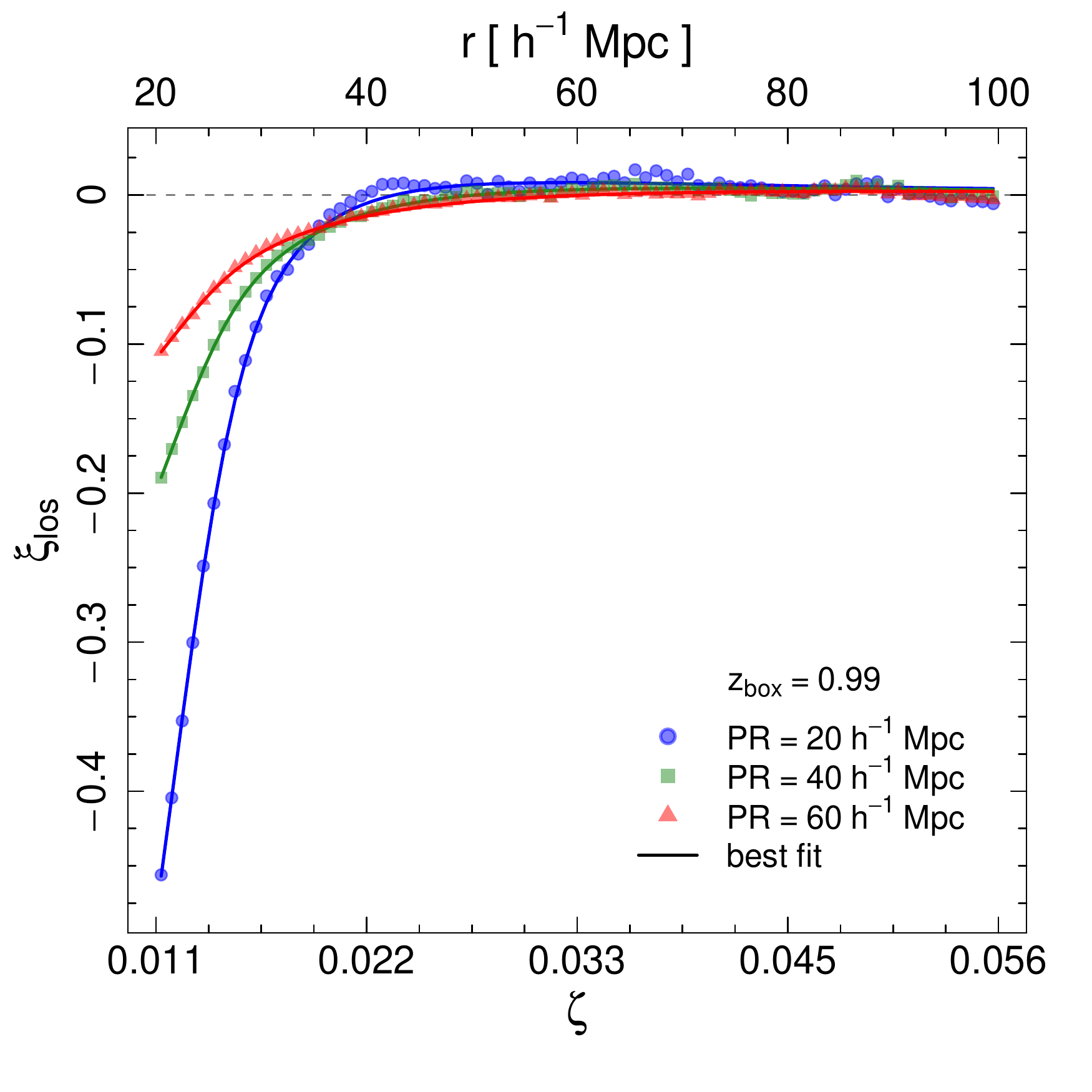}
    \caption[
    POS and LOS projections of the void-halo cross-correlation function for a sample of voids selected from the MXXL simulation.
    ]{
    Plane-of-sky (\textit{left-hand panel}) and line-of-sight (\textit{right-hand panel}) void-halo cross-correlation functions corresponding to the void sample taken from the MXXL snapshot $\zbox=0.99$.
    Here are shown the cases $\mathrm{PR}/\hmpc = 20$ (blue circles), $40$ (green squares) and $60$ (red triangles).
    Both the observable-space (below) and real-space (above) axes are shown for comparison.
    The solid curves are the theoretical predictions obtained after the application of the model from Section~\ref{sec:cosmotest_model} with the best fitted parameters obtained in Section~\ref{sec:cosmotest_test}.
    }
    \label{fig:correlations_ospace}
\end{figure}

To perform the analysis, we used the void catalogues defined in Section~\ref{sec:data_voids1} and summarised in Table~\ref{tab:catalogues1}.
Specifically, for each MXXL snapshot, we selected a void sample with sizes between $20 \leq R_\mathrm{v}/\hmpc \leq 25$, as indicated in Figure~\ref{fig:rad_dist}.
For better clarity in the discussion of the results, we will express the projection ranges in units of $\hmpc$ using the MXXL cosmology.
In these units, we took equal projection ranges in both directions, so that we will refer to both with the common notation PR.
We analysed $8$ different projection ranges: $\mathrm{PR}/\hmpc = 1$, $5$, $10$, $20$, $30$, $40$, $50$ and $60$.
As an example, Figure~\ref{fig:correlations_ospace} shows the projected correlation functions $\xiposo$ and $\xiloso$ corresponding to the void sample taken from the snapshot\footnote{In this chapter, $\zbox$ will denote the redshift of the MXXL snapshots.} $\zbox=0.99$, for the cases $\mathrm{PR}/\hmpc=20$ (blue circles), $40$ (green squares) and $60$ (red triangles).
The POS and LOS binning steps were selected in such a way that each step has an approximate length of $1~\hmpc$.
The error bars, taken from the diagonal of the associated covariance matrices (see Section~\ref{subsec:cosmotest_test_mcmc}), are not shown because they are smaller than the data points symbols.
It can be seen that the profiles flatten as the PR increases, and that the LOS projection is more affected by RSD than the POS projection.
The remaining snapshots show a similar behaviour.
By way of comparison, we show both the observable (below) and real-space (above) axes.
We also show the theoretical predictions (solid curves) obtained after the application of the model from Section~\ref{sec:cosmotest_model} with the best fitting parameters obtained from the likelihood analysis of Section~\ref{sec:cosmotest_test}.

\section{Model}
\label{sec:cosmotest_model}

In this section, we present a physical model for the observable-space correlation function for a general cylindrical binning scheme, in which each bin is characterised by the quantities $\theta_\mathrm{int}$, $\theta_\mathrm{ext}$, $\zeta_\mathrm{low}$ and $\zeta_\mathrm{up}$.
This model takes into account the RSD and AP effects, and also the mixture of scales due to the geometry and sizes of the bins.
In this way, the projected correlation functions can be treated as special cases with the appropriate bin limits: $\theta_\mathrm{int}$, $\theta_\mathrm{ext}$, $\zeta_\mathrm{low}=0$ and $\zeta_\mathrm{up} = \mathrm{PR}_\zeta$ for the POS projection; $\theta_\mathrm{int}=0$, $\theta_\mathrm{ext}=\mathrm{PR}_\theta$, $\zeta_\mathrm{low}$ and $\zeta_\mathrm{up}$ for the LOS projection.
Firstly, we will give the full treatment, the one that must be applied when working with a spectroscopic survey.
Then, we will turn to the simplifications used for our simulation study.

The first step is to model the geometry of an observable-space cylindrical bin $(i,j)$ and to map the redshift-space scales $\sigma$ and $\pi$ that are involved.
First, note that for a given void centre at redshift $z'$, there are two possible values of the redshift $z$ for each limit in $\zeta$ that match the criterion $\zeta=|z-z'|$.
Specifically, tracers inside the bin will have angular coordinates between $\theta_\mathrm{int}$ and $\theta_\mathrm{ext}$, and their redshifts $z$ will belong to either one of these two disjoint intervals: $(z'+\zeta_\mathrm{low}, z'+\zeta_\mathrm{up})$ or $(z'-\zeta_\mathrm{up}, z'-\zeta_\mathrm{low})$.
Therefore, each cylindrical bin corresponds to two volumes on the data. 
Given a set of cosmological parameters, these two regions correspond to two different volumes in redshift space.

Taking these considerations into account, the expected number of void-tracer pairs, $DD_{ij}$, can be calculated by expressing Eq.~(\ref{eq:estimator_voids}) in differential form and then integrating over the volume of the slice taken from the catalogue and the volume of the bin, obtaining the following expression:
\begin{equation}
\begin{aligned}
	DD_{ij} =& 2\,\hat{\pi}\int_{z'_\mathrm{min}}^{z'_\mathrm{max}} dz'~\chi^2(z')~n_\mathrm{v}(z')~V_\mathrm{slice}\\
    &\left[
    \int_{z'+\zeta_\mathrm{low}}^{z'+\zeta_\mathrm{up}}dz~\frac{d\chi}{dz}(z)~n_\mathrm{t}(z)
    \int_{\theta_\mathrm{int}}^{\theta_\mathrm{ext}}d\theta~\theta~\left[1 + \xi(\sigma, \pi)\right]
    \right.+\\
    &\left.\int_{z'-\zeta_\mathrm{up}}^{z'- \zeta_\mathrm{low}}dz~\frac{d\chi}{dz}(z)~n_\mathrm{t}(z)
    \int_{\theta_\mathrm{int}}^{\theta_\mathrm{ext}}d\theta~\theta~[1 + \xi(\sigma, \pi)]\right],\\
\end{aligned}
    \label{eq:data_pairs}
\end{equation}
where $n_\mathrm{v}(z')$ is the number density distribution of voids in the slice $z'_\mathrm{min} \leq z' \leq z'_\mathrm{max}$, $V_\mathrm{slice}$ the volume of this slice, and $n_\mathrm{t}(z)$ the number density distribution of tracers in the bin.
Here, $\xisp$ is a theoretical correlation function defined in redshift space that must be modelled considering the RSD effect.
Its arguments, $\sigma = \sigma(\theta,z')$ and $\pi = \pi(z,z')$, depend on the observable-space coordinates by means of Eqs.~(\ref{eq:sigma_ap2}) and (\ref{eq:pi_ap2}).
The expected number of pairs in a uniform distribution, $RR_{ij}$, is given in a similar fashion:
\begin{equation}
\begin{aligned}
	RR_{ij} =& \hat{\pi} \left(\theta^2_\mathrm{ext}-\theta^2_\mathrm{int}\right)\int_{z'_\mathrm{min}}^{z'_\mathrm{max}} dz'~\chi^2(z')~n_\mathrm{v}(z')~V_\mathrm{slice}\\
        &\left[
        \int_{z'+\zeta_\mathrm{low}}^{z'+\zeta_\mathrm{up}}dz~\frac{d\chi}{dz}(z)~n_\mathrm{t}(z)
        + \int_{z'-\zeta_\mathrm{up}}^{z'- \zeta_\mathrm{low}}dz~\frac{d\chi}{dz}(z)~n_\mathrm{t}(z) \right].
\end{aligned}
	\label{eq:random_pairs}
\end{equation}
In this way, combining Eqs.~(\ref{eq:data_pairs}) and (\ref{eq:random_pairs}) into Eq.~(\ref{eq:estimator_voids}), we get a theoretical estimation for $\hat{\xi}(\theta_i, \zeta_j)$.
This is a general expression that takes into account the AP effect and all possible mixture of scales due to the binning scheme.

The scope of this work is to present a novel cosmological test focusing the analysis on the effects of the different types of distortions that arise in measurements.
In order to test these effects, we used simplified mock catalogues taken from the MXXL simulation.
When applied to observational data, the full treatment developed above must be considered.
For the moment, for our simulation study, some simplifications can be done.
On the one hand, $n_\mathrm{t}$ is a constant function.
On the other hand, we have assumed a unique redshift for void centres, the one corresponding to the MXXL snapshot, $z_\mathrm{box}$.
In this way, $n_\mathrm{v}(z')$ can be thought of as a Dirac-delta distribution.
Therefore, according to Eqs.~(\ref{eq:data_pairs}) and (\ref{eq:random_pairs}), the estimator of the correlation function simplifies to the following expression:
\begin{equation}
\begin{aligned}
	\hat{\xi}(\theta_i, \zeta_j) = & -1 + 2~\left(\theta^2_\mathrm{ext}-\theta^2_\mathrm{int}\right)^{-1}  \\ 
    &\left[\int_{z_\mathrm{box}+\zeta_\mathrm{low}}^{z_\mathrm{box}+\zeta_\mathrm{up}} \frac{dz}{H(z)}
    \int_{\theta_\mathrm{int}}^{\theta_\mathrm{ext}}d\theta~\theta~\left[1 + \xi(\sigma, \pi)\right]\right. +\\
    &\left.\left.\int_{z_\mathrm{box}-\zeta_\mathrm{up}}^{z_\mathrm{box}- \zeta_\mathrm{low}}\frac{dz}{H(z)}
    \int_{\theta_\mathrm{int}}^{\theta_\mathrm{ext}}d\theta~\theta~[1 + \xi(\sigma, \pi)]\right]\right/\\
    &\left[\int_{z_\mathrm{box}+\zeta_\mathrm{low}}^{z_\mathrm{box}+\zeta_\mathrm{up}} \frac{dz}{H(z)} +\int_{z_\mathrm{box}-\zeta_\mathrm{up}}^{z_\mathrm{box}- \zeta_\mathrm{low}} \frac{dz}{H(z)}~\right],
\end{aligned}
	\label{eq:xi_gd}
\end{equation}
where Eq.~(\ref{eq:dcom_z}) was used to express the derivative of the comoving distance.

The next step is to model $\xisp$, the redshift-space correlation function, needed in Eqs.~(\ref{eq:data_pairs}) or (\ref{eq:xi_gd}), that takes into account the RSD effect.
We chose the GS model developed by \citeonline{clues2} and discussed in Section~\ref{subsec:voids_rsd_gsm}.
Specifically, what we need is Eq.~(\ref{eq:gsm0}), which we rewrite here as it is a fundamental result for the development of the test:
\begin{equation}
    1 + \xi(\sigma, \pi) = 
    \int_{-\infty}^{\infty} [1 + \xi(r)] \frac{1}{\sqrt{2\hat{\pi}}\sigma_\mathrm{v}}
    \mathrm{exp} \left[- \frac{(v_\parallel - v(r)\frac{r_\parallel}{r})^2}{2\sigma_\mathrm{v}^2} \right]
    \mathrm{d}v_\parallel.
    \label{eq:gsm}
\end{equation}
This relation indicates that the computation of $\xisp$ needs a recipe for the density profile, $\xi(r)$, the velocity profile, $v(r)$, and the velocity dispersion, $\sigma_\mathrm{v}$.


\subsection{Density profile}
\label{subsec:cosmotest_model_densdiff}

Before proceeding to a model for the real-space density profile, it is instructive to first study the characteristics of this profile by measuring it directly from the simulation.
This analysis will help us to build such a model.

\begin{figure}
    \centering
    \includegraphics[width=79mm]{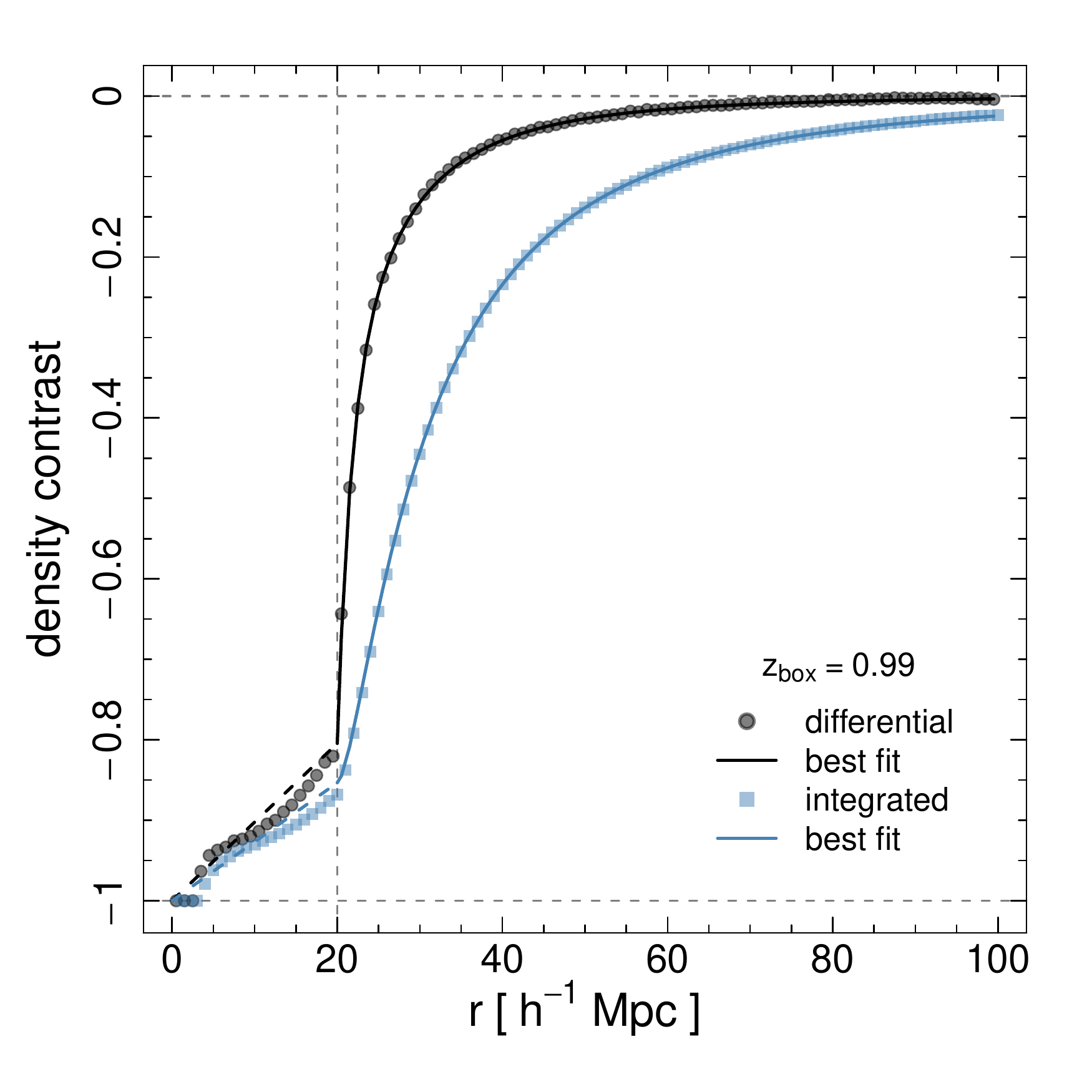}
    \includegraphics[width=79mm]{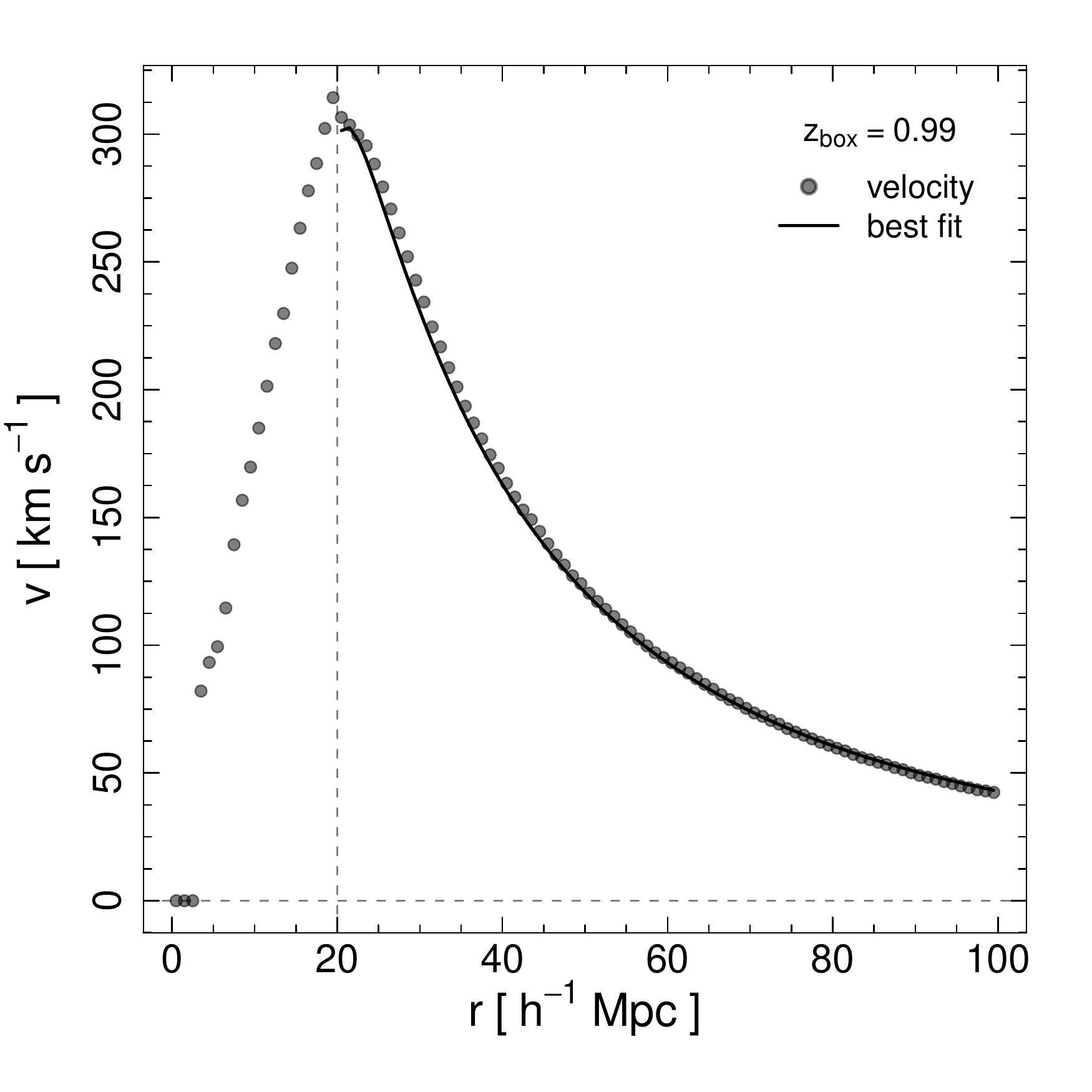}
    \caption[
    Density contrast and velocity profiles for a sample of voids selected from the MXXL simulation.
    ]{
    \textit{Left-hand panel.}
    Density contrast profile (black circles) and its integration (light-blue squares) corresponding to the void sample taken from the MXXL snapshot $\zbox=0.99$.
    \textit{Right-hand panel.}
    The corresponding radial velocity profile (black circles).
    In both panels, the solid curves are the theoretical predictions obtained after the application of the model from Section~\ref{sec:cosmotest_model} with the best fitted parameters obtained in Section~\ref{sec:cosmotest_test} for the case $\mathrm{PR}=40\hmpc$.
    }
    \label{fig:densvel}
\end{figure}

The left-hand panel of Figure~\ref{fig:densvel} shows, with black circles, the density profile corresponding to the void sample taken from the snapshot $\zbox = 0.99$.
This profile was measured using the prescription of Eq.~(\ref{eq:estimator_natural}) by counting void-halo pairs within a spherical binning scheme.
In this geometry, a bin is a void-centric spherical shell with internal radius $r_\mathrm{int}$, external radius $r_\mathrm{ext}$, and a binning step $\delta r := r_\mathrm{ext} - r_\mathrm{int}$, equal for all bins.
Here, we used a binning step of $\delta r = 1~\hmpc$.
The remaining snapshots show a similar behaviour.
Note the three reference dashed lines in the figure: (i) the horizontal $\xi = -1$ line, which indicates total emptiness, as is the case near the void centres, (ii) the horizontal $\xi = 0$ line, which is the mean value of the Universe, and (iii) the vertical $r = r_\mathrm{cut}$ line, which indicates the minimum radius of the sample and can be thought as a representative border between the inner parts of the voids and their environment.
The same panel of the figure also shows, with light-blue squares, the corresponding integrated density profile, $\Delta(r)$.
As can be seen, the sample is represented by an increasing profile that tends to the mean density of the Universe at large distances, that is, the sample is mostly composed of R-type voids.
One of the reasons that R-type voids are more convenient to perform the cosmological test is because they are less sensitive to non-linear effects compared to the S-type ones.

In Section~\ref{sec:voids_density}, we discussed that there is not a first-principle model for $\xi(r)$, but there are many empirical and parametric approaches in the literature.
In our case, we developed an own parametric model suitable for our R-type void samples:
\begin{equation}
	\xi(r) =
	\left\{
		\begin{array}{ll}
			Ar - 1 & \mathrm{if} ~ r < r_\mathrm{cut}\\
            -\xi_0 \left[ \left( \frac{r}{r_0} \right)^{-3} + \left( \frac{r}{r_0} \right)^{-\alpha} \right]  &
            \mathrm{if} ~ r \geq r_\mathrm{cut},
		\end{array}
	\right.
	\label{eq:dens_model}
\end{equation}
where $\xi_0$, $r_0$ and $\alpha$ are the three parameters of the model.
Note that this model is not fixed, in the sense that it depends on the global cosmology as $r_\mathrm{cut}$ and $r$ must be calculated from the relations given by Eqs.~(\ref{eq:sigma_rsd}) and (\ref{eq:pi_rsd}), which in turn are given by Eqs.~(\ref{eq:sigma_ap2}) and (\ref{eq:pi_ap2}).
Three ranges can be identified: (i) the \textit{void-inner zone} $r<r_\mathrm{cut}$, (ii) the \textit{void-environment zone} $r_\mathrm{cut} \le r \le 100~\hmpc$, and (iii) the \textit{void-outer zone} $r > 100~\hmpc$.

Let us start with the void-environment zone.
This range is quantified by a double power law with slopes $-3$ and $\alpha$.
The first exponent describes the behaviour near the void walls, whereas the second one, in the more remote areas.
The other two parameters are an amplitude $\xi_0$ and a pivot scale $r_0$, where the slope changes.
The model fails to describe the void-inner zone, since this function tends to $-\infty$ as $r$ tends to zero.
It also fails to describe the outer zone beyond $100~\hmpc$ because of the BAO feature.

We now describe the remaining zones.
The void-inner zone, on the one hand, is not relevant in terms of correlation signal and is not trivial to model.
Moreover, the condition $\Delta(R_\mathrm{v})~=~\Delta_\mathrm{id}$ imposed by the spherical void finder has a direct impact on the shape of the profile in this zone.
This is evident in Figure~\ref{fig:densvel}, where the curves break at $r = r_\mathrm{cut}$.
For these reasons, we decided to measure correlations only in the void-environment zone, as Figure~\ref{fig:correlations_ospace} reflects.
Nevertheless, the contributions from the inner and outer zones when reproducing RSD are in fact significant, specially the inner zone.
This is because some scales from the inner zone can be shifted into the environment zone in redshift space, making a significant contribution to RSD.
In the same way, some scales from the environment zone can be shifted into the outer zone. 
Hence, these ranges must be modelled, even though data from there are not used.
For the inner zone, we found that it is sufficient a segment connecting the points $(r=0, \xi=-1)$ (void centre) and $(r_\mathrm{cut}, \xi_\mathrm{cut})$ (void wall), where $\xi_\mathrm{cut} := \xi(r_\mathrm{cut})$, in such a way that the segment and the rest of the curve are connected.
The slope of this segment is
\begin{equation}
	A = \frac{1}{r_\mathrm{cut}}~(\xi_\mathrm{cut} + 1).
	\label{eq:slope}
\end{equation}
For the outer zone, on the other hand, it is sufficient to extend the scope of validity of the double power-law model, finding no significant deviations.
Although the slope $A$ is not a known quantity, since $\xi_\mathrm{cut}$ can only be truly known in real space, in Section~\ref{subsec:cosmotest_model_densint} we will give an approximate theoretical value for it.


\subsection{Velocity profile}
\label{subsec:cosmotest_model_vel}

In this subsection, we continue the analysis with the velocity profile of the void sample, shown in the right-hand panel of Figure~\ref{fig:densvel}.
As in the case of the density profile, the remaining snapshots show a similar behaviour.
Each data point represents the average of the void-centric radial component of the peculiar velocities of the haloes around voids using the same binning scheme as in the case of the density profile.
There are two reference dashed lines: (i) the horizontal $v = 0$ line, which represents the global mean value, and (ii) the vertical $r = r_\mathrm{cut}$ line.
As can be seen, the sample is represented by a profile that shows only expansion velocities, as expected for R-type voids.
The characteristic distance $r_\mathrm{cut}$ represents the change between increasing (inside the voids) and decreasing (outside the voids) expansion.

In Section~\ref{sec:voids_velocity}, we showed that $v(r)$ can be analytically modelled following linear perturbation theory.
What we need is Eq.~(\ref{eq:velocity0}), which we rewrite here as it is a fundamental result for the development of the test and for the analysis of the following chapters:
\begin{equation}
    v(r) = - \frac{1}{3} \frac{H(z)}{(1+z)} \beta(z) r \Delta(r).
    \label{eq:velocity}
\end{equation}
The RSD parameter, $\beta(z)$, is thus a fundamental cosmological parameter in the test. 
The dynamical information of our void sample regarding its type is contained then in the integrated density contrast profile, $\Delta(r)$, which we discuss below.


\subsection{Integrated density profile}
\label{subsec:cosmotest_model_densint}

The integrated density contrast profile can be modelled by combining Eqs.~(\ref{eq:delta_int}) and (\ref{eq:dens_model}):
\begin{equation}
	\Delta(r) =
	\left\{
		\begin{array}{ll}
			\frac{3}{4}Ar - 1 & \mathrm{if} ~ r < r_\mathrm{cut} \\
            \frac{3}{r^3} \left[ \frac{Ar_\mathrm{cut}^4}{4} - \frac{r_\mathrm{cut}^3}{3} + I(r) - I(r_\mathrm{cut}) \right] &
            \mathrm{if} ~ r \geq r_\mathrm{cut},
		\end{array}
	\right.
	\label{eq:deltaint_model}
\end{equation}
where $I(r)$ is a primitive function for the integral given in Eq.~(\ref{eq:delta_int}) (without the prefactor $3/r^3$) and with Eq.~(\ref{eq:dens_model}) as integrand:
\begin{equation}
    I(r) = -\xi_0 \left[ r_0^3 \mathrm{ln}(r)
         + \frac{r_0^\alpha}{3-\alpha} r^{3-\alpha} \right] + \mathrm{constant}.
	\label{eq:integral}
\end{equation}

To complete the model, we need an approximate value for the slope $A$.
For this, we realised that the true value $\Delta_\mathrm{cut}:=\Delta(r_\mathrm{cut})$ can be approximated by the identification-method value $\Delta_\mathrm{id}$ shown in Table~\ref{tab:catalogues1}.
Then, from Eq.~(\ref{eq:deltaint_model}) for $r<r_\mathrm{cut}$,
\begin{equation}
    A \approx \frac{4}{3 r_\mathrm{cut}} (\Delta_\mathrm{id} + 1).
	\label{eq:slope2}
\end{equation}
From Eqs.~(\ref{eq:slope}) and (\ref{eq:slope2}), we can also give an approximate value for $\xi_\mathrm{cut}$:
\begin{equation}
    \xi_\mathrm{cut} \approx \frac{4}{3}(\Delta_\mathrm{id} + 1) - 1.
	\label{eq:xicut_value}
\end{equation}
The validity of these approximations can be corroborated in Figure~\ref{fig:densvel} visually, where the approximated values for $\Delta_\mathrm{cut}$ and $\xi_\mathrm{cut}$ (where the dashed segments and solid curves match) are near the corresponding true values (data points at $r=r_\mathrm{cut}$).

In the next section, we will evaluate in detail the performance of the models that we have developed for the correlation function and the real-space density and velocity profiles.
For the moment, we just mention that the solid curves in Figure~\ref{fig:densvel} are the predictions of Eqs.~(\ref{eq:dens_model}), (\ref{eq:velocity}), and (\ref{eq:deltaint_model}) for the density and velocity profiles using the best fitted parameters obtained from the likelihood analysis of Section~\ref{sec:cosmotest_test} applied to the measurements of the projected correlation functions.
The good agreement between the predictions and the data points shows that our method is capable of recovering these profiles.
This is part of the calibration of the test.


\section{Testing the method}
\label{sec:cosmotest_test}

In this section, we test the performance of our model in reproducing the observed features on the projected POS and LOS correlation functions, as well as on the corresponding real-space density and velocity profiles.
The aim is to extract cosmological information from the parameters involved in the model.
For the implementation of the test, we adopted a flat-$\Lambda$CDM cosmology.
Moreover, we considered a fixed $H_0$ value, the one corresponding to the MXXL.
When applied to a real data set, $H_0$ can be extracted from a different method.
In view of this, the parameters involved can be summarised in two sets: (i) the cosmological set $\lbrace \Omega_m, \beta \rbrace$ (from Eqs.~\ref{eq:sigma_ap2}, \ref{eq:pi_ap2} and \ref{eq:velocity}), and (ii) the nuisance set $\lbrace \sigma_\mathrm{v}, \xi_0, r_0, \alpha \rbrace$ (from Eqs.~\ref{eq:gsm} and \ref{eq:dens_model}).
We will focus our analysis mainly on the cosmological set.

A comment worth mentioning about the cosmological set.
Within the standard model, $\Omega_m$ and $\beta=f/b$ are not independent, they are connected via Eqs.~(\ref{eq:fz}) and (\ref{eq:beta}).
As we mentioned in Section~\ref{sec:voids_cosmology}, theories of Modified Gravity predict that deviations from the predictions of General Relativity should be more pronounced in unscreened low-density environments, making voids a powerful tool for detecting them.
Moreover, a simple linear bias relation may not be a suitable choice for the case of voids.
In such cases, these relations must be modified.
Therefore, in order to detect any tensions with these standard assumptions, it is instructive to take $\beta$ as a free parameter of the model, and not to incorporate the explicit dependence on $\Omega_m$ or $b$.
In this way, we keep the method more general.

In order to constrain these parameters, we implemented a Bayesian analysis based on the Markov Chain Monte Carlo (MCMC) technique using a Metropolis-Hastings sampler \cite{metropolis_mcmc_1953,hastings_mcmc_1970}.
Although we adopted a standard flat-$\Lambda$CDM cosmology, the method can be easily generalised to incorporate other models as well.
In fact, we implemented in our code the analytical and numerical expressions for the comoving distance in the general non-flat case given in terms of elliptical functions by \citeonline{cosmo_rollin}.

Given that AP distortions are sensitive to the redshift of void identification (Eqs.~\ref{eq:sigma_ap2} and \ref{eq:pi_ap2}), we analysed the performance of our test with the redshift using the three MXXL snapshots: $\zbox=0.51$, $0.99$ and $1.50$.
This is important in view of the new generation of galaxy spectroscopic surveys,
which will in general cover a volume with a median redshift larger than $0.5$, a significant improvement with respect to available surveys.


\subsection{Likelihood analysis and covariance matrices}
\label{subsec:cosmotest_test_mcmc}

Let us denote the likelihood function with $\mathcal{L}(\boldsymbol{\theta} | \boldsymbol{x})$, where $\boldsymbol{\theta}=\lbrace \Omega_m, \beta, \sigma_\mathrm{v}, \xi_0, r_0, \alpha \rbrace$ represents the parameter space, and $\boldsymbol{x}$, the measured data.
The MCMC chains explore $\mathcal{L}(\boldsymbol{\theta} | \boldsymbol{x})$ until they reach the equilibrium distribution near its maximum, that is, until they find the $\boldsymbol{\theta}$-values that make the data most probable with the corresponding confidence regions.
We took the \citeonline{gelman_rubin} convergence criterion, which compares the spread of the distribution of the means of the chains with the variance of the target distribution.
Once the chains satisfy this criterion, the unburned parts are discarded and the remaining ones are used to sample the likelihood function.

The $\mathcal{L}$-function is obtained by computing the differences between the measured and theoretical correlations for a given set of parameters:
\begin{equation}
    \mathrm{ln}(\mathcal{L}) = - \boldsymbol{\Delta\xi}^T \boldsymbol{C}^{-1} \boldsymbol{\Delta\xi} + \mathrm{constant}.
	\label{eq:likelihood}
\end{equation}
Here, both the measured and theoretical correlations are stored in vectors of the form $\boldsymbol{\xi}:=(\xi_\mathrm{pos},\xi_\mathrm{los})$ containing the correlation values of each bin, whereas $\boldsymbol{\Delta\xi}$ denotes the corresponding difference vector.
$\boldsymbol{C}$ denotes the associated covariance matrix.
Each element $C_{ij}$ is computed on the data by a jackknife resampling using the multivariate generalisation of \citeonline{efron_jackknife_1982}:
\begin{equation}
    C_{ij} = \frac{n-1}{n} \sum_{k=1}^n \left[ \xi_{(k)} - \xi_{(.)} \right]_i \left[ \xi_{(k)} - \xi_{(.)} \right]_j,
	\label{eq:cov}
\end{equation}
where $n$ is the number of jackknife realisations, $\xi_{(k)}$ the correlation function for the $k^\mathrm{th}$ jackknife realisation, and $\xi_{(.)}$ the average of $\xi_{(k)}$ over the $n$ realisations.

If $2m$ is the number of bins used to measure correlations ($m$ for the POS projection and $m$ for the LOS projection), then $2m$ is the length of the correlation vector containing the measurements: $\boldsymbol{\xi}[m]$, and therefore $2m~\times~2m$ is the dimension of the associated covariance matrix: $\boldsymbol{C}[2m \times 2m]$.
This is by far much smaller than in the traditional case, where the correlation measurements are contained in a matrix $\boldsymbol{\xi}[m \times m]$, and the associated covariance is a matrix $\boldsymbol{C}[m^2 \times m^2]$.
This is a key aspect of our method, first because the estimation of the inverse of a smaller matrix is numerically more stable, and second and more important, because the propagation of covariance errors into the likelihood estimates are substantially reduced, allowing us to use a smaller number of mock catalogues to estimate covariances \cite{{Taylor13,Dodelson13}}.
In this work, we used $2\cdot80=160$ bins, which implies correlation vectors of length $\boldsymbol{\xi}[160]$, with associated covariance matrices of dimension $\boldsymbol{C}[160 \times 160]$.

\begin{figure}
    \centering
    \includegraphics[width=\textwidth]{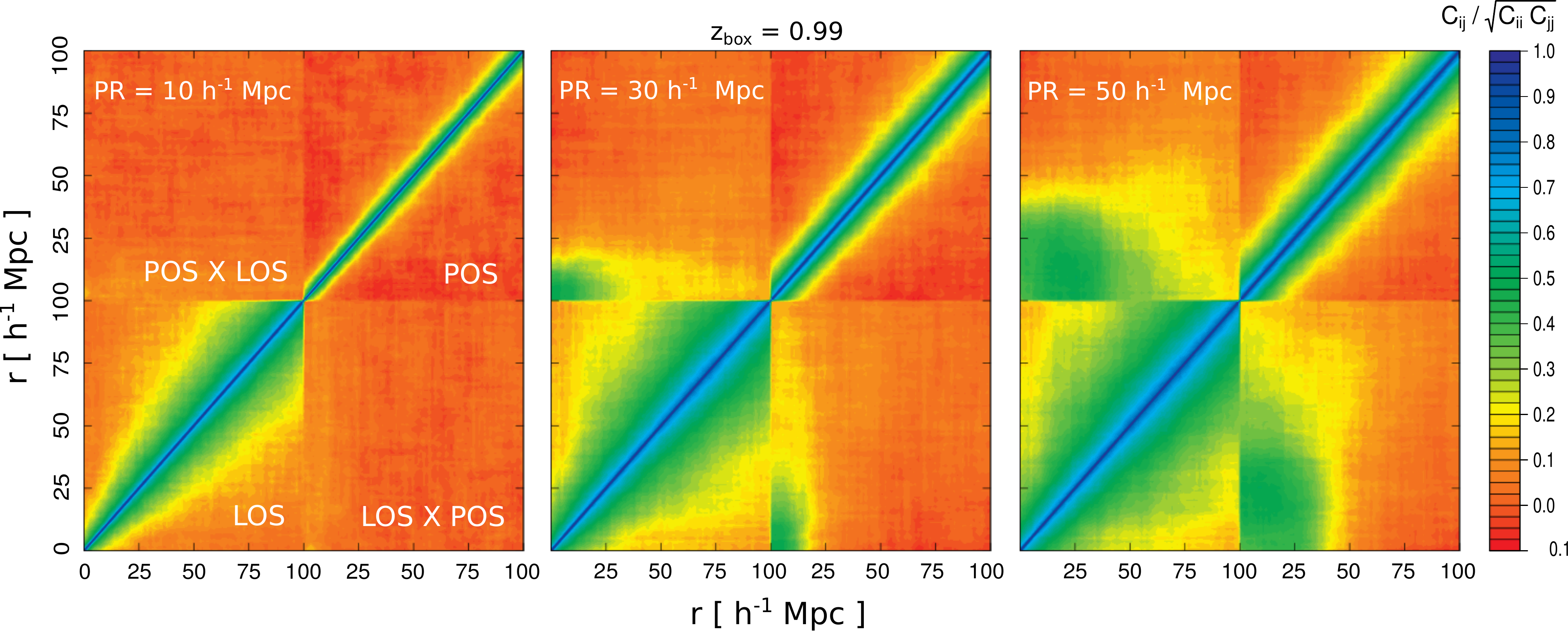}
    \caption[
    Normalised covariance matrices associated with the projected correlations method, measured with the MXXL simulation.
    ]{
    Normalised covariance matrices, shown as coloured contour maps, corresponding to the void sample taken from the MXXL snapshot $\zbox=0.99$ for the cases $\mathrm{PR}/\hmpc=10$, $30$ and $50$.
    Absolute values range from $0$ (red) to $1$ (blue).
    The axes are expressed in terms of the real-space coordinates.
    Four distinct quadrants can be distinguished on each matrix: the covariance for the LOS projection (bottom left), the covariance for the POS projection (top right), and the symmetric and cross $\mathrm{LOS} \times \mathrm{POS}$ covariance (bottom right and top left).
    The diagonal indicates the global variance and makes up the implicit error bars in Figure~\ref{fig:correlations_ospace}.
    Clear off-diagonal patterns arise as the PR increases.    
    }
    \label{fig:cov}
\end{figure}

Figure~\ref{fig:cov} shows the covariance matrices corresponding to the void sample taken from the snapshot $\zbox=0.99$ for the cases $\mathrm{PR}/\hmpc~=~10$, $30$ and $50$.
The remaining snapshots show a similar behaviour.
Technically, the so-called correlation matrices are shown: $C_{ij}/\sqrt{C_{ii}C_{jj}}$, which acquire absolute values from $0$ to $1$ encoded as a coloured contour map from red to blue.
To avoid confusion of nomenclature with the correlation function, we will refer to them simply as normalised covariance matrices.
The x- and y-axes are expressed in real-space coordinates for a better comparison.
Note that the matrices are not diagonal since the independence of the correlation values for bins at different scales can not be guaranteed.
In fact, they show clear patterns.
Focusing on an individual matrix, four distinct quadrants can be seen: (i) the bottom left quadrant is the covariance submatrix for the LOS  projection, (ii) the top right quadrant, the covariance submatrix for the POS projection, whereas (iii) and (iv) the bottom right and top left quadrants are the cross and symmetric LOS $\times$ POS covariance submatrices.
The diagonal of the entire matrix is the global variance.
The square root of the values from these diagonals make up the implicit error bars in Figure~\ref{fig:correlations_ospace}.
Comparing now the three matrices, it can be seen that if the PR is small, the covariance matrices tend to be diagonal, whereas as the PR increases, off-diagonal values become more prominent.
For instance, note the increase of the cloud pattern observed on the LOS $\times$ POS quadrants.

\subsection{Cosmological constraints}
\label{subsec:cosmotest_test_constraints}

We now present the results of the likelihood analysis.
Since $\Omega_m$ and $\beta$ are the two parameters of interest, we will focus mainly on them.
The goal is to calibrate the method.
This is achieved by retrieving at the end of the process the values of these parameters inherent to the MXXL.
On the one hand, $\Omega_m=0.25$ (Section~\ref{subsec:data_sim}).
On the other hand, the target $\beta(\zbox)$ values were directly inferred by fitting the real-space velocity profiles measured in the simulation boxes with Eq.~(\ref{eq:velocity}).

\begin{figure}
    \centering
    \includegraphics[width=79mm]{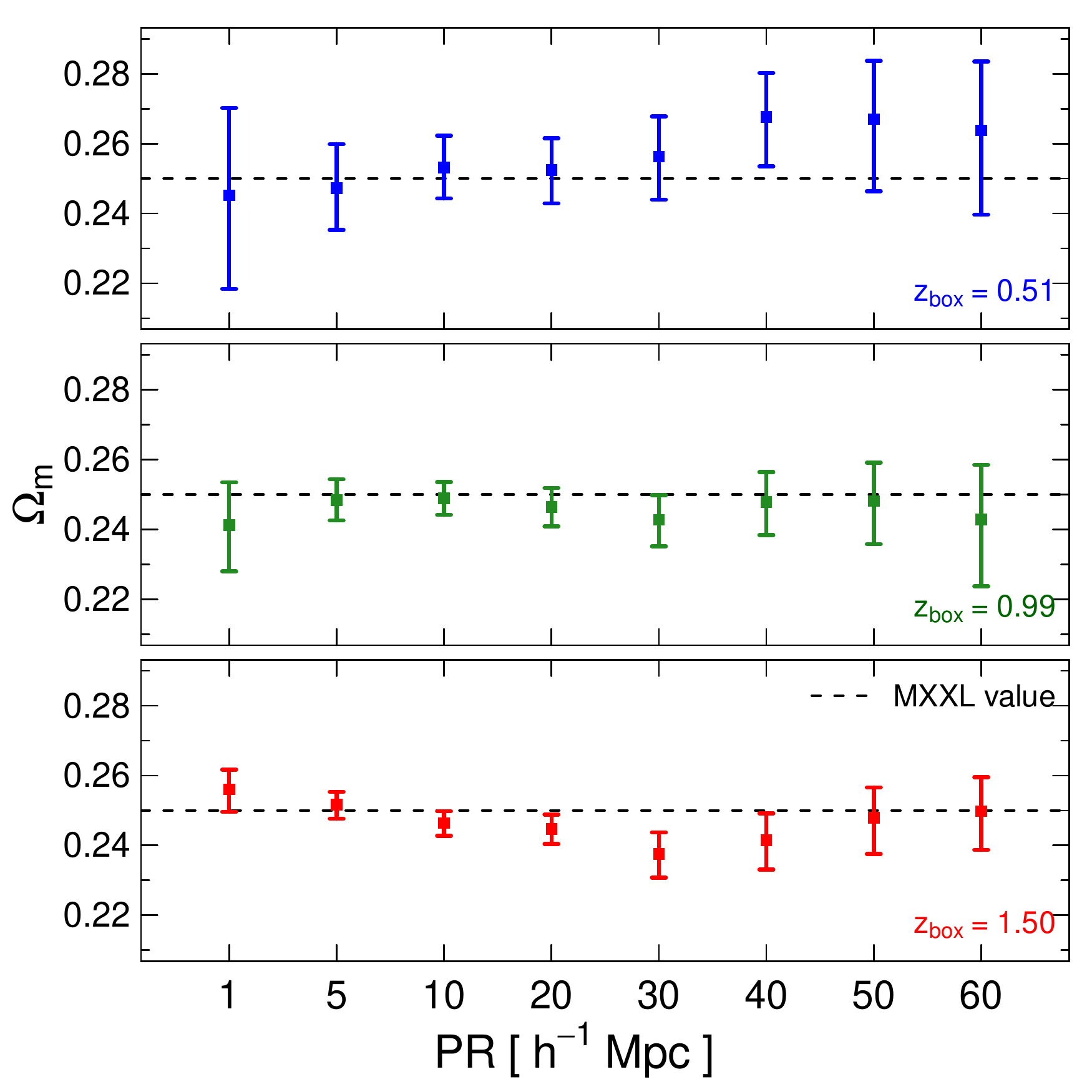}
    \includegraphics[width=79mm]{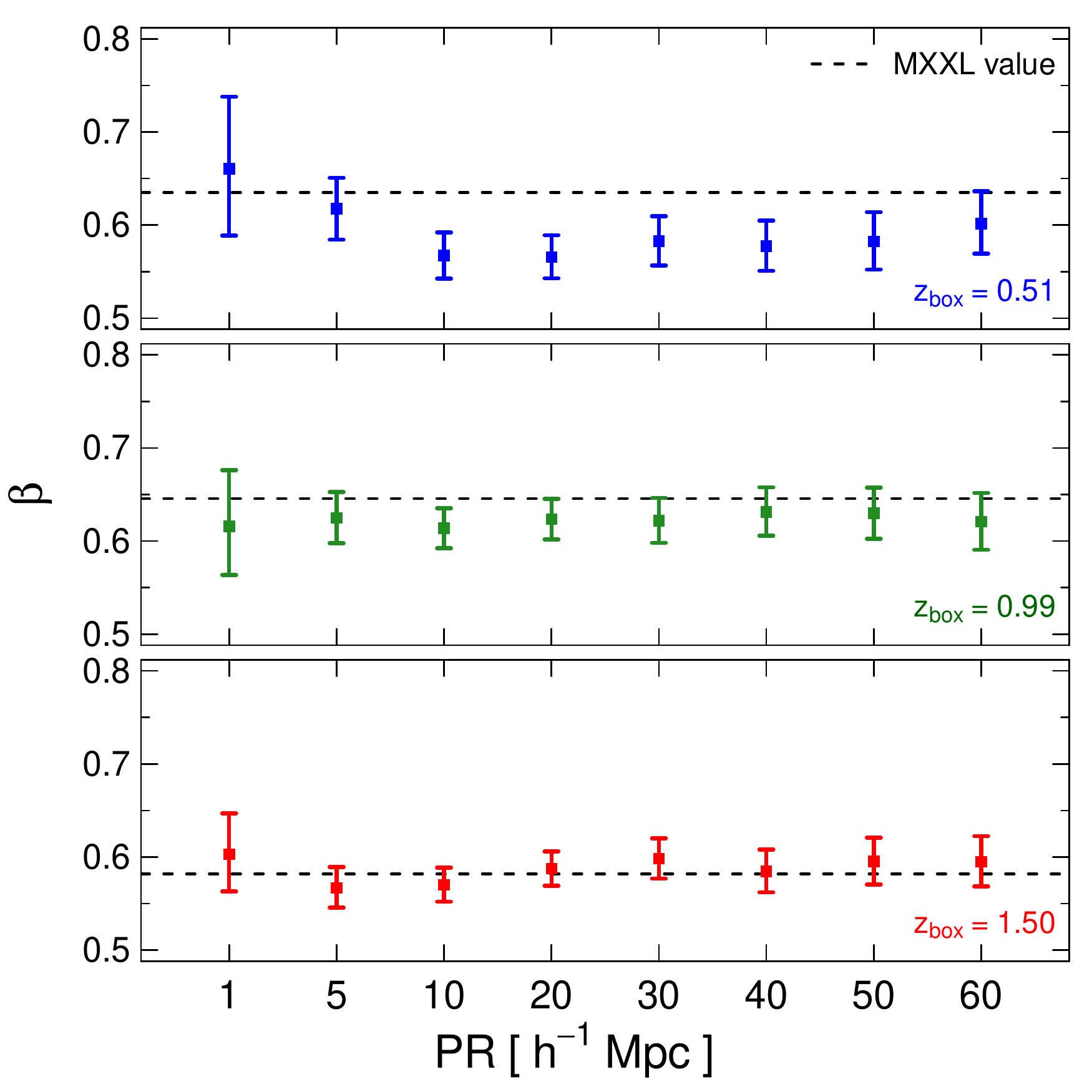}
    \caption[
    Evaluation of the cosmological test: marginalised likelihood distributions over the cosmological parameters $\Omega_m$ and $\beta$.
    ]{
    Marginalised likelihood distributions over the parameters $\Omega_m$ (\textit{left-hand panel}) and $\beta$ (\textit{right-hand panel}) for each MXXL snapshot and for each projection range, shown as $1\sigma$ ($68.3\%$) error bars.
    The dashed horizontal lines indicate the respective MXXL target values.
    The test is robust with the redshift and the projection range.
    }
    \label{fig:constraints}
\end{figure}

Figure~\ref{fig:constraints} shows the marginalised likelihood distributions over the parameters $\Omega_m$ and $\beta$ for each MXXL snapshot and for each projection range as $1\sigma$ ($68.3\%$) error bars, since these distributions show a Gaussian shape.
The dashed horizontal lines indicate the MXXL target values.
As can be seen, the target values fall inside the $1\sigma$ error bars in most cases, and fall inside $3\sigma$ ($99.7\%$) of error in all of them, which is the consistency check we were looking for to calibrate the method.
This is a consequence of the ability of our model to reproduce the RSD and AP distortions as well as the mixture of scales.
This can also be corroborated by inspecting Figures~\ref{fig:correlations_ospace} and \ref{fig:densvel}, where the theoretical profiles (solid curves) obtained with the best fitted parameter values from this likelihood analysis match very well the data points.

Inspecting Figure~\ref{fig:constraints} in more detail, we can make some conclusions from the behaviour of the constraints.
First, note that the true value of $\beta$ is slightly dependent on the redshift.
Note also that the error bars for $\beta$ are almost constant, nearly independent of the PR and $\zbox$.
In the case of $\Omega_m$, by contrast, the error bars reach a minimum at a PR between $10$ to $20~\hmpc$, which points out that there is an optimal range to perform the test.
Moreover, they generally decrease from lower to higher $z_\mathrm{box}$, which elucidates that tighter confidence regions are obtained by performing the test at higher redshifts.
This is due to the fact that the model is more sensitive to the AP effect at higher redshifts.
Note also that, although the MXXL values fall inside the error bars in most cases, there is an appreciable deviation in $\beta$ for the case $z_\mathrm{box}=0.51$ and $\mathrm{PR}~\geq~10~\hmpc$ (upper right-hand panel).
This is possibly due to a deficiency in our model for RSD, since this effect prevails over the AP effect at lower redshifts.
Incidentally, \citeonline{rsd_achitouv1} and \citeonline{rsd_nadathur} present an improved RSD model for voids, analysing non-linearities and second order effects.
Alternatively, since $\Omega_m$ and $\beta=f/b$ are not independent, and the constraints are unbiased for the former, the tensions detected for the latter may come from the assumption of a linear bias relation.
In other words, one should expect the constraints on $f$ to be unbiased too, as the constraint on $\Omega_m$ should translate analytically into a constraint on $f$, the only additional uncertainty to consider is the constraint on $b$.
We leave this problematic for a future investigation.

Two aspects worth mentioning.
First, we are using a high-density halo sample in a large volume.
Therefore, the confidence levels on the estimated parameters must be understood in a precision limit framework.
When applied to real data, the confidence regions will be larger.
Second, for a fixed snapshot, the error bars do not represent independent estimates, since it is about measuring the same correlation function for the same void sample but merely adding more void-halo pairs when increasing the projection range.
Having said this, Figure~\ref{fig:constraints} is a robust confirmation that the test can be applied with a wide variety of projection ranges.

\begin{figure}
    \centering
    \includegraphics[width=\textwidth]{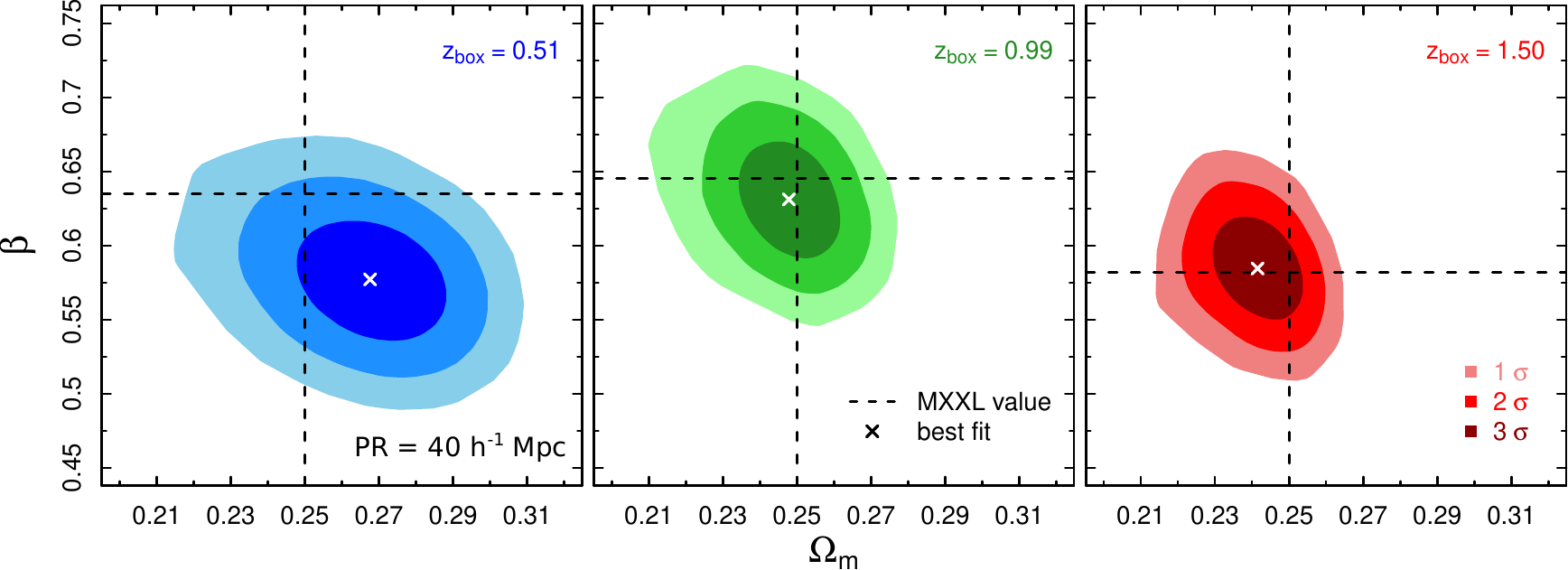}
    \caption[
    Evaluation of the cosmological test: marginalised likelihood distributions onto the plane $\Omega_m-\beta$.
    ]{
    Two-dimensional marginalised likelihood distributions onto the plane $\Omega_m-\beta$ for each MXXL snapshot and for the case $\mathrm{PR}=40\hmpc$.
    From the inner to the outermost, the coloured contour levels enclose $1\sigma$ ($68.3\%$), $2\sigma$ ($95.5\%$) and $3\sigma$ ($99.7\%$) confidence regions.
    The dashed lines indicate the respective MXXL target values, whereas the white crosses, the best fitted values.    
    }
    \label{fig:bananas}
\end{figure}

Figure~\ref{fig:bananas} shows, as an example, the two-dimensional likelihood marginalisations onto the $\Omega_m-\beta$ plane for the case $\mathrm{PR}=40~\hmpc$ for the three MXXL snapshots.
From the inner to the outermost, the coloured contour levels enclose $1\sigma$ ($68.3\%$), $2\sigma$ ($95.5\%$) and $3\sigma$ ($99.7\%$) confidence regions.
The dashed lines indicate the respective MXXL values, whereas the white crosses, the best fit values.
Note that the target values fall inside the $1\sigma$ confidence region for medium and high redshifts, whereas for low redshift the deviation of $\beta$ explained before can be appreciated.
While the best results (tightest constraint and smallest deviation) were obtained for the case $\mathrm{PR}=5~\hmpc$ case, we decided to show the case $\mathrm{PR}=40~\hmpc$ to highlight the robustness of the test with the projection range, which is important because the wider it is, then more data pairs are counted, and therefore, the measured signal increases.
In this sense, this is a more realistic case applicable to real data.

\begin{figure}
    \centering
    \includegraphics[width=\textwidth]{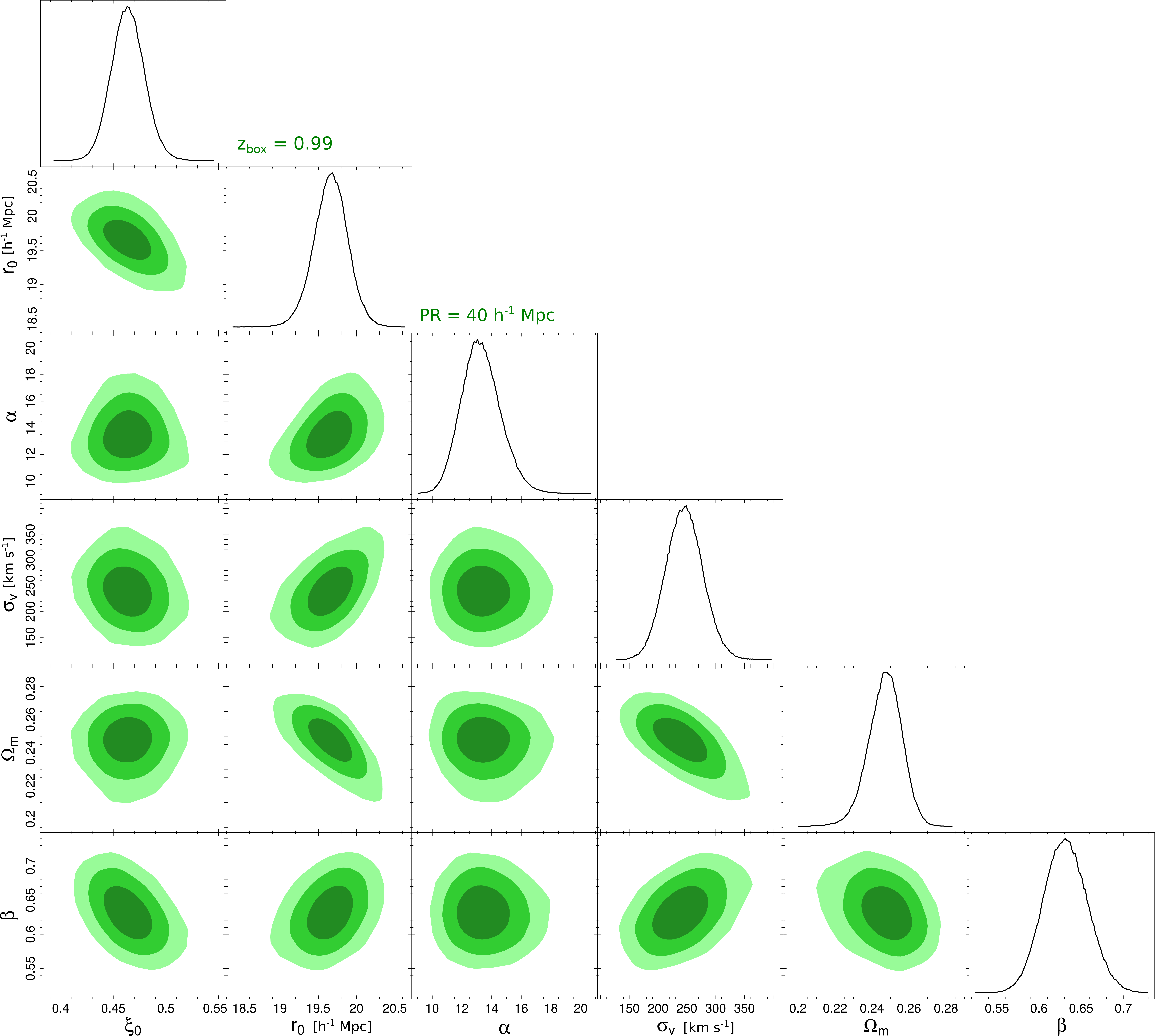}
    \caption[
    Evaluation of the cosmological test: analysis of the full parameter space.
    ]{
    Marginalised likelihood distributions of the full parameter space for the case $\zbox=0.99$ and $\mathrm{PR} = 40\hmpc$.
    The $\Omega_m-\beta$ panel is the same as the central panel of Figure~\ref{fig:bananas}.
    No degeneracies are observed.
    The individual distributions follow a Gaussian behaviour.
    }
    \label{fig:bananas2}
\end{figure}

For completeness, Figure~\ref{fig:bananas2} shows the two-dimensional likelihood marginalisations of the full parameter space for the case $\zbox = 0.99$ and $\mathrm{PR} = 40\hmpc$.
The constraints are tight, showing no degeneracies between the parameters.
Moreover, the one-dimensional marginalisations show a Gaussian shape in all cases.
The remaining snapshots and projection ranges show a similar behaviour.

\chapter{Redshift-space effects in voids}
\label{chp:zeffects}
\section*{Abstract}

Voids are promising cosmological probes.
Nevertheless, every cosmological test based on voids must necessarily employ methods to identify them in redshift space.
The dynamical and geometrical distortions have an impact on the void identification process itself, affecting statistical properties of voids such as their number, size and spatial distribution.
This generates additional distortion patterns in observations.
Using the spherical void finder, we developed a statistical and theoretical framework to physically describe the connection between the identification in real and redshift space.
We found that redshift-space voids above the shot-noise level have a unique real-space counterpart spanning the same region of space.
In this context, it is valid to assume void number conservation under the redshift-space mapping.
Moreover, they are systematically bigger and their centres are preferentially shifted along the line-of-sight direction.
The \textit{expansion effect} is a by-product of the RSD induced by tracer dynamics at scales around the void radius (t-RSD), whereas the \textit{off-centring effect} constitutes a different class of RSD induced at larger scales by the global dynamics of the whole region containing the void (v-RSD).
The volume of voids is also altered by the fiducial cosmology assumed to measure distances, this is the \textit{AP change-of-volume effect}.
We developed a theoretical framework to model these effects, which was tested using the MXXL simulation.
This description depends strongly on dynamical and cosmological considerations.
This part of the work has been published in the Monthly Notices of the Royal Astronomical Society Journal \cite{zvoids_correa}.


\section{The standard picture of RSD around voids}
\label{sec:zeffects_rsd}

In Section~\ref{sec:lss_rsd}, we studied the standard picture of RSD on large and linear scales, represented in Figure~\ref{fig:rsd2} schematically.
However, recent studies have demonstrated that this is a description too simplified for the case of voids.
\citeonline{rsd_nadathur} and \citeonline{reconstruction_nadathur} noticed that there are four commonly assumed hypotheses in all RSD models for voids, which are violated when voids are identified in redshift space, i.e. from observations.

\begin{enumerate}

\item 
\textit{Conservation of the number of void-galaxy pairs}, under the mapping from real-space into redshift-space.
This condition is expressed in Eq.~(\ref{eq:number_consv}).

\item
\textit{Invariability of void-centre positions}.
This condition is expressed in Eq.~(\ref{eq:rsd}).
This means that the centres do not suffer any RSD, and it is only the velocity of galaxies that are relevant when quantifying the change from $\mathbf{r}$ to $\mathbf{s}$.

\item
\textit{Isotropy of the velocity field around voids.}
This fundamental hypothesis is expressed in Eq.~(\ref{eq:vel_rad}).
Motivated by the two previous items, this condition assumes that, from the point of view of the observational centres (identified in redshift space), the average radial outflow of the galaxies is still radially directed.

\item
\textit{Isotropy of the density field around voids.}
As in the previous item, this condition assumes that, from the point of view of the observational centres, the surrounding density field in real space has spherical symmetry.

\end{enumerate}
The failure of these hypotheses finds an explanation on the RSD and AP effects.
Traditionally, we have focused only on the spatial distribution of the galaxies around voids.
The truth is that these effects also have a direct impact on the identification process itself, affecting global properties of voids such as their number, size and spatial distribution.
For this reason, it is not clear if there is a relation between both void populations: the one identified from observational data (in redshift space) and the true underlying one (in real space).
The validity of these hypotheses is also discussed in \citeonline{aprsd_hamaus2020}.

This problematic is important when designing cosmological tests, since these void systematicities generate additional deviations and anisotropic patterns in the measurements of the void size function and the correlation function.
Given the precision achievable with modern surveys nowadays, it is extremely important to detect and model all these effects in order to obtain unbiased cosmological constraints.
This is the reason why we have used real-space voids in the previous chapter in order to develop and calibrate our cosmological test.

\citeonline{reconstruction_nadathur} propose as a solution to use the reconstruction technique, which has been first applied to the case of BAO analyses \cite{reconstruction_eisenstein}.
Reconstruction is an algorithm to approximately recover the real-space position of galaxies from redshift space based on the Zel'dovich approximation.
The idea is to apply reconstruction before performing the void finding step.
As this is a cosmology-dependent procedure, it can be used to measure the growth rate factor parameter if reconstruction plus the void finding step are applied iteratively.
This method has been recently applied to data \cite{aprsd_nadathur, aprsd_nadathur2020}, showing a robust power in constraining the cosmological parameters and recovering the statistical properties of voids.
However, it also presents some disadvantages.
One the one hand, it is computationally expensive, since it is an iterative process.
On the other hand, it can be quite redundant, since it requires combining it with a model for RSD around the recovered voids (after this effect was previously removed) in order to obtain accurate constraints.
Finally, it does not take advantage of the valuable physical information about the structure and dynamics of voids contained in these additional systematicities, which manifest when voids are directly identified from observations.

In the present work, we propose an alternative approach: to analyse the void finding process in order to find a physical connection between the resulting void populations in real and redshift space, and in this way, be able to detect the physical effects responsible for the observed differences between their statistical properties.
This is the goal of this chapter.
Some conventions before we start.
The real space and redshift space will be referred to as $r$-space and $z$-space, respectively.
Moreover, we will distinguish between real and redshift-space void radii with the notation $\rrs$ and $\rzs$, respectively.


\section{Bijective mapping}
\label{sec:zeffects_map}

We begin the analysis with a visual inspection of $r$-space and $z$-space voids, returning back to Figure~\ref{fig:mxxl_slice}.
This figure shows two slices of the MXXL simulation box using the TC void catalogues.
The aim of this section is to study the impact of RSD alone, postponing the analysis of the combined RSD and AP effects until the next chapter, where we will use the FC void catalogues.
In Section~\ref{sec:data_voids2}, we discussed that the left-hand panel of this figure is a representation of the POS distribution of haloes and voids, whereas the right-hand panel shows the corresponding LOS distribution.
There, $r$-space void centres are represented with blue dots, whereas $z$-space centres, with red squares.
From the figure, it is clear that both types of voids approximately span the same regions of space.

In order to link $r$-space and $z$-space voids, we looked for a correspondence between them by cross-correlating catalogues TC-rs-f and TC-zs-f.
Specifically, for each $z$-space centre, we picked the nearest $r$-space centre with the condition that it must lay inside $1~\rzs$.
Then, we filtered those voids if no partner could be found.
Note that this mapping from $z$-space into $r$-space is a well defined function, since the condition of the nearest $r$-space neighbour assigns only one object to each $z$-space void.
Moreover, this mapping is also injective, since the non-overlapping condition imposed by the void finder implies that each $r$-space void can only be reached by a single $z$-space one.
Furthermore, the filtering condition guarantees then a one-to-one relationship between $z$-space and $r$-space voids.
For this reason, these voids constitute what we call the \textit{bijective catalogues}.
In Table~\ref{tab:catalogues2}, these catalogues were denoted by TC-rs-b and TC-zs-b, the former for the bijective voids in $r$-space, the latter for the bijective voids in $z$-space.
Note that, by construction, both catalogues have the same number of elements: $318784$.
Moreover, it is ensured in this way that a void and its associated counterpart span the same region of the simulated universe.
In order to distinguish the bijective catalogues from the original ones, we will refer to the latter as the \textit{full catalogues}, denoted by TC-rs-f and TC-zs-f in Table~\ref{tab:catalogues2}.
Going back to Figure~\ref{fig:mxxl_slice}, bijective voids are represented with circles around their centres, which correspond to the intersections of their volumes with the midplane of the slice.
The rest are voids of the full catalogues without a partner in the other space.

\begin{figure}
    \centering
    \includegraphics[width=79mm]{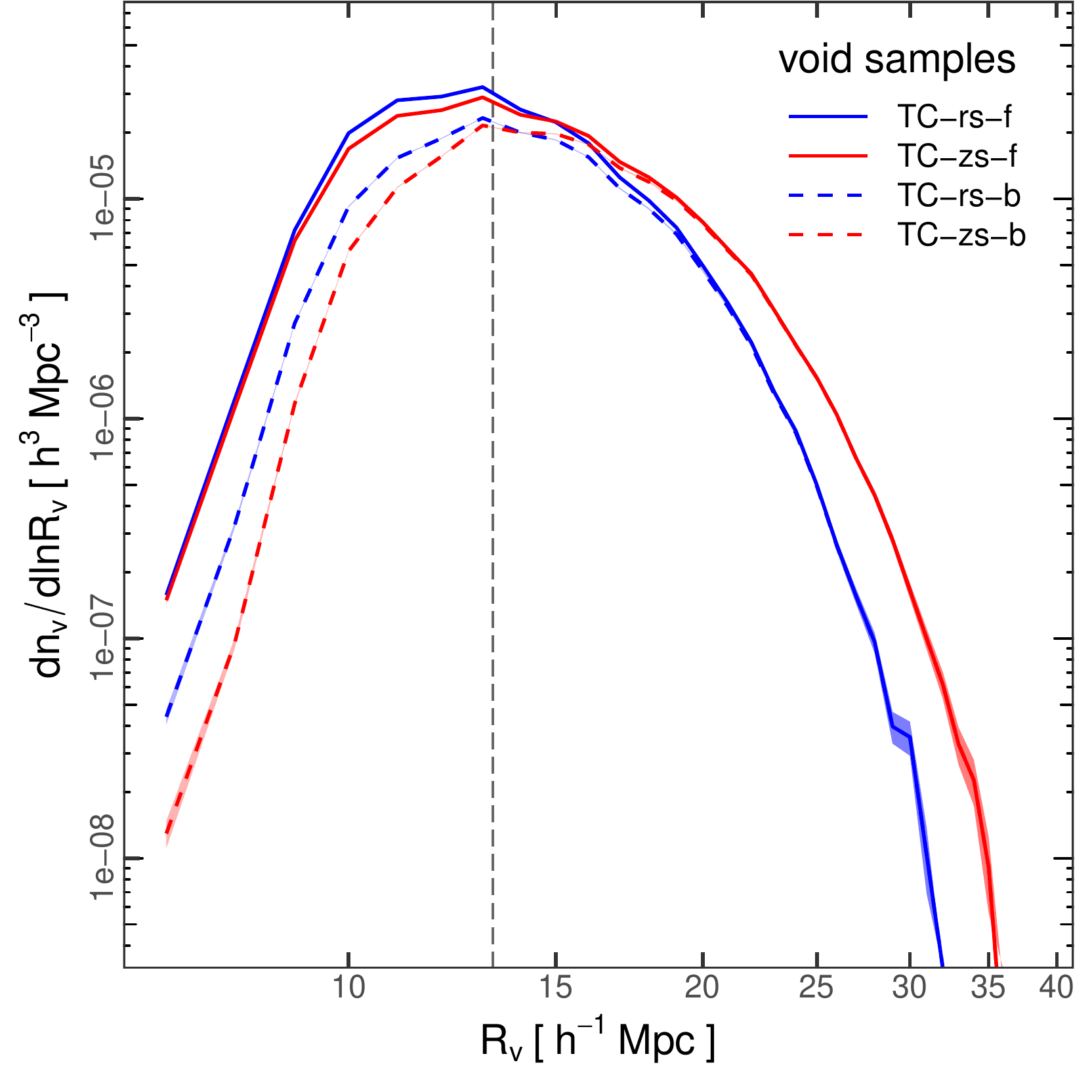}
    \includegraphics[width=79mm]{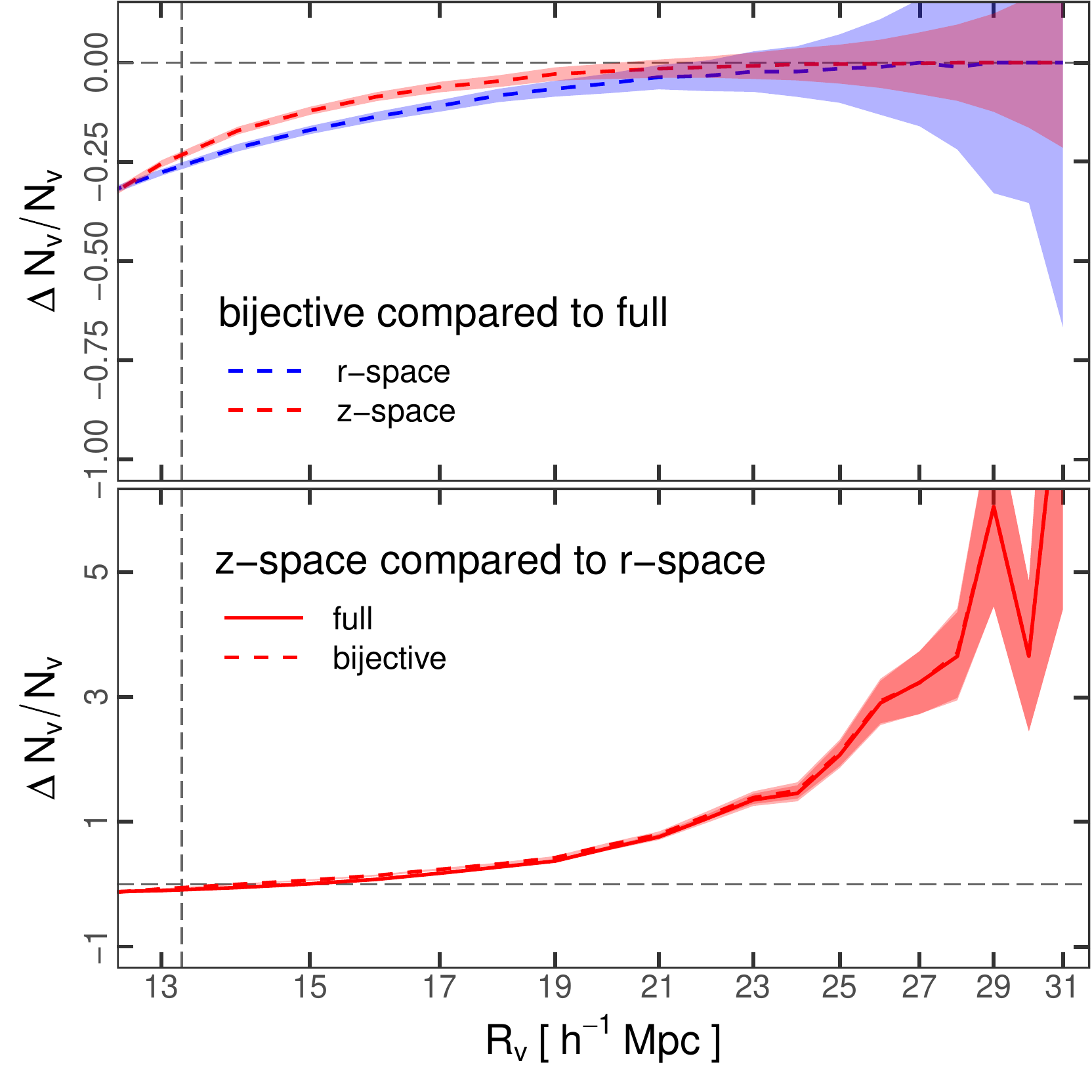}
    \caption[
    Void size functions of the TC void catalogues (Table~\ref{tab:catalogues2}).
    ]{
    \textit{Left-hand panel.}
    Void size functions of the TC void catalogues.
    The solid curves correspond to the full catalogues, both in $r$-space (blue) and $z$-space (red).
    The dashed curves correspond to the bijective catalogues.
    The vertical dashed line is the median of the $z$-space full distribution (TC-zs-f), which separates the small voids dominated by shot noise (on the left) from the genuine ones relevant for cosmological analyses (on the right).
    \textit{Upper right-hand panel.}
    Fractional differences of void counts between the bijective and full catalogues.
    It quantifies the void loss under the mapping between $z$-space and $r$-space.
    Voids above the shot-noise level are almost conserved.
    \textit{Lower right-hand panel.}
    Fractional differences of void counts between the $z$-space and $r$-space catalogues.
    }
    \label{fig:TC_VSF}
\end{figure}

In order to enquire deeper into the relation between $r$-space and $z$-space voids, the left-hand panel of Figure~\ref{fig:TC_VSF} shows the void size functions of the four TC catalogues (in $r$-space and $z$-space, with their respective full and bijective versions).
The abundances were computed from the radius distribution of each catalogue, expressing the void counts as comoving differential number densities, $dn_\mathrm{v}$, and normalising them by the logarithmic sizes of the radius bins, $d\mathrm{ln}R_\mathrm{v}$.
The solid curves represent the abundances of the full catalogues, both in $r$-space (blue) and $z$-space (red), whereas the dashed curves, the abundances of the bijective catalogues.
In all cases, the error bands were calculated from Poisson errors in the void counting process.
The vertical dashed line represents the median of the $z$-space full catalogue (TC-zs-f), which will serve as a reference line throughout this chapter.
This value is equal to $13.26~\hmpc$ ($2.28$ in units of the mean interparticle separation).
The qualitative behaviour of the VSFs is consistent with previous studies, where we can distinguish two main behaviours separated by the vertical line: (i) on the left, small voids dominated by shot noise, and (ii) on the right, genuine voids decreasing their number as the radius increases with a functional shape similar to those predicted by the theory (see Figure~\ref{fig:abundance_models} and the corresponding discussion in Section~\ref{sec:voids_abundance}).
Small voids dominated by shot noise, in this sense, are not reliable for any statistical and cosmological analyses, hence we will mainly focus throughout this work on larger voids delimited by the reference line.

Note that the full (solid curves) and bijective (dashed curves) abundances tend to the same values as the radius increases.
This means that the loss of voids under the mapping between both spatial configurations is only significant in the region dominated by shot noise, whereas the larger and relevant voids are almost conserved.
This can be better appreciated in the upper right-hand panel of Figure~\ref{fig:TC_VSF}, where we present the corresponding fractional differences of void counts between the full and bijective catalogues: $\Delta N_\mathrm{v}/N_\mathrm{v} = (N_\mathrm{v}^\mathrm{bij} - N_\mathrm{v}^\mathrm{full})/N_\mathrm{v}^\mathrm{full}$.
For all radii of interest, the loss of voids decreases as the radius increases, being less than $25\% ~ (\Delta N_\mathrm{v}/N_\mathrm{v} < 0.25)$ in the worst case.
We arrive here at the first and one of the most important conclusions of this chapter: voids identified from an observational catalogue above the shot-noise level are true voids, i.e. they have a real-space counterpart.
Therefore, it is valid to treat the full and bijective catalogues indistinctly in their statistical properties.

With this in mind, let us now compare the $r$-space (blue curves) and $z$-space (red curves) abundances.
It is clear that they are very different from each other.
In particular, the corresponding fractional differences $\Delta N_\mathrm{v}/N_\mathrm{v} = (N_\mathrm{v}^\mathrm{zs} - N_\mathrm{v}^\mathrm{rs})/N_\mathrm{v}^\mathrm{rs}$ (lower right-hand panel) increase as the radius increases, and they can be very high at the largest sizes.
However, in the context of the bijective mapping, as both types of voids are in fact the same entities, this means that these differences can only be attributed to some physical effect that voids suffer when they are mapped from $r$-space into $z$-space.
Incidentally, note that for each radius, $z$-space voids are systematically bigger than their $r$-space counterparts.
This hints of an expansion effect.


\section{Theoretical description}
\label{sec:zeffects_theo}

The goal of this section is to theoretically study the possible physical mechanisms responsible for the transformation of $r$-space voids into their associated $z$-space counterparts.
We will do this in the context of the four hypotheses commonly assumed to model RSD around voids noticed in Section~\ref{sec:zeffects_rsd} and the bijective mapping of the last section.
In the next section, we will provide the corresponding statistical evidence of the framework developed here.


\subsection{Void number conservation}
\label{subsec:zeffects_theo_num}

Since the galaxy number is conserved under the redshift-space mapping, condition (1) concerning the conservation of void-galaxy pairs can be inspected by analysing the corresponding void number conservation.
This is not trivial, since unlike galaxies, which can be considered as particles that are totally conserved under this mapping (only their position changes), some voids can be destroyed while new artificial voids can be created in this process.

Strictly speaking, void number conservation is violated.
This is evident when analysing the full catalogues, since the two versions in $r$-space and $z$-space have different number of elements (compare the number of voids between the TC-rs-f and TC-zs-f entries in Table~\ref{tab:catalogues2}) and different abundances (compare the blue and red curves in Figure~\ref{fig:TC_VSF}).
Nevertheless, this condition is not violated in the context of the bijective mapping that we defined.
This is supported by two reasons.
First, bijective voids are, by definition, the same entities spanning the same regions of space.
This is why the bijective catalogues have the same number of voids (compare the number of voids between the TC-rs-b and TC-zs-b entries in Table~\ref{tab:catalogues2}).
Second, voids above the shot-noise level (characterised by the median of the radius distribution) are practically conserved under this mapping, as we have demonstrated in the previous section.
For these sizes, the full and bijective catalogues are equivalent regarding their statistical properties.

Therefore, any difference detected in the statistical properties between $z$-space and $r$-space voids can only be attributed to some physical effect that impacts on voids when they are mapped from $r$-space into $z$-space.
Such an effect, then, must be associated with the distortions in the observed spatial distribution of galaxies.
Hence, a physical description must find its bases on the large-scale dynamics of galaxies.


\subsection{Expansion effect}
\label{subsec:zeffects_theo_trsd}

\begin{figure}
    \centering
    \includegraphics[width=50mm]{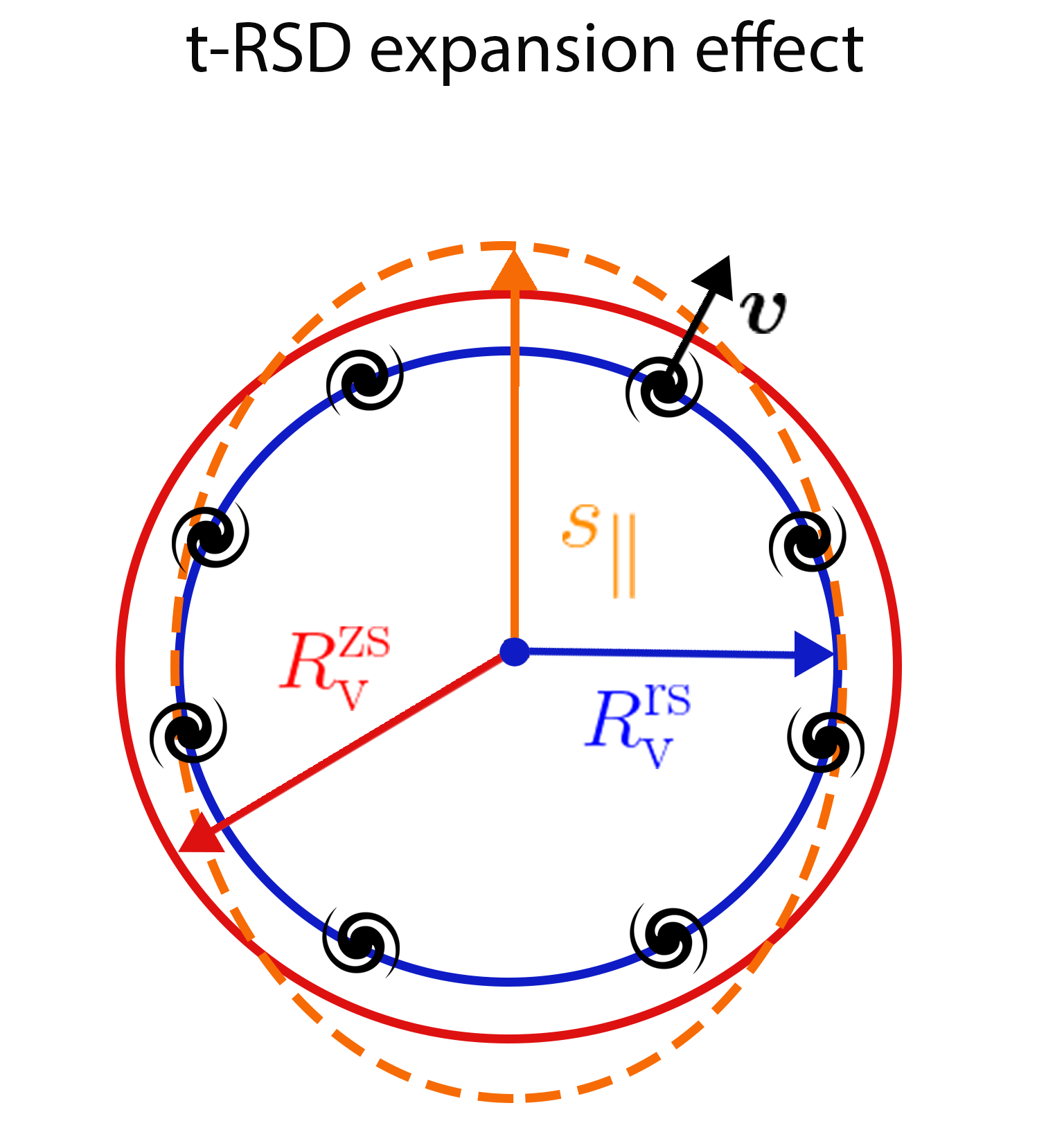}
    \includegraphics[width=50mm]{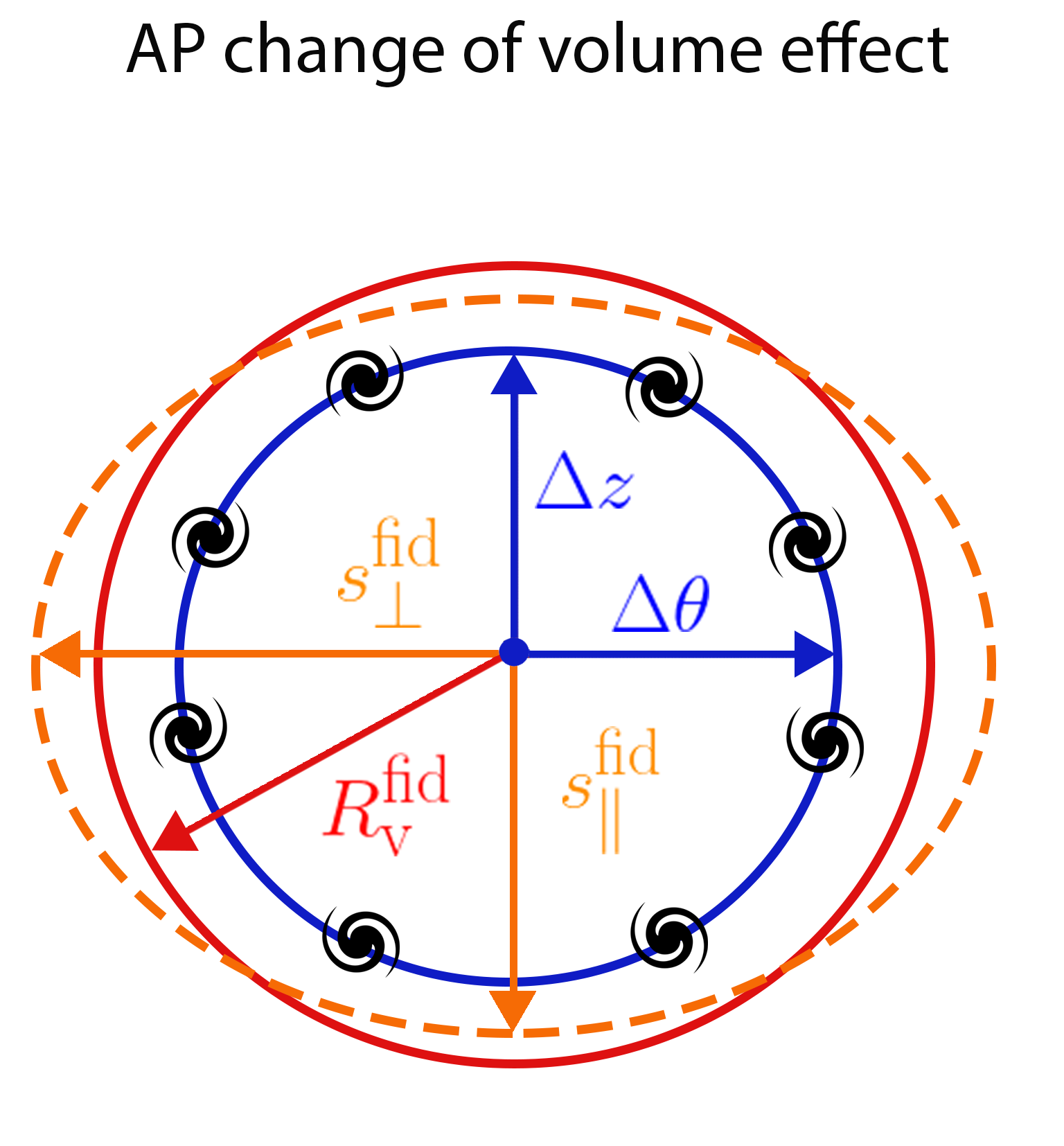}
    \includegraphics[width=50mm]{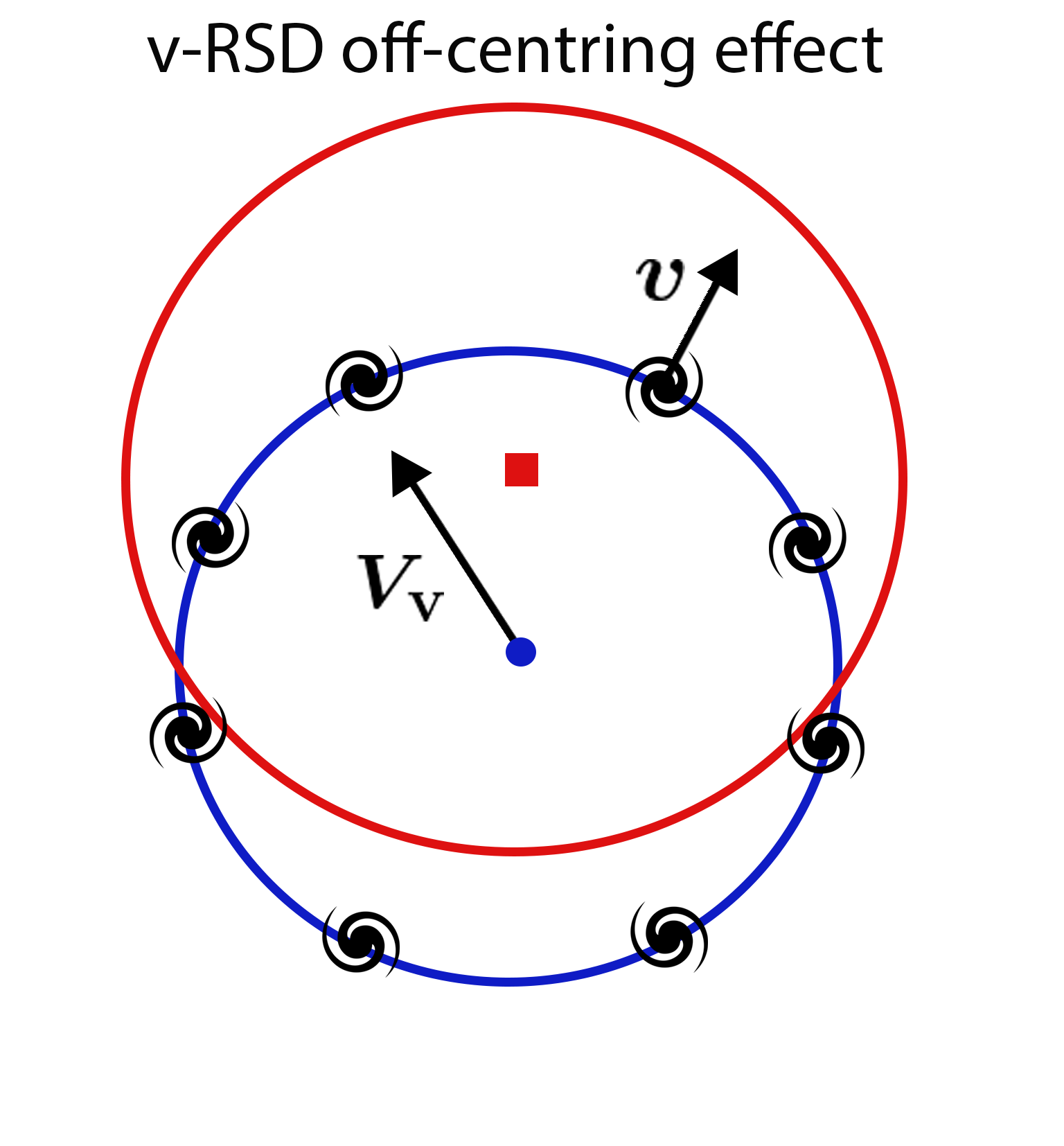}
    \caption[
    Redshift-space effects in voids: expansion (t-RSD), AP change-of-volume and off-centring (v-RSD).
    ]{
    Schematic illustration of the three $z$-space effects in voids.
    In all cases, a hypothetical spherical $r$-space void of radius $\rrs$ is represented with a blue solid circle with some galaxies around.
    The LOS direction is assumed vertical.
    \textit{Left-hand panel. Expansion effect.}
    In $z$-space, the void appears elongated along the LOS (orange dashed ellipse) due to the LOS component of the peculiar velocity of the galaxies that surround it.
    The void finder identifies an equivalent and bigger sphere of radius $\rzs=\qrsd\rrs$ (red solid circle).
    This effect is a by-product of tracer dynamics at scales around the void radius (t-RSD).
    \textit{Central panel. AP change-of-volume effect.}
    The dimensions of the void are distorted in both directions due to the AP effect, manifesting itself again as an ellipsoid (orange dashed ellipse).
    The void finder identifies an equivalent sphere of radius $R_\mathrm{v}^\mathrm{fid}=\qap\rrs$ (red solid circle).
    \textit{Right-hand panel. Off-centring effect.}
    In $z$-space, the void centre appears shifted along the LOS due to the LOS component of the void net velocity.
    This effect is a by-product of the global void dynamics (v-RSD).
    The three effects can be treated independently.
    }
    \label{fig:effects}
\end{figure}

In Section~\ref{sec:zeffects_map}, we showed that $z$-space voids are systematically bigger than their $r$-space counterparts.
This suggests that voids expand when they are mapped from $r$-space into $z$-space.
The left-hand panel of Figure~\ref{fig:effects} depicts this \textit{expansion effect} schematically.
There, a spherical $r$-space void of radius $\rrs$ (represented with a blue solid circle with some galaxies) appears elongated along the LOS direction in $z$-space due to the RSD induced by the LOS component of the peculiar velocities of the tracers surrounding it.
The $r$-space spherical void has been transformed into a $z$-space ellipsoid (orange dashed ellipse in the figure) with semi-axes $s_\perp$ and $s_\parallel$, where $s_\perp$ is the POS semi-axis (equal for both $x_1$ and $x_2$ directions), and $s_\parallel$, the LOS semi-axis.

We now derive analytical expressions for the semi-axes.
We will assume that RSD do not affect the void dimensions across the POS.
Hence, we can consider that $s_\perp = \rrs$.
An expression for $s_\parallel$, on the other hand, can be obtained by means of Eq.~(\ref{eq:pi_rsd}), the expression for the $z$-space void-centric distance along the LOS.
In this case, $\pi$ and $r_\parallel$ must be replaced by $s_\parallel$ and $\rrs$, respectively.
The next ingredient is an expression for $v_\parallel$, which can be obtained from Eq.~(\ref{eq:velocity}), the expression for the void-centric radial velocity profile characterising the peculiar velocity field around voids.
Here, $v_\parallel = v(r_\parallel) = v(\rrs)$, for which $\Delta(r)$ must be evaluated at $r = \rrs$, which in turn is equal to the threshold of void identification: $\Delta(\rrs) = \Delta_\mathrm{id}$.
In this way, combining Eqs.~(\ref{eq:pi_rsd}) and (\ref{eq:velocity}) with the mentioned replacements, we get an expression for $s_\parallel$:
\begin{equation}
    s_\parallel = \rrs \left( 1 - \frac{1}{3} \beta(\zsim) \did \right).
	\label{eq:q_elip}
\end{equation}
Note that here, we have assumed the validity of hypotheses (3) and (4) concerning the isotropy of the density and velocity fields in $r$-space in order to explain a $z$-space phenomenon, even if this isotropy is no longer valid in this last spatial configuration.

Our void finder identifies spherical regions instead of ellipsoidal ones.
Hence, as a first ansatz, we will assume that $z$-space spherical voids enclose the same volume of the corresponding ellipsoidal regions.
This is depicted in Figure~\ref{fig:effects} with a red solid circle.
Denoting the radius of this equivalent sphere by $\rzs$, equating both volumes, and using Eq.~(\ref{eq:q_elip}), we finally get an expression for $\rzs$:
\begin{equation}
    \rzs = \qrsds \rrs,
	\label{eq:q1_rsd}
\end{equation}
where
\begin{equation}
    \qrsds = \sqrt[3]{1 - \frac{1}{3} \beta(\zsim) \did}.
	\label{eq:q1_rsd_factor}
\end{equation}
Note that $\qrsds$, which we will call the \textit{RSD factor}, is simply a scale-independent proportionality constant.
Moreover, it has a strong cosmological dependence, since it depends on $\beta$.
To get the explicit value of $\qrsds$ inherent to the MXXL simulation, we need the corresponding values of $\did$ and $\beta(\zsim)$.
As specified in Table~\ref{tab:catalogues1}, $\did = -0.853$ for $\zsim = 0.99$.
The corresponding value of $\beta$ can be obtained from the analysis of Section~\ref{subsec:cosmotest_test_constraints}: ${\beta = 0.65}$ (see the central panel of Figure~\ref{fig:constraints}).
With these two quantities, we get $\qrsds = 1.058$.
Note that $\qrsds > 1$, hence $\rzs > \rrs$, which is in agreement with our assumption that voids expand when they are mapped from $r$-space into $z$-space.

In reality, the actual value of $\rzs$ is expected to vary between $\rrs$ and $s_\parallel$, i.e. $\rrs \leq \rzs \leq s_\parallel$.
In order to test for possible deviations in the predictions of Eq.~(\ref{eq:q1_rsd}), we also considered a more general approach by introducing a new variable $\dR$ that quantifies the variation in radius:
\begin{equation}
    \dR := \frac{\rzs - \rrs}{s_\parallel - \rrs}.
    \label{eq:delta_R}
\end{equation}
Considering that $\qrsd := \rzs/\rrs$, and combining Eqs.~(\ref{eq:q_elip}) and (\ref{eq:delta_R}), we obtain the following linear relation:
\begin{equation}
    \qrsd = 1 - \frac{1}{3} \dR \beta(\zsim) \did.
    \label{eq:q_rsd}
\end{equation}
Note that $\qrsd \geq 1$ always, since $\did < 0$.
Therefore, $\dR$ is expected to vary in the range $[0, 1]$.
One limit case is $\dR = 0$ ($\qrsd = 1$), which corresponds to the unlikely case of no expansion at all, i.e. $\rzs = \rrs$.
The other limit case is $\dR = 1$ ($\qrsd = 1.185$), which corresponds to the case where the $z$-space void is characterised by a sphere of radius $\rzs=s_\parallel$.
This last case is also unlikely because it would mean that RSD affect the dimensions of voids equally in all directions.
In particular, the theoretical prediction of Eq.~(\ref{eq:q1_rsd}) corresponds to the case $\dRs = 0.315$.

In Section~\ref{subsec:zeffects_stat_rad}, we will show that the theoretical prediction $\qrsds$ fits very well the median of the overall $\rzs/\rrs$ ratio distribution (see Figure~\ref{fig:cor_radius}).
However, we found that voids well above the shot-noise level respond better to the value $\dRl = 0.5$, i.e. to a $z$-space radius that is the mean between $\rrs$ and $s_\parallel$.
We will discuss these aspects in more detail in Sections~\ref{sec:zeffects_stat} and in the next chapter.
In this way, we get a prediction slightly different from that of Eq.~(\ref{eq:q1_rsd}):
\begin{equation}
    \rzs = \qrsdl \rrs,
	\label{eq:q2_rsd}
\end{equation}
where
\begin{equation}
    \qrsdl = 1 - \frac{1}{6} \beta(\zsim) \did,
	\label{eq:q2_rsd_factor}
\end{equation}
with an explicit value of $\qrsdl = 1.092$ inherent to the MXXL simulation at $\zsim = 0.99$.

The discrepancies between Eqs.~(\ref{eq:q1_rsd}) and (\ref{eq:q2_rsd}) can be attributed to the way in which the void finder performs the average spherical integration of the density field in an ellipsoidal underdense region.
Hence, the optimal value of $\qrsd$ will depend on the shape and slope of the real-space density profiles in the inner parts of voids.
We leave for a future investigation a deeper analysis about the derivation of $\qrsd$ considering these aspects.


\subsection{Alcock-Paczyński change-of-volume effect}
\label{subsec:zeffects_theo_ap}

Up to here, a true distance scale was implicitly assumed.
Note, however, that the only information available from observational catalogues are angular positions and redshifts of astrophysical objects like galaxies.
These observable quantities must be transformed into a physical distance scale, which involves the assumption of a fiducial cosmology.
A deviation between the true and fiducial cosmologies will lead to additional distortions in the spatial distribution of galaxies.
Hence, the AP effect will also affect the volume of voids.
To understand the impact of this effect, we will consider for the following analysis the distribution of galaxies in $r$-space, free of RSD.

In principle, the size of a spherical void can be quantified by two directly measurable quantities: an angular radius\footnote{We use $\Delta \theta$ instead of $\Delta \phi$ to be in agreement with the notation used in \citeonline{zvoids_correa}, from which this part of the work is based.} on the plane of the sky, $\Delta \theta$, and a redshift radius along the line of sight, $\Delta z$.
According to Eqs.~(\ref{eq:sigma_ap}) and (\ref{eq:pi_ap}), these observable quantities are related to their respective physical dimensions $R_\mathrm{v \perp}$ and $R_\mathrm{v \parallel}$ by the following transformation equations:
\begin{equation}
    R_\mathrm{v \perp} = D_\mathrm{M}(\zsim) \Delta \theta
	\label{eq:void_size_pos}
\end{equation}
and
\begin{equation}
    R_\mathrm{v \parallel} = \frac{c}{H(\zsim)} \Delta z.
	\label{eq:void_size_los}
\end{equation}
Note that if one knew the true cosmology, then it would not be necessary to distinguish between the POS and LOS dimensions.
Both would be equal to the $r$-space void radius: $\rrs = R_\mathrm{v \parallel} = R_\mathrm{v \perp}$.
However, assuming a fiducial cosmology leads to possible discrepancies between both quantities, therefore a spherical void will appear again as an ellipsoid in the underlying coordinate system.
Nevertheless, unlike the RSD-ellipsoids from the expansion effect, the AP-ellipsoids are distorted in both the POS and LOS directions.
Furthermore, the net result is not necessarily an expansion, it can also be a contraction, it all depends on the chosen cosmology.
This additional \textit{AP change-of-volume effect} is schematically depicted in the central panel of Figure~\ref{fig:effects}.

We can describe this AP-volume effect following a similar approach to that used for the expansion effect.
Considering that the ellipsoid has semi-axes $s_\perp^\mathrm{fid}$ and $s_\parallel^\mathrm{fid}$, given by Eqs.~(\ref{eq:void_size_pos}) and (\ref{eq:void_size_los}) with fiducial values $H_\mathrm{fid}$ and $D_{M}^\mathrm{fid}$, then a direct comparison with its true dimension, $\rrs$, also given by the same expressions but with the true values $H_\mathrm{true}$ and $D_{M}^\mathrm{true}$, leads to the following relations: 
\begin{equation}
    s_\perp^\mathrm{fid} = \qap^\perp \rrs
    \label{eq:q_ap_pos2}
\end{equation}
and
\begin{equation}
    s_\parallel^\mathrm{fid} = \qap^\parallel \rrs,
    \label{eq:q_ap_los2}
\end{equation}
where $\qap^\perp$ and $\qap^\parallel$ are given by Eqs.~(\ref{eq:qap_perp}) and (\ref{eq:qap_parallel}), respectively.
Here, we adopted the index ``true'' to refer to quantities based on the true underlying cosmology, although unknown.
Finally, considering the equivalent sphere with the same volume of the ellipsoid, and calling this new radius by $R_\mathrm{v}^\mathrm{fid}$, we get an expression similar to Eq.~(\ref{eq:q1_rsd}):
\begin{equation}
    R_\mathrm{v}^\mathrm{fid} = \qap \rrs,
	\label{eq:q_ap}
\end{equation}
where
\begin{equation}
    \qap = \sqrt[3]{(\qap^\perp)^2 \qap^\parallel}.
	\label{eq:q_ap_factor}
\end{equation}

Like the RSD factor, the \textit{AP factor} is also a proportionality constant, scale-independent and cosmology-dependent.
However, there is an interesting difference between both.
On the one hand, $\qap$ depends only on the background cosmological parameters, such as $\Omega_m$, $\Omega_\Lambda$ and $H_0$, hence it encodes information about the expansion history and geometry of the Universe.
On the other hand, $\qrsd$ depends only on $\beta$, hence it is related to the dynamics and growth rate of cosmic structures.

For the development of the next chapter, we will need the explicit values of the AP factor for the two FC void catalogues defined in Section~\ref{sec:data_voids2}.
Catalogue FC-l, which assumes a fiducial value of $\Omega_m^l = 0.20$, has a corresponding value of $\qap^l = 1.046$.
Catalogue FC-u, with $\Omega_m^u = 0.30$, has a corresponding value of $\qap^u = 0.960$.
Note that $\qap^l > 1$ for the former, hence according to Eq.~(\ref{eq:q_ap}), it is expected an expansion of these fiducial voids.
Conversely, $\qap^u < 1$ for the latter, hence a contraction is expected.


\subsection{AP and RSD combined contributions}
\label{subsec:zeffects_theo_aprsd}

The volume of a void will be affected by the combined contributions of the AP and RSD effects, which are indistinguishable in observations.
A priori, it is not trivial to ensure that both effects can be treated independently as we did.
However, in the next chapter we will provide statistical evidence of this.
From a theoretical point of view, the fact that the factors $\qrsd$ and $\qap$ encode different cosmological information is a good sign of this assumption.

Assuming this independence, we can relate the $z$-space and $r$-space void radii making a two-step correction: first, we apply Eq.~(\ref{eq:q_ap}) to correct for the AP effect, and then, we apply Eq.~(\ref{eq:q1_rsd}) (or Eq.~\ref{eq:q2_rsd}) to correct for the expansion effect.
Note that this correction can be performed in reverse order as well.
Therefore, the final ratio between the radii in both spatial configurations is obtained by combining both expressions:
\begin{equation}
    \rzs = \qap ~ \qrsd ~ \rrs.
    \label{eq:q_ap_rsd}
\end{equation}


\subsection{Off-centring effect}
\label{subsec:zeffects_theo_vrsd}

A simple visual inspection of Figure~\ref{fig:mxxl_slice} shows that $z$-space void centres are shifted with respect to their $r$-space counterparts.
This \textit{off-centring effect} is a direct consequence of the failure of hypothesis (2) concerning the invariability of centre positions when voids are mapped from $r$-space into $z$-space.
\citeonline{reconstruction_nadathur} remark that this hypothesis is equivalent to assuming that void positions do not suffer any RSD themselves.
On the other hand, \citeonline{lambas_sparkling_2016,ceccarelli_sparkling_2016} and \citeonline{lares_sparkling_2017} demonstrated that voids move through space as whole entities with a net velocity $\velv$.
Inspired by these results, then the off-centring effect can be simply understood as a new kind of RSD induced by the global dynamics of voids, and therefore it is expected that their centres appear preferentially shifted along the LOS when they are identified in $z$-space, in the same way as tracers do.
The right-hand panel of Figure~\ref{fig:effects} depicts this effect schematically.
We will provide a solid statistical evidence of this effect in Section~\ref{subsec:zeffects_stat_dispvel}.

We can make an analytical prediction of this effect by considering it as a dynamical phenomenon.
The void finder used in this work provides the position $\posv = (\posvx, \posvy, \posvz)~[\hmpc]$ and peculiar velocity $\velv = (\velvx, \velvy, \velvz)~[\kms]$ of void centres (see the next section for more details about how these velocities are calculated).
Therefore, in order to account for the LOS shifting of the centres, it is only necessary to write an expression equivalent to Eq.~(\ref{eq:halo_zspace}) but applied to voids:
\begin{equation}
    \Tilde{X}_\mathrm{v 3} = \posvz + \frac{1 + \zsim}{H(\zsim)} \velvz,
	\label{eq:void_zspace}
\end{equation}
where $\Tilde{X}_\mathrm{v3}$ denotes the shifted $\posvz$-coordinate.
As before, it is not trivial to know if the volume and off-centring effects are independent from each other.
In Section~\ref{subsec:zeffects_stat_cross} we will provide evidence of this.
In this case, Eq.~(\ref{eq:pi_rsd}) is still valid provided that $r_\parallel = |\posvz - x_3|$ and $v_\parallel = |\velvz - v_3|$.

Before ending this section, a brief reflection about the $z$-space effects in voids described here.
One the one hand, the expansion effect is a by-product of the RSD induced by \textit{tracer dynamics} at scales around the void radius. 
At these scales, the velocity field of tracers responds to a divergence originated in the local density minimum that defines the void.
On the other hand, the off-centering effect is also a result of the RSD effect, but induced at larger scales.
Its source is the bulk motion of galaxy tracers in the whole region containing the void following the large-scale dynamics of the gravitational field \cite{lares_sparkling_2017}.
This last aspect resembles a \textit{void dynamics}, that presents itself as a different kind of RSD.
Therefore, it is expected that both effects leave a footprint on the cosmological statistics, such as the void size function and the void-galaxy cross-correlation function.

Hereinafter, we will refer to the expansion effect with the acronym t-RSD, whereas we will refer to the off-centring effect with the acronym v-RSD.
This is motivated by the fact that, although both effects are a consequence of the dynamical distortions present in the spatial distribution of galaxies, the net effect manifests at different scales: the prefix t- refers to tracer dynamics itself (classic RSD effect), whereas the prefix v- refers to void dynamics.
This will be particularly useful to distinguish the different contributions to the deviations and anisotropic patterns observed on the void size function and the correlation function.


\section{Statistical analysis}
\label{sec:zeffects_stat}

The statistical analysis of this section aims to provide evidence about the expansion and off-centring effects postulated in the last section.
In the next chapter, we will complete the analysis by incorporating the additional AP change-of-volume effect.
For this reason, we will continue using the TC catalogues, specifically the bijective versions (TC-rs-b and TC-zs-b of Table~\ref{tab:catalogues2}).

This analysis is based on looking for correlations between three statistics that characterise the volume alteration and movement of a void:
(i) the $z$-space to $r$-space radius ratio $\q$, (ii) the $z$-space displacement of the centre $\disp = (\dispx,\dispy,\dispz)$, and (iii) its $r$-space net velocity $\velv = (\velvx,\velvy,\velvz)$.
Specifically, $\disp$ is calculated as the displacement of a void centre in going from $r$-space into $z$-space normalised to the $r$-space radius:
\begin{equation}
    \disp := \frac{\Tilde{\mathbf{X}}_\mathrm{v} - \posv}{\rrs}.
    \label{eq:disp_centres}
\end{equation}
The net velocity, on the other hand, is computed as the sum of all the individual velocities of the haloes that fall inside a spherical shell with dimensions $0.8 \leq r/\rrs \leq 1.2$.
This is an unbiased and fair estimation of the bulk flow velocity of the void, as was demonstrated in \citeonline{lambas_sparkling_2016} (see their Fig.~1).


\subsection{Correlations between \texorpdfstring{$r$}{r}-space and \texorpdfstring{$z$}{z}-space void radii}
\label{subsec:zeffects_stat_rad}

\begin{figure}
    \centering
    \includegraphics[width=79mm]{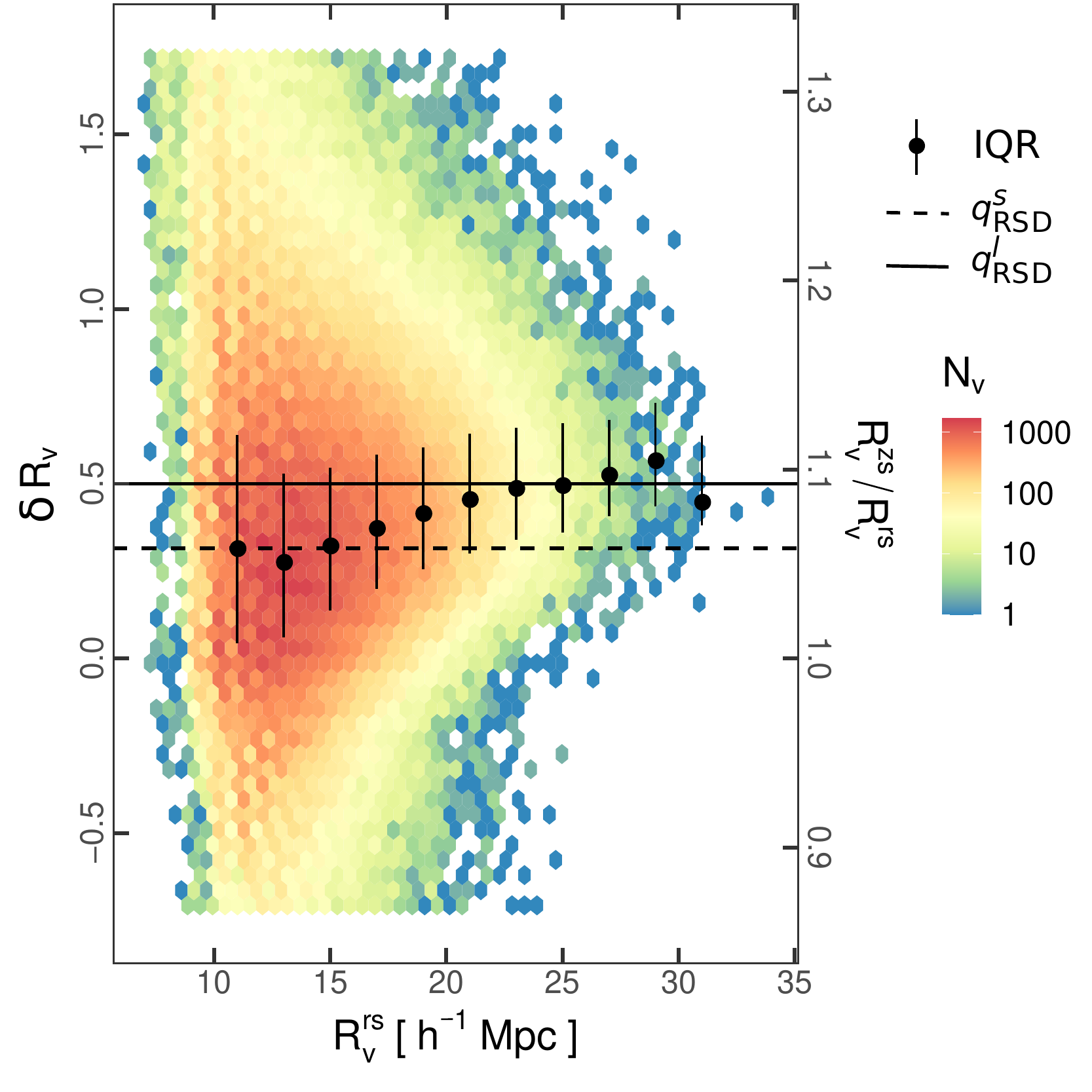}
    \includegraphics[width=79mm]{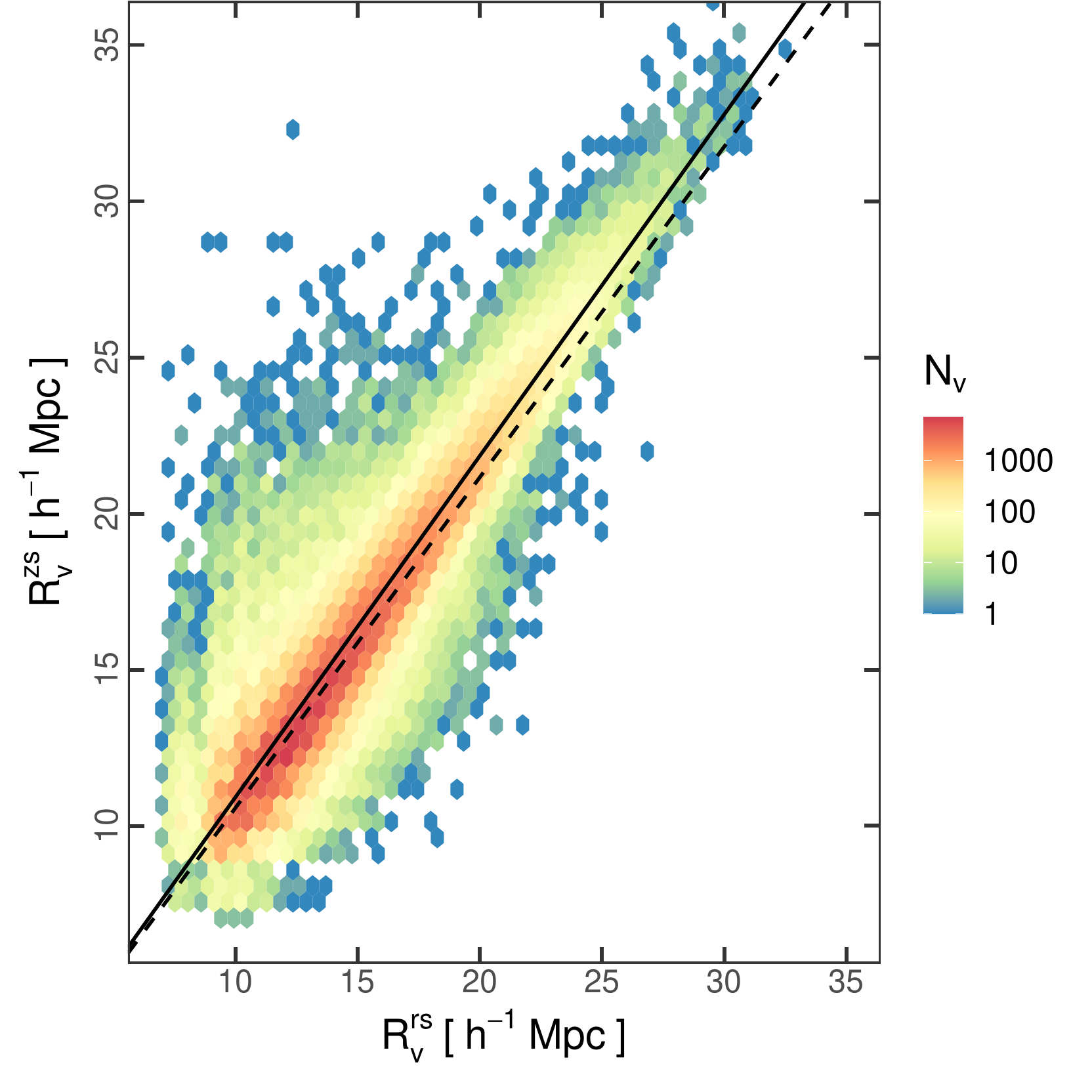}
    \caption[
    Statistical demonstration of the expansion effect.
    ]{
    \textit{Left-hand panel.}
    Two-dimensional distribution $(\rrs,\dR)$ shown as a heat map.
    From blue to red, the colours span from low to high number of void counts in a logarithmic scale.
    The right-hand axis shows the equivalent scale based on the ratio $\q$, related to $\dR$ via Eq.~(\ref{eq:q_rsd}).
    The black dots with error bars show the evolution of the median and interquartile range of this distribution with void radius.
    The horizontal lines indicate the predictions of Eqs.~(\ref{eq:q1_rsd}) ($\dRs = 0.315$, $\qrsds = 1.058$, dashed line) and (\ref{eq:q2_rsd}) ($\dRl = 0.5$, $\qrsdl = 1.092$, solid line).
    \textit{Right-hand panel.}
    Two-dimensional distribution $(\rrs,\rzs)$.
    There is a linear trend, whose slope is correctly described by the factors $\qrsds$ (dashed line) and $\qrsdl$ (solid line).
    In both panels, it is clear that $\qrsds$ is a better predictor of the median for smaller voids, whereas $\qrsdl$ is more suitable for the larger ones.
    This analysis constitutes a statistical demonstration of the t-RSD expansion effect.
    }
    \label{fig:cor_radius}
\end{figure}

The left-hand panel of Figure~\ref{fig:cor_radius} shows the two-dimensional distribution $(\rrs, \dR)$ as a heat map.
From blue to red, the colours span from low to high void counts $N_\mathrm{v}$.
These counts are presented in a logarithmic scale in order to highlight the patterns of the distribution at different scales.
The right-hand axis shows the equivalent scale based on the ratio $\q$, related to $\dR$ via Eq.~(\ref{eq:q_rsd}).
In order to study the evolution of this distribution with void radius, we computed the median and interquartile range (IQR) taking bins of width $2~\hmpc$ in the range $10 \leq \rrs/\hmpc \leq 32$, represented in the figure with black dots and error bars.
The horizontal lines indicate the predictions of Eqs.~(\ref{eq:q1_rsd}) ($\dRs = 0.315$, $\qrsds = 1.058$, dashed line) and (\ref{eq:q2_rsd}) ($\dRl = 0.5$, $\qrsdl = 1.092$, solid line).
Note that $\qrsds$ is a better predictor of the median for smaller voids, whereas $\qrsdl$ is more suitable for larger voids, the ones more relevant for cosmological studies.
The right-hand panel of the figure shows the two-dimensional distribution $(\rrs, \rzs)$.
There is a clear linear trend between both radii, whose slope can be correctly described by the RSD factors $\qrsds$ (dashed line) and $\qrsdl$ (solid line).
As before, $\qrsdl$ is more suitable for larger voids.

From this analysis, we arrive at the second important conclusion of this chapter: voids expand when they are mapped from $r$-space into $z$-space, and this expansion can be statistically quantified as an increase of void radius by a factor $\qrsd$ ($\qrsds$ for smaller voids or $\qrsdl$ for larger voids).
These results give support to the t-RSD expansion effect postulated in Section~\ref{subsec:zeffects_theo_trsd}.

\subsection{Correlations between displacement of centres and net velocity}
\label{subsec:zeffects_stat_dispvel}

\begin{figure}
    \centering
    \includegraphics[width=\textwidth/2]{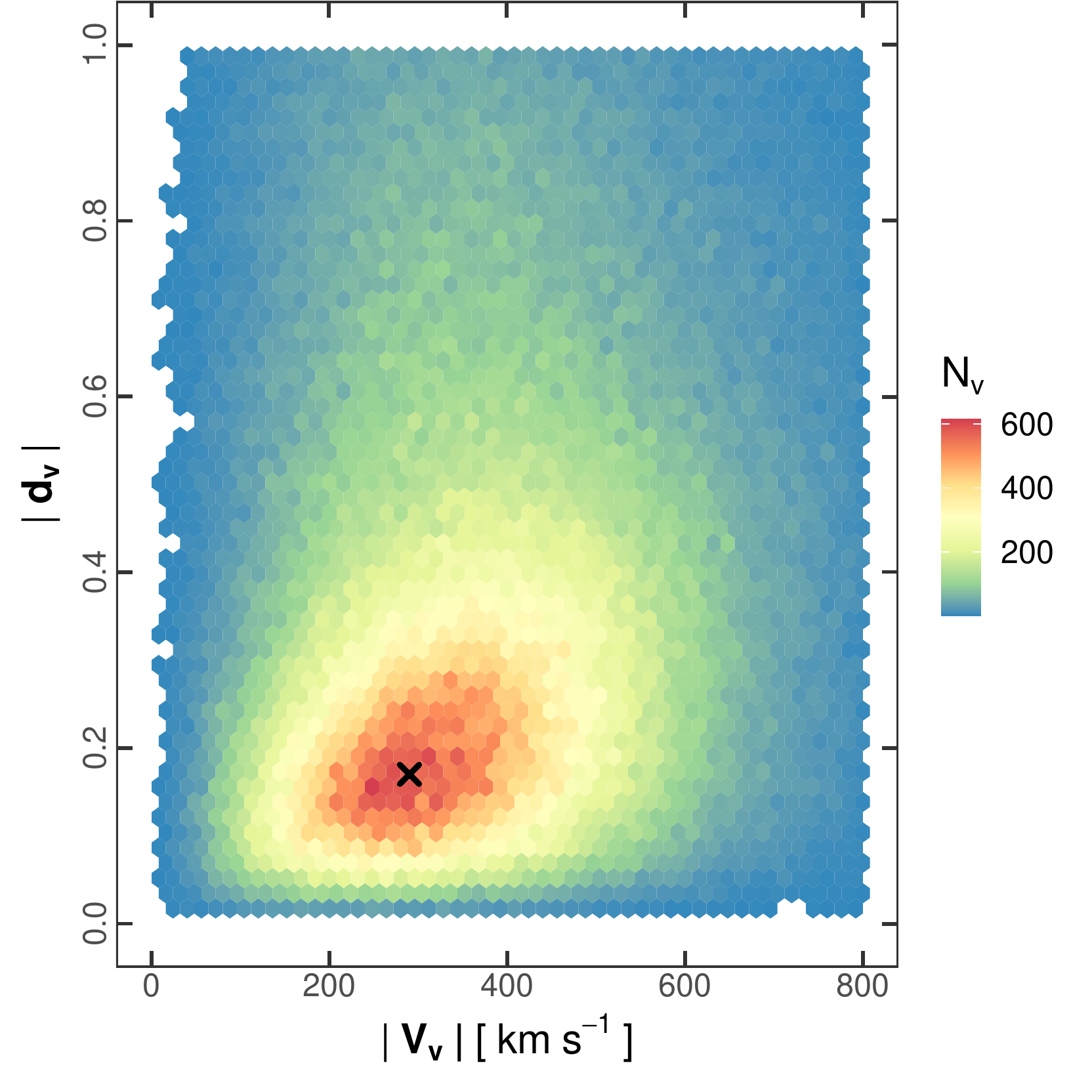}
    \caption[
    Characteristics of the global movement of voids.
    ]{
    Two-dimensional distribution $(|\velv|, |\disp|)$.
    The cross indicates the two-dimensional mode, which shows that voids tend to move with a speed of $290~\kms$, and their centres tend to shift an amount of $0.17~\rrs$.
    }
    \label{fig:cor_disp_vel}
\end{figure}

Figure~\ref{fig:cor_disp_vel} shows the two-dimensional distribution $(|\velv|, |\disp|)$, where the module of these vectors were taken.
This distribution contains information about the dynamics of voids as whole entities.
The cross in the figure indicates the mode of the distribution, which shows that voids tend to move with a speed of $290~\kms$, and their centres tend to displace an amount of $0.17~\rrs$.
It is clear then, that voids cannot be considered at rest.

\begin{figure}
    \centering
    \includegraphics[width=\textwidth/2]{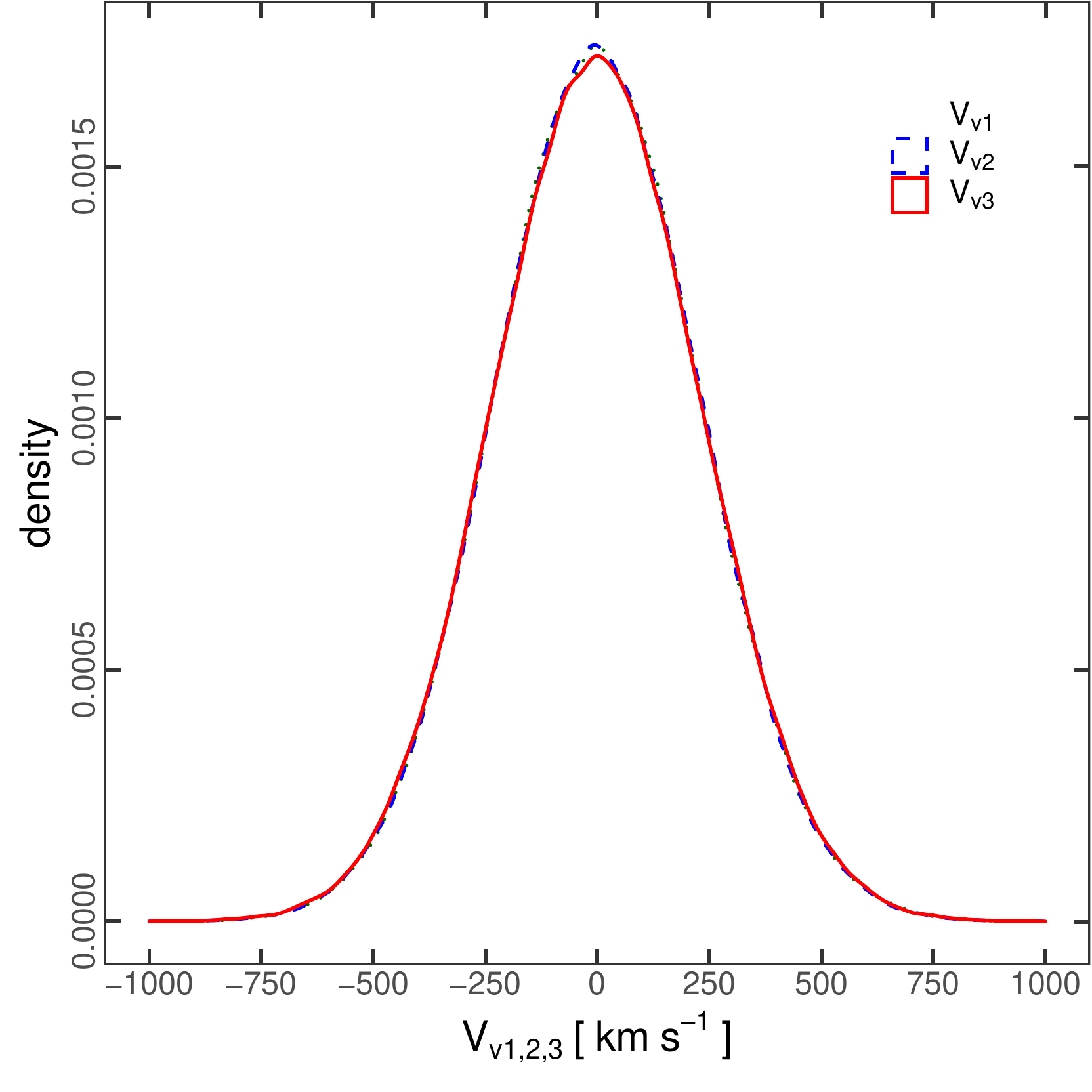}
    \caption[
    Isotropy in the global movement of voids.
    ]{
    Distribution of the components of $\velv$ along the three directions of the simulation box.
    They show Gaussian shapes, centred at $0~\kms$, with a dispersion of $231~\kms$.
    This is a manifestation of the isotropic movement of voids.
    }
    \label{fig:hist_vel}
\end{figure}

Concerning the velocities, Figure~\ref{fig:hist_vel} shows the distribution of the components of $\velv$ along the three directions of the simulation box.
They all show Gaussian shapes, centred at $0~\kms$ with a dispersion of $231~\kms$.
This was expected, since there must not be any privileged direction of motion for voids according to the cosmological principle.

\begin{figure}
    \centering
    \includegraphics[width=79mm]{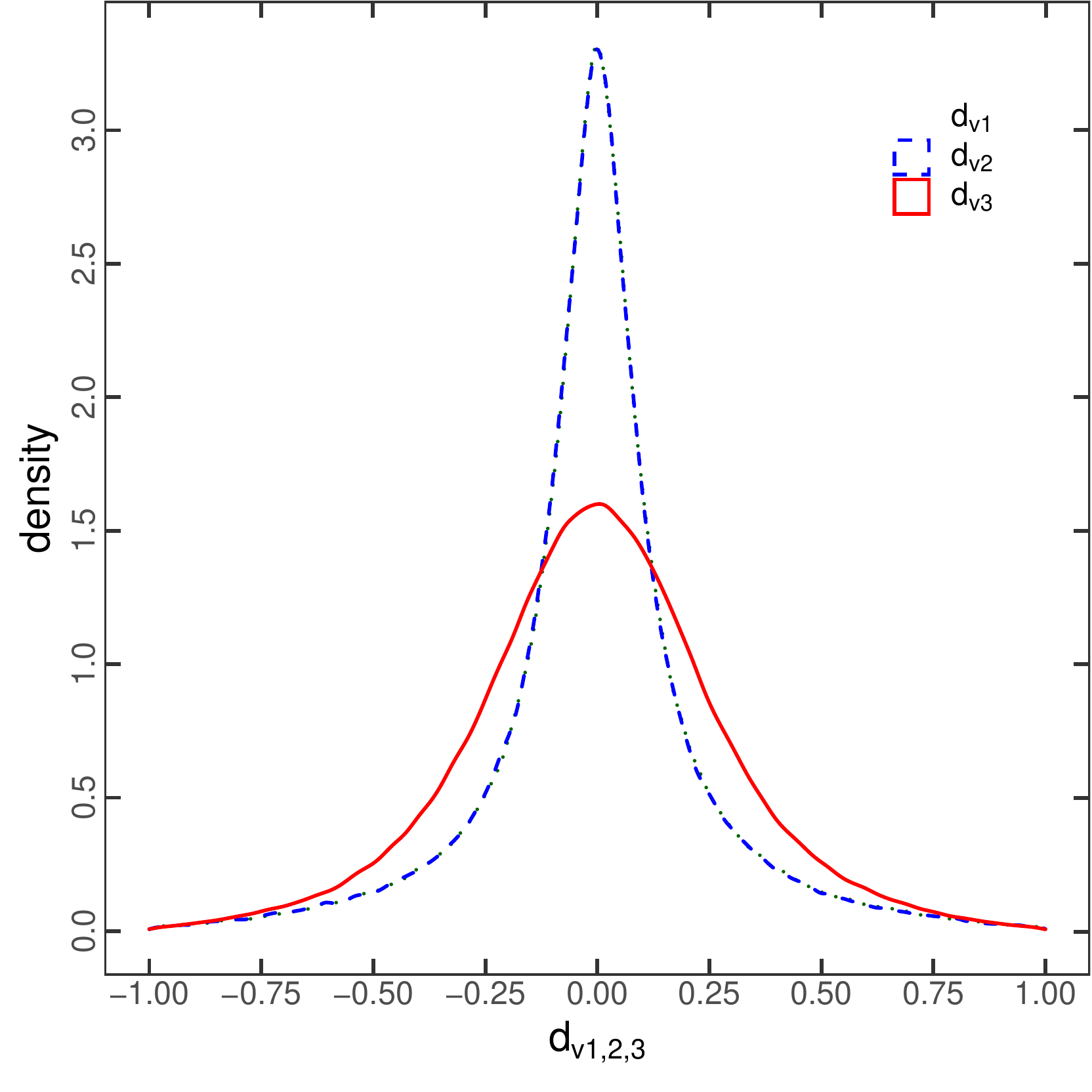}
    \includegraphics[width=79mm]{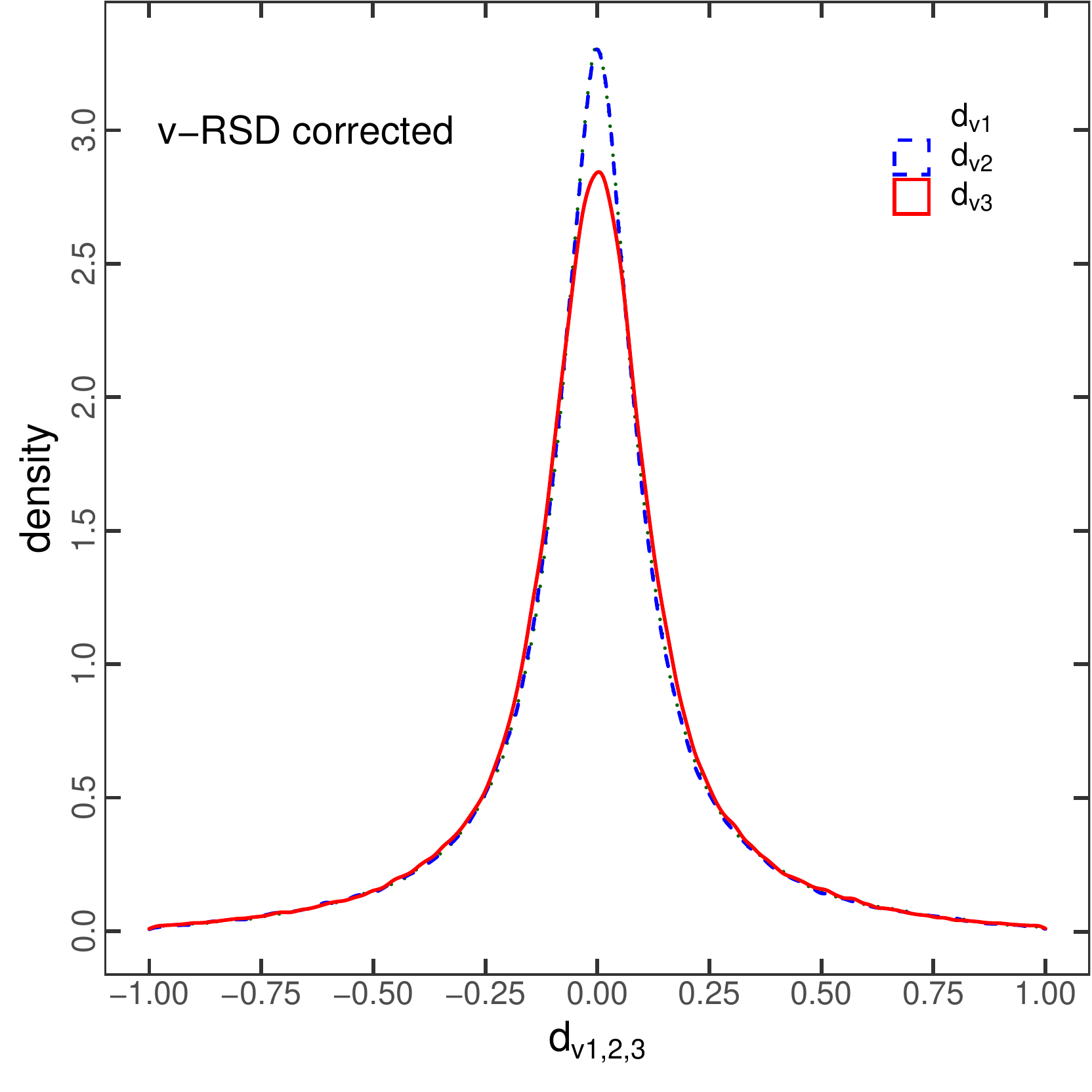}
    \caption[
    Statistical demonstration of the off-centring effect: displacement.
    ]{
    \textit{Left-hand panel}.
    Distribution of the components of $\disp$ along the three directions of the simulation box.
    They show Gaussian shapes centred at $0$.
    The POS distributions (green dotted and blue dashed curves) are almost identical with a dispersion of $0.25$, whereas the LOS distribution (red solid curve) is very different, with a dispersion of $0.3$.
    \textit{Right-hand panel.}
    The three distributions are almost identical after correcting the LOS displacements with Eq.~(\ref{eq:void_zspace}).
    This is a statistical demonstration of the v-RSD off-centring effect.
    }
    \label{fig:hist_disp}
\end{figure}

Concerning the displacements, the left-hand panel of Figure~\ref{fig:hist_disp} shows the distribution of the components of $\disp$ along the three directions of the simulation box.
They all show Gaussian shapes centred at $0$.
However, unlike the velocities, displacements manifest a difference depending on the direction.
On the one hand, the POS distributions (green dotted curve and blue dashed curve) are almost identical, as expected, with a dispersion of $0.25$.
On the other hand, the LOS distribution (red solid curve) has a dispersion of $0.3$, different to the other two.
However, after correcting the LOS displacements with Eq.~(\ref{eq:void_zspace}), a distribution that practically coincides with the previous ones is obtained, as shown in the right-hand panel.
Nevertheless, the three still show a residual and isotropic displacement.
This can be attributed to Poisson noise when the void finder tries to localise the optimum centre.
This is step (4) in the void identification process (Section~\ref{sec:data_id}).

\begin{figure}
    \centering
    \includegraphics[width=79mm]{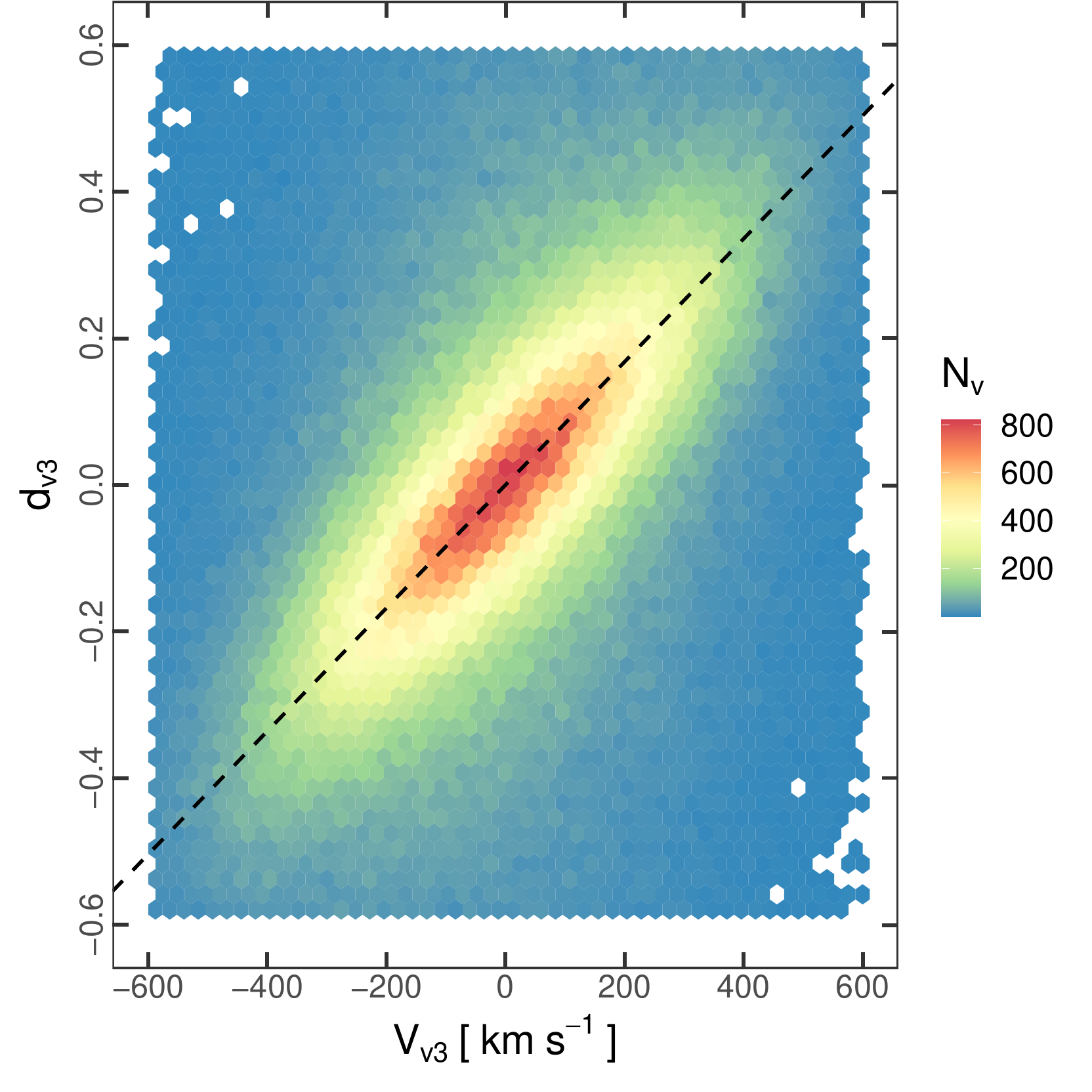}
    \includegraphics[width=79mm]{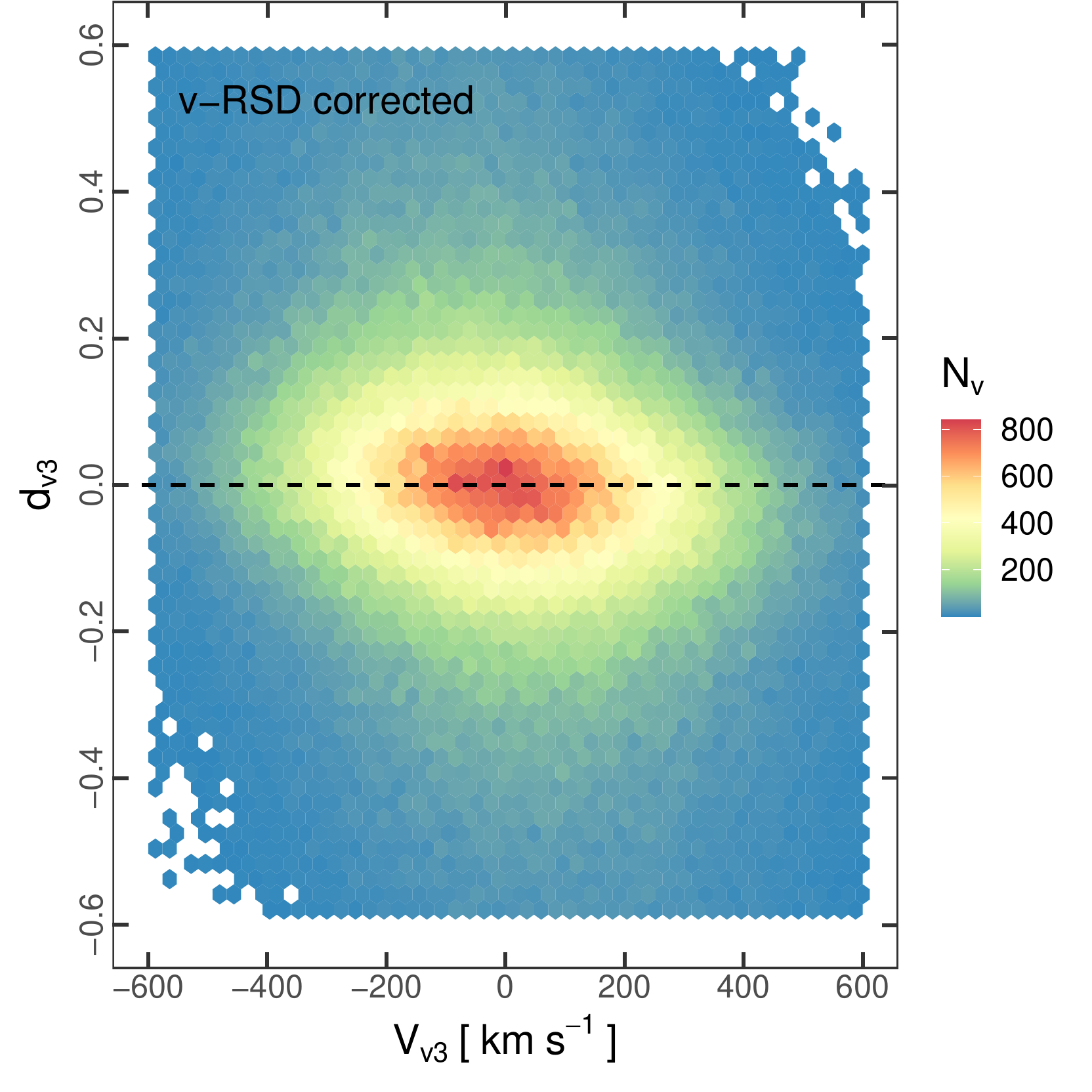}
    \caption[
    Statistical demonstration of the off-centring effect: displacement and velocity.
    ]{
    \textit{Left-hand panel}.
    Two-dimensional distribution $(\velvz, \dispz)$.
    There is a linear trend that is correctly described by the linear relation given by Eq.~(\ref{eq:void_zspace}) (dashed line).
    \textit{Right-hand panel}.
    The correlation disappears after correcting the LOS displacements with this equation.
    This is a statistical demonstration of the v-RSD off-centring effect.    
    }
    \label{fig:cor_disp3_vel3}
\end{figure}

The phenomenon described in the last paragraph is more evident in the left-hand panel of Figure~\ref{fig:cor_disp3_vel3}, where the two-dimensional distribution $(\velvz, \dispz)$ is shown, i.e. for the LOS components of the velocity and displacement.
There is a linear trend between both quantities, which is correctly described by Eq.~(\ref{eq:void_zspace}), represented by the dashed line.
Specifically, the slope of this line is given by the term $(1+\zsim)/H(\zsim)$.
The right-hand panel of the figure shows that, after correcting the LOS displacements with this equation, the correlation disappears, leading to a two-dimensional distribution that is almost identical to the corresponding ones for the POS components: $(\velvx, \dispx)$ and $(\velvy, \dispy)$ (not shown here).

From this analysis, we arrive at the third important conclusion of this chapter: void centres shift systematically along the line of sight when they are mapped from $r$-space into $z$-space.
This is an RSD-type displacement that can be statistically quantified by means of Eq.~(\ref{eq:void_zspace}).
These results give support to the v-RSD off-centring effect postulated in Section~\ref{subsec:zeffects_theo_vrsd}.


\subsection{Cross correlations}
\label{subsec:zeffects_stat_cross}

\begin{figure}
    \centering
    \includegraphics[width=79mm]{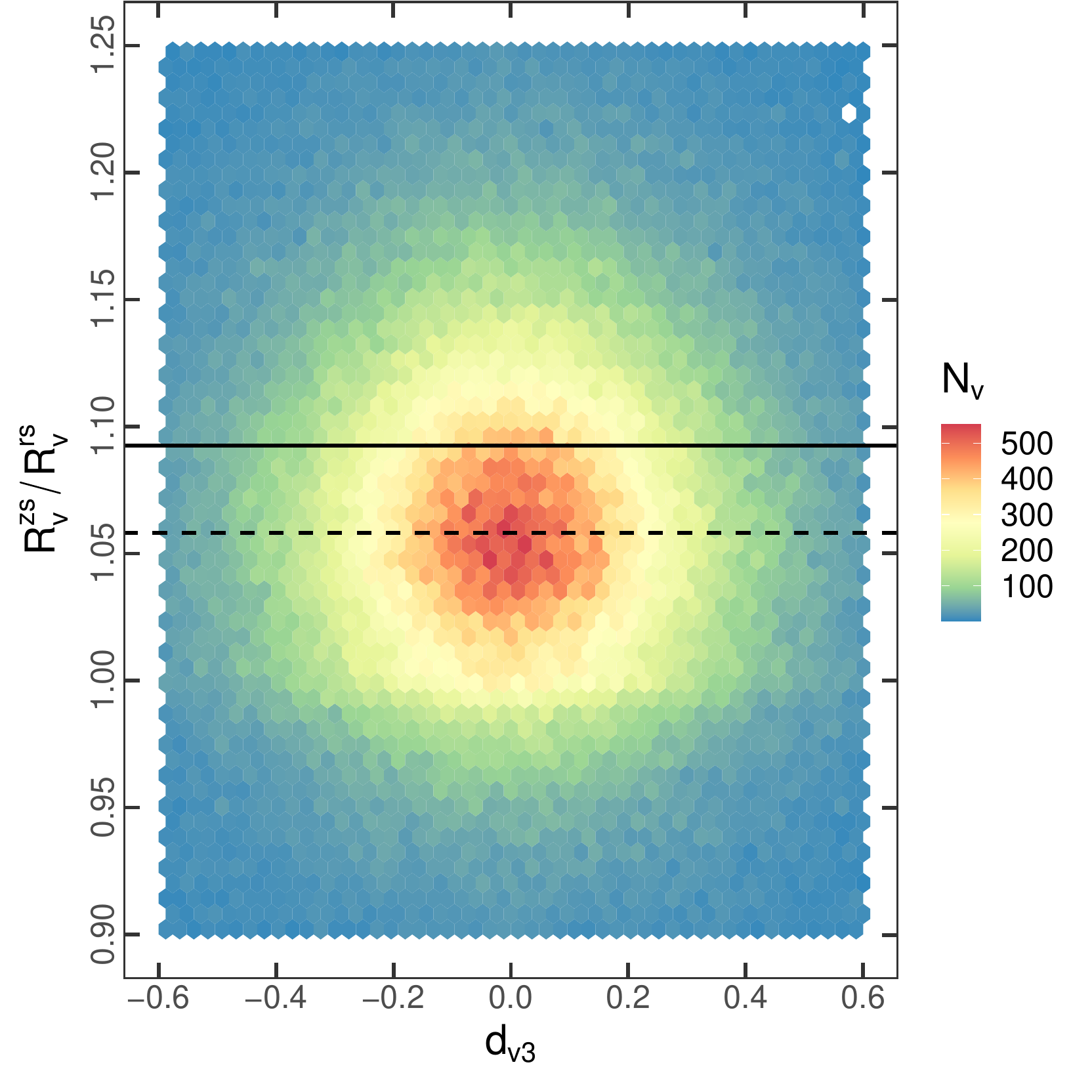}
    \includegraphics[width=79mm]{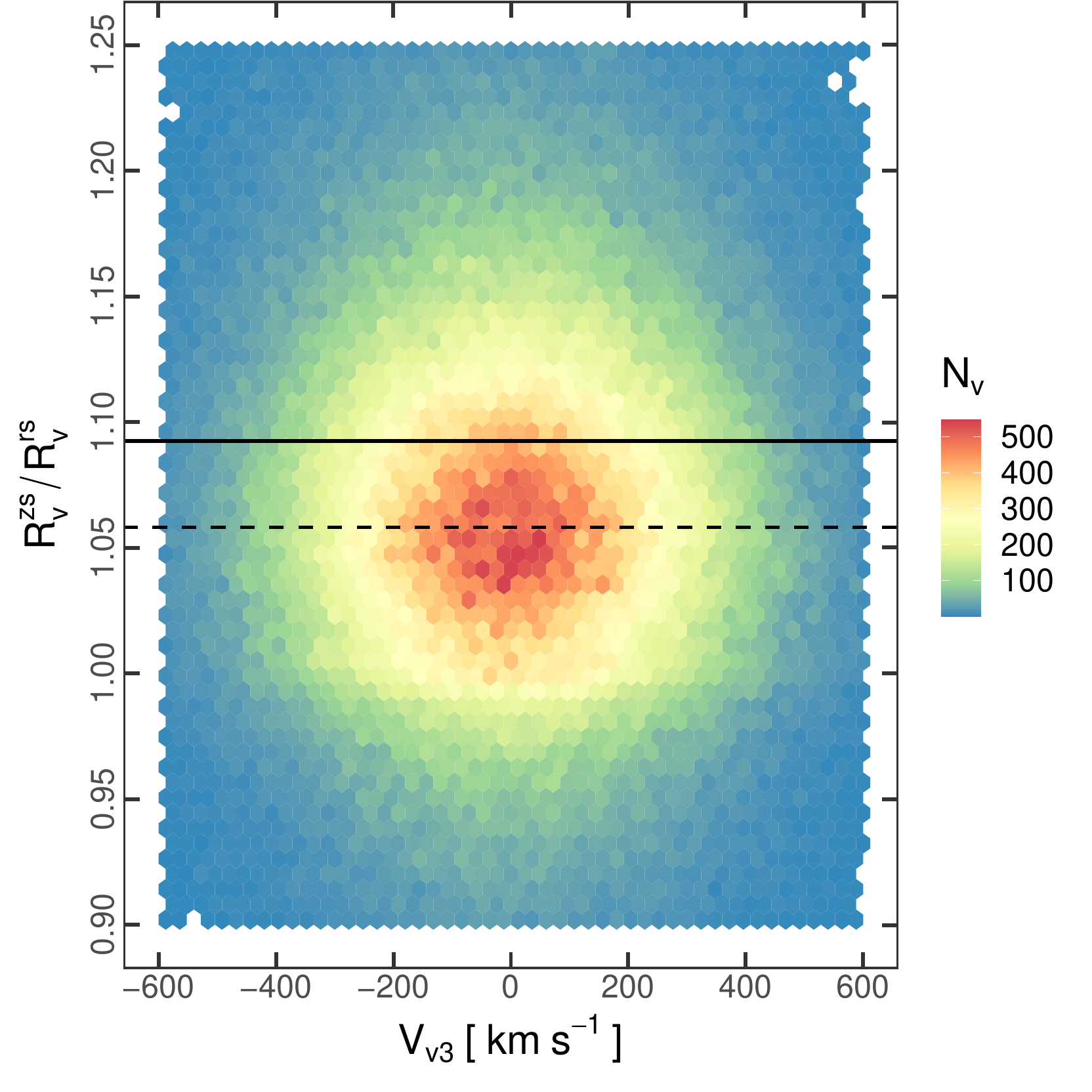}
    \caption[
    Statistical demonstration of the independence between the expansion and off-centring effects.
    ]{
    \textit{Left-hand panel.}
    Two-dimensional distribution $(\dispz, \q)$.
    \textit{Right-hand panel.}
    Two-dimensional distribution $(\velvz, \q)$.
    In both panels, the horizontal lines are the theoretical predictions $\qrsds$ (Eq,~\ref{eq:q1_rsd}, dashed line) and $\qrsdl$ (Eq.~\ref{eq:q2_rsd}, solid line).
    No correlations are observed, suggesting that the t-RSD expansion effect and the v-RSD off-centring effect are statistically independent.
    }
    \label{fig:cor_cross}
\end{figure}

Since the ratio $\q$ (or equivalently $\dR$) characterises the change of volume in voids, the statistical analysis of Section~\ref{subsec:zeffects_stat_rad} gives support to the t-RSD expansion effect postulated in Section~\ref{subsec:zeffects_theo_trsd}.
On the other hand, as the displacement $\disp$ and net velocity $\velv$ characterise the movement of voids, the statistical analysis of Section~\ref{subsec:zeffects_stat_dispvel} gives support to the v-RSD off-centring effect postulated in Section~\ref{subsec:zeffects_theo_vrsd}.
It only remains to test if both effects are statistically independent by looking for cross correlations between these quantities.
Figure~\ref{fig:cor_cross} shows the two-dimensional distributions $(\dispz, \q)$ (left-hand panel) and $(\velvz, \q)$ (right-hand panel).
The horizontal lines are the theoretical predictions given by $\qrsds$ (dashed line) and $\qrsdl$ (solid line).
No correlations are observed, giving support to the postulated independence.
Although we have only presented the distributions corresponding to the LOS components of the displacement and velocity, it is worth mentioning that the analogue POS distributions, namely $(\dispx, \q)$, $(\dispy, \q)$, $(\velvx, \q)$ and $(\velvy, \q)$, show a similar behaviour.

The results presented in this section allow us to make the following interpretation: the large-scale dynamics of the region containing a void (v-RSD) is decoupled from the dynamics of the tracers at scales of the void radius (t-RSD or classic RSD).
This also suggests that the potential distortion patterns in observations due to these two effects can be treated separately.
This is the fourth important conclusion of this chapter.

\chapter{Impact on the void size function}
\label{chp:impact_vsf}
\section*{Abstract}

The three redshift-space systematicities that affect the void identification process: the expansion effect, the off-centring effect and the AP change-of-volume effect, have an impact on the cosmological statistics, since they generate additional distortion patterns in the measurements.
This leads to biased cosmological constraints if they are not taken into account appropriately.
In this chapter, we focus on the impact that they have on the void size function.
We will tackle the consequences on the correlation function in the next chapter.
Using the theoretical framework developed in the previous chapter to model these effects, we were able to recover the statistical properties of the abundance of voids in real space.
We found that the VSF is only affected by the t-RSD and AP effects, while it is free of the v-RSD effect.
In this way, we lay the foundations for improvements in current models for the abundance of voids in order to obtain unbiased cosmological constraints from modern spectroscopic surveys.
This part of the work can be found in the same published paper that contains the analysis of the previous chapter \cite{zvoids_correa}.


\section{A comment about the VSF modelling}
\label{sec:impact_vsf_general}

We start this chapter by reviewing some generalities concerning the modelling of the abundance of voids.
As we discussed in Section~\ref{sec:voids_abundance}, the real-space VSF can be modelled using the excursion set formalism in combination with the spherical evolution of matter underdensities derived from perturbation theory.
This model is analogous to the mass function used to describe the abundance of dark-matter haloes.
We also saw that there are two main approaches: the SvdW model \cite{svdw} and the Vdn model \cite{abundance_jennings}.

It is important to highlight that both models are only applicable to voids identified from the matter distribution.
In simulations, this corresponds to voids identified from dark-matter particles.
Voids identified from matter tracers like haloes or galaxies, by contrast, can be substantially different regarding their statistical properties.
Nevertheless, as we discussed in Section~\ref{sec:data_id}, many authors claim that both types of voids can still be related to each other with a linear bias approach, and hence, the SvdW and Vdn models are still valid in the case of tracers.
Such a model should fit the $r$-space abundances of Figure~\ref{fig:TC_VSF} (blue curves).

In practice, however, it is only possible to identify voids from the observed spatial distribution of galaxies.
Therefore, the $z$-space effects studied in the previous chapter are expected to have a strong impact on the VSF.
We tackle this problematic in this chapter using the theoretical framework developed in Section~\ref{sec:zeffects_theo} to describe these effects.
This description has a strong cosmological dependence, hence it must be combined with the excursion set formalism in order to obtain unbiased cosmological constraints from spectroscopic surveys.
In this way, we lay the foundations for a complete treatment of the VSF modelling, leaving for a future investigation a full analysis combining both developments.


\section{Alcock-Paczyński correction}
\label{sec:impact_vsf_ap}

This chapter has a double intention.
On the one hand, we will finish the analysis of the previous chapter by incorporating the additional AP change-of-volume effect not treated yet.
On the other hand, we will study the impact of all the $z$-space effects in voids (t-RSD, v-RSD and AP) on the VSF as a cosmological test.
For this reason, we now turn to the FC void catalogues, fully affected by $z$-space systematicities (see Table~\ref{tab:catalogues2}).

\begin{figure}
    \centering
    \includegraphics[width=79mm]{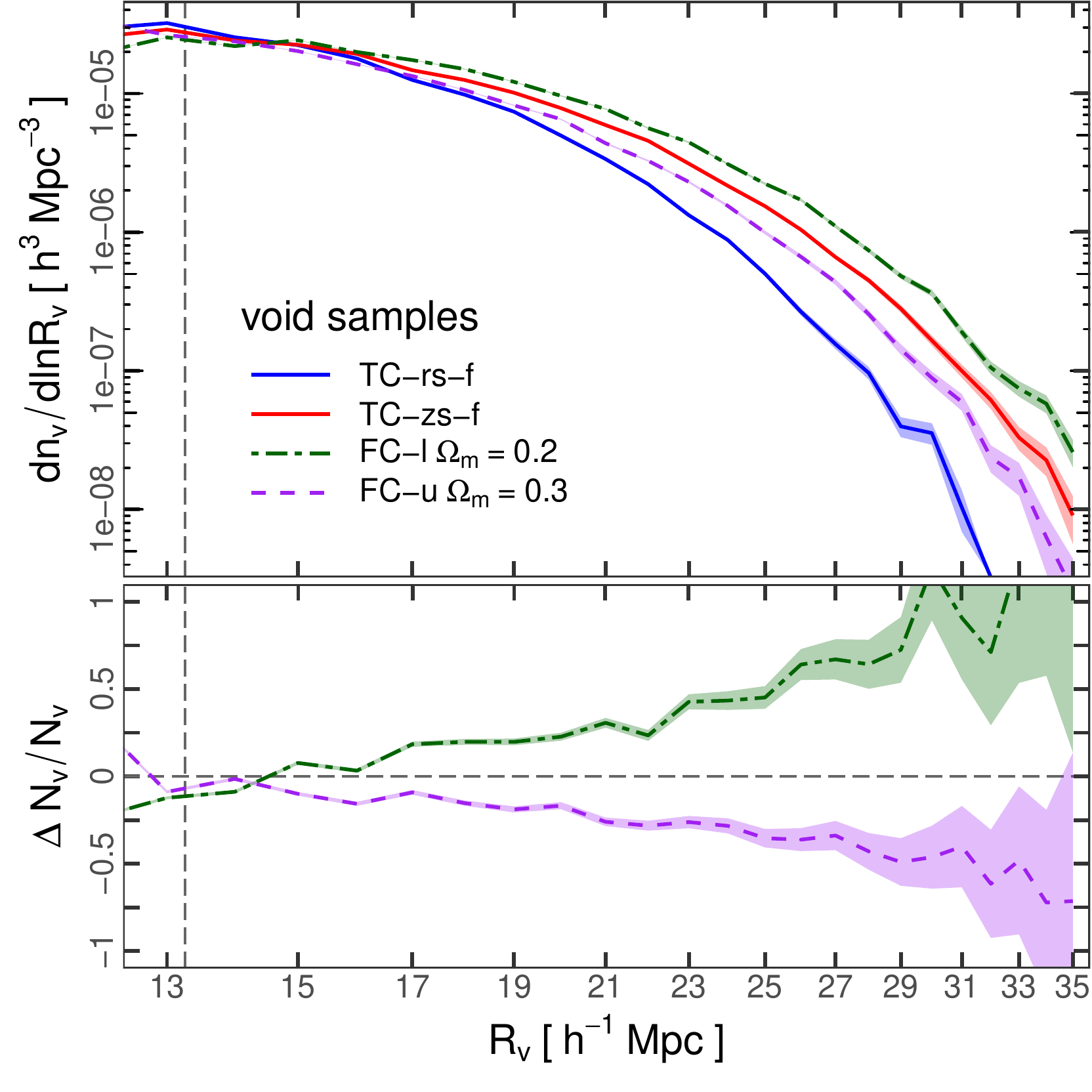}
    \includegraphics[width=79mm]{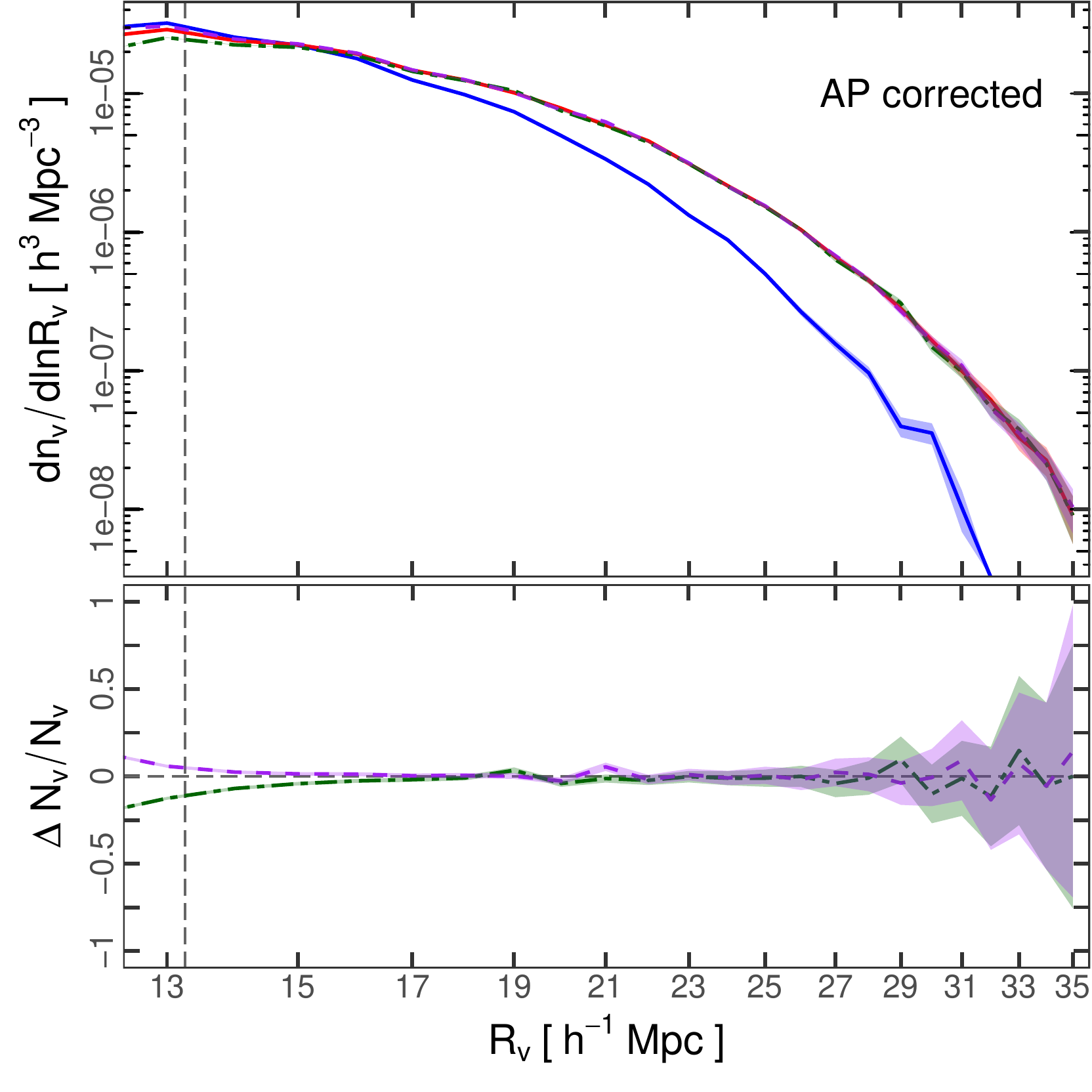}
    \caption[
    Void size functions of the FC void catalogues (Table~\ref{tab:catalogues2}). Alcock-Paczy\'nski correction.
    ]{
    Alcock-Paczyński correction of void abundance.
    \textit{Left-hand panel.}
    Abundances of the FC catalogues.
    The VSF of the FC-l catalogue (with a fiducial value of $\Omega_m^l = 0.20$) is represented by the green dot-dashed curve, whereas the VSF of the FC-u catalogue (with a fiducial value of $\Omega_m^u = 0.30$) is represented by the purple dashed curve.
    By way of comparison, the abundances of catalogues TC-rs-f (blue solid curve) and TC-zs-f (red solid curve) are also shown as references.
    The corresponding fractional differences of void counts between the FC and TC-zs-f catalogues are shown in the lower panel.
    \textit{Right-hand panel.}
    The same as the left-hand panel, but after performing the correction over the FC abundances.
    }
    \label{fig:FC_VSF_AP}
\end{figure}

The left-hand panel of Figure~\ref{fig:FC_VSF_AP} shows the void abundances of the two FC catalogues.
Since they are fully affected by RSD and AP distortions, they mimic two possible observational measurements.
The VSF of the FC-l catalogue, which assumes a fiducial value of $\Omega_m^l = 0.20$, lower than the true MXXL value, is represented by the green dot-dashed curve, whereas the VSF of the FC-u catalogue, which assumes a higher fiducial value of $\Omega_m^u = 0.30$, by the purple dashed curve.
The goal of this section is to correct these abundances using the theoretical framework developed in Section~\ref{sec:zeffects_theo} in order to recover the true underlying $r$-space abundance.
Graphically, we aim to recover the blue solid curve of the figure, corresponding to the TC-rs-f catalogue.
To do this, it is sufficient to correct each void radius by applying Eq.~(\ref{eq:q_ap_rsd}) and using the values of the AP and RSD factors that we have derived for the MXXL simulation: $\qrsdl=1.092$, $\qap^l=1.046$ and $\qap^u=0.960$.

Instead of performing this simple correction directly, we will split it in a two-step procedure in order to discuss the different physical mechanisms involved.
In this section, we discuss the first step, the correction for the AP change-of-volume effect with the AP factors.
In the next section, we will discuss the second step, the correction for the t-RSD expansion effect with the RSD factor.
The goal here is to recover the $z$-space VSF of reference, which is affected by RSD but unaffected by the AP effect.
Graphically, we aim to recover the red solid curve of the figure, corresponding to the TC-zs-f catalogue.
This correction is shown in the right-hand panel of Figure~\ref{fig:FC_VSF_AP}.

The first aspect clearly seen when comparing the abundances of the FC-l and FC-u catalogues with respect to the $z$-space VSF of reference, is that a higher abundance is obtained when a lower value of $\Omega_m$ is assumed, whereas the opposite behaviour occurs when a higher value of $\Omega_m$ is assumed.
In the context of the bijective mapping, this means that FC-l voids are systematically bigger, whereas FC-u voids are systematically smaller.
This is in excellent agreement with our discussion of Section~\ref{subsec:zeffects_theo_ap}, where we expected an AP-expansion for the former, since $\qap^l > 1$, and an AP-contraction for the latter, since $\qap^u < 1$.
Note that after correcting the radii for the AP change-of-volume effect (right-hand panel), both curves coincide with the $z$-space VSF of reference remarkably well at all scales of interest.
This is better appreciated by looking at the lower panels, where we show the corresponding fractional differences of void counts, $\Delta N_\mathrm{v}/N_\mathrm{v} = (N_\mathrm{v}^\mathrm{FC}-N_\mathrm{v}^\mathrm{zs})/N_\mathrm{v}^\mathrm{zs}$, as an indicator of the quality of the correction.
Note that after the correction, the differences between the fiducial and the $z$-space abundances are notably reduced, being $\Delta N_\mathrm{v}/N_\mathrm{v} < 0.2$ in the worst case.
Furthermore, this is a clear signature that this effect is independent of the others.

We arrive here at the first important conclusion of this chapter: the volume of voids is also affected by the fiducial cosmology assumed to measure distances, which manifests as an overall expansion or contraction depending on the chosen fiducial parameters.
Moreover, this effect is independent of the other $z$-space systematicities, and can be statistically quantified as a variation of the radius by a factor $\qap$.
These results give support to the AP change-of-volume effect postulated in Section~\ref{subsec:zeffects_theo_ap}.

\section{Expansion-effect correction}
\label{sec:impact_vsf_trsd}

\begin{figure}
    \centering
    \includegraphics[width=79mm]{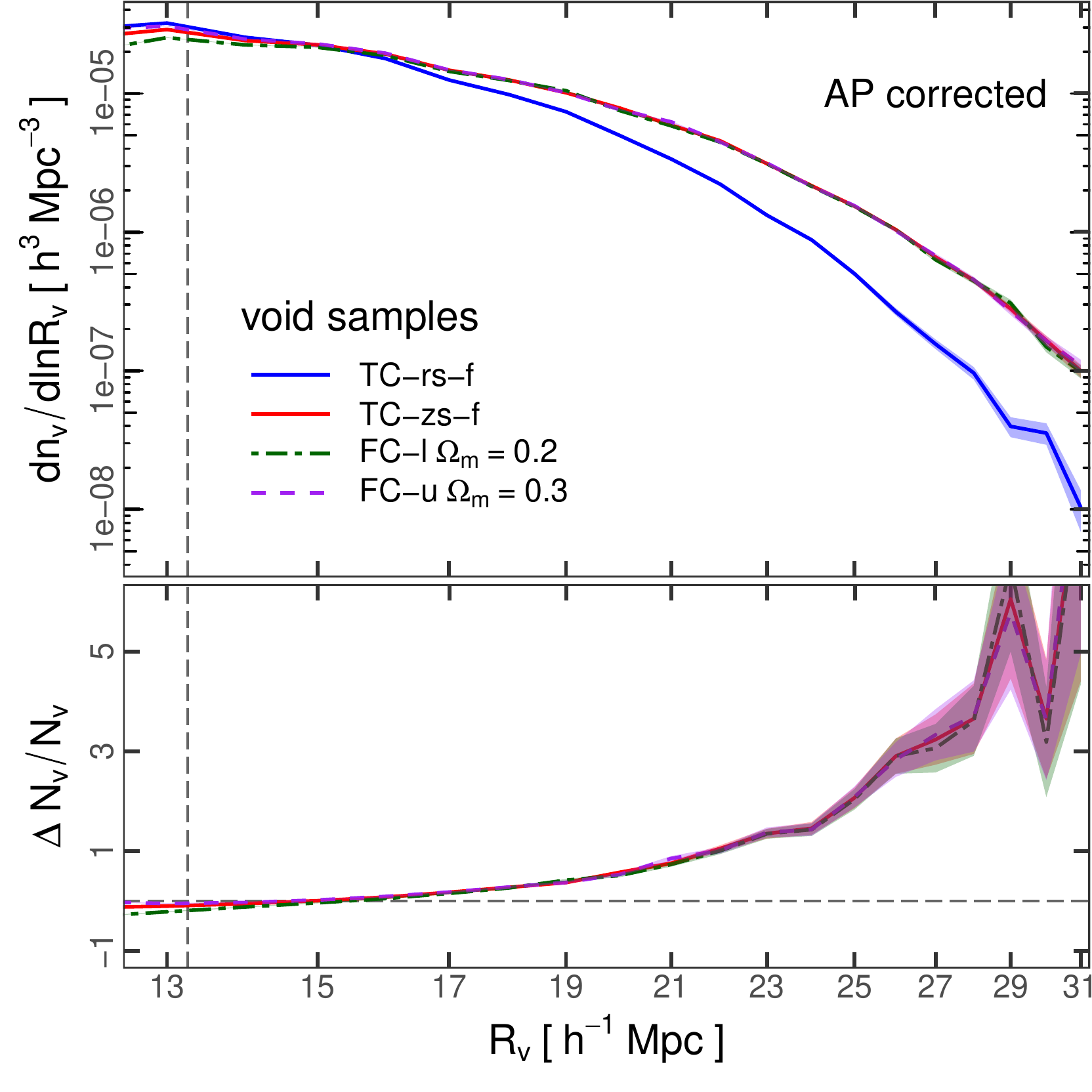}
    \includegraphics[width=79mm]{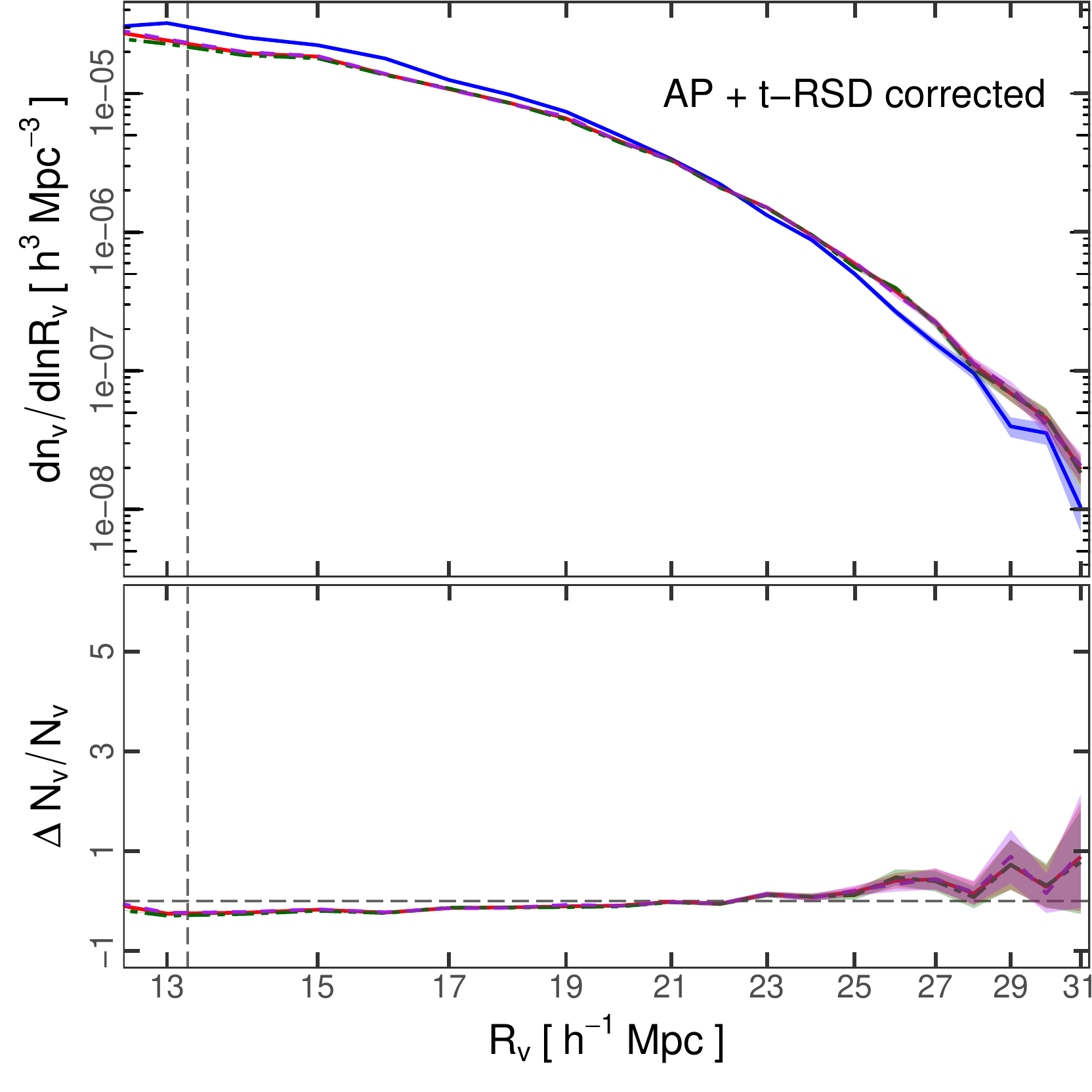}
    \caption[
    Expansion-effect correction of void abundances.
    ]{
    Expansion-effect correction of void abundance.
    \textit{Left-hand panel.}
    The same as the right-hand panel of Figure~\ref{fig:FC_VSF_AP}, except for the fact that the fractional differences are taken with respect to the TC-rs-f catalogue.
    \textit{Right-hand panel.}
    The same as the left-hand panel, but after performing the t-RSD correction.
    }
    \label{fig:FC_VSF_AP_RSD}
\end{figure}

In this section, we discuss the second step of the correction: the t-RSD expansion effect.
The goal now is to recover the $r$-space VSF of reference.
This is shown in Figure~\ref{fig:FC_VSF_AP_RSD}.
The left-hand panel is the same as the right-hand panel of Figure~\ref{fig:FC_VSF_AP}, except for the fact that the fractional differences are now referred to the $r$-space catalogue:
$\Delta N_\mathrm{v}/N_\mathrm{v} = (N_\mathrm{v}^\mathrm{FC}-N_\mathrm{v}^\mathrm{rs})/N_\mathrm{v}^\mathrm{rs}$.
The right-hand panel shows the correction per se.
This is satisfactory for all radii of interest, although there are some appreciable deviations at the smallest scales near to the threshold imposed by the median of the distribution.
Note that the large differences between $z$-space and $r$-space voids, already noted in Figure~\ref{fig:TC_VSF}, which can be as high as $\Delta N_\mathrm{v}/N_\mathrm{v} > 4$ at the largest scales, have been reduced to $\Delta N_\mathrm{v}/N_\mathrm{v} < 0.8$ in all cases.

\begin{figure}
    \centering
    \includegraphics[width=\textwidth/2]{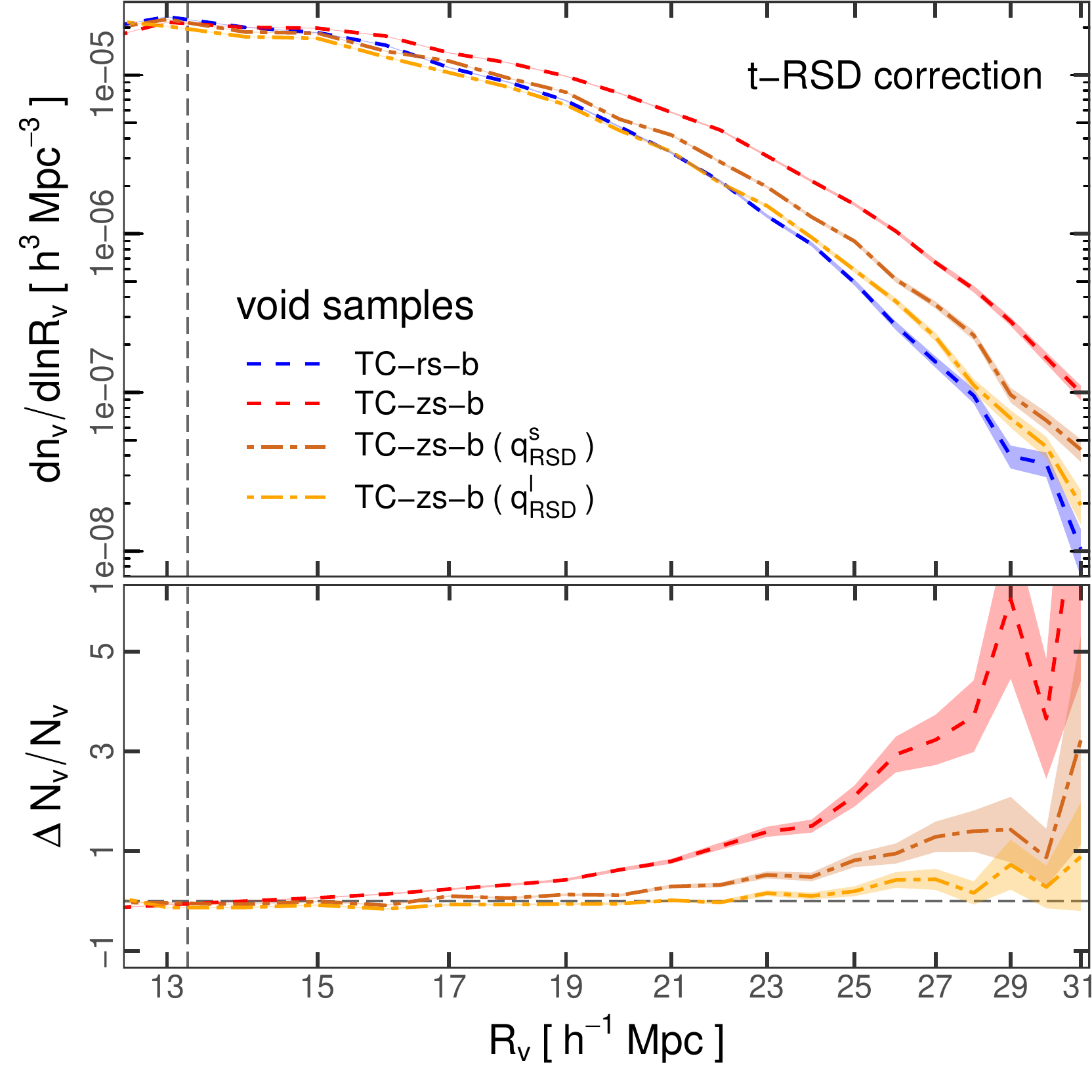}
    \caption[
    Expansion-effect correction: comparative analysis between the full and bijective catalogues, and between the factors $\qrsds$ and $\qrsdl$.
    ]{
    Expansion-effect correction using the bijective catalogues instead of the full ones.
    \textit{Upper panel.}
    The red and blue dashed curves represent the VSF of catalogues TC-zs-b and TC-rs-b, the same as in Fig.~\ref{fig:TC_VSF}.
    The brown dot-dashed curve represents the correction using Eq.~(\ref{eq:q1_rsd}) with the factor $\qrsds$.
    The orange dot-dashed curve, the correction using Eq.~(\ref{eq:q2_rsd}) with the factor $\qrsdl$.
    \textit{Lower panel.}
    Fractional differences of void counts between the uncorrected and corrected $z$-space abundances with respect to the $r$-space one (TC-rs-b).
    Note that $\qrsdl$ performs better than $\qrsds$.
    A comparison with Figure~\ref{fig:FC_VSF_AP_RSD} demonstrates that the deviations at small radii have been attenuated.
    They are due to the contamination of non-bijective voids in the sample, which typically have these sizes.
    }
    \label{fig:TC_VSF_RSD}
\end{figure}

For the analysis up to here, we have used as references the full catalogues in their two versions: TC-rs-f for $r$-space, and TC-zs-f for $z$-space.
However, we could have used the bijective catalogues instead.
This was motivated by the fact that, in the spirit of the bijective mapping analysis, the full and bijective catalogues can be treated indistinctly.
On the other hand, we have used Eq.~(\ref{eq:q2_rsd}) (with $\qrsdl$) to correct the void radii for the expansion effect instead of using Eq.~(\ref{eq:q1_rsd}) (with $\qrsds$).
This was motivated by the fact that the former is more suitable for those voids above the shot-noise level.
Therefore, we repeated the analysis, using this time the bijective catalogues as references: TC-rs-b for $r$-space, and TC-zs-b for $z$-space.
This allows us to test the impact of the impurity due to non-bijective voids in the full catalogues.
Moreover, we evaluated the performance of both RSD factors, $\qrsds$ and $\qrsdl$.
Given that the AP correction works well at all scales, we exclusively focused on the correction for the expansion effect.
Specifically, starting from the $z$-space VSF corresponding to the TC-zs-b catalogue, we set out to retrieve the $r$-space VSF corresponding to the TC-rs-b catalogue.
This is shown in Figure~\ref{fig:TC_VSF_RSD}.
The red and blue curves are the same as in Figure~\ref{fig:TC_VSF}.
The brown dot-dashed curve represents the correction made with the factor $\qrsds$, whereas the orange dot-dashed curve, the analogue correction made with the factor $\qrsdl$.
Two conclusions can be made.
First, $\qrsdl$ performs better than $\qrsds$, specially at larger scales.
This confirms our suggestion that $\qrsdl$ is more suitable than $\qrsds$ for describing large voids, the ones of interest for cosmological analyses.
Second, unlike what happens in Figure~\ref{fig:FC_VSF_AP_RSD}, there are not any appreciable deviations at small scales.
These deviations can then be attributed to the contamination of non-bijective voids in the full catalogues at these scales.
The analysis carried out in this section reinforces the conclusions formulated in Section~\ref{subsec:zeffects_stat_rad}.

\section{Free of off-centring effect and cosmological relevance}
\label{sec:impact_vsf_vrsd}

The achievements of the correction process prove another important fact: the VSF is not affected by the v-RSD off-centring effect.
This was implicitly assumed in the two-step correction formula, since Eq.~(\ref{eq:q_ap_rsd}) only includes the t-RSD and AP effects contained in the factors $\qrsd$ and $\qap$, respectively.
This constitutes the second important conclusion of this chapter.

In summary, the only two ingredients necessary to correct an observational VSF are the AP and RSD factors, which relate void radii in $r$-space and $z$-space.
These factors are simply two proportionality constants since they are scale-independent.
Moreover, they depend strongly on cosmology, encoding different information in a decoupled way.
On the one hand, $\qap$ depends only on the background cosmological parameters, such as $\Omega_m$, $\Omega_\Lambda$ and $H_0$, hence it is related to the expansion history and geometry of the Universe.
On the other hand, $\qrsd$ depends only on $\beta$, hence it is related to the dynamics and growth rate of cosmic structures.
Therefore, the framework developed in this work must be combined with the excursion set theory used to model void abundances in order to obtain unbiased cosmological constraints from spectroscopic surveys.
This is the third important conclusion of this chapter.

\chapter{Impact on the correlation function}
\label{chp:impact_vgcf}
\section*{Abstract}

In this chapter, we focus on the impact that the AP, t-RSD and v-RSD effects have on the void-galaxy cross-correlation function, specifically, on the projected versions that we developed in Chapter~\ref{chp:cosmotest}.
We found that, unlike the case of the void size function, the correlation function is affected by the three redshift-space systematicities; the t-RSD effect being the most important.
We also found a fourth source of distortions not previously considered in the literature: the intrinsic ellipticity of voids (e-RSD).
This is the first time that the v-RSD and e-RSD effects are detected and quantified from measurements in simulations.
With a simple preliminary test, we verified that the Gaussian streaming model is still a robust model provided that all these effects are taken into account.
In this way, we lay the foundations for improvements in current models for the correlation function in order to obtain unbiased cosmological constraints from modern spectroscopic surveys.
The analysis carried out here is not only important from this practical point of view, but also to deepen our understanding about the structure and dynamics of voids, and more generally, of the Universe at the largest scales.
Moreover, some of the effects studied here can constitute cosmological tests by themselves, as is the case of the ellipticity of voids.
This part of the work will be submitted soon to be considered for publication (Correa et al., in preparation).


\section{Void sample}
\label{sec:impact_vgcf_sample}

In this chapter, we will not consider AP distortions.
In Section~\ref{sec:impact_vgcf_ap}, we will justify this choice.
For this reason, we now return to the TC void catalogues.
As the analysis of this chapter complements the two previous ones, we will continue using the MXXL snapshot $\zsim = 0.99$, keeping in mind that the remaining snapshots evince the same general results.

\begin{figure}
    \centering
    \includegraphics[width=79mm]{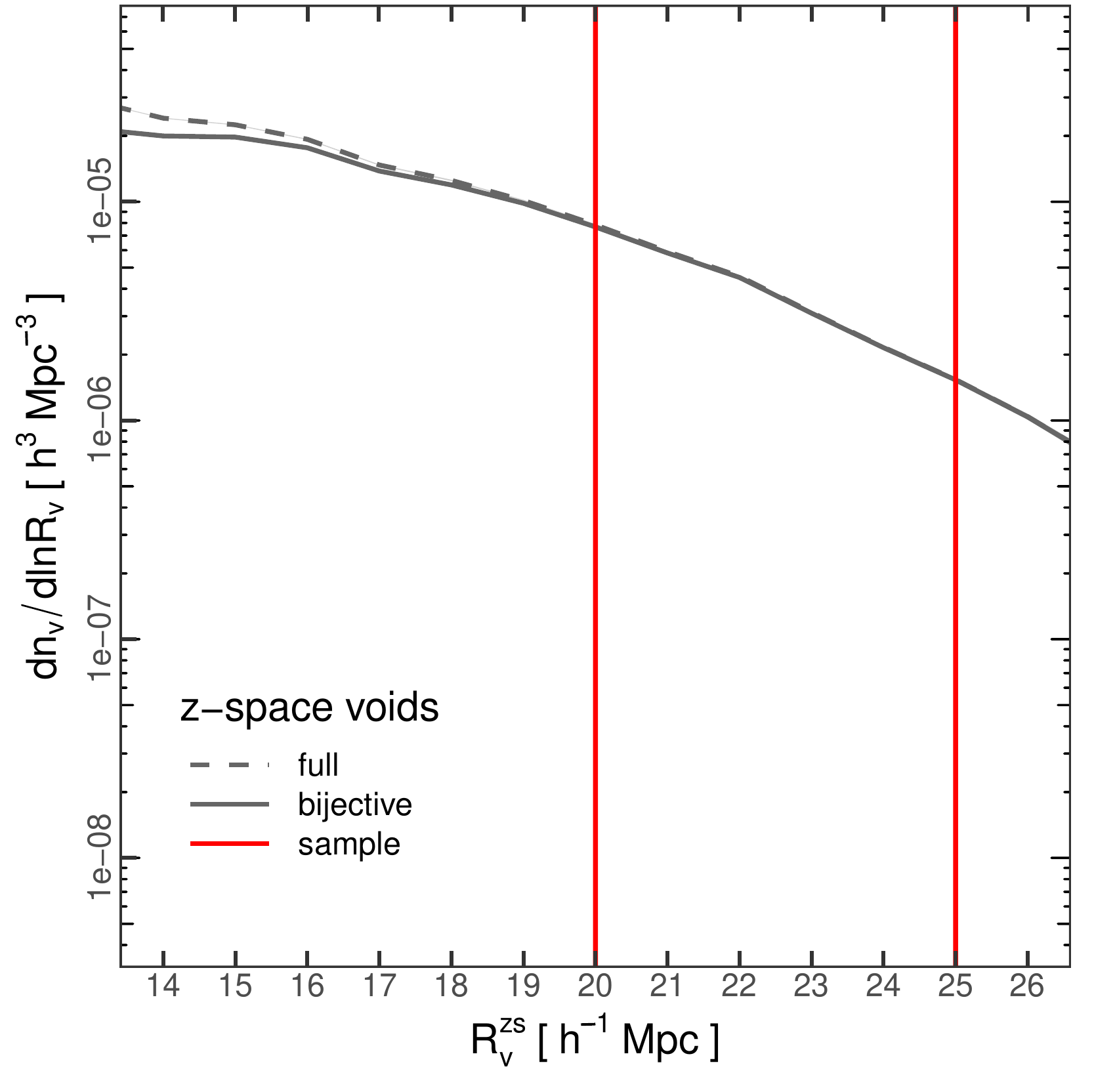}
    \includegraphics[width=79mm]{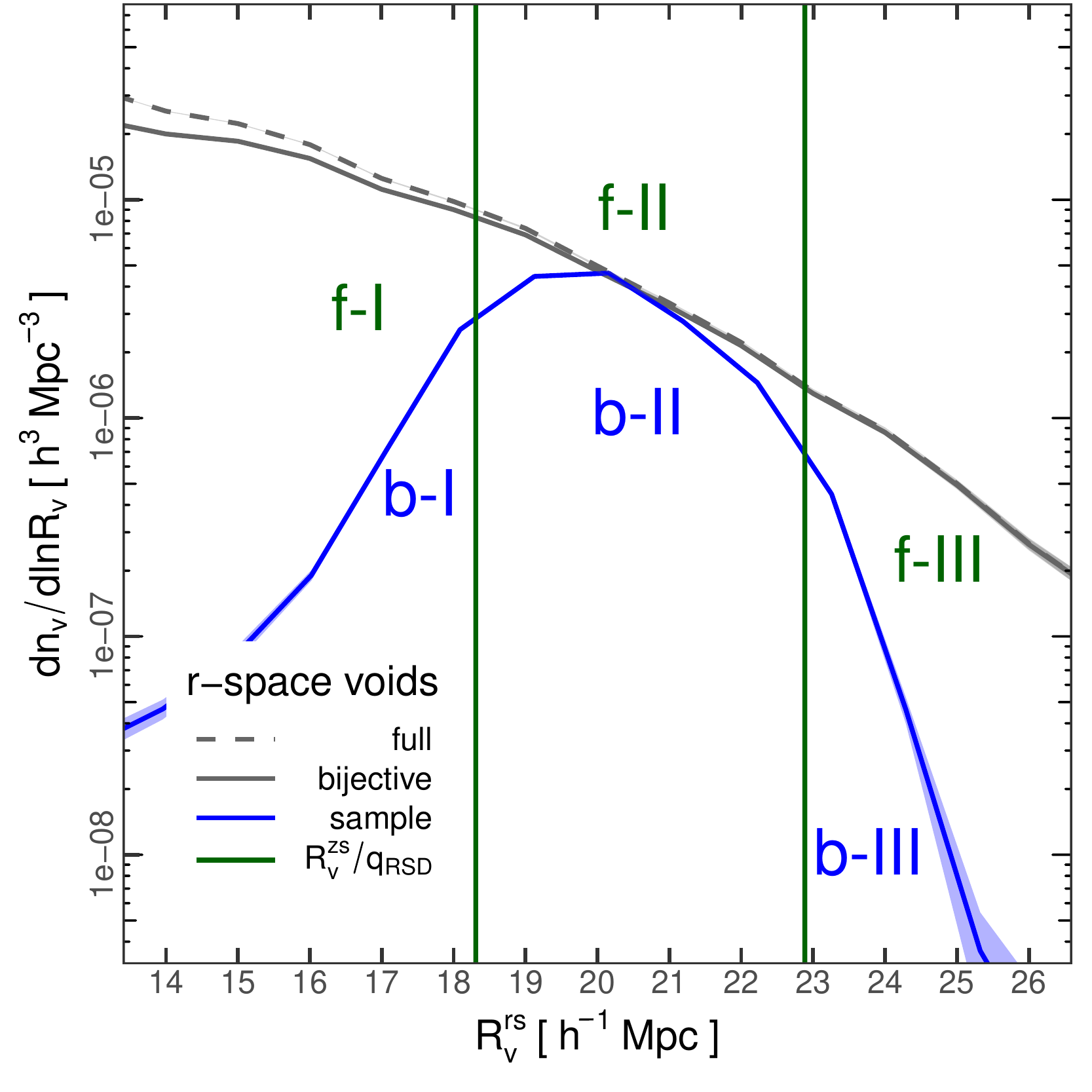}
    \caption[
    Samples of voids used to study the impact of the AP, t-RSD, v-RSD and e-RSD effects on the correlation function.
    Manifestations in real and redshift space.
    ]{
    Void sample used in this chapter to analyse the correlation function.
    \textit{Left-hand panel.}
    Void size function in $z$-space.
    The full catalogue (TC-zs-f) is represented with the grey dashed curve; the bijective catalogue (TC-zs-b), with the grey solid curve.
    The red vertical lines delimit the sample.
    \textit{Right-hand panel.}
    Void size function in $r$-space.
    The corresponding full (TC-rs-f) and bijective (TC-rs-b) catalogues have the same representation.
    The blue solid curve represents the radius distribution of the $r$-space counterparts of the voids in the sample.
    The green vertical lines indicate the t-RSD correction of radius using Eq.~(\ref{eq:q2_rsd}), and constitutes a robust prediction of the completeness region where the bulk of the $r$-space counterparts must be found.
    Subsamples f-I, f-II, f-III, b-I, b-II and b-III are used to analyse the ellipticity effect of voids.
    }
    \label{fig:VSF_zoom}
\end{figure}

The left-hand panel of Figure~\ref{fig:VSF_zoom} shows a zoom to the void abundances curves presented previously in Figure~\ref{fig:TC_VSF} for the full and bijective catalogues in $z$-space (TC-zs-f and TC-zs-b, respectively).
The representation has been interchanged for a better comprehension of the discussion that we will develop throughout this chapter.
Specifically, the full VSF is represented with the grey dashed curve, whereas the bijective VSF, with the grey solid curve.
In Section~\ref{sec:zeffects_map}, we demonstrated that the full and bijective catalogues are statistically equivalent at these scales.
For this reason, we will not distinguish between them hereinafter, unless otherwise stated.

In order to measure the correlation function, we selected a void sample with sizes between $20 \leq \rzs/\hmpc \leq 25$.
The red vertical lines in the figure indicate this cut.
We have verified that the general results do not depend on this cut as long as this is done in the bijective range (above the shot-noise level).

The right-hand panel of the figure shows the corresponding abundances in $r$-space.
The representation is the same: the grey dashed curve is the full abundance (corresponding to catalogue TC-rs-f), whereas the grey solid curve is the bijective abundance (corresponding to catalogue TC-rs-b).
In particular, the blue solid curve describes the radius distribution of the $r$-space counterparts of the voids from the original $z$-space sample.
That is, it is the same sample of voids, but as manifested in $r$-space according to the bijective mapping.
Note that, unlike the $z$-space voids, the $r$-space ones are not confined to a defined band, but they follow a more complex distribution covering a very extended radius range, although notice the log-scale of the y-axis.
This is a central aspect in the analysis of this chapter, so we will come back to this figure later to explain the meaning of the remaining elements represented in it.

\section{The projected correlation functions revisited}
\label{sec:impact_vgcf_cor}

We will follow the methodology developed in Chapter~\ref{chp:cosmotest} in order to measure the correlation function.
It is worth noting that the original method relies on measuring correlations directly in terms of void-centric angles and redshifts.
The cosmological dependence of these quantities with a physical distance scale is contained in the model, thus allowing us to evaluate different fiducial cosmologies without assuming any one in particular.
In this way, the AP effect is taken into account naturally.
However, as we will not consider AP distortions in the following analysis, we will treat correlations in the comoving coordinate system defined by the simulation, and concentrate on RSD exclusively.

We recall the basics here in the context of these considerations.
$\xisp$ denotes the $z$-space void-galaxy cross-correlation function, where $\sigma$ represents the void-centric comoving distance on the plane of the sky, and $\pi$, the analogue distance along the line of sight.
Projecting $\xisp$ towards the plane of the sky (towards the $\sigma$-axis) in a given $\pi$-range of scales, $\mathrm{PR}_\pi$, we get the POS correlation function, denoted by $\xipos$, a one-dimensional function that depends only of the POS-variable $\sigma$.
Conversely, projecting $\xisp$ towards the line of sight (towards the $\pi$-axis) in a given $\sigma$-range of scales, $\mathrm{PR}_\sigma$, we get the LOS correlation function, $\xilos$, a one-dimensional function that depends only of the LOS-variable $\pi$.

The measurement of the $z$-space correlation function is based on counting void-galaxy pairs in a cylindrical binning scheme.
In this geometry, a bin is a cylindrical shell oriented along the LOS, with internal radius $\sigma_\mathrm{int}$, external radius $\sigma_\mathrm{ext}$, a lower height $\pi_\mathrm{low}$, and an upper height $\pi_\mathrm{up}$.
Following Eq.~(\ref{eq:estimator_voids}), the correlation function can be simply estimated as the ratio between the number of counted void-tracer pairs, $DD_{ij}$, and the expected number of pairs in a homogeneous distribution, $RR_{ij}$, where the subindices allude to the bin $(i,j)$:
\begin{equation}
    \hat{\xi}(\sigma_i, \pi_j) = \frac{DD_{ij}}{RR_{ij}} - 1.
    \label{eq:estimator_voids3}
\end{equation}
In turn, $RR_{ij}$ can be analytically estimated as the product of the density of tracers, the total number of voids and the volume of the bin.
Here, $(\sigma_i, \pi_j)$ denotes the coordinates of the geometrical centre of the bin, to which the measured correlation value is assigned.

The projected correlation functions can be considered special cases of this binning scheme.
The scheme for the POS correlation involves bins with dimensions $\sigma_\mathrm{int}$, $\sigma_\mathrm{ext}$, $\pi_\mathrm{low}=0$ and $\pi_\mathrm{up}=\mathrm{PR_\pi}$.
In this way, $\delta \sigma := \sigma_\mathrm{ext} - \sigma_\mathrm{int}$ is the POS binning step.
Similarly, the scheme for the LOS correlation involves bins with dimensions $\sigma_\mathrm{int}=0$, $\sigma_\mathrm{ext}=\mathrm{PR_\sigma}$, $\pi_\mathrm{low}$ and $\pi_\mathrm{up}$.
In a similar way, $\delta \pi := \pi_\mathrm{up} - \pi_\mathrm{low}$ is the LOS binning step.
For the development of this chapter, we took equal projection ranges in both directions: $\mathrm{PR_\sigma} = \mathrm{PR_\pi} = 40~\hmpc$, which for simplicity, we will refer to both with the common notation $\mathrm{PR}$.
Moreover, we also took equal binning steps in both directions: $\delta \sigma = \delta \pi = 1~\hmpc$.


\subsection{Different configurations}
\label{subsec:impact_vgcf_cor_space}

We measured correlations in different configurations of the spatial distribution of voids and haloes.
The measurements made with $z$-space voids and $z$-space haloes correspond to what we call the $\zxz$-space configuration.
Similarly, the measurements made with $r$-space voids and $z$-space haloes correspond to what we call the hybrid $\rxz$-space configuration.
Finally, the measurements made with $r$-space voids and $r$-space haloes correspond to what we call the $\rxr$-space configuration.
This notation will also apply for measurements of the density and velocity fields.

\begin{figure}
    \centering
    \includegraphics[width=79mm]{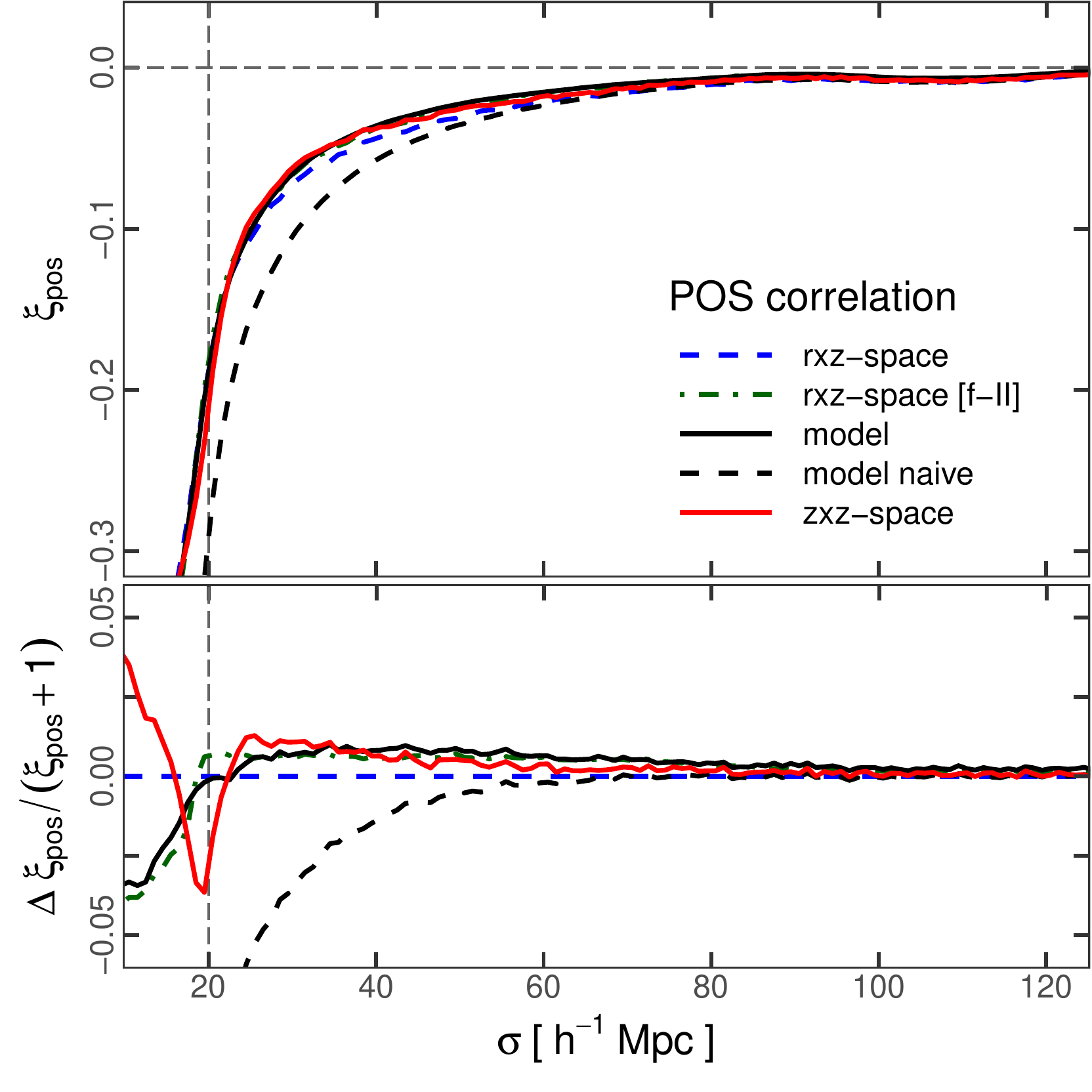}
    \includegraphics[width=79mm]{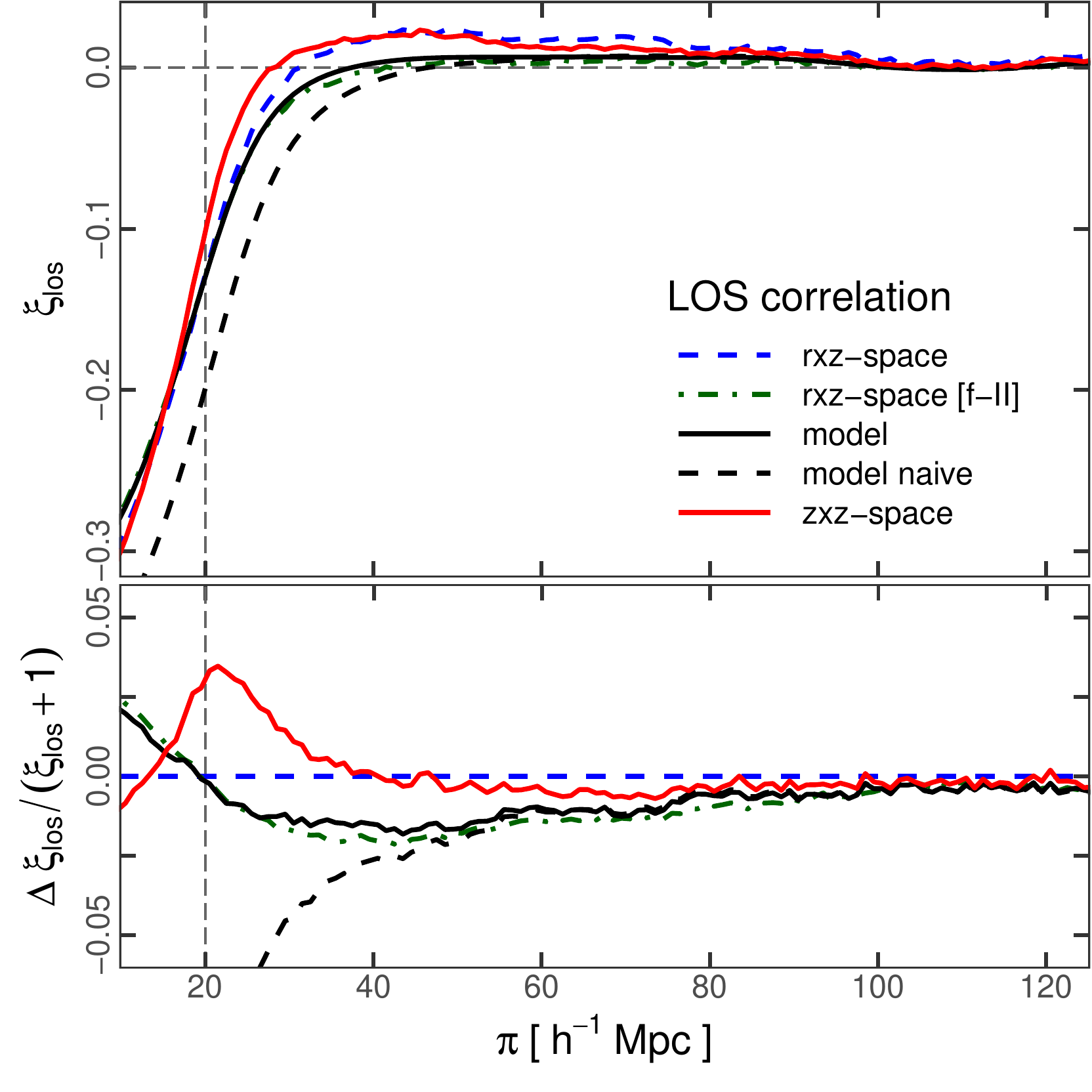}
    \caption[
    POS and LOS projections of the correlation function measured in the $\zxz$- and $\rxz$-space configurations.
    Comparison with model predictions.
    ]{
    Plane-of-sky (\textit{left-hand panel}) and line-of-sight (\textit{right-hand panel}) projections of the correlation function measured in different configurations.
    The red solid curves represent the measurement performed in the $\zxz$-space configuration using the $z$-space sample of voids defined in Section~\ref{sec:impact_vgcf_sample} and shown in Figure~\ref{fig:VSF_zoom}.
    It mimics a possible observational measurement.
    The blue dashed curves represent the measurement performed in the hybrid $\rxz$-space configuration using the $r$-space counterparts of the voids in the sample.
    It allows us to interpret the RSD-model predictions.
    The black curves represent two theoretical predictions: without applying (``naive'', dashed line) and after applying (solid line) the expansion-effect correction.
    The green dot-dashed curves represent the $\rxz$-space correlation function using subsample f-II.
    It allows us to test the predictions of the t-RSD correction.
    In all cases, a projection range of $40~\hmpc$ was used.
    \textit{Lower panels.}
    Corresponding fractional differences taking the $\rxz$-space configuration as a reference.
    }
    \label{fig:correlations_systematics}
\end{figure}

Figure~\ref{fig:correlations_systematics} shows the projected POS (left-hand panel) and LOS (right-hand panel) void-halo correlation functions measured in different spatial configurations.
Specifically, the red solid curves represent the result of measuring correlations in the  $\zxz$-space configuration using the void sample defined in Section~\ref{sec:impact_vgcf_sample} (Figure~\ref{fig:VSF_zoom}), i.e. they are the result of measuring correlations from the spatial distribution of both voids and haloes in $z$-space, and hence mimic a possible observational measurement.
The blue dashed curves, on the other hand, represent the result of measuring correlations in the hybrid $\rxz$-space configuration, i.e. they are obtained by taking the associated $r$-space voids of the sample but keeping the haloes in $z$-space.
This distinction is important, since current models for RSD around voids are defined to operate on this hybrid configuration. \cite{rsd_nadathur,reconstruction_nadathur}, and the aim of this chapter is to compare model predictions with observations.
For the moment, we will focus on these two configurations.
We will explain later the meaning of the remaining curves.

It is more useful to compare two correlations measured in different configurations by means of their fractional differences: $\Delta \xi / (\xi + 1) := [(\xi_\mathrm{tar}+1)-(\xi_\mathrm{ref}+1)]/(\xi_\mathrm{ref}+1)$, where $\xi_\mathrm{tar}$ is the target correlation that we want to compare, and $\xi_\mathrm{ref}$ the one used as reference.
These differences are shown in the lower panels of the figure.
We chose the hybrid $\rxz$-space configuration as a reference (blue dashed line).

\subsection{Simplification of the model}
\label{subsec:impact_vgcf_cor_model}

We adapted the model developed in Section~\ref{sec:cosmotest_model} under the assumptions made at the beginning of this chapter.
Specifically, we adapted Eq.~(\ref{eq:xi_gd}) for the case of our simulation study and the initial hypothesis of working in a comoving coordinate system without considering AP distortions:
\begin{equation}
    \hat{\xi}(\sigma_i, \pi_j) = -1 + 2 
    \frac{ \int_\mathrm{\pi_\mathrm{low}}^{\pi_\mathrm{up}} d\pi \int_{\sigma_\mathrm{int}}^{\sigma_\mathrm{ext}} \sigma [1 + \xi(\sigma,\pi)] d\sigma }
    { (\sigma^2_\mathrm{ext}-\sigma^2_\mathrm{int}) (\pi_\mathrm{up}-\pi_\mathrm{low}) }.
    \label{eq:xi_gd2}
\end{equation}
The rest of the modelling is identical.
An important aspect to keep in mind is that this model is defined to work in the hybrid $\rxz$-space configuration.
Hence, in Eq.~(\ref{eq:gsm}), the quantities $r_\perp$ and $r_\parallel$ are the $\rxr$-space analogues of the $\rxz$-space quantities $\sigma$ and $\pi$.
In a similar way, $\xi(r)$ and $v(r)$ must be understood as the $\rxr$-space density and velocity profiles of a void sample, respectively.

\begin{figure}
    \centering
    \includegraphics[width=79mm]{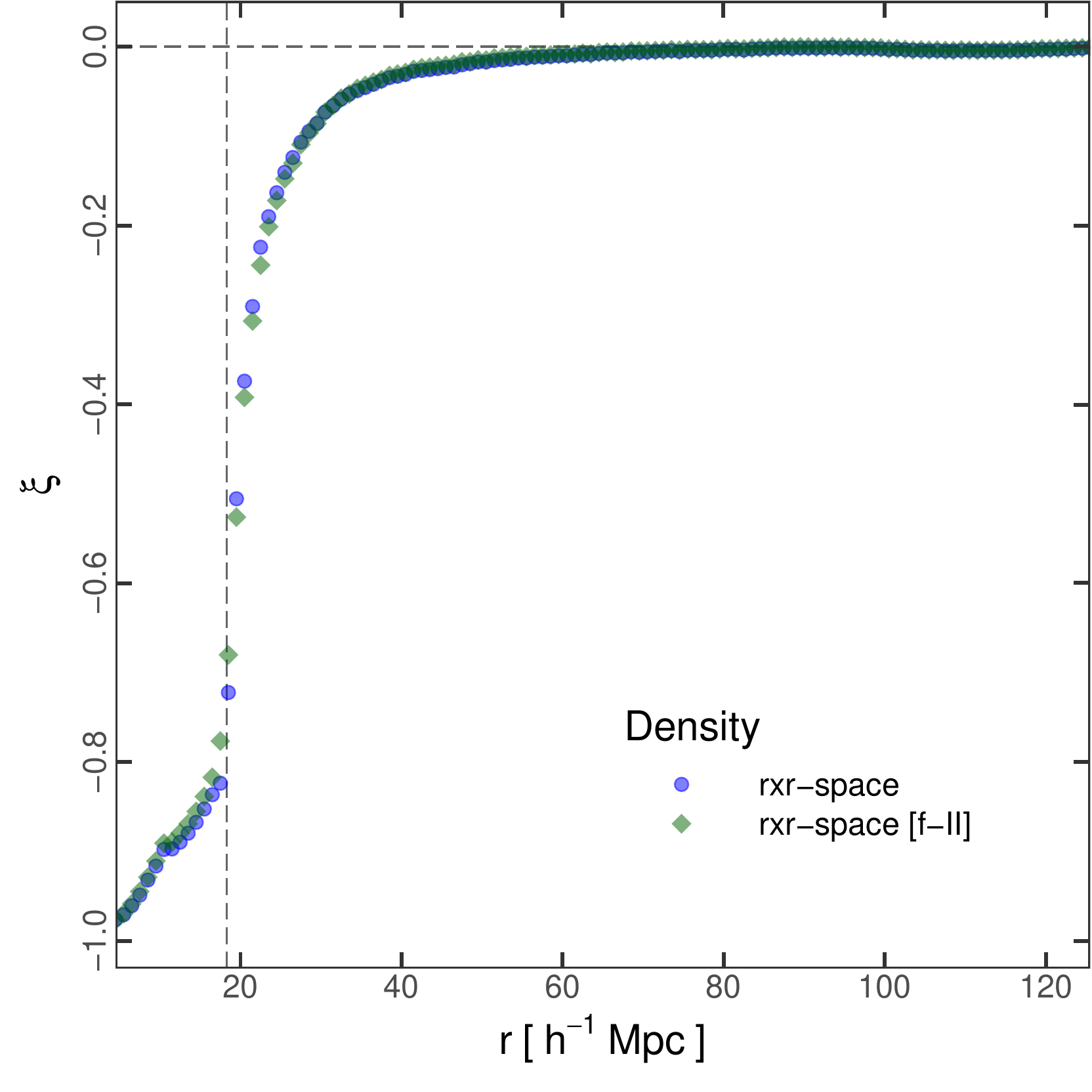}
    \includegraphics[width=79mm]{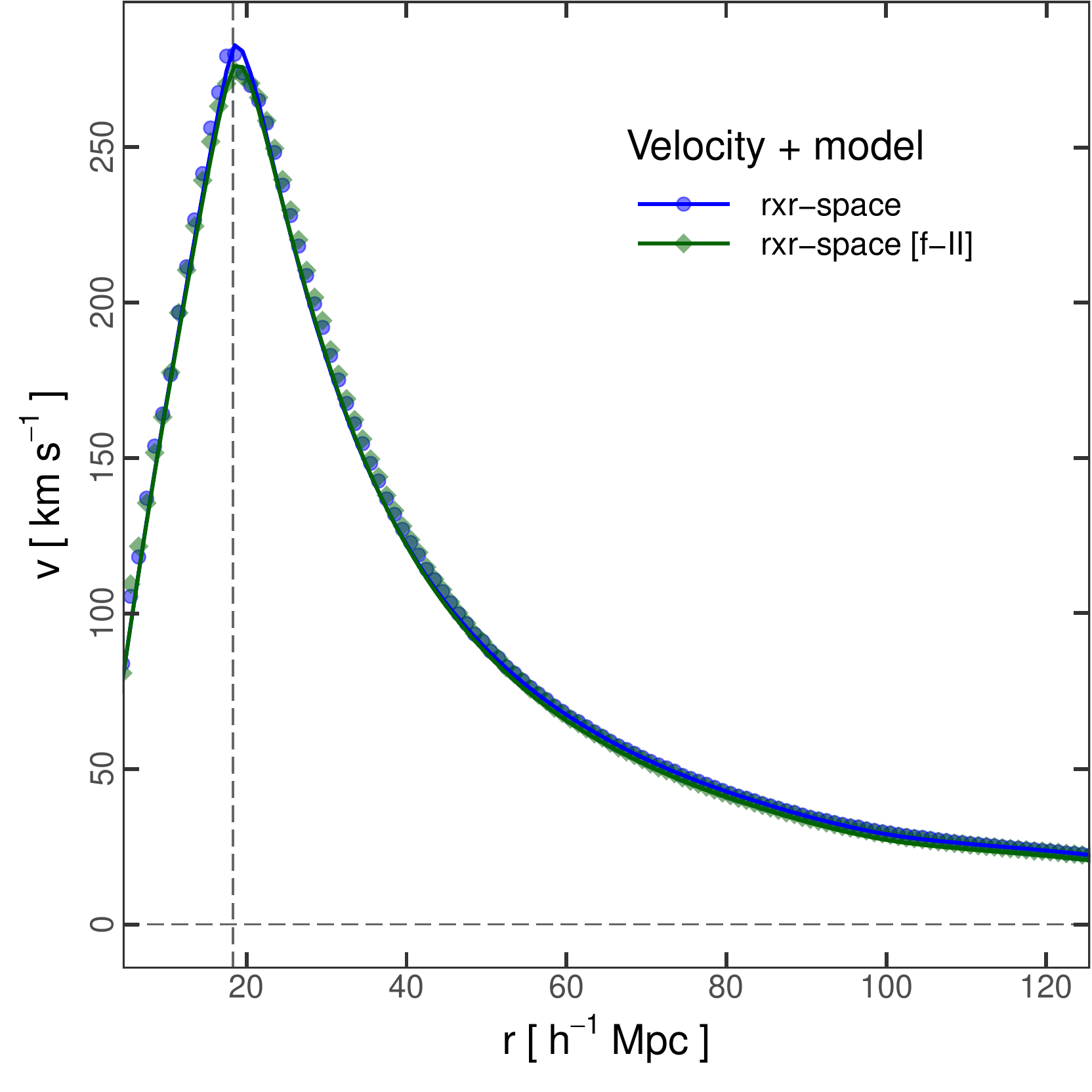}
    \caption[
    Density contrast and velocity profiles: analysis of the $\rxr$-space configuration.
    ]{
    Density contrast (\textit{left-hand panel}) and velocity (\textit{right-hand panel}) profiles in $\rxr$-space.
    The blue circles correspond to the $r$-space counterparts of the voids in the original sample, whereas the green diamonds, to subsample f-II.
    The results for both samples are almost identical, which demonstrates the potential of the t-RSD correction.
    The blue and green curves are the corresponding velocity predictions of Eq.~(\ref{eq:velocity}) for each case, in excellent agreement with the measured data at all scales.
    }
    \label{fig:rrspace_profiles}
\end{figure}

In Section~\ref{subsec:cosmotest_model_densdiff}, we discussed that, unlike the case of the velocity profile, it is not possible to obtain a satisfactory model for the density profile from first principles, thus a common practice is to use parametric and empirical approaches.
In particular, we developed our own model suitable for R-type voids.
Nevertheless, we will not use it here, nor any other model.
Instead, we will use the $\rxr$-space density profile of the sample directly measured from the simulation as input in the model.
This will allow us to figure out with precision all the $z$-space systematic effects that affect the correlation function, the aim of this chapter, since we will not introduce spurious effects from the performance of density models in this way.
This profile is shown in the left-hand panel of Figure~\ref{fig:rrspace_profiles} with blue circles.
The right-hand panel shows the corresponding velocity profile, also with blue circles.
The blue solid curve is the prediction of Eq.~(\ref{eq:velocity}), which works remarkably well at all scales.
We will explain the meaning of the remaining profiles displayed in the figure (represented with green diamonds) in Section~\ref{sec:impact_vgcf_trsd}.

\section{Impact of the impurity of the sample}
\label{sec:impact_vgcf_impurity}

The goal of this section and the following ones is to figure out all the sources of distortion patterns that lead to the observed features in the $\zxz$-space configuration, the red solid curves in Figure~\ref{fig:correlations_systematics}, which mimic a possible observational measurement.
The analysis will be based on the theoretical framework developed in Chapter~\ref{chp:zeffects}.
Although we use the projected version of the correlation function, the conclusions we will reach are general, also applicable to the traditional way of measuring correlations.
Moreover, the same results are observed with different projection ranges.
We chose to show the case $\mathrm{PR} = 40~\hmpc$ as this constitutes a realistic case applicable to data (see the discussion of Section~\ref{subsec:cosmotest_test_constraints}).

\begin{figure}
    \centering
    \includegraphics[width=\textwidth/2]{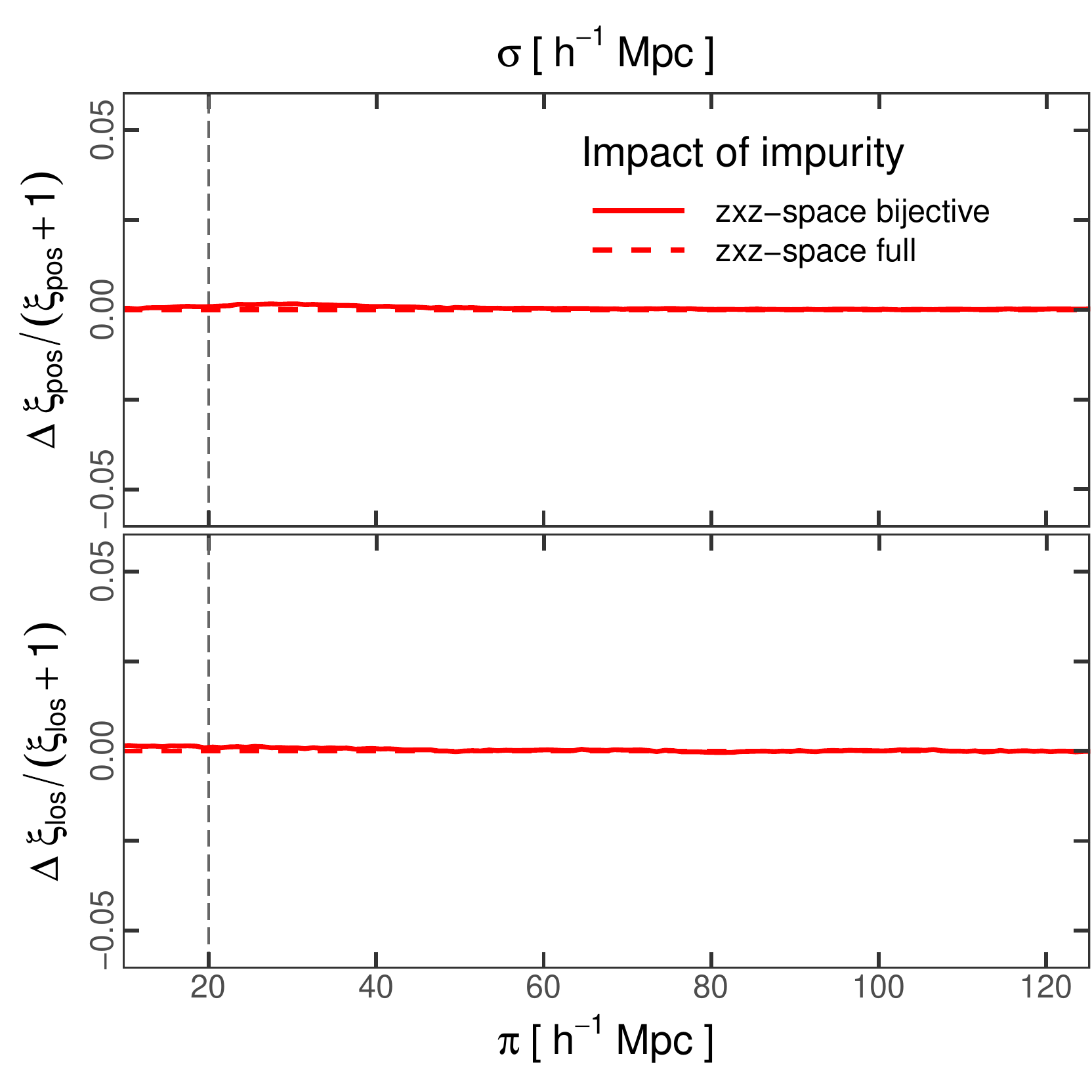}
    \caption[
    Impact of the impurity of a sample on the measurement of the correlation function due to non-bijective voids.
    ]{
    Impact of the impurity of a sample on the measurement of the correlation function due to non-bijective voids, quantified by means of the fractional differences of correlation between the original sample, selected from the bijective catalogue (red solid curves), and an analogue sample selected from the full catalogue (red dashed lines).
    The differences are practically negligible at all scales.
    }
    \label{fig:impurity}
\end{figure}

Although we are not distinguishing between the full and bijective catalogues since they are statistically equivalent, we make an exception here to reinforce this concept in the context of the correlation function.
In Section~\ref{sec:impact_vgcf_sample}, we defined the void sample to be used throughout this chapter.
It is made up of voids identified in $z$-space with sizes in the range $20 \leq \rzs/\hmpc \leq 25$.
Technically, this cut was applied on the bijective catalogue (TC-zs-b).
In view of this, we repeated the selection of the sample, but applying the cut on the full catalogue (TC-zs-f) this time.
As can be appreciated visually in the left-hand panel of Figure~\ref{fig:VSF_zoom}, both samples are almost identical.
With this new sample, we measured the projected correlation functions under the same conditions as in Section~\ref{sec:impact_vgcf_cor}.
Figure~\ref{fig:impurity} shows the fractional differences between the correlations measured with both samples.
The upper panel corresponds to the POS projection, whereas the lower panel to the LOS projection.
The impact of impurity due to non-bijective voids in the sample taken from the full catalogue is determined by quantifying the deviations observed in this figure.
To make the comparison, we have kept the scale used in Figure~\ref{fig:correlations_systematics}.
The differences are practically negligible, being less than $0.2\%$ at all scales.

\section{Expansion-effect correction}
\label{sec:impact_vgcf_trsd}

We are in conditions to make a first model prediction.
Recall that the model needs the $\rxr$-space density profile $\xi(r)$ as input.
A first and naive ansatz is to consider the density profile of an $r$-space void sample with radii in the same range as the original $z$-space sample: $20 \leq \rrs/\hmpc \leq 25$.
The model prediction using this profile is shown in Figure~\ref{fig:correlations_systematics} by means of the black dashed curves.
It is clear that it completely fails to reproduce the $\zxz$-space measurement: compare the black dashed curves with the red solid curves.
However, this was expected, since the model is defined to operate in the hybrid $\rxz$-space configuration, as it does not take into account the global dynamics of voids.
Nevertheless, it also fails to reproduce the correlations in this configuration too, as evident in the figure if we now compare the black dashed curves with the blue dashed curves.
Note that the relative differences can be as high as $8\%$, specially near the void walls.

In reality, the failed prediction of the last case was also expected, we showed it on purpose to highlight the importance of providing to models the correct $\rxr$-space density and velocity information.
The mistake was to assume that the $r$-space counterparts of the $z$-space voids from the original sample fall in the same range in the radius distribution, i.e. that their sizes do not change.
The right-hand panel of Figure~\ref{fig:VSF_zoom} clearly shows that this does not happen, in fact, the $r$-space distribution is more complex, covering a very extended range.
In view of this, we repeated the analysis using this time the $\rxr$-space density profile measured with the $r$-space counterparts of the voids from the original sample.
This is the profile shown in Figure~\ref{fig:rrspace_profiles} (left-hand panel with blue circles).
The corresponding model prediction is shown in Figure~\ref{fig:correlations_systematics} by means of the black solid curves.
Note that the deviations previously observed have now been noticeably reduced, although not completely.
There is a residual deviation of order $2\%$ at scales near the void walls.

The immediate question that arises is how to describe the complex radius distribution that the voids in the sample exhibit in $r$-space.
The answer is Eq.~(\ref{eq:q2_rsd}), namely, the t-RSD expansion effect (hereinafter, we will drop the superscript $l$ of factor $\qrsdl$ for simplicity).
According to this effect, the $r$-space counterparts of the voids in the sample should fall in the range $20/\qrsd \leq \rrs/\hmpc \leq 25/\qrsd$.
This is shown in the right-hand panel of Figure~\ref{fig:VSF_zoom} by means of the green vertical lines.
We arrive here at a very important conclusion: the thus-delimited band captures the bulk of the voids that constitute the sample.
To understand this better, let us take these voids inside the band, which will be referred to as subsample f-II for reasons that will be clarified later.
Note that this subsample approximates the true sample with respect to its $r$-space statistical properties.  
This is evident in Figure~\ref{fig:rrspace_profiles}, which shows the corresponding density and velocity profiles, represented with green diamonds.
Note that they are almost identical to the profiles corresponding to the true sample (represented with blue circles).
This demonstrates the ability of the t-RSD correction to recover the $r$-space statistical properties of voids, even if they have been identified in $z$-space.

Up to here, we can assert that the most important source of distortion patterns in the correlation function is the expansion effect (t-RSD), which can be modelled, in first approximation, by incorporating Eq.~(\ref{eq:q2_rsd}) to the GS model.
The remaining deviations between the observations (red solid curves) and the model prediction (black solid curves) can be separated into two components.
The hybrid $\rxz$-space configuration (blue dashed curves) will serve as a mediator between both of them.
This is the subject of study of the rest of this chapter.


\section{Alcock-Paczyński correction}
\label{sec:impact_vgcf_ap}

Before moving on, a brief comment about the impact of the AP change-of-volume effect on the correlation function, although we are not considering it in this chapter.
Here, we are making use of the TC void catalogues, affected by RSD, but not affected by AP distortions.
The AP distortions on tracers per se were tackled thoroughly in Chapter~\ref{chp:cosmotest}.
The AP distortions on voids, though, are in fact important for the current analysis.
However, the net effect is completely analogous to the expansion effect.
Specifically, repeating the analysis using the FC void catalogues, affected by both RSD and AP distortions, the radius distribution behaves similarly to Figure~\ref{fig:VSF_zoom}.
The only difference is that the bulk of $r$-space voids are found using Eq.~(\ref{eq:q_ap_rsd}) instead of Eq.~(\ref{eq:q2_rsd}), which combines the contribution of both effects.
Concretely, they are found in the range $20/(\qap~\qrsd) \leq \rrs/\hmpc \leq 25/(\qap~\qrsd)$.


\section{Off-centring-effect correction}
\label{sec:impact_vgcf_vrsd}

The first component of the remaining deviations is related to the differences between the correlations measured in the $\zxz$- and $\rxz$-space configurations (red solid curves and blue dashed curves in Figure~\ref{fig:correlations_systematics}, respectively).
In the context of the bijective mapping, these differences can only be attributed to the displacement of centres when voids are mapped from $r$-space into $z$-space, i.e. to the off-centring effect.
We highlight the fact that this effect is responsible for an additional distortion pattern in the correlation function, different from the classic distortions due to tracer dynamics (t-RSD), but due to the global dynamics of voids (v-RSD).
It is also true that this deviation is smaller than that produced by the expansion effect, of order $3\%$ at scales near the void walls.

\begin{figure}
    \centering
    \includegraphics[width=\textwidth/2]{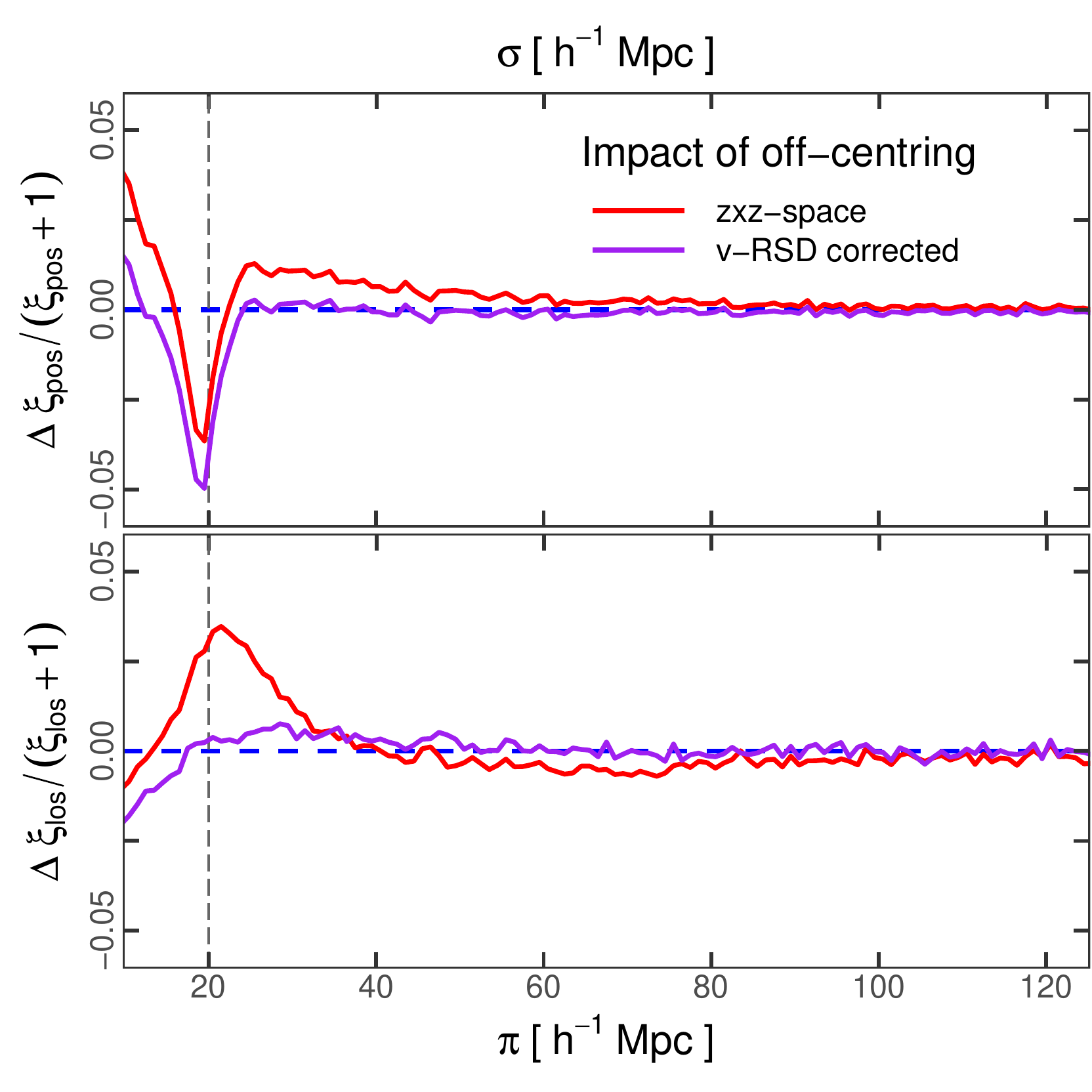}
    \caption[
    Impact of the off-centring effect on the measurement of the correlation function.
    ]{
    Impact of the off-centring effect on the measurement of the correlation function, quantified by means of the fractional differences of correlation between the $\zxz$-space (red solid curves) and $\rxz$-space (blue dashed lines) configurations.
    This is an additional distortion pattern, not previously taken into account, due to the global dynamics of voids (v-RSD).
    The deviations diminish notably after correcting the position of void centres with Eq.~(\ref{eq:void_zspace}) (purple solid curves).
    }
    \label{fig:vrsd}
\end{figure}

We can quantify the impact of v-RSD distortions by means of Eq.~(\ref{eq:void_zspace}), an RSD-displacement expression applied to the void-centre positions.
Figure~\ref{fig:vrsd} shows what happens when the $z$-space centre positions are corrected with this expression (purple solid curves).
The deviations diminish notably, with residual differences well below $1\%$ at all scales, thus recovering the correlation function in the hybrid $\rxz$-space configuration.
This is the first time that these types of distortions are detected and quantified.

The correction performed here was possible because we are working with a simulation.
The void finder can compute the velocity of a void from the individual velocity of tracers, hence Eq.~(\ref{eq:void_zspace}) can be applied to correct the centre position.
Nevertheless, this is not possible in practice.
One feasible solution is to incorporate the void net velocity distribution (Figure~\ref{fig:hist_vel}) into the GS model.
We leave this topic for a future investigation.

\section{Void ellipticity}
\label{sec:impact_vgcf_ersd}

The second component of the remaining deviations is related to the differences between the correlation measured in the hybrid $\rxz$-space configuration and the theoretical prediction of the model (blue dashed curves and black solid curves in Figure~\ref{fig:correlations_systematics}, respectively), which are of order $2\%$ at scales near the void walls, as noted previously in Section~\ref{sec:impact_vgcf_trsd}.

Given the fact that subsample f-II (obtained by means of the t-RSD correction) is a fair approximation of the original sample in $r$-space, it is expected that the correlation function measured with both (in $\rxz$-space) would be almost identical.
However, this is not the case.
The green dot-dashed curves of Figure~\ref{fig:correlations_systematics} represent the projected correlations measured with the f-II subsample.
A comparison with the blue dashed curves shows that there are appreciable deviations, specially in the LOS projection.
This was not expected, since both the f-II subsample and the original sample have almost identical $\rxr$-space density and velocity profiles, as shown in Figure~\ref{fig:rrspace_profiles}.
Evidently, the tails in the $r$-space radius distribution of the original sample have an appreciable effect on correlations.
Note however that the model predicts remarkably well the f-II correlations: compare the black solid curves with the green dot-dashed ones.
This result ratifies the methodology developed in Chapter~\ref{chp:cosmotest}.

To solve this question, we need to go back to the right-hand panel of Figure~\ref{fig:VSF_zoom} and define some additional subsamples.
We have already defined one of them: subsample f-II, made up of those voids inside the radius band delimited by the t-RSD correction using the factor $\qrsd$ (green vertical lines).
Analogously, the subsample of voids that also fall inside this band, but below the distribution of the original sample (below the blue curve), will constitute subsample b-II.
Note that both subsamples f-II and b-II are very similar because of the t-RSD correction.
Now, those voids belonging to the left tail will constitute subsample b-I, whereas those belonging to the right tail will be referred to as subsample b-III.
In a similar way, all voids to the left of the t-RSD band covering the same radius range as subsample b-I will constitute subsample f-I, whereas all voids to the right of the band covering the same range as subsample b-III will be referred to as subsample f-III.
Note that the subsets f-I, f-II and f-III contain the subsets b-I, b-II and b-III, respectively.

\begin{figure}
    \centering
    \includegraphics[width=\textwidth]{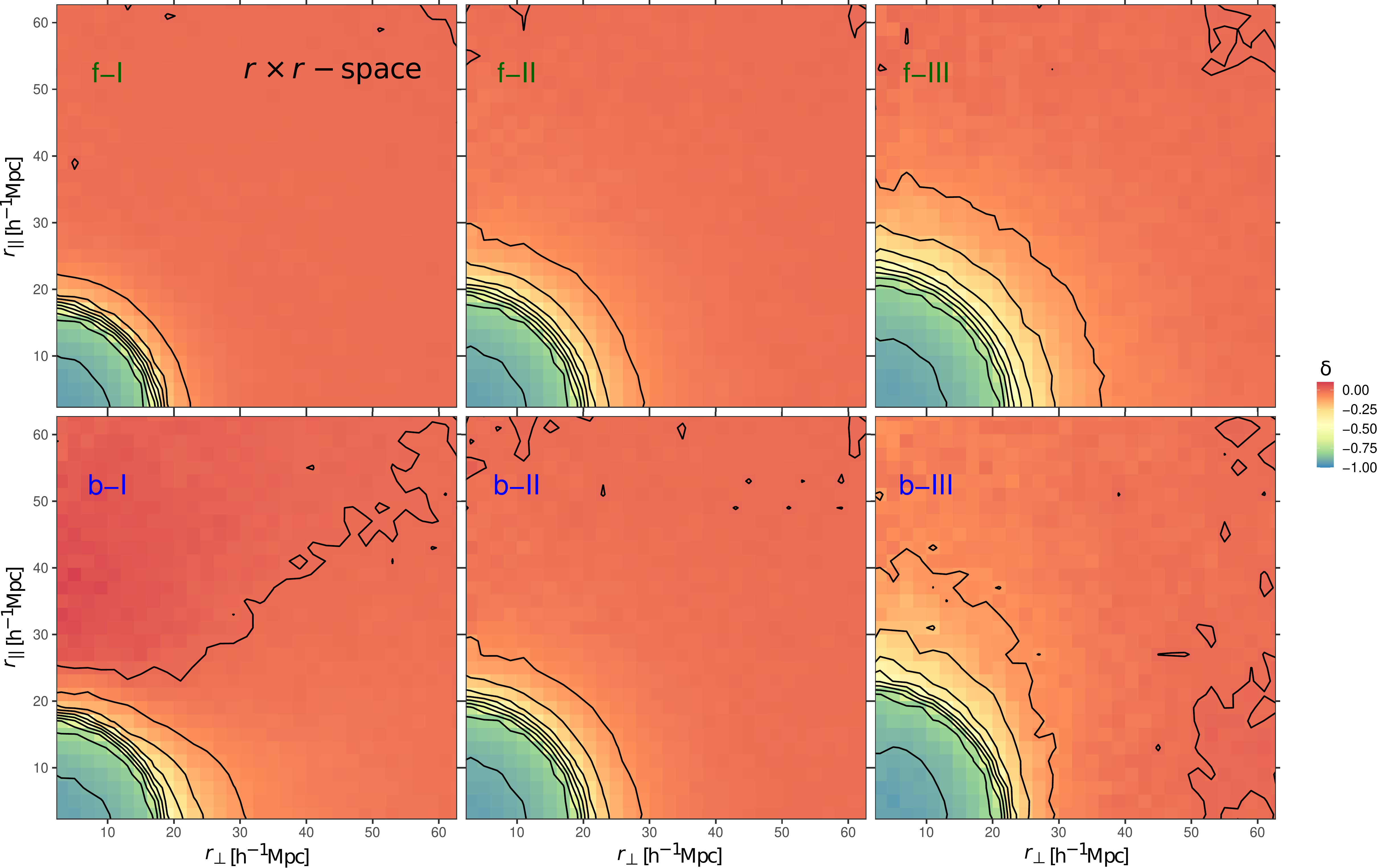}
    \caption[
    Density contrast maps measured in the $\rxr$-space configuration: anisotropies in real space.
    ]{
    Density contrast maps in $\rxr$-space corresponding to the different subsamples defined from Figure~\ref{fig:VSF_zoom}.
    Subsamples f-I, f-II and f-III (\textit{upper panels}) exhibit circular contours with no signs of anisotropies.
    The only difference between them relies on the size of their emptiest regions, in agreement with the sizes of their voids.
    Subsample b-II (\textit{lower central panel}) behaves similarly to subsample f-II.
    This highlights the power of the t-RSD correction.
    The remaining subsamples, b-I and b-II, exhibit prominent and opposite anisotropic patterns.
    This is a manifestation of the ellipsoidal nature of voids.
    }
    \label{fig:map2d_dens}
\end{figure}

Figure~\ref{fig:map2d_dens} shows the $\rxr$-space density field map, $\xi(r_\perp,r_\parallel)$, for each subsample defined above.
In each map, the bluer zones describe the emptier regions of space, whereas the redder ones describe the more crowded regions.
All maps were colour-coded in order to have the same scale.
The first relevant aspect to be outlined here is that subsamples f-I, f-II and f-III (upper panels) behave as expected: there are no anisotropies since they exhibit circular contours.
The only difference between them can be found in the size of the empty regions, in agreement with the fact that, from I to III, they are composed of voids of increasing radii.
Note that subsample b-II (lower central panel) also behaves as expected, in particular, it is very similar to subsample f-II.
However, subsamples b-I and b-III (lower left-hand panel and lower right-hand panel) behave unexpectedly.
They exhibit prominent anisotropic patterns.
This is a curious result, since we are analysing the density field in the $\rxr$-space configuration, expecting spherical symmetry.
Moreover, the anisotropies are opposite: b-I voids are elongated along the POS axis, whereas b-III voids are elongated along the LOS direction.

\begin{figure}
    \centering
    \includegraphics[width=\textwidth]{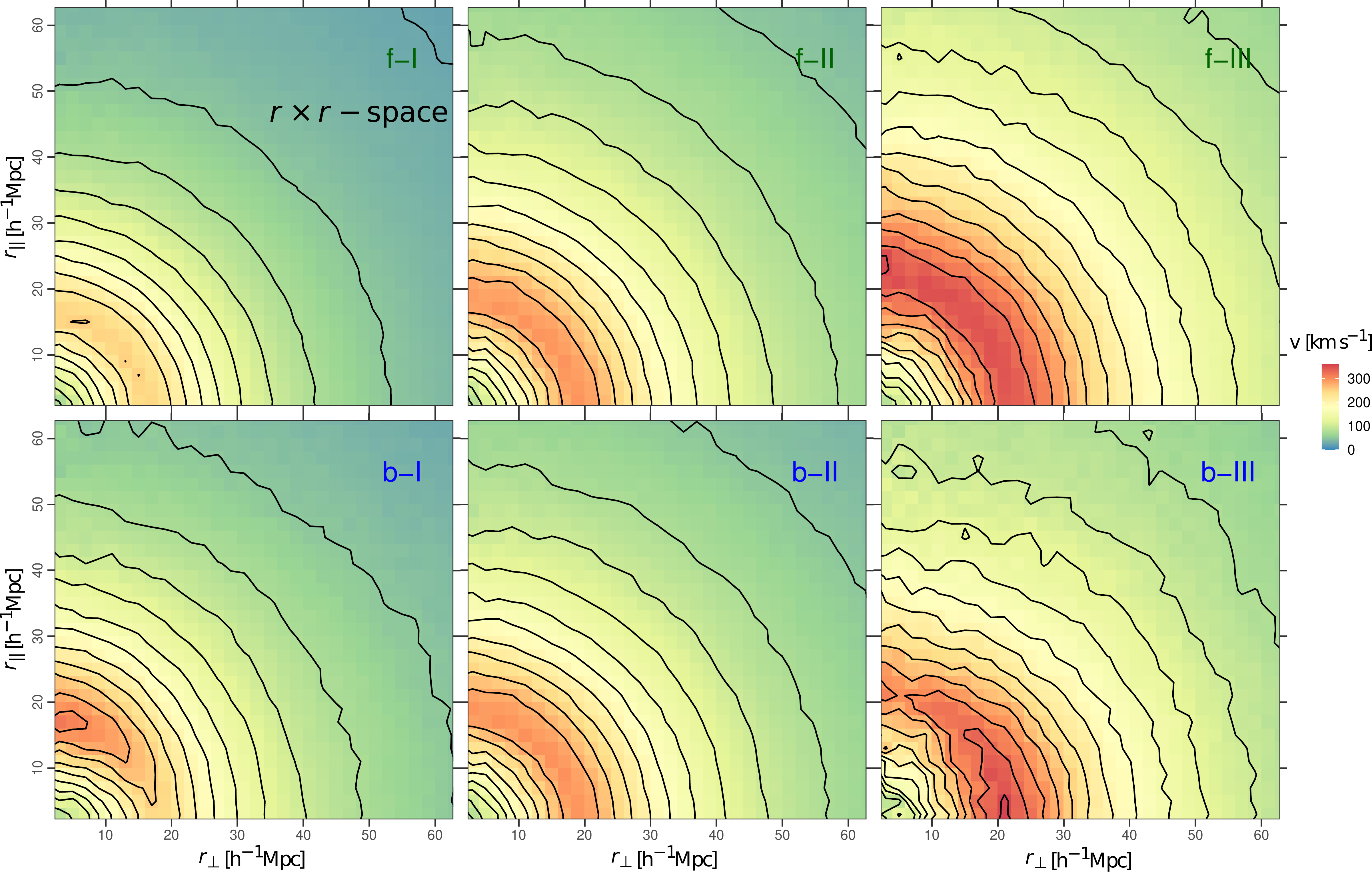}
    \caption[
    Velocity maps measured in the $\rxr$-space configuration: over-expanding and under-expanding voids.
    ]{
    Velocity maps in $\rxr$-space, supplementary information to Figure~\ref{fig:map2d_dens}.
    The same qualitative behaviour is observed.
    Voids from subsample b-I show an over-expanding behaviour, since their radii are greater than the prediction of Eq.~(\ref{eq:q2_rsd}), whereas the voids from subsample b-III show an under-expanding behaviour, since their radii are smaller.
    }
    \label{fig:map2d_vel}
\end{figure}

Figure~\ref{fig:map2d_vel} shows the associated two-dimensional $\rxr$-space velocity field, $v(r_\perp,r_\parallel)$, for each subsample.
As in the case of the density, each map was colour-coded so that they all have the same scale.
The same results are observed: subsamples f-I, f-II and f-III exhibit circular contours, subsample b-II behaves identically to subsample f-II, whereas subsamples b-I and b-III exhibit prominent and opposite anisotropic patterns.

We interpret these results in the following way.
Individual voids are typically ellipsoidal.
However, they are oriented randomly in space (there is not any privileged direction).
Therefore, the ellipticity has not any significant impact on the statistical properties of a complete sample of voids.
This is the case of subsamples f-I, f-II and f-III.
This is also the case of subsample b-II, since it is very similar to subsample f-II.
In this sense, the expansion-effect correction is a good predictor of the region of completeness in the radius distribution: f-II and b-II voids are almost the same.
Voids from subsamples b-I and b-III, however, are not complete, but they constitute a special selection of voids.
They do not follow the velocity and expansion predictions of Eqs.~(\ref{eq:velocity}) and (\ref{eq:q2_rsd}), respectively.
On the one hand, b-I voids are \textit{over-expanding voids}, since they fall in the selection range when they are identified in $z$-space, but their radii $\rzs$ are greater than the prediction given by the factor $\qrsd$.
Conversely, b-III voids are \textit{under-expanding voids}, since they also fall in the selection range, but their radii are lower than the corresponding prediction.

\begin{figure}
    \centering
    \includegraphics[width=79mm]{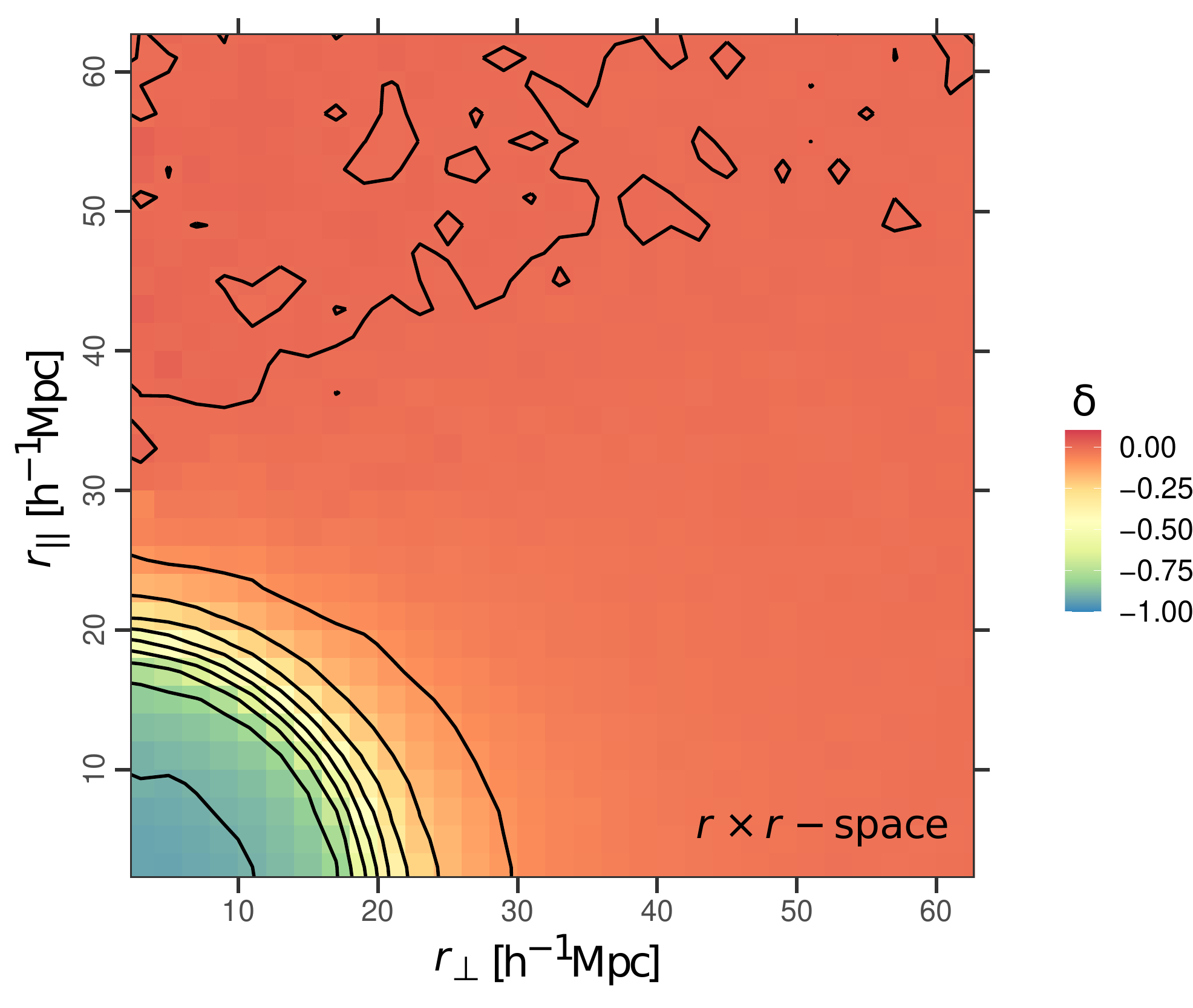}
    \includegraphics[width=79mm]{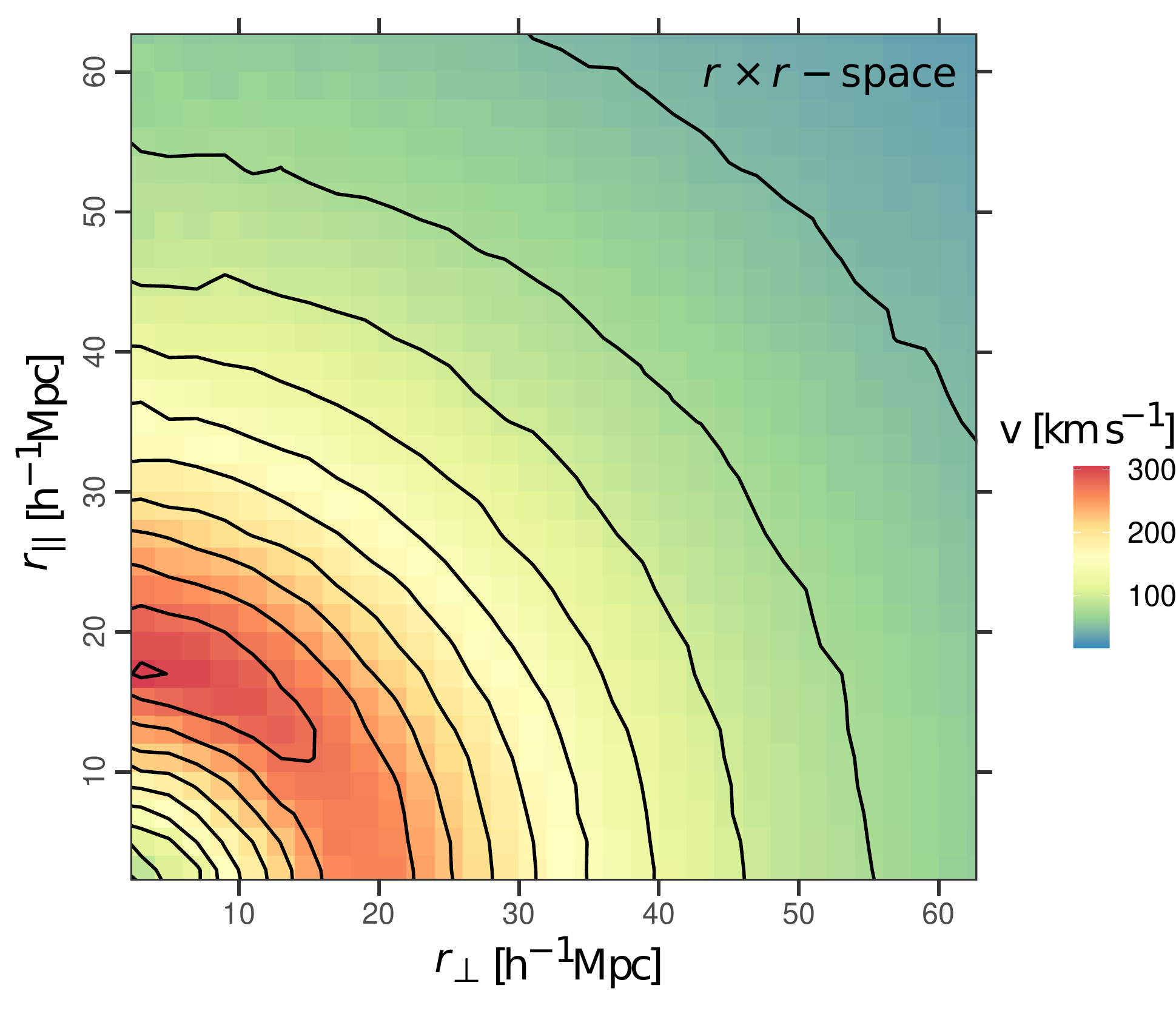}
    \caption[
    Impact of the ellipticity of voids on the measurement of the correlation function (e-RSD).
    ]{
    Density contrast (\textit{left-hand panel}) and velocity (\textit{right-hand panel}) maps in $\rxr$-space corresponding to the original sample of voids.
    The opposite behaviour observed in Figures~\ref{fig:map2d_dens} and \ref{fig:map2d_vel} do not cancel because b-I voids are many more than b-III voids.
    The tails in the radius distribution of the sample detected in Figure~\ref{fig:VSF_zoom} are the responsible for an additional distortion pattern observed on the correlation function, not previously taken into account, which is due to the intrinsic ellipticity of voids (e-RSD).
    }
    \label{fig:map2d}
\end{figure}

Figure~\ref{fig:map2d} shows the $\rxr$-space density (left-hand panel) and velocity (right-hand panel) maps corresponding to the original sample (b-I $\cup$ b-II $\cup$ b-III).
Note that the anisotropies are still present, the opposite behaviour of the tails does not cancel.
Particularly, the behaviour of the left tail is the prevailing one.
This is because subsample b-I has many more elements than subsample b-III.
Therefore, although the spherical averaged statistics erase the elliptical features of voids when estimating the density and velocity profiles, as Figure~\ref{fig:rrspace_profiles} shows, it manifests when calculating correlations.
This is because the tails of the distribution contribute independently, a fact that is more evident when distinguishing the behaviour along the POS and LOS directions, as is the case of the projected correlations.
This is the reason why the model cannot correctly reproduce the correlations even in the $\rxz$-space configuration.
We arrive here at an important conclusion: besides the t-RSD and v-RSD types of distortions, the intrinsic ellipticity of voids is another source of distortions in the correlation function.
We will refer to this effect, particularly the type of distortions that generates, with the acronym e-RSD.

\section{Towards an improved model}
\label{sec:impact_vgcf_egsm}

In order to appropriately describe the observations, current models for the void-galaxy cross-correlation function must be improved by taking into account the $z$-space effects that affect void properties.
The most important aspect is the t-RSD + AP correction over the volume of voids, i.e. the application of Eq.~(\ref{eq:q_ap_rsd}) to relate an observational sample of voids with their $r$-space counterparts, and in this way, be able to model their $\rxr$-space statistical properties correctly.
This is particularly important in order to characterise the density and velocity fields around voids.

The remaining deviations are smaller and can be attributed to two sources of distortions: the off-centring effect (v-RSD) and the intrinsic ellipticity of voids (e-RSD).
To model the former, the net peculiar velocity of voids must be taken into account statistically, for instance, in Eq.~(\ref{eq:gsm}).
A hint is to incorporate a void velocity distribution (see Figure~\ref{fig:hist_vel}).
For the latter, an option is to review the GS model by taking into account the elliptical symmetry that the $\rxr$-space density and velocity fields impose, for which it will be necessary to study deeper the connection between the two with a generalisation of Eq.~(\ref{eq:velocity0}).
Incidentally, there are some works about the ellipticity of voids and its cosmological importance in the literature \cite{ellipticity_park_lee, ellipticity_bos}.
We leave for a future investigation to tackle these topics in order to improve our model.

\begin{figure}
    \centering
    \includegraphics[width=79mm]{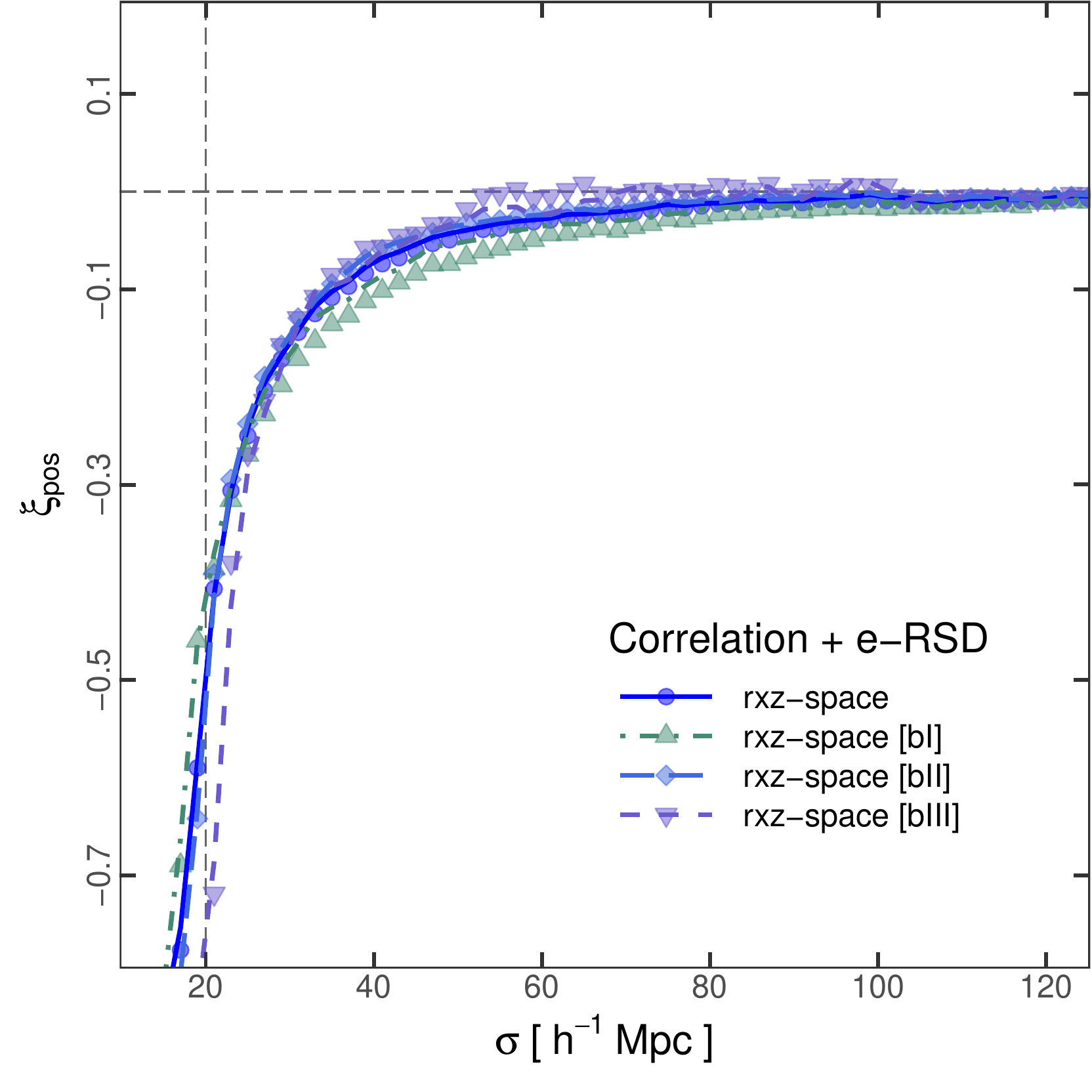}
    \includegraphics[width=79mm]{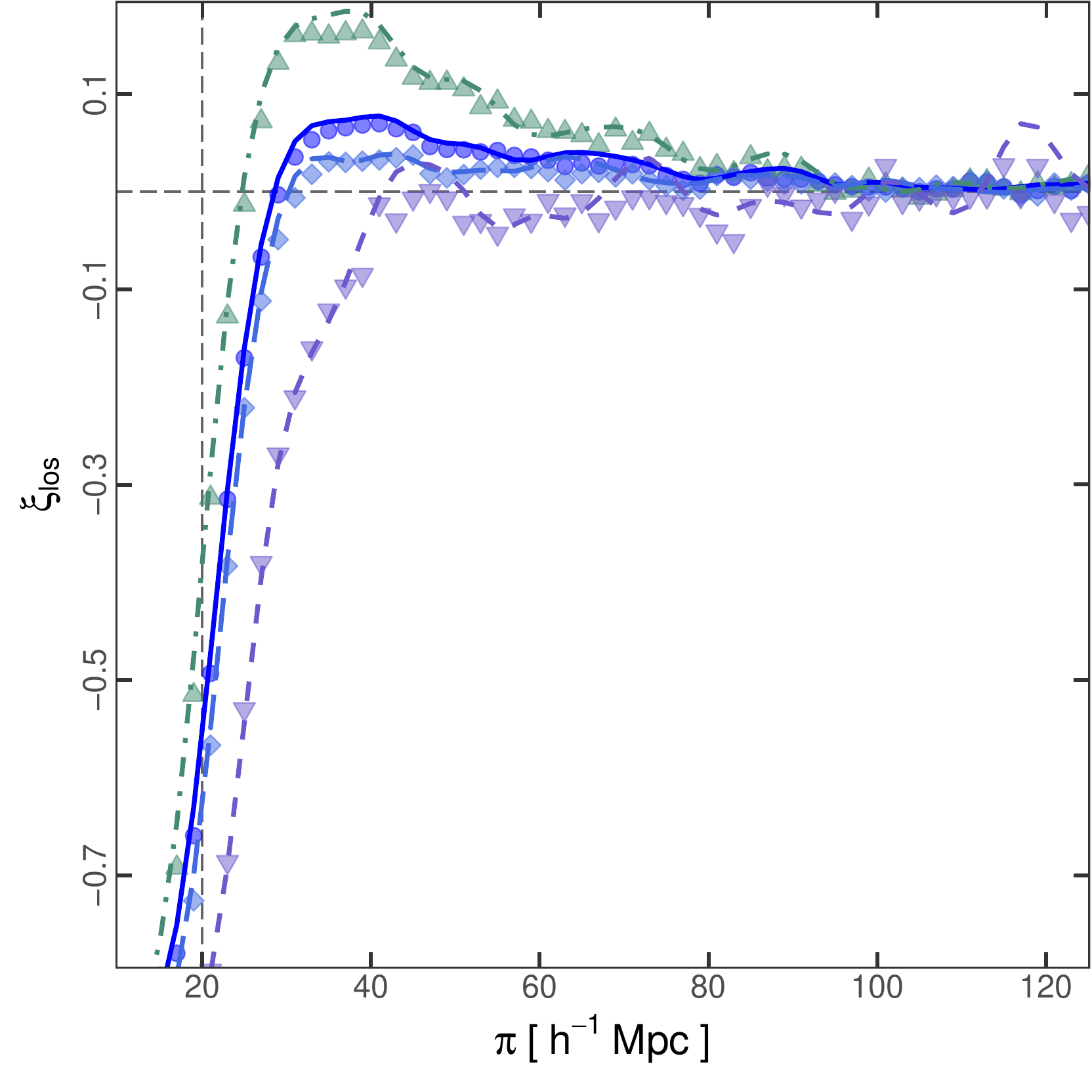}
    \caption[
    Testing the validity of the Gaussian streaming model by incorporating information about the ellipticity of voids.
    ]{
    Preliminary test of the Gaussian streaming model by incorporating information about the ellipticity of voids from the density and velocity maps from Figures~\ref{fig:map2d_dens} to \ref{fig:map2d}.
    The projected correlations were measured using a thinner projection range: $10~\hmpc$.
    The dots represent the measurements, whereas the curves, the corresponding theoretical predictions.
    The samples involved are the following: original sample (in blue, circles + solid curves), subsample b-I (in green, upward triangles + dot-dashed curves), subsample b-II (in light-blue, diamonds + long-dashed curves) and subsample b-III (in purple, downward triangles + short-dashed curves).
    }
    \label{fig:egsm}
\end{figure}

In the meantime, and as a first approach, we tested the GS model by incorporating the information provided by the $\rxr$-space density $\xi(r_\perp,r_\parallel)$ and velocity $v(r_\perp,r_\parallel)$ maps of Figure~\ref{fig:map2d}.
We measured again the $\rxz$-space projected correlation functions, but using this time a thinner projection range: $\mathrm{PR}=10~\hmpc$, which allows us to effectively capture the behaviour of the two fields along both directions.
This is shown in Figure~\ref{fig:egsm} with blue circles.
The binning step used here is $\delta \sigma = \delta \pi = 2~\hmpc$.
Instead of using a single $\rxr$-space density profile $\xi(r)$ as input in the model, we used two profiles: one suitable for the POS correlation, $\xi_\mathrm{pos}(r_\perp)$, and one suitable for the LOS correlation, $\xi_\mathrm{los}(r_\parallel)$.
Both were obtained by projecting $\xi(r_\perp,r_\parallel)$ towards the POS and LOS axes, respectively, using the same PR.
The corresponding model prediction is represented by means of the blue solid curves in the figure.
They match the observed data points at all scales.

For completeness, we repeated the analysis for the subsamples defined previously: b-I (in green, up-triangles + dot-dashed curves), b-II (in light-blue, diamonds + long-dashed curves) and b-III (in purple, down-triangles + short-dashed curves).
This is also shown in Figure~\ref{fig:egsm}.
As before, the model recovers the observations remarkably well at all scales.
This preliminary analysis demonstrates that the GS model is still a robust model.
In this way we lay the foundations for a full modelling.

\chapter{Analysis of BOSS DR12 data}
\label{chp:boss}
This is the last chapter of the thesis.
We present a preliminary analysis of the void-galaxy cross-correlation function measured from observational data, namely, from the BOSS DR12 spectroscopic survey introduced in Chapter~\ref{chp:data} (see Sections~\ref{sec:data_boss} to \ref{sec:data_voids_boss}).
The main goal is to understand the origin of the anisotropic patterns observed in the measurements within the context of the redshift-space effects in voids studied in Chapter~\ref{chp:zeffects} and the impact they have on this statistic studied in Chapter~\ref{chp:impact_vgcf}.
We will continue using the projected versions of the correlation function developed in Chapter~\ref{chp:cosmotest}.
We will modify our model with a simple correction that incorporates the expansion and AP-volume effects, as this is the main correction needed.
We also aim to study the impact of the remaining two sources of distortions, namely, the off-centring and void ellipticity effects.
However, they deserve a deeper and more detailed analysis in order to model them properly, hence we leave this topic for a future investigation.


\section{The projected correlation functions in BOSS}
\label{sec:boss_correlations}

We begin the analysis by returning back to Figure~\ref{fig:boss_abundance}, which shows the radius distribution of the BOSS DR12 voids.
This is also a representation of the void abundance in BOSS, but without the proper normalisation.
The shape of this curve is consistent with our previous analyses with the MXXL simulation in Chapters~\ref{chp:zeffects} and \ref{chp:impact_vsf}, see for instance Figure~\ref{fig:TC_VSF}.
The different radii range spanned in both figures is due to the different populations of tracers used to identify voids: dark-matter haloes from the MXXL and galaxies from BOSS DR12.
According to the analysis of Chapter~\ref{chp:impact_vsf}, this observational VSF is affected by the t-RSD and AP effects that impact on the volume of voids, hence it must be corrected with Eq.~(\ref{eq:q_ap_rsd}).
We recall that this correction must be incorporated in current models for the void abundance, such as the SvdW and Vdn models, in order to properly constrain the cosmological parameters involved.

To analyse the void-galaxy cross-correlation function in BOSS, we used the void sample defined in Section~\ref{sec:data_voids_boss}, with sizes between $30 \leq \rzs/\mathrm{Mpc} \leq 35$.
In order to measure the projected correlation functions, we followed the procedure developed in Section~\ref{sec:cosmotest_pcorrelations}, taking the following projection ranges: $\mathrm{PR}_\theta = 0.02320994$ (expressed in radians) and $\mathrm{PR}_\zeta = 0.013105$.
These values were chosen so that the projections ranges were approximately equal to $30~\mathrm{Mpc}$ according to the cosmology of the Patchy mocks, which is the minimum radius of the sample.
Such a value allows us to effectively capture the anisotropic patterns in the correlation function in both directions.
Regarding the binning steps, we chose the following values: $\delta \theta = 0.00617$ and $\delta \zeta = 0.00350$.
For this preliminary observational study, we used the \citeonline{estimator_davis} estimator given by Eq.~(\ref{eq:estimator_dp}), also explained in \citeonline{clues2} for the void-galaxy case.
To estimate $DR$, we used the galaxy randoms provided by BOSS in the public DR12 data.

\begin{figure}
    \centering
    \includegraphics[width=79mm]{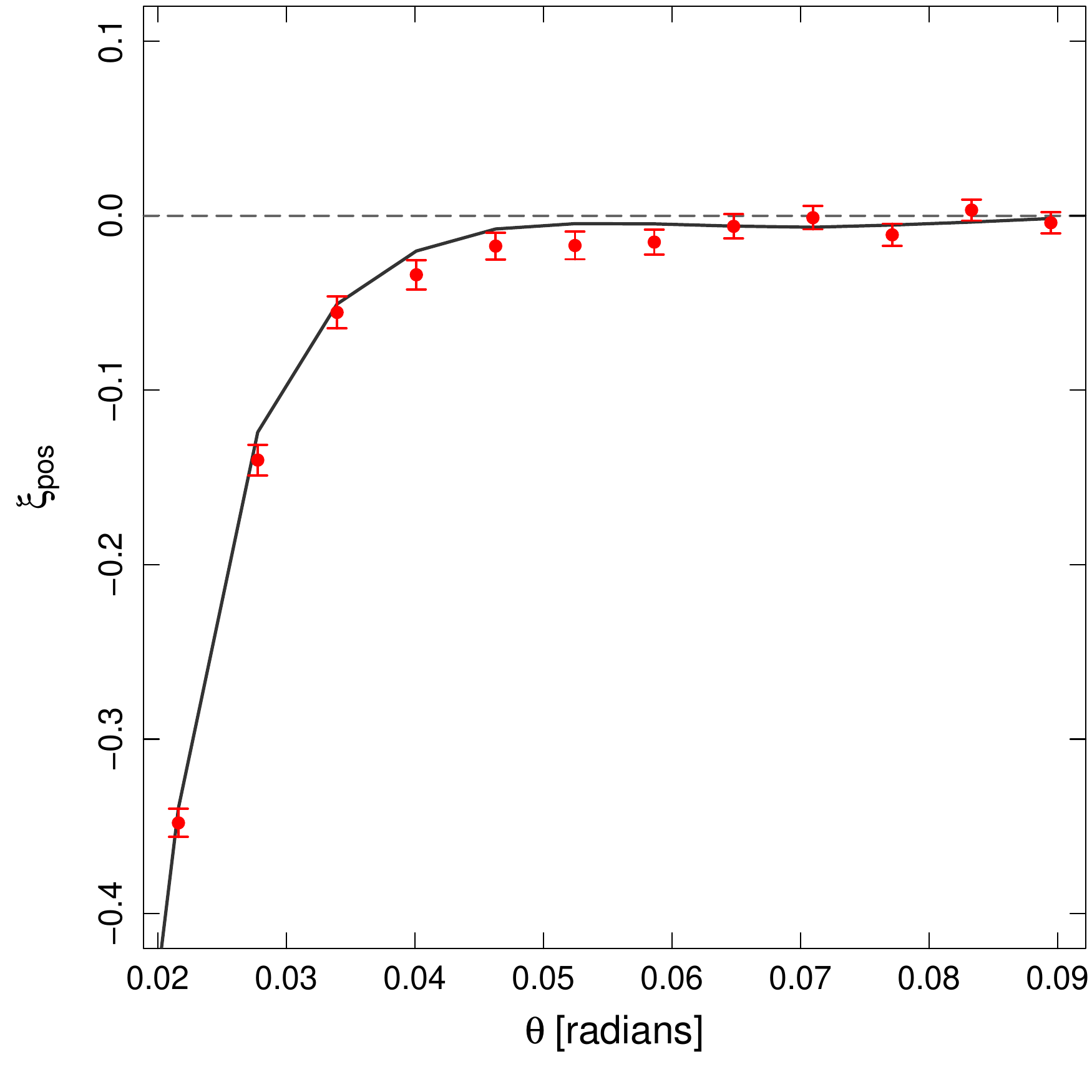}
    \includegraphics[width=79mm]{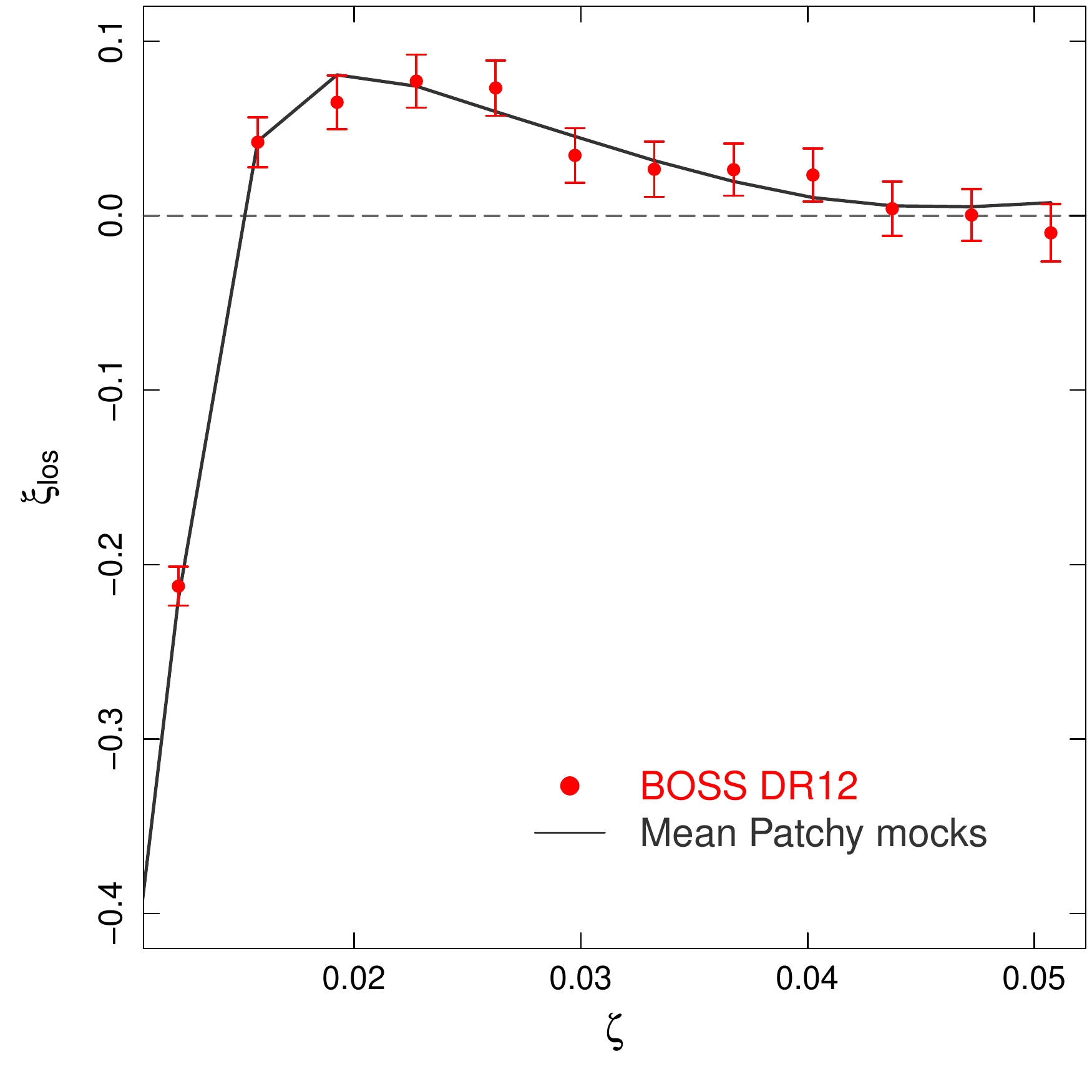}
    \caption[
    POS and LOS projections of the void-galaxy cross-correlation function for a sample of voids selected from the BOSS DR12 survey.
    Comparison with the Patchy mocks.
    ]{
    Plane-of-sky (\textit{left-hand panel}) and line-of-sight (\textit{right-hand panel}) void-galaxy cross-correlation functions corresponding to a void sample selected from BOSS DR12 data, with sizes in the range $30 \leq \rzs/\mathrm{Mpc} \leq 35$ (red circles with error bars).
    It is also shown the mean correlation obtained with the Patchy mocks (grey solid curves).
    The following projection ranges were used: $\mathrm{PR}_\theta = 0.023$ and $\mathrm{PR}_\zeta = 0.013$.
    The error bars were taken from the diagonal of the covariance matrix measured with the Patchy mocks (Figure~\ref{fig:patchy_covariance}).
    }
    \label{fig:boss_correlations}
\end{figure}

\begin{figure}
    \centering
    \includegraphics[width=\textwidth/2]{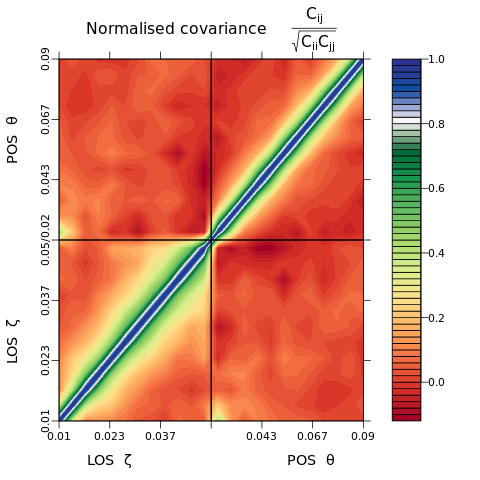}
    \caption[
    Normalised covariance matrix measured with the Patchy mocks used to analyse BOSS data.
    ]{
    Normalised covariance matrix measured with the Patchy mocks used to analyse the BOSS projected correlation functions.
    }
    \label{fig:patchy_covariance}
\end{figure}

Figure~\ref{fig:boss_correlations} shows the POS (left-hand panel) and LOS (right-hand panel) projections measured in this way.
The red circles with error bars correspond to BOSS data, whereas the grey solid curves, to the mean correlation obtained with the Patchy mocks.
The measurements are consistent with our previous analyses using the MXXL simulation.
Note, for instance, that the profiles exhibit similar features as the ones presented in Figures~\ref{fig:correlations_ospace} and \ref{fig:correlations_systematics}.
Moreover, note that the BOSS and Patchy measurements are consistent with each other.
Figure~\ref{fig:patchy_covariance} shows the associated (normalised) covariance matrix, calculated from the Patchy mocks, and from which the error bars of the previous figure were obtained.
It is also consistent with our previous results, showing similar features as the matrices presented in Figure~\ref{fig:cov} using the MXXL simulation.

\section{Analysis of distortions}
\label{sec:boss_distortions}

In this section, we analyse the anisotropic patterns observed in the POS and LOS correlation functions measured from BOSS data.
In Chapter~\ref{chp:impact_vgcf}, we explained that, in order to reproduce these features on the correlation function, the t-RSD expansion effect together with the AP change-of-volume effect constitute the most important correction to be done.
This is because both effects alter the void sizes under the $r$-space into $z$-space mapping, affecting the population of voids considered when selecting a sample and the scales involved in the correlation function directly.
In our model developed in Section~\ref{sec:cosmotest_model}, this can be simply accomplished by means of the two-step correction of Eq.~(\ref{eq:q_ap_rsd}) by making the following replacement: $r_\mathrm{cut} \longrightarrow r_\mathrm{cut}/(\qap~\qrsd)$.
This correction allows us to refer better to the $\rxr$-space statistical properties of the sample, namely, the density and velocity fields.
As we mentioned at the beginning of this chapter, we leave for a future investigation the question of modelling the remaining v-RSD and e-RSD effects.

\begin{figure}
    \centering
    \includegraphics[width=\textwidth/2]{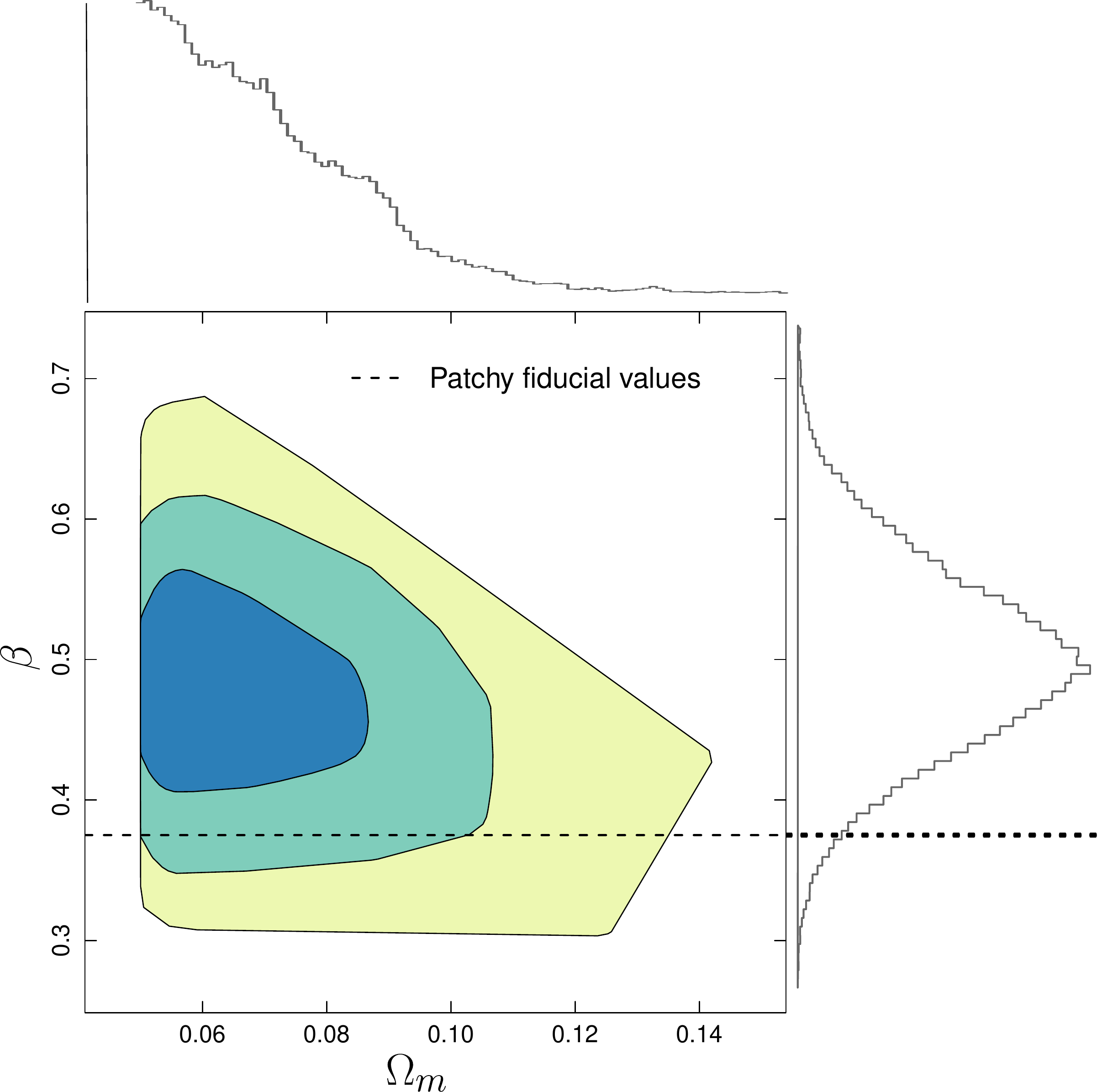}
    \caption[
    Marginalised likelihood distribution onto the plane $\Omega_m-\beta$ from the analysis of BOSS data.
    ]{
    Two-dimensional marginalised likelihood distributions onto the plane $\Omega_m-\beta$ obtained from the analysis of BOSS.
    From the inner to the outermost, the coloured contour levels enclose $1\sigma$ ($68.3\%$), $2\sigma$ ($95.5\%$) and $3\sigma$ ($99.7\%$) confidence regions.
    The dashed line indicates the fiducial value of $\beta$ corresponding to the Patchy mocks.
    }
    \label{fig:boss_mcmc}
\end{figure}

With this simple modification, we followed the procedure described in Section~\ref{sec:cosmotest_test} in order to constrain the cosmological parameters involved in the model.
Unfortunately, the constraints are not good, since $\Omega_m$ and the velocity dispersion $\sigma_\mathrm{v}$ tend towards very low values.
This is shown in Figure~\ref{fig:boss_mcmc}, where we present the marginalised likelihood distribution onto the plane $\Omega_m-\beta$.
Note however, that we still obtain a fair value for $\beta$ if we compare it with its fiducial value obtained from the Patchy mocks: $\beta_\mathrm{fid}=0.375$.

\begin{figure}
    \centering
    \includegraphics[width=79mm]{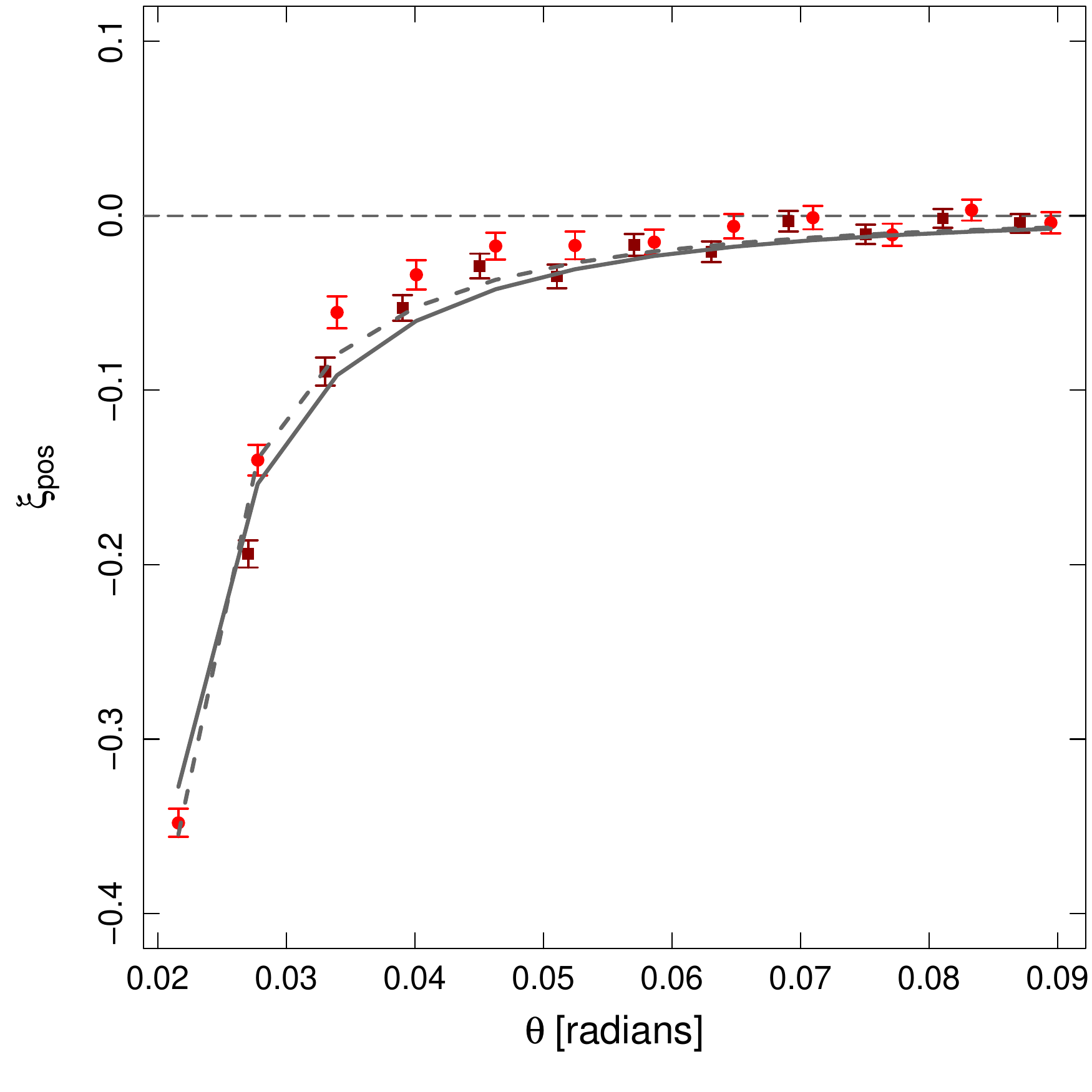}
    \includegraphics[width=79mm]{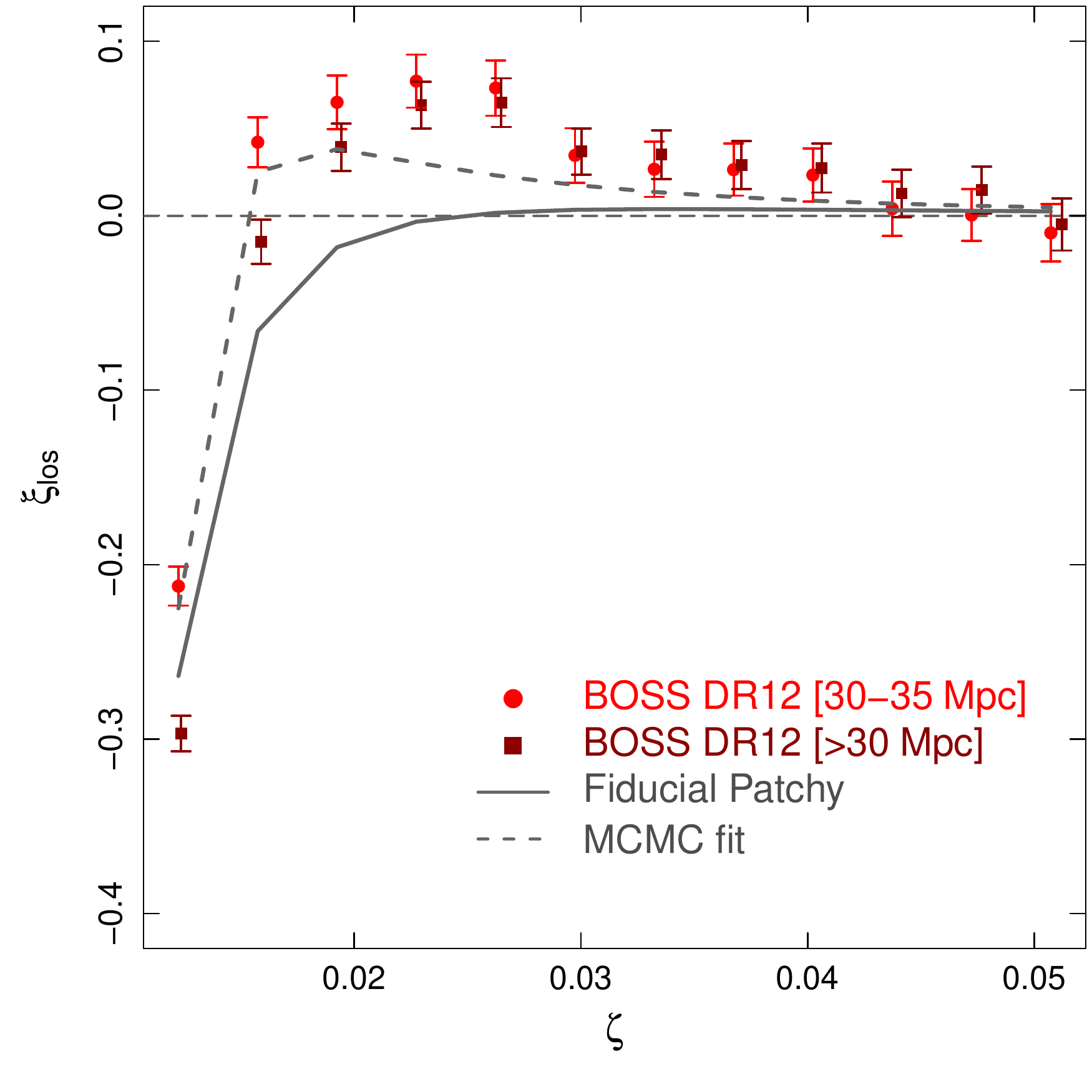}
    \caption[
    Complementary analysis of Figure~\ref{fig:boss_correlations}.
    Comparison between measurements and model predictions.
    Manifestation of the v-RSD and e-RSD effects in observations.
    ]{
    Complementary analysis of Figure~\ref{fig:boss_correlations}, where the measurements for an additional sample with sizes $\rzs \geq 30~\mathrm{Mpc}$ were added (dark red squares with error bars).
    The grey solid curves show the model predictions using the fiducial parameters of the Patchy mocks, whereas the grey dashed curves, the predictions using the inferred parameters obtained from an MCMC exploration.
    Both predictions describe well the POS projection, but they fail to describe the LOS counterpart due to an enhancement of correlation at intermediate scales.
    This is a manifestation of the v-RSD and e-RSD effects.
    This is the first time that these new types of distortion patterns are detected from observational data.
    }
    \label{fig:boss_fit}
\end{figure}

The key aspect here is that this bad behaviour can be totally explained in terms of the $z$-space effects in voids studied throughout this work.
This highlights the importance of developing a complete model that takes into account all the intervening systematicities.
Figure~\ref{fig:boss_fit} is a complement of Figure~\ref{fig:boss_correlations} that shows, in addition, two theoretical predictions.
The dashed curves, on the one hand, represent the model prediction using the fiducial parameters taken from the Patchy mocks.
The solid curves, on the other hand, represent the model prediction using the best fitted parameters from an MCMC likelihood exploration.
Although both describe the POS correlation function correctly, they are unable to reproduce the LOS correlation function mainly due to a pronounced bump located at intermediate scales.
We see that the model is forcing $\Omega_m$ and $\sigma_\mathrm{v}$ to get very low values in order to achieve this.
However, we have already noticed a deviation of this type in our analysis of Chapter~\ref{chp:impact_vgcf}.
In particular, Figure~\ref{fig:correlations_systematics} shows that this deviation is of a similar order.
What is happening here is that we are detecting two new types of anisotropic patterns: v-RSD and e-RSD, a manifestation of two physical effects that originate under the $z$-space mapping of voids.
This is the first time that these new types of distortion patterns are detected from observational measurements.

The pronounced bump observed on the LOS correlation function is partly due to the e-RSD effect when selecting a void sample.
As we have seen in Section~\ref{sec:impact_vgcf_ersd}, the two tails of the radius distribution of the $r$-space counterparts of the voids in a sample are the responsible of this feature (see the right-hand panel of Figure~\ref{fig:VSF_zoom}).
Hence, an elliptical model for void dynamics seems to be mandatory in order to obtain reliable cosmological constraints.
In order to see if this bump can be alleviated, we repeated the analysis using this time a sample with sizes in the range $\rzs \geq 30~\mathrm{Mpc}$, i.e. where the right-hand tail of the $r$-space sample distribution is not present, since all voids are considered in this range.
This is also shown in Figure~\ref{fig:boss_fit}, where the measurements are represented with dark red squares and error bars.
Note that the results do not change significantly.
This is because the left-hand tail has many more voids than the right-hand tail, and hence, it is the main contributor to the bump.
This has already been noticed in the analysis using the MXXL, in particular, it is clearly evident in Figure~\ref{fig:egsm}.
Therefore, the e-RSD distortions cannot be avoided in observations, since a left-hand cut is always necessary to select a sample of voids above the shot-noise level, and hence relevant for statistical analyses, as we discussed in Section\ref{sec:zeffects_map}.

We took a big step forward in describing and modelling the distortion patterns observed in the void-galaxy cross-correlation function.
Our model reproduces three of these systematicities: the t-RSD effect, the AP effect and the mixture of scales.
The first two do not only affect the spatial distribution of galaxies around voids, they also alter their sizes, significantly affecting the selection of voids in a sample and the scales involved in the correlation function.
However, in the context of high-precision cosmological measurements, it is also necessary to take into account the v-RSD and e-RSD effects.
The positive aspect, though, is that both effects are intimately related to the intrinsic structure and dynamics of voids.
Observational data show that the signal is good enough to detect and study them.
Therefore, their modelling is not only important to obtain unbiased cosmological constraints, but it is also important for large-scale structure studies per se, since these effects encode valuable information about the structure and dynamics of the Universe at the largest scales.
Even more, they encode additional cosmological information.
For instance, \citeonline{ellipticity_park_lee} show that the ellipticity distribution of voids constitutes a cosmological probe by itself.

\chapter{Conclusions}
\label{chp:conclusions}
One of the major challenges of modern Cosmology is to understand the nature of the accelerated expansion of the Universe.
The standard model postulates a flat-$\Lambda$CDM Universe, in which this cosmic acceleration is explained by introducing a new component called dark energy, whose density and pressure induce a net effect equivalent to a repulsive gravitational force.
From a theoretical point of view, this phenomenon is consistent with the cosmological constant of Einstein's field equations, $\Lambda$, which can also be associated with the vacuum energy.
Alternatively, this could be a hint that General Relativity needs to be reviewed, leading to the field of Modified Gravity theories.
There is a wide variety of dark-energy models, hence, in order to constrain them, it is important to apply several complementary statistical methods.
The three most important probes of dark energy to date are the following: the Hubble diagram using distant SNe Ia, the study of the anisotropies in the CMB, and the AP test using the BAO scale as a standard ruler.

Cosmic voids are vast underdense regions of the Universe.
Since they are the largest observable structures, they encode key information about the expansion history and geometry of the Universe.
Hence, voids emerge as natural alternative candidates to test dark energy.
Moreover, their potential has increased recently in view of the new generation of galaxy spectroscopic surveys, which will probe our Universe covering a volume and redshift range without precedents.
Furthermore, the study of voids offers two special advantages over the high density regime.
On the one hand, void dynamics can be treated linearly in good approximation, allowing us to assume spherical symmetry for the density and velocity fields, and therefore, it is easier to model systematic effects such as redshift-space distortions. 
On the other hand, Modified Gravity theories predict that potential deviations from the predictions of General Relativity must be more pronounced in these unscreened low density environments.

There are two primary statistics in void studies: the void size function and the void-galaxy cross-correlation function.
The void size function, on the one hand, describes the abundance of voids and can be modelled using the excursion set formalism in combination with the spherical evolution of density perturbations.
The correlation function, on the other hand, characterises the density contrast field around voids when considered at small to intermediate scales.
Although a way to model the density from first principles has not yet been found, there are some parametric and empirical models in the literature.

Both statistics are affected by distortions in the observed spatial distribution of galaxies, which translate into anisotropic patterns or deviations in the measurements.
Nevertheless, these distortion patterns can be modelled from physical principles, hence they encode valuable cosmological information.
The two main sources of distortions are the following: (i) the redshift-space distortions (RSD), a dynamical effect caused by the peculiar velocities of galaxies, and (ii) the Alcock-Paczyński (AP) effect, a geometrical effect caused by the selection of a fiducial cosmology necessary to transform the angles and redshifts provided by a survey into a distance scale expressed in physical units.

Below, we summarise and discuss the main results of our research about cosmic voids as cosmological laboratories.
Our tests and models were calibrated using the Millennium XXL simulation in combination with the spherical void finder developed by \citeonline{clues3}.
This is a large and high-resolution simulation, ideal to make feasibility cosmological analyses for modern galaxy surveys.
At the end, we provide the main conclusions of our preliminary results using observational data from the Baryon Oscillation Spectroscopic Survey Data Release 12.
To interpret these results, we used the Patchy mocks obtained from the Big MultiDark simulation.


\section{Designing a new cosmological test}
\label{sec:conclusions_cosmotest}

In the first part of the work, we designed a new cosmological test based on the void-galaxy cross-correlation function.
The main characteristics, virtues and limitations of the method can be summarised in the following statements.

\begin{enumerate}

\item 
\textit{Free of fiducial cosmology.}
We treat correlations directly in terms of void-centric angular distances, $\theta$, and redshift differences, $\zeta$, between void-galaxy pairs, hence it is not necessary to assume a fiducial cosmology.
In this way, the AP effect is taken into account naturally.

\item
\textit{Mixture of scales.}
Besides the RSD and AP distortions, there is a third type of systematic effect that affects the cosmological inference when contrasting theoretical predictions and observational measurements.
In order to perform a measurement, a binning scheme is assumed, and hence, several scales are mixed in this process.
Increasing the bin sizes improves the signal, but then, theoretical models must take into account their geometry and volume carefully.

\item
\textit{Projected correlation functions.}
Our method takes into account the mixture of scales due to the binning scheme.
This allows us to work with bins of arbitrary sizes, furthermore, this allows us to work with full projections of the correlation function.
This variant of measuring correlations constitutes the main statistic of the cosmological test.
Projecting the observable-space correlation function, $\xispo$, towards the plane of the sky ($\theta$-axis) in a given range of redshift separations $\mathrm{PR}_\zeta$, we get the POS correlation function, $\xiposo$, a function that depends only on the angular coordinate $\theta$.
Conversely, projecting $\xispo$ towards the line of sight ($\zeta$-axis) in a given angular range $\mathrm{PR}_\theta$, we get the LOS correlation function, $\xiloso$, a function that depends only on the redshift-difference coordinate $\zeta$.

\item
\textit{Model.}
We developed a physical model for the correlation function on observable space for a general cylindrical binning scheme $(\theta_\mathrm{int}, \theta_\mathrm{ext}, \zeta_\mathrm{low}, \zeta_\mathrm{up})$.
This model takes into account the RSD and AP effects, together with the scale mixing due to the geometry and sizes of the bins.
The projected POS and LOS correlation functions constitute two special cases in this scheme.
The model can be summarised in the following equations: (i) Eqs.~(\ref{eq:sigma_ap2}) and (\ref{eq:pi_ap2}) to map the observable quantities $\theta$ and $\zeta$ into comoving distances, (ii) Eqs.~(\ref{eq:estimator_voids}), (\ref{eq:data_pairs}) and (\ref{eq:random_pairs}) for the correlation estimator (Eq.~\ref{eq:xi_gd} is a simplification for the case of our simulation study), (iii) Eq.~(\ref{eq:gsm}) for the Gaussian streaming model, (iv) Eq.~(\ref{eq:dens_model}) for the real-space density contrast profile, and (v) Eq.~(\ref{eq:velocity}) for the real-space velocity profile.

\item
\textit{Density contrast profile.}
We developed our own parametric model suitable for R-type void samples.
This is Eq.~(\ref{eq:dens_model}), which is basically a double power law.
The first exponent is fixed and is equal to $-3$.
It describes the behaviour near the void walls.
The second one, $\alpha$, describes the more remote areas.
The other two parameters are: an amplitude $\xi_0$, and a pivot scale $r_0$ where the slope changes.

\item
\textit{Covariance matrices.}
The data covariance matrices associated with the method are dimensionally much smaller than in the traditional case.
This is a key aspect, first because the estimation of the inverse of a smaller matrix is numerically more stable, and second and more important, because the propagation of covariance errors into the likelihood estimates are then substantially reduced, allowing us to use a smaller number of mock catalogues to estimate covariances.

\item
\textit{Cosmological constraints.}
The parameters of the model can be summarised in two sets: the cosmological set $\lbrace \Omega_m, \beta \rbrace$, and the nuisance set $\lbrace \sigma_\mathrm{v}, \xi_0, r_0, \alpha \rbrace$.
The ultimate goal is to perform an AP test to constrain $\Omega_m$ and $\beta$.
$\Omega_m$ is more sensitive to AP distortions, whereas $\beta$ is more sensitive to RSD.
We implemented a likelihood exploration using an MCMC technique.
The constraints are tight, showing no degeneracies between the parameters, and showing Gaussian distributions in all cases.
The main results are presented in Figure~\ref{fig:constraints}, which shows the likelihood marginalisations over $\Omega_m$ and $\beta$ as $1\sigma$ ($68.3\%$) error bars.
The inherent MXXL values fall inside the error bars in almost all cases, which is the consistency check of the reliability of the method.
The test is also robust with the projection range.
The success of the method calibration is further enhanced by the fact that the model is capable of reproducing the POS and LOS projected correlation functions (Figure~\ref{fig:correlations_ospace}), as well as the real-space density and velocity profiles (Figure~\ref{fig:densvel}), showing an excellent agreement between the measurements and the theoretical predictions.
The error bars in the case of $\Omega_m$ show that there is an optimum PR to perform the test, and that better confidence regions are obtained at higher redshifts, where the model is more sensitive to the AP effect.

\item
\textit{A comment about the validity of the model.}
The model was tested with real-space voids of the MXXL simulation.
This is because RSD models are defined to work in the hybrid $\rxz$-space configuration.
One option is to use the reconstruction technique.
We propose an alternative approach: to incorporate to our method the framework developed in Chapters~\ref{chp:zeffects} and \ref{chp:impact_vgcf} to model the non-trivial redshift-space effects that affect the void identification process.

\end{enumerate}


\section{Redshift-space effects in voids}
\label{sec:conclusions_zeffects}

There are different types of void finders.
Despite the intrinsic differences between them, they are generally based on the spatial distribution and/or dynamics of matter tracers.
Therefore, the RSD and AP effects have a direct impact on the void identification process itself, affecting global properties of voids, such as their number, size and spatial spatial distribution.
This generates additional distortion patterns in the measurements of the cosmological statistics, such as the void size function and the void-galaxy cross-correlation function.
Given the precision achievable nowadays with modern galaxy spectroscopic surveys, these patterns can be detected with high precision, hence it is extremely important to model them correctly in order to obtain unbiased cosmological constraints.

One approach is to use the reconstruction technique to approximately recover the $r$-space position of tracers before applying the void finding step.
While this method has proved to be accurate in recovering the $r$-space void statistics and in extracting cosmological information from them, it also has some disadvantages.
For instance, it is computationally expensive since it is an iterative process.
Moreover, it is also redundant, since it is necessary to model the RSD effect around the recovered $r$-space voids in order to obtain accurate constraints.
Finally, the method does not take advantage of the valuable physical and cosmological information contained in the mentioned additional systematicities, which only manifest themselves by identifying voids in $z$-space.
These effects provide key clues about the structure and dynamics of voids.

In this work, we explored an alternative approach: to analyse the void finding process in order to understand physically the underlying effects that manifest in $z$-space.
Using our spherical void finder, we performed a statistical comparison between the resulting voids identified in real space and redshift space in the context of the four commonly assumed hypotheses to model RSD around voids, which are only valid in $r$-space, i.e. they are violated if voids are identified in $z$-space: (1) void number conservation, (2) invariability of centre positions, (3) isotropy of the velocity field, and (4) isotropy of the density field.

The main conclusions of this part of the work can be summarised in the following statements.

\begin{enumerate}

\item
\textit{Bijective mapping.}
There is a one-to-one relationship between $z$-space and $r$-space voids at scales not dominated by shot noise.
This means that each $z$-space void has a unique $r$-space counterpart and vice versa, in such a way that both span the same region of space.
In this context, condition (1) concerning void number conservation is still valid.

\item
\textit{Expansion effect.}
Voids in $z$-space are systematically bigger than their $r$-space counterparts.
This can be understood as an expansion effect and can be statistically quantified as an increase in void radius by a constant factor $\qrsds$ (Eq.~\ref{eq:q1_rsd}).
Actually, the slightly modified factor $\qrsdl$ (Eq.~\ref{eq:q2_rsd}) has proved to be more suitable for voids well above the shot-noise level, the ones of interest for cosmological studies.
For this analysis, we assumed the validity of hypotheses (3) and (4) concerning the isotropy of the density and velocity fields in $r$-space in order to explain a $z$-space phenomenon, even if this isotropy is no longer valid for $z$-space voids.
This effect is a by-product of the RSD induced by tracer dynamics at scales around the void radius (t-RSD).

\item
\textit{Off-centring effect.}
Void centres are systematically shifted along the LOS direction when they are identified in $z$-space.
This is a direct consequence of the violation of hypothesis (2) concerning the invariability of centre positions.
This off-centring can be statistically quantified as an RSD-displacement by means of Eq.~(\ref{eq:void_zspace}).
Hence, it constitutes a different class of RSD induced by large-scale flows in the matter distribution.
Interpreting voids as whole entities moving through space with a net velocity, this effect can be thought as a by-product of the RSD induced by the global dynamics of voids (v-RSD).

\item
\textit{AP change-of-volume effect.}
The volume of voids is also altered by the fiducial cosmology assumed to transform angular positions and redshifts into physical distances, which manifests itself as an overall expansion or contraction depending on the chosen fiducial parameters.
This effect can be statistically quantified as a variation of void radius by a constant factor $\qap$ (Eq.~\ref{eq:q_ap}).

\item
\textit{Independence of the effects.}
The three effects: AP, t-RSD and v-RSD, are statistically independent, and therefore, can be treated separately.

\item
\textit{Potential of the spherical void finder.}
The simplicity of the spherical void finder allows us to explain these $z$-space effects naturally.
This is because the method returns spherical non-overlapping voids with a well defined centre and radius.

\end{enumerate}


\section{Impact on the void size function}
\label{sec:conclusions_vsf}

Regarding the impact of the redshift-space effects in voids on the void size function, we highlight the following two conclusions.

\begin{enumerate}

\item 
\textit{AP + t-RSD correction.}
The void size function is affected by the effects that alter the volume of voids, namely, the expansion effect and the AP change-of-volume effect, but it is free of the off-centring effect.
Therefore, an observational VSF can be corrected in order to recover the true underlying $r$-space abundance by the simple two-step correction of void radius given by Eq.~(\ref{eq:q_ap_rsd}).

\item
\textit{Cosmological relevance.}
The only two necessary ingredients to correct the void size function are the factors $\qap$ and $\qrsd$.
These factors are simply two scale-independent proportionality constants, but strongly cosmology-dependent.
There is, however, an interesting difference between them.
On the one hand, $\qap$ depends only on the background cosmological parameters, such as $\Omega_m$, $\Omega_\Lambda$ and $H_0$, hence it is related to the expansion history and geometry of the Universe.
On the other hand, $\qrsd$ depends only on $\beta$, hence it is related to the dynamics and growth rate of cosmic structures.
The framework developed here must be combined with the excursion set theory used to model void abundances in order to obtain unbiased cosmological constraints from spectroscopic surveys.
    
\end{enumerate}


\section{Impact on the correlation function}
\label{sec:conclusions_vgcf}

Regarding the impact of the redshift-space effects in voids on the void-galaxy cross-correlation function, we highlight the following conclusions.

\begin{enumerate}

\item
\textit{Impurity of a sample.}
The impurity of a sample due to non-bijective voids has a negligible impact on correlation measurements.
This reinforces the fact that void number conservation under the $z$-space mapping is a valid assumption.

\item
\textit{Configurations.}
We measured correlations, densities and velocities in different configurations of the spatial distribution of haloes and voids.
Measurements made with $z$-space voids and $z$-space haloes correspond to the $\zxz$-space configuration, mimicking possible observational measurements.
Measurements made with $r$-space voids and $z$-space haloes correspond to the hybrid $\rxz$-space configuration, where current models for RSD around voids are defined to work.
Measurements made with $r$-space voids and $r$-space haloes correspond to the $\rxr$-space configuration, free of the RSD and AP effects.

\item
\textit{AP + t-RSD correction.}
It is fundamental to provide models with the correct $\rxr$-space statistical properties of a void sample, particularly the density contrast and velocity fields.
This can largely be achieved by considering the t-RSD and AP effects on the volume of voids, correcting void radii with Eq.~(\ref{eq:q_ap_rsd}).
The remaining deviations between the observations and model predictions are smaller and caused by the following two sources.

\item
\textit{v-RSD correction.}
One source is the off-centring effect.
This effect is responsible for an additional distortion pattern due to the global void dynamics (v-RSD), different from the classic anisotropic patterns due to tracer dynamics (t-RSD).
These distortions are noticeably reduced after correcting the centre positions with Eq.~(\ref{eq:void_zspace}).
This is the first time that this type of distortions are detected and quantified.

\item
\textit{Void ellipticity.}
The other source is the intrinsic $\rxr$-space ellipticity of voids.
Voids are typically ellipsoidal, although they are oriented randomly in space.
Therefore, this ellipticity has not a significant impact when considering a complete sample of voids.
However, when the sample is selected in $z$-space from a radius bin, the $r$-space counterparts are distributed in a complex way, covering a very extended radius range.
The t-RSD correction is a good predictor of the completeness region, where the ellipticity is not important.
Nevertheless, the remaining tails of this distribution have an appreciable impact on the measurement of the correlation function.
They are composed of special voids: over-expanding voids elongated along the POS direction, and under-expanding voids elongated along the LOS direction.
They are responsible for an additional distortion pattern not previously taken into account in the literature (e-RSD).

\item
\textit{Towards an improved model.}
The AP + t-RSD correction of void radius is the most important aspect to be considered in models.
Although the remaining deviations due to the v-RSD and e-RSD effects are smaller, they have an appreciable impact when trying to fit the data.
The v-RSD effect can be treated in models by incorporating information about the void net velocity distribution.
This is the connection needed between the $\rxz$- and $\zxz$-space configurations.
Regarding the e-RSD effect, current models must be reviewed by taking into account the ellipsoidal structure of voids and a proper connection between the $\rxr$-space density and velocity fields.
We leave for a future investigation to tackle these issues.
With a simplified test, we showed that the GS model can still be a robust model even in this case.

\item
\textit{Comparing the void size function and the correlation function.}
The former is affected by two types of systematic effects: t-RSD and AP, whereas the latter, by five types of systematic effects: t-RSD, AP, mixture of scales, v-RSD and e-RSD.

\end{enumerate}


\section{Analysis of BOSS DR12 data}
\label{sec:conclusions_boss}

We finish this work with the main aspects of the preliminary analysis of the void-galaxy cross-correlation function measured from BOSS DR12 data.

\begin{enumerate}

\item 
\textit{Projected correlation functions in BOSS.}
The measurements are consistent with our previous analyses using the MXXL simulation, exhibiting similar features.
Moreover, the results from BOSS are in excellent agreement with the results from the Patchy mocks.
The associated covariance matrix is also consistent with the previous results using the MXXL.

\item
\textit{Analysis of distortions.}
We modified our model to incorporate the AP + t-RSD correction for void sizes.
However, it is not enough.
The remaining distortion patterns due to the v-RSD and e-RSD effects have a strong impact and must be modelled properly in order to obtain unbiased constraints.
In particular, the LOS projection is severely affected by them.
The POS projection, by contrast, is well described by the model.
This is the first time that these new types of distortions are detected in observational measurements.

\end{enumerate}

As a final reflection, we have taken a big step forward in our initial goal of establishing cosmic voids as reliable cosmological probes to address the dark-energy problem.
The new generation of spectroscopic surveys, such as BOSS, HETDEX, DESI and Euclid, enabled us to enter into a new era of high-precision cosmological measurements without precedents.
Therefore, it is fundamental to detect and model all the systematic effects that affect observational measurements in order to obtain unbiased cosmological constraints.
The present work has demonstrated that this is particularly important in the case of voids.
On the one hand, this is the first time that redshift-space systematicities are considered in the abundance of voids.
The community has concentrated its efforts on modelling the true underlying real-space abundance with the excursion set formalism.
On the other hand, we have detected and described all types of distortion patterns present in the void-galaxy cross-correlation function, discovering new effects not previously taken into account.
In this way, we lay solid foundations for a full and proper modelling of the void size function and the correlation function with the aim of designing new and reliable cosmological tests to be applied in modern surveys.
In particular, we have clearly established the validity framework of our test using the projected POS and LOS correlation functions.
We will seek to improve it by incorporating the v-RSD and e-RSD effects.
There is an excellent signal to detect and study them from observational data.
Finally, it should be noted that, in addition to its practical and immediate cosmological importance, the analysis of all the redshift-space effects in voids treated here is also important for large-scale structure studies per se, since it encodes valuable information about the structure and dynamics of voids, and more generally, about the Universe at the largest scales.
Even more, some of these effects constitute cosmological probes by themselves, as is the case of the void ellipticity.


\citeoption{abnt-etal-list=3}

\bibliographystyle{abntex2-cite-min}

\providecommand{\abntreprintinfo}[1]{%
 \citeonline{#1}}
\setlength{\labelsep}{0pt}


\phantompart
\printindex

\end{document}